\documentclass[letterpaper,12pt,leqno]{article}
\usepackage{paper}
\usepackage{tikz}
\usepackage{hyperref}
\usepackage{url}
\usepackage{tikz}

\newcommand{\circledgraphfour}[9]{%
\begin{tikzpicture}
\node[anchor=south west, inner sep=0] (img) at (0,0)
{\includegraphics[width=#1]{#2}};
\draw[red, very thick] (#3) circle (#4);
\draw[red, very thick] (#5) circle (#6);
\draw[red, very thick] (#7) circle (#8);
\draw[red, very thick] (#9) circle (0.25);
\end{tikzpicture}%
}

\hypersetup{pdftitle={Paper Title}}

\usepackage{float}
\usepackage{graphicx}
\usepackage{caption,subcaption}
\usepackage[font=small]{caption}
\usepackage{atbegshi}
\AtBeginDocument{\AtBeginShipoutNext{\AtBeginShipoutDiscard}}
\begin{document}
\title{Beveridgean Unemployment Gap with Part-time Employment}
\author{Rongjin Zhang}
\thanks{Department of Economics, University of California, Santa Cruz.\\ I am deeply indebted to Pascal Michaillat for his invaluable advice, continuous guidance, encouragement, and support. I am extremely grateful to Kenneth Kletzer and Alonso Villacorta who contributed valuable comments and suggestions at different stages of this project. The paper has also benefited from suggestions made at the macroeconomics workshop at UCSC.} 
\date{June 2026}     

\begin{titlepage}\maketitle


This paper extends the sufficient-statistics formula for efficient unemployment developed by \citet{MS21b} to account for part-time employment. I introduce two additional sufficient statistics that measure the share of part-time employment and part-time hours relative to full-time hours. Applying the framework to the United States (1951--2026) and Japan (1970--2025), I compare the effects of total part-time employment and involuntary part-time employment on efficient unemployment. Total part-time employment has substantially larger effects than involuntary part-time employment. While involuntary part-time employment provides information about labor-market slack, the main change in efficient unemployment comes from part-time work itself because part-time workers supply fewer market hours than full-time workers. Under the total part-time calibration, efficient unemployment averages 4.7\% in the United States before COVID and 4.2\% after COVID. In the Japanese application, the full-sample average is 2.7\%. The distinction is especially important in Japan, where part-time employment is widespread and often reflects flexible work arrangements. These findings suggest that aggregate labor input, rather than involuntary part-time employment alone, is an important determinant of labor-market efficiency. 

\end{titlepage}\section{Introduction}\label{s:introduction}

Workers who are classified as employed can differ substantially in the amount of labor they supply. Part-time workers typically work fewer hours than full-time workers and therefore contribute less market production. As a result, the same unemployment rate can correspond to different levels of aggregate labor input depending on the composition of employment. This distinction matters for welfare-based measures of labor-market efficiency, which depend on labor input rather than employment status alone.

Part-time employment has become increasingly common in many advanced economies. Figure~\ref{fig:oecd_pt} illustrates this trend across several OECD countries, showing a persistent rise in the part-time employment share over the past several decades. In several countries, part-time workers now account for a substantial fraction of total employment. The growing prevalence of part-time work therefore makes the distinction between employment counts and labor input empirically relevant.

Part-time employment is generally defined as employment involving fewer than 35 hours of work per week. According to the U.S. Bureau of Labor Statistics, part-time workers can be classified as working part time for non-economic reasons or for economic reasons. The latter group consists of workers who would prefer additional hours of work but are unable to obtain full-time employment.

Interest in involuntary part-time employment has grown substantially because it is closely linked to business-cycle conditions. The cyclical behavior of part-time employment differs across worker groups. Total part-time employment combines voluntary and involuntary part-time work and therefore reflects both structural labor-supply choices and business-cycle conditions. Involuntary part-time employment is more directly linked to business-cycle conditions than total part-time employment. As shown in Figure~\ref{fig:us_pt}, involuntary part-time employment in the United States rises sharply during recessions and often remains elevated well into subsequent recoveries \citep{vbl20}. These patterns suggest that involuntary part-time employment contains valuable information about labor-market fluctuations over the business cycle.

\setcounter{figure}{0}
\begin{figure}
    \centering
    \includegraphics[width=1\linewidth]{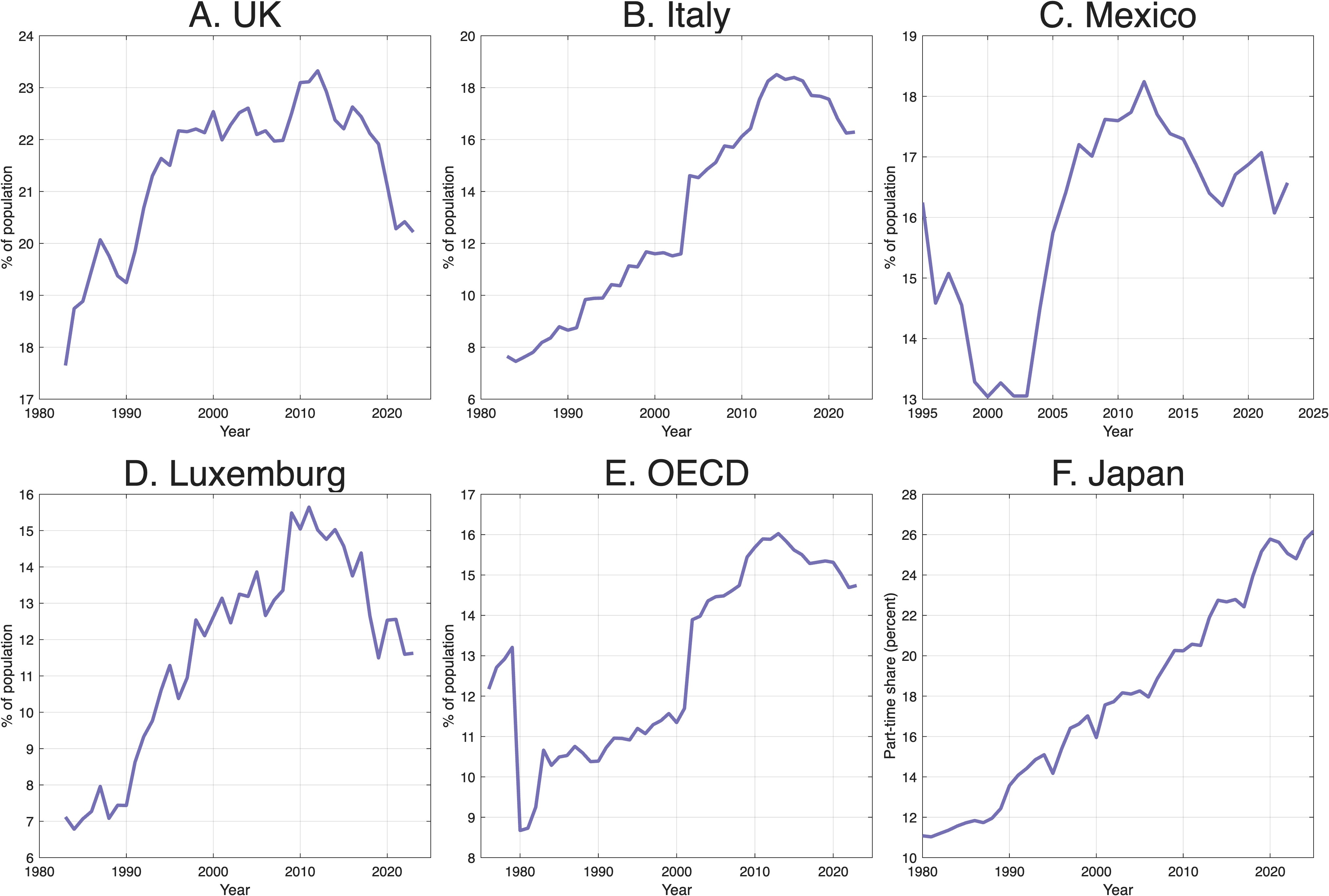}
    \caption{Part-time employment in some OECD countries}
    \label{fig:oecd_pt}
    \note{\centering\textit{Source}: OECD Database.}
\end{figure}

\begin{figure}[H]
    \centering
    \includegraphics[width=0.7\linewidth]{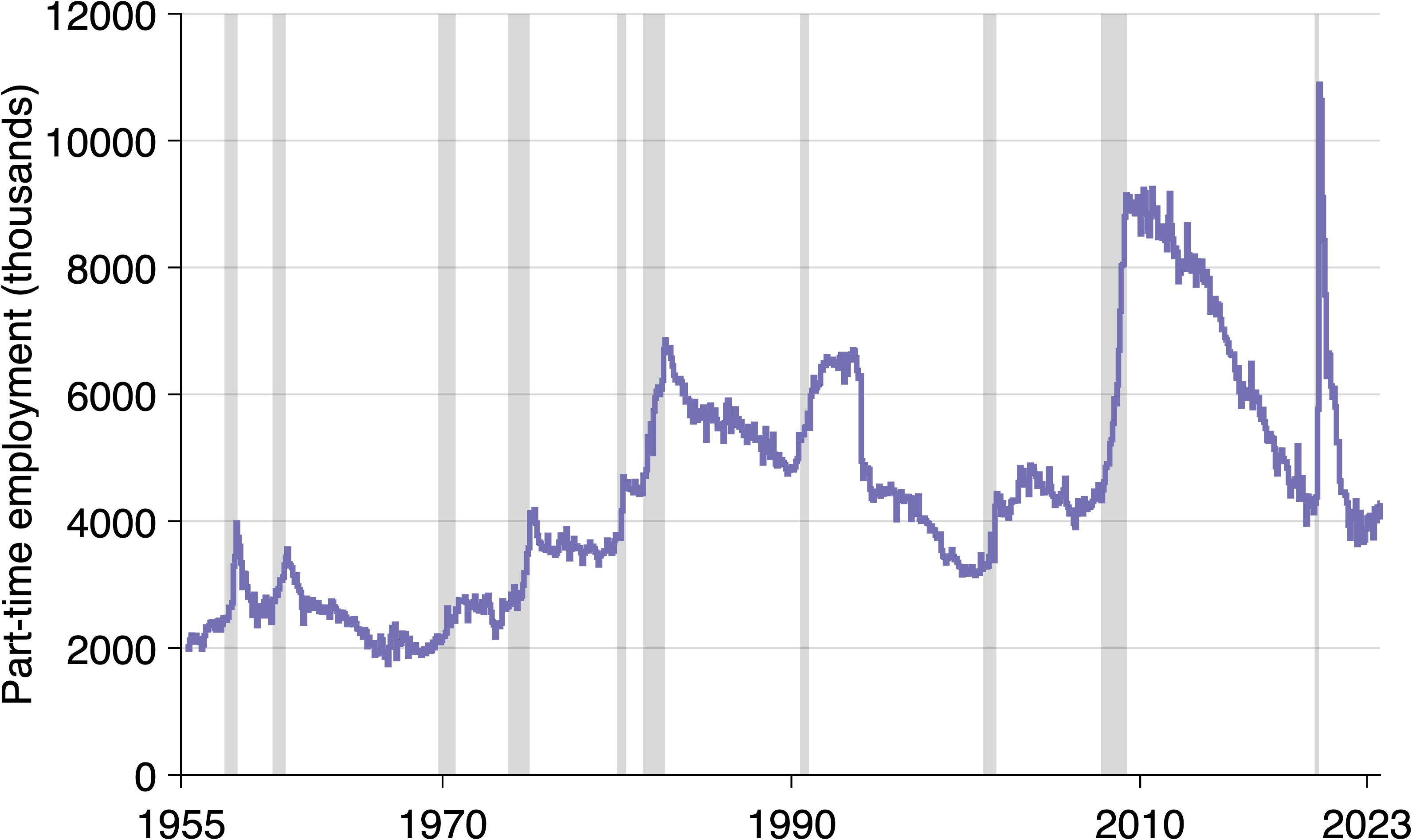}
    \caption{Involuntary part-time employment in the U.S.  \textit{Source}: U.S. Bureau of Labor Statistics}
    \label{fig:us_pt}
\end{figure}

However, from a welfare perspective, labor-market conditions depend not only on whether workers are employed, but also on how much labor they supply. Even voluntarily chosen part-time employment contributes fewer market hours than full-time employment and therefore reduces effective labor input. As a result, two economies with identical unemployment rates may differ substantially in market production and labor-market efficiency if their employment structures differ.

\begin{figure}[H]
    \centering
    \includegraphics[width=1\linewidth]{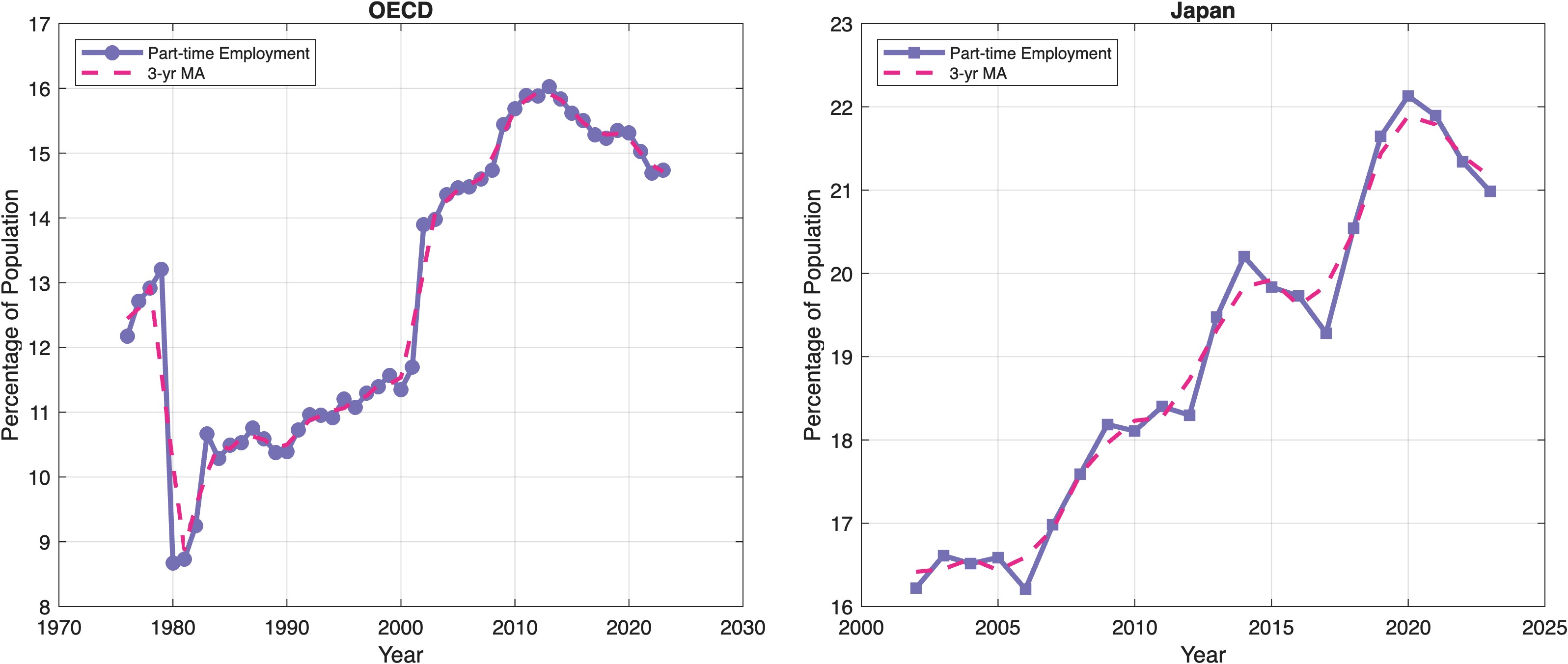}
    \caption{OECD part-time employment and Japanese part-time employment.\\[0.2em]
    \centering\textit{Source}: OECD Database.}
    \label{fig:oecd_jpn}
\end{figure}

To account for this dimension of employment heterogeneity, I introduce part-time employment into the Beveridgean unemployment-gap framework of \citet{MS21b}. In the extended framework, part-time workers contribute less market production than full-time workers but generate greater social value than unemployed workers. This extension yields a generalized measure of efficient unemployment that incorporates differences in hours worked among employed workers. I apply the framework to data from the United States and Japan to examine how part-time employment changes estimates of efficient unemployment and unemployment gaps. Figure~\ref{fig:oecd_jpn} motivates the Japan application by showing that part-time employment is much more prevalent in Japan than in the OECD average. While the OECD part-time employment rate rises after the early 2000s and stabilizes below 16 percent, Japan's part-time employment rate increases steadily from about 16 percent in 2002 to above 22 percent around 2020 before declining slightly. This pattern suggests that part-time employment is an important feature of Japanese labor utilization and should be incorporated into the measurement of efficient unemployment.

I also separate involuntary part-time employment from total part-time employment and repeat the analysis to distinguish the effect of reduced labor input from the effect of underemployment. 

\section{Background and existing measures of labor-market slack}
As discussed by \citet{MS21b}, two approaches are commonly used to assess unemployment gaps. The first measures deviations of the unemployment rate from its long-run trend, while the second compares the unemployment rate to the non-accelerating inflation rate of unemployment (NAIRU). Although widely used, neither approach is derived from a social welfare framework and therefore neither directly measures the distance between observed labor-market outcomes and the welfare-maximizing allocation.

Discussions of part-time employment often focus on involuntary part-time employment, which is commonly interpreted as a form of underemployment. Statistical agencies such as the U.S. Bureau of Labor Statistics (BLS) measure underemployment using indicators such as the involuntary part-time rate (IPTR), defined as the share of workers employed part-time for economic reasons. Another widely used measure is the U-6 unemployment rate, which includes unemployed workers, marginally attached workers, and workers employed part-time for economic reasons. Because U-6 extends U-5 by incorporating involuntary part-time workers, the gap between the two measures is often interpreted as reflecting the contribution of involuntary part-time employment to labor-market slack.
While informative, these indicators have important limitations. First, they identify the number of involuntary part-time workers but do not measure the intensity of underemployment, such as the extent to which these workers would like to increase their hours. Second, because they are based on worker counts rather than hours worked, they may not fully capture differences in labor utilization across economies. As emphasized by \citet{BB19}, comparisons based solely on employment status can be misleading when working-time arrangements differ substantially across labor markets.

This paper takes a different approach by extending the efficient-unemployment framework of \citet{MS21b}. The original framework uses the official unemployment rate, U-3, which measures total unemployed persons as a share of the civilian labor force. While U-3 is the standard measure of unemployment, it treats all employed workers as equivalent and therefore does not distinguish between full-time and part-time employment. As a result, two labor markets with the same unemployment rate may differ substantially in their effective labor input if one has a larger share of part-time employment.

One possible alternative is U-6, which includes not only unemployed workers but also marginally attached workers and persons employed part time for economic reasons \citep{BLS_underutilization}. However, U-6 remains a worker-count measure and does not account for differences in hours worked across employed workers. Consequently, simply replacing U-3 with U-6 does not provide a welfare-based measure of labor-market efficiency.

The distinction between full-time and part-time employment is particularly important from a welfare perspective. Part-time workers contribute fewer market hours on average than full-time workers and therefore generate less market production. For a given unemployment rate, an increase in part-time employment reduces effective labor input and may alter firms' incentives to post vacancies. Thus, labor markets with identical unemployment rates may differ in their degree of efficiency depending on the composition of employment.

To capture this dimension of labor-market heterogeneity, I extend the sufficient-statistics framework of \citet{MS21b} by decomposing employment into full-time and part-time workers and incorporating differences in working hours directly into the welfare analysis. The extension introduces two additional sufficient statistics: the employment composition between full-time and part-time workers and the relative hours worked by part-time workers. As a result, the efficient unemployment rate and the unemployment gap depend not only on the number of employed workers but also on the effective labor input they provide. This approach recognizes that two economies with identical unemployment rates may differ substantially in labor-market efficiency if their employment structures and working-time arrangements differ.

I first extend the U.S. analysis of \citet{MS21b} through 2026, thereby incorporating the COVID-19 period. Because the U.S. Beveridge curve shifts unusually during and after the pandemic, I present the results separately for the pre-pandemic and post-pandemic periods. The results show that incorporating involuntary part-time employment has only a modest effect on estimates of efficient unemployment, whereas incorporating total part-time employment leads to somewhat larger changes in the measured unemployment gap and efficient unemployment rate. Nevertheless, the unemployment gap remains strongly countercyclical, and the labor market continues to exhibit periods of inefficiency. These findings suggest that the efficient-unemployment framework of \citet{MS21b} is largely robust to the inclusion of involuntary part-time employment in the United States because workers employed part time for economic reasons account for only a small share of total labor input.

I then apply the extended framework to Japan, where part-time employment represents a substantially larger share of total employment. In contrast to the United States, incorporating total part-time employment has quantitatively important implications for the measurement of labor-market efficiency. Accounting for heterogeneity in working hours raises the efficient unemployment rate and materially alters the assessment of labor-market conditions. By contrast, incorporating involuntary part-time employment produces only small changes in the results. This finding reflects the fact that involuntary part-time employment constitutes a relatively small and declining share of total part-time employment in Japan. More broadly, the results indicate that the welfare implications of part-time work are driven primarily by differences in effective labor input rather than by the involuntary nature of part-time employment.

\section{Review of the Beveridgean unemployment gap}\label{s:section}

The Beveridgean unemployment-gap framework of \citet{MS21b} measures labor-market efficiency using three sufficient statistics: the Beveridge elasticity, the social value of nonwork, and the recruiting cost. These statistics jointly determine the welfare-maximizing point on the Beveridge curve and yield closed-form expressions for efficient labor-market tightness and efficient unemployment, which serve as benchmarks for measuring unemployment gaps.

\subsection{Three sufficient statistics in \citet{MS21b} }
The measurement of unemployment proposed by \citet{MS21b} is grounded in the theory of efficiency within modern labor market models, which are governed by the Beveridge curve—a negative relationship between unemployment and vacancies. This implies that vacancies and unemployment cannot be simultaneously reduced, making the resolution of the unemployment-vacancy tradeoff, the elasticity of the Beveridge curve, central to their analysis.

As in \citet{MS21b}, the Beveridge elasticity is the elasticity of the vacancy rate with respect to the unemployment rate along the Beveridge curve, normalized to be positive: 

\begin{equation}\label{my_first_eqn}
	\epsilon = -\frac{d ln(v(u))}{d ln(u)}=-\frac{u}{v} \cdot v'(u) 
\end{equation}

Both unemployment and vacancies induce welfare costs. Therefore, consider a flow social welfare that is given by the linear function:

\begin{equation}\label{my_first_eqn}
	W(n, u, v) = p(n+zu-cv)L
\end{equation}
where $n$ is the employment rate, the number of employed workers divided by the size of the labor force. L denotes the total number of workers in the labor force. Unemployed workers engage in home production or other nonmarket activities valued at productivity $zp$, where $0<z<1$. Thus, $z$ measures the social value of nonwork relative to market production.\\

A more general form can be expressed as:

\begin{equation}\label{my_first_eqn}
	\widehat{W} (u, v) = W (1-u, u, v)
\end{equation}
where $\widehat{W}$ is strictly decreasing in u and v.\\

 Employed workers have a productivity level $p>0$, which is greater than that of home production, where $pz<p$.  Consequently, from the welfare analysis equation (2), the social value of nonwork $z$ is defined as the marginal rate of substitution between unemployment and employment:

	$$ \frac{\partial W}{\partial u}/\frac{\partial W}{\partial n} = z <1 $$ 

Therefore, the social cost of unemployment is the marginal contribution of unemployed workers to welfare, relative to the marginal contribution of employed workers:

$$ (\frac{\partial W}{\partial u}-\frac{\partial W}{\partial n} )/\frac{\partial W}{\partial n} = 1- z >0 $$

The last sufficient statistic is the recruiting cost, which equals c in equation (2), representing the marginal rate of substitution between vacancies and employment in the welfare function:
$$  - \frac{\partial W}{\partial v}/\frac{\partial W}{\partial n} = c >0$$

According to Proposition 1 in (\citeauthor{MS21b}, 2021), efficiency is achieved at the point at which the Beveridge curve is tangent to an isowelfare curve; therefore, the efficient unemployment rate is implicitly defined by:

 \begin{equation}\label{my_first_eqn}
	 v'(u) = -\frac{\frac{\partial W}{\partial u}-\frac{\partial W}{\partial n}}{\frac{\partial W}{\partial v}}=-\frac{1-z}{c}
\end{equation}
\noindent\textit{Note:} Appendix ~\ref{app:notation} shows the correspondence between the notation used in \citet{MS21b} and the notation adopted in this paper, and derives the part-time extension.

A typical limitation for sufficient statistics as pointed out in \citet{MS21b} is that the parameters may depend on each other (\citeauthor{c9}, 2009). Thus, a key assumption to address the endogeneity of the sufficient statistics as shown above to compute the efficient unemployment rate (assumption 3 in \citet{MS21b}):\\
The Beveridge elasticity $\epsilon$, social value of nonwork $z$, and recruiting cost $c$ do not depend on the unemployment and vacancy rates.

This assumption implies an isoelastic Beveridge curve, $v(u) = \hat{a} \cdot u^{-\epsilon}$, then an efficient tightness from the elasticity of Beveridge curve can be derived:

 \begin{equation}\label{my_first_eqn}
	\theta^{*} = \frac{1-z}{c \epsilon}
\end{equation}

\begin{figure}[h]
    \centering
    \includegraphics[width=0.7\linewidth]{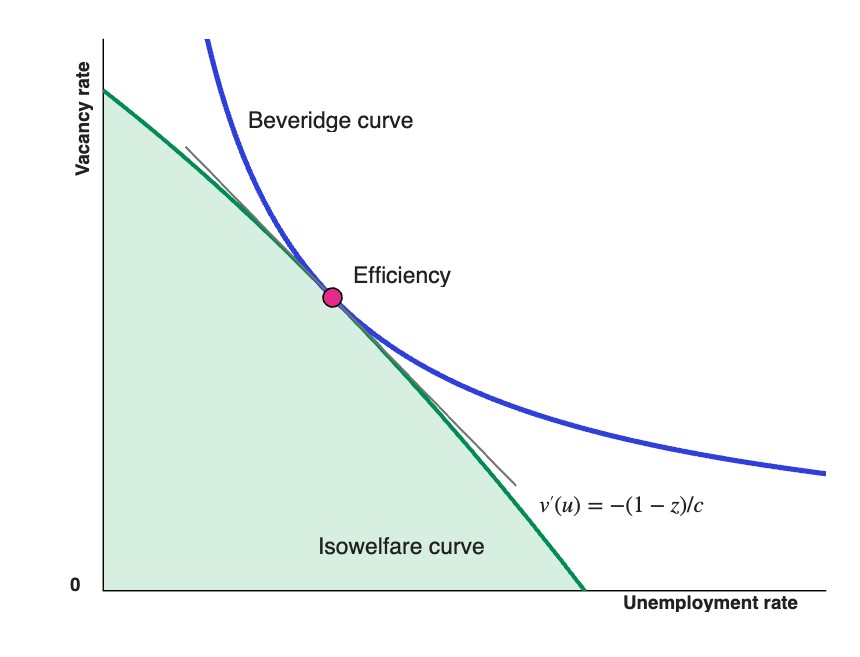}
    \caption{Efficient unemployment rate. \textit{Notes:} The Beveridge curve has slope $v'(u)$. The isowelfare curve has slope $-\frac{1-z}{c}$, where $z$ is the social value of nonwork and $c$ is the recruiting cost. The tangency of the Beveridge and isowelfare curves gives the efficient labor-market allocation. \textit{Source:} \citet{MS21b}}
    \label{fig:ustar_A}
\end{figure}

Hence, the efficient unemployment can be expressed in sufficient statistics:
 \begin{equation}\label{my_first_eqn}
u^{*} = (\frac{c \epsilon}{1-z}\cdot \frac{v}{u^{-\epsilon}})^{1/(1+\epsilon)}
\end{equation}
Figure~\ref{fig:ustar_A} illustrates how the three sufficient statistics jointly determine labor-market efficiency in the Beveridgean framework of \citet{MS21b}. The Beveridge elasticity governs the slope of the Beveridge curve, while the social value of nonwork and recruiting cost determine the slope of the isowelfare curve. Labor-market efficiency is achieved at the tangency point between the Beveridge curve and an isowelfare curve. At this point, the marginal welfare gain from reducing unemployment is exactly offset by the marginal welfare cost of creating additional vacancies. This tangency condition forms the basis for the sufficient-statistics formula used to compute efficient labor-market tightness and efficient unemployment.

\section{Incorporating part-time employment}\label{s:section}

This section extends the Beveridgean unemployment-gap framework by distinguishing between full-time and part-time employment. In the standard framework, all employed workers are treated identically regardless of hours worked. Yet part-time workers supply fewer market hours than full-time workers and therefore contribute less market production. Ignoring this distinction may overstate effective labor utilization in economies where part-time employment represents a substantial share of total employment.

To account for differences in labor input, I decompose employment into full-time and part-time workers. Part-time workers are assumed to generate a combination of market and nonmarket production, implying a productivity level between that of full-time workers and unemployed workers. This extension introduces two additional sufficient statistics: the share of full-time employment in total employment, $\alpha$, and the share of working hours supplied by part-time workers relative to full-time workers, $\gamma$. Together, these statistics allow the efficient-unemployment measure to account for both the composition of employment and differences in hours worked.

\subsection{Flow social welfare with part-time employment}

Following the social-welfare analysis of \citet{MS21b}, I decompose employment into full-time employment, $n_F$, and part-time employment, $n_P$, with unemployment rate $u$ and vacancy rate $v$. Full-time workers have productivity $p>0$. A part-time worker allocates a fraction $\gamma$ of time to market production and the remaining fraction $1-\gamma$ to nonmarket activities with productivity $zp$, where $z$ denotes the social value of nonwork relative to market production. The effective productivity of a part-time worker is therefore
\[
p\gamma + pz(1-\gamma)
=
p[\gamma+(1-\gamma)z].
\]

Since part-time workers supply fewer market hours than full-time workers but more than unemployed workers engaged in home production,
\[
pz < p[\gamma+(1-\gamma)z] < p.
\]

The flow social welfare function becomes
\begin{equation}
W(n_F,n_P,u,v)
=
p\Big(n_F+[\gamma+(1-\gamma)z]n_P-cv+zu\Big)L.
\label{eq:welfare_pt}
\end{equation}

Equation~\eqref{eq:welfare_pt} extends the Beveridgean framework by allowing different employment states to contribute different amounts of effective labor input. As shown below, the extension introduces two additional sufficient statistics: the employment share of full-time workers, $\alpha$, and the relative hours supplied by part-time workers, $\gamma$. In the model, $\gamma$ denotes the fraction of time that part-time workers allocate to market production, with the remaining fraction $1-\gamma$ allocated to nonmarket production valued at $zp$. In the empirical implementation, I measure this fraction using the ratio of average weekly hours worked by part-time and full-time workers. This mapping assumes that differences in weekly hours reflect differences in market time relative to full-time workers. Under this assumption, the remaining time of part-time workers is valued consistently with the social value of nonwork used elsewhere in the model.

\subsection{Two additional sufficient statistics}

The extended framework requires information on both the composition of employment and the relative labor input supplied by part-time workers. These dimensions are summarized by two additional sufficient statistics, $\alpha$ and $\gamma$.

\subsubsection{Population share of full-time and part-time employment}

Let $\alpha^{P}$ denote the share of full-time employment in total employment:
\begin{equation}
\alpha^{P}
=
\frac{n_F}{n_F+n_P},
\end{equation}
where $n_F$ and $n_P$ denote the employment rates of full-time and part-time workers, respectively. The share of part-time employment is therefore
\begin{equation}
1-\alpha^{P}
=
\frac{n_P}{n_F+n_P}.
\end{equation}
The aggregate unemployment rate satisfies
\begin{equation}
u
=
1-(n_F+n_P).
\end{equation}
In the baseline analysis, I classify workers as part-time if their actual hours worked (ahrsworkt) are fewer than 35 hours per week and as full-time if their actual hours worked are at least 35 hours per week. The 35-hour threshold is consistent with the conventional distinction between full-time and part-time employment used by the U.S. Bureau of Labor Statistics \citep{BLS_parttime}, although the BLS applies the threshold to usual rather than actual hours worked. I use actual hours in the baseline to avoid known comparability problems in usual-hours (uhrswork1) reporting in the ASEC microdata (notably in 1998) as shown in Appendix Figure~\ref{f:actual_usual_pt}, the part-time employment share constructed from usual hours exhibits a large spike in 1998 that is absent from the actual-hours series and difficult to reconcile with broader labor-market developments. Outside this episode, the two measures generate very similar part-time employment shares. Using actual hours therefore avoids spurious variation in the classification of workers while maintaining consistency with the construction of the relative-hours statistic $\gamma$.

As a robustness check, I also construct full-time and part-time employment shares using usual weekly hours, excluding the anomalous 1998 observation. The resulting employment shares are close to those obtained from actual hours (for example, mean $\alpha$ is 0.75 under actual hours and 0.78 under usual hours over 1994--2025), and the efficient unemployment rate and unemployment gap change by at most 0.13 percentage point relative to the baseline. Appendix Figure~\ref{f:actual_usual_robustness} demonstrates that the choice between actual and usual hours has little effect on the results. Panels (a) and (b) show that efficient unemployment rates constructed from the two measures are almost indistinguishable in both the pre-COVID and post-COVID samples. Likewise, panels (c) and (d) show that unemployment gaps are virtually identical under the two classifications. Therefore, the use of actual hours worked in the baseline analysis is driven by data consistency rather than by any material effect on the estimated efficient unemployment rate or unemployment gap.

A rise in voluntary part-time employment need not indicate greater labor-market slack. Nevertheless, even voluntarily chosen part-time work reduces market hours supplied and therefore affects effective labor utilization. Consequently, two economies with identical unemployment rates may differ substantially in aggregate labor input and market production. The purpose of the extension is not to classify voluntary part-time work as inefficient, but rather to account for differences in labor input when measuring efficient unemployment.

Total part-time employment includes both voluntary and involuntary part-time workers. The involuntary group is especially relevant for measuring labor-market slack because these workers are employed part time for economic reasons and would prefer to work more hours at the prevailing wage \citep{BLS_PTER}. I therefore also examine involuntary part-time employment separately from total part-time employment to distinguish between the role of part-time employment more broadly and the role of underemployment.

Let $n_{IP}$ denote involuntary part-time employment. When focusing on this margin, the full-time employment share is defined as
\begin{equation}
\alpha^{IP}
=
\frac{n_F}{n_F+n_{IP}},
\end{equation}
where $n_F$ denotes full-time employment. The corresponding involuntary part-time employment share is
\begin{equation}
1-\alpha^{IP}
=
\frac{n_{IP}}{n_F+n_{IP}}.
\end{equation}

\begin{figure}[H]
    \centering
    \includegraphics[width=0.8\linewidth]{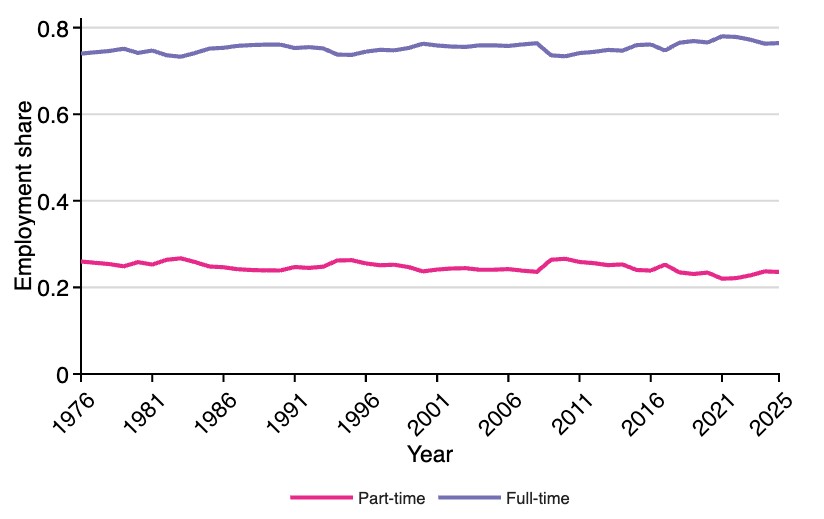}
    \caption{Share of full-time and part-time employment in the United States, 1976--2025. \textit{Notes:} The figure reports the share of full-time employment, $\alpha^{P}$, and the share of part-time employment, $1-\alpha^{P}$, in total employment. Full-time workers are defined as employed individuals with actual weekly hours worked of at least 35 (\texttt{ahrsworkt} $\geq 35$); part-time workers are defined as employed individuals with actual weekly hours worked of fewer than 35 (\texttt{ahrsworkt} $< 35$). The sample includes employed workers (\texttt{EMPSTAT} = 10, 12) with valid actual hours and positive person weights. Employment shares are computed using \texttt{ASECWT}, with the 2014 ASEC split-sample adjustment (\texttt{ASECWT} $\times 5/8$ if \texttt{HFLAG}=0; $\times 3/8$ if \texttt{HFLAG}=1). Population counts are $n^F_t=\sum_i w_{it}\mathbf{1}\{\text{FT}\}$ and $n^P_t=\sum_i w_{it}\mathbf{1}\{\text{PT}\}$; $\alpha^{P}_t=n^F_t/(n^F_t+n^P_t)$. \textit{Source:} IPUMS CPS ASEC.}

    \label{fig:alpha-wholept}
\end{figure}

Figure~\ref{fig:alpha-wholept} plots the evolution of full-time and total part-time employment shares in the United States from 1976 to 2025. The share of part-time employment fluctuates over time but remains within a relatively narrow range. Table~\ref{tab:us_pt_shares} shows that full-time employment accounts for approximately 75 percent of total employment on average, while part-time employment accounts for the remaining 25 percent. This stability motivates calibrating the employment-share parameter at its sample average, $\alpha=0.75$.

\begin{figure}[H]
    \centering
    \includegraphics[width=0.8\linewidth]{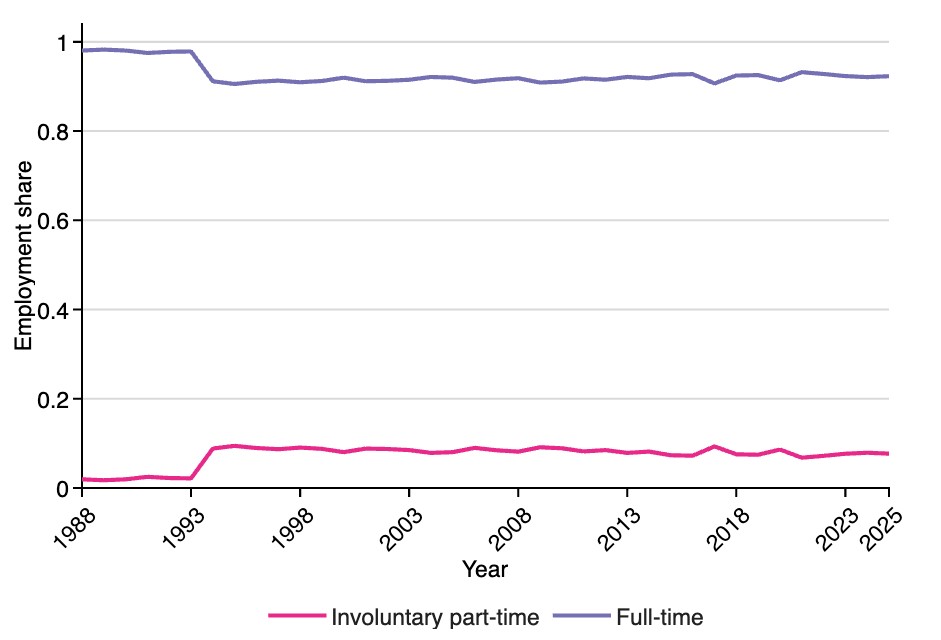}
    \caption{Share of full-time and involuntary part-time employment in the United States, 1988--2025. \textit{Notes:} The figure reports the share of full-time employment, $\alpha^{IP}$, and the share of involuntary part-time employment, $1-\alpha^{IP}$, among full-time and involuntary part-time workers. Full-time workers are defined as employed individuals with actual weekly hours worked of at least 35 (\texttt{ahrsworkt} $\geq 35$). Involuntary part-time workers are defined as employed individuals with \texttt{WKSTAT} = 12 or 21 (part-time for economic reasons). The sample includes employed workers (\texttt{EMPSTAT} = 10, 12) with positive person weights. Employment shares are computed using \texttt{ASECWT}, with the 2014 ASEC split-sample adjustment (\texttt{ASECWT} $\times 5/8$ if \texttt{HFLAG}=0; $\times 3/8$ if \texttt{HFLAG}=1). Population counts are $n^F_t=\sum_i w_{it}\mathbf{1}\{\text{FT}\}$ and $n^{IP}_t=\sum_i w_{it}\mathbf{1}\{\text{IPT}\}$; $\alpha^{IP}_t=n^F_t/(n^F_t+n^{IP}_t)$. The series begins in 1988, when \texttt{WKSTAT} codes for involuntary part-time are available. \textit{Source:} IPUMS CPS ASEC.}

    \label{fig:alpha}
\end{figure}

For involuntary part-time and full-time employment, the overall composition of employment remains remarkably stable from 1988 to 2025, as shown in Figure~\ref{fig:alpha}. Table~\ref{tab:us_ipt_shares} confirms this pattern quantitatively: full-time employment accounts for approximately 93 percent of the relevant employment measure on average, while involuntary part-time employment accounts for the remaining 7 percent.

For the historical periods in which direct measures are unavailable, I use the sample-average employment shares as calibration values. For total part-time employment, the average full-time employment share is

\begin{equation}
\alpha^{P} = 0.75,
\end{equation}

implying that approximately three-quarters of employed workers are full-time workers and one-quarter are part-time workers. This value is used to calibrate the pre-1976 historical period. For the period in which CPS ASEC data are available, beginning in 1976, I incorporate the observed time-series values of $\alpha_t$.

For involuntary part-time employment, the corresponding average full-time employment share is

\begin{equation}
\alpha^{IP}=0.93,
\end{equation}

implying that approximately 93 percent of workers are employed full time, while 7 percent are employed part time for economic reasons. Since the involuntary part-time series is available beginning in 1988, I use $\alpha^{IP}=0.93$ to calibrate earlier historical periods and use the observed time-series values of $\alpha^{IP}_t$ after 1988.

\begin{table}[H]
\centering
\caption{Summary Statistics of Full-Time and Total Part-Time Employment Shares}
\label{tab:us_pt_shares}
\begin{tabular}{lcccc}
\toprule
 & Full-time ($n^F$) & Part-time ($n^P$) & $\alpha^{P}$ & $1-\alpha^{P}$ \\
\midrule
Mean  & 93.75 million & 30.52 million & 0.753 & 0.247 \\
SD    & 16.70 million &  4.55 million & 0.011 & 0.011 \\
Min   & 61.01 million & 21.42 million & 0.733 & 0.220 \\
Max   & 120.82 million & 37.25 million & 0.780 & 0.267 \\
$N$   & 50 & 50 & 50 & 50 \\
\bottomrule
\end{tabular}

\medskip
\begin{minipage}{\linewidth}
\footnotesize
\textit{Notes:} Annual employment counts from the CPS Annual Social and Economic Supplement (ASEC), 1976--2025 ($N=50$).
The sample includes employed workers (\texttt{empstat} = 10, 12) with positive person weights (\texttt{asecwt}).
In 2014, weights are adjusted for the ASEC split sample (\texttt{asecwt} $\times$ 5/8 if \texttt{hflag}=0; $\times$ 3/8 if \texttt{hflag}=1).
Full-time is defined by actual weekly hours worked $\geq 35$ (\texttt{ahrsworkt}); total part-time by actual hours $<35$ (workers with valid actual hours only).
Population counts are $n^F_t=\sum_i w_{it}\mathbf{1}\{\text{FT}\}$ and $n^P_t=\sum_i w_{it}\mathbf{1}\{\text{PT}\}$;
$\alpha^{P}_t=n^F_t/(n^F_t+n^P_t)$ and $1-\alpha^{P}_t=n^P_t/(n^F_t+n^P_t)$.
For years before 1976 in the quarterly figures, we set $\alpha$ to the sample mean reported above (0.75) and use the annual series $\alpha^{P}_t$ from 1976 onward.
\end{minipage}
\end{table}

\begin{table}[H]
\centering
\caption{Summary Statistics of Full-Time and Involuntary Part-Time Employment Shares}
\label{tab:us_ipt_shares}
\begin{tabular}{lcccc}
\toprule
 & Full-time ($n^F$) & Invol.\ part-time ($n^{IP}$) & $\alpha^{IP}$ & $1-\alpha^{IP}$ \\
\midrule
Mean  & 101.02 million &  8.16 million & 0.927 & 0.073 \\
SD    &  11.56 million &  2.78 million & 0.023 & 0.023 \\
Min   &  82.33 million &  1.64 million & 0.906 & 0.019 \\
Max   & 120.82 million & 11.24 million & 0.981 & 0.094 \\
$N$   & 38 & 38 & 38 & 38 \\
\bottomrule
\end{tabular}

\medskip
\begin{minipage}{\linewidth}
\footnotesize
\textit{Notes:} Same CPS ASEC sample and weighting as Table~\ref{tab:us_pt_shares}, 1988--2025 ($N=38$).
Full-time: actual weekly hours worked $\geq 35$ (\texttt{ahrsworkt}).
Involuntary part-time (IPT): \texttt{wkstat} = 12 or 21 (at work part-time for economic reasons, or usually part-time for economic reasons); no additional hours restriction.
Counts: $n^F_t$, $n^{IP}_t$; $\alpha^{IP}_t=n^F_t/(n^F_t+n^{IP}_t)$.
The IPT series is meaningful from 1988 when \texttt{wkstat} is available; before 1988 we calibrate $\alpha^{IP}$ to the sample mean above (0.93) in the quarterly figures and use observed $\alpha^{IP}_t$ thereafter.
\end{minipage}
\end{table}

\subsubsection{Share of working hours}

\begin{figure}[H]
\centering
\includegraphics[width=0.9\linewidth]{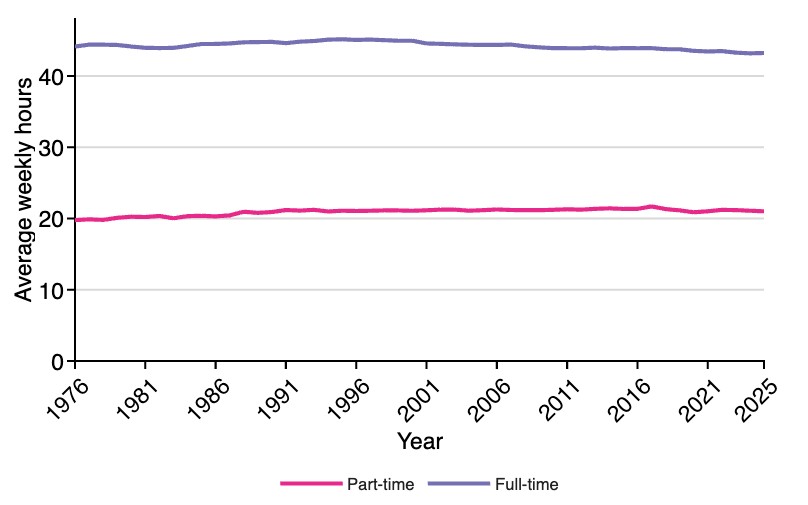}
\caption{Average weekly hours worked by full-time and part-time workers in the United States, 1976--2025. \textit{Notes:} The figure reports annual average actual weekly hours worked by full-time and part-time workers. Workers are classified using actual hours worked last week (\texttt{ahrsworkt}): full-time if \texttt{ahrsworkt} $\geq 35$, part-time if \texttt{ahrsworkt} $< 35$ (valid actual hours only). The sample includes employed workers (\texttt{EMPSTAT} = 10, 12) with positive person weights. Averages are population-weighted means of \texttt{ahrsworkt} among full-time and part-time workers, using \texttt{ASECWT} with the 2014 ASEC split-sample adjustment (\texttt{ASECWT} $\times 5/8$ if \texttt{HFLAG}=0; $\times 3/8$ if \texttt{HFLAG}=1). \textit{Source:} IPUMS CPS ASEC.}
\label{fig:working-hours-pt}
\end{figure}

The second sufficient statistic is the relative-hours contribution of part-time workers. I construct two versions of this statistic. The first uses total part-time employment, while the second uses involuntary part-time employment. The distinction is useful because total part-time employment captures the aggregate-hours contribution of part-time work, whereas involuntary part-time employment isolates labor underutilization among workers who would like to work more hours. For total part-time employment, I define the relative-hours parameter as

\begin{equation}
\gamma^{P}_{t}=\frac{H^{P}_{t}}{H^{F}_{t}},
\end{equation}

where $H^{P}_{t}$ and $H^{F}_{t}$ denote the average weekly hours worked by part-time and full-time workers, respectively. Table~\ref{tab:us_pt_hours} shows that part-time workers work 20.92 hours per week on average, while full-time workers work 44.26 hours. Figure~\ref{fig:working-hours-pt} plots the corresponding time series of average weekly hours for full-time and part-time workers from 1976 to 2025. The implied relative-hours ratio averages 0.47 and is highly stable over time, with a standard deviation of only 0.012. Therefore, for the historical period in which annual hours data are not available, I use the sample average $\gamma^{P}=0.47$ as the calibration value. For the period in which the data are available, I use the observed time-series value $\gamma^{P}_{t}$.

\begin{figure}[H]
\centering
\includegraphics[width=0.9\linewidth]{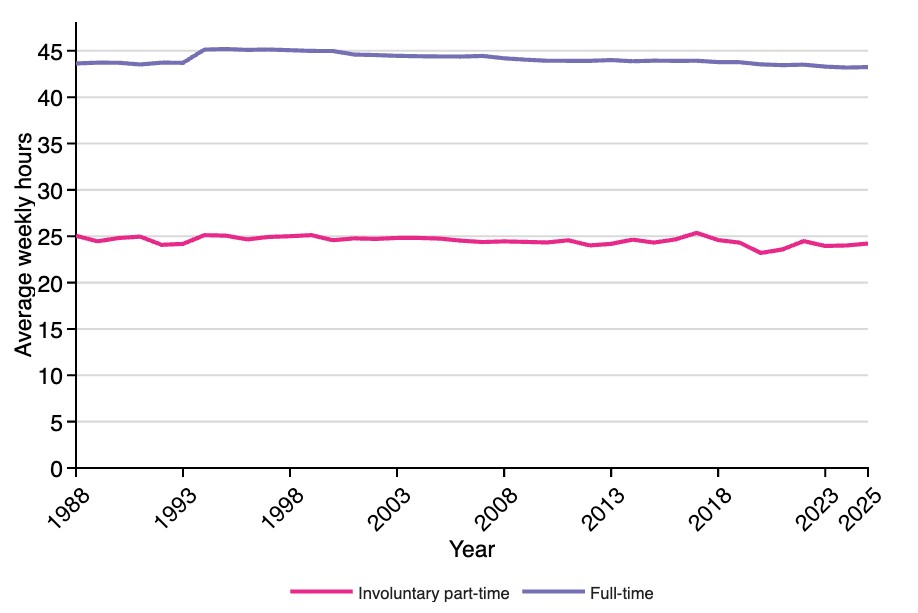}
\caption{Average weekly hours worked by full-time and involuntary part-time workers in the United States, 1988--2025. \textit{Notes:} The figure reports annual average actual weekly hours worked by full-time and involuntary part-time workers. Full-time workers are defined as employed individuals with actual weekly hours worked of at least 35 (\texttt{ahrsworkt} $\geq 35$). Involuntary part-time workers are defined as employed individuals with \texttt{WKSTAT} = 12 or 21 (part-time for economic reasons). The sample includes employed workers (\texttt{EMPSTAT} = 10, 12) with valid actual hours and positive person weights. Averages are population-weighted means of \texttt{ahrsworkt} among full-time and involuntary part-time workers, using \texttt{ASECWT} with the 2014 ASEC split-sample adjustment (\texttt{ASECWT} $\times 5/8$ if \texttt{HFLAG}=0; $\times 3/8$ if \texttt{HFLAG}=1). The series begins in 1988, when \texttt{WKSTAT} codes for involuntary part-time are available. \textit{Source:} IPUMS CPS ASEC.}
\label{fig:working-hours-ipt}
\end{figure}

For involuntary part-time employment, I define

\begin{equation}
\gamma^{IP}_{t}=\frac{H^{IP}_{t}}{H^{F}_{t}},
\end{equation}

where $H^{IP}_{t}$ denotes the average weekly hours worked by involuntary part-time workers. Figure~\ref{fig:working-hours-ipt} reports average weekly hours for full-time and involuntary part-time workers in the United States from 1988 to 2025. Full-time workers work approximately 44 hours per week on average, while involuntary part-time workers work approximately 25 hours per week.

As shown in Table~\ref{tab:us_ipt_hours} and Figure~\ref{fig:working-hours-ipt}, the relative-hours ratio for involuntary part-time workers averages 0.56, with a standard deviation of only 0.008. This implies that involuntary part-time workers supply approximately 56 percent as many market hours as full-time workers. As with total part-time employment, I use the sample average $\gamma^{IP}=0.56$ for the historical period without available annual data and use the observed time-series value $\gamma^{IP}_{t}$ for the period in which the data are available.

\begin{table}[htbp]
\centering
\caption{Summary Statistics of Average Weekly Hours: Total Part-Time ($\gamma^{P}$)}
\label{tab:us_pt_hours}
\begin{tabular}{lccc}
\toprule
 & Full-time hours ($H^F$) & Part-time hours ($H^P$) & $\gamma$ \\
\midrule
Mean  & 44.26 & 20.92 & 0.473 \\
SD    &  0.52 &  0.47 & 0.012 \\
Min   & 43.18 & 19.77 & 0.446 \\
Max   & 45.16 & 21.69 & 0.494 \\
$N$   & 50 & 50 & 50 \\
\bottomrule
\end{tabular}

\medskip
\begin{minipage}{\linewidth}
\footnotesize
\textit{Notes:} Annual averages from the CPS ASEC, 1976--2025 ($N=50$), using the same sample and weights as Table~\ref{tab:us_pt_shares}.
Workers are classified into full-time and total part-time by actual hours worked (\texttt{ahrsworkt} $\geq 35$ vs.\ $<35$).
$H^F_t$ and $H^P_t$ are population-weighted means of actual weekly hours worked (\texttt{ahrsworkt}) among full-time and part-time workers, respectively (valid actual hours only).
$\gamma_t = H^P_t/H^F_t$.
For years before 1976 in the quarterly figures, we set $\gamma^{P}$ to the sample mean (0.47) and use the annual series from 1976 onward.
\end{minipage}
\end{table}

\begin{table}[htbp]
\centering
\caption{Summary Statistics of Average Weekly Hours: Involuntary Part-Time ($\gamma^{IP}$)}
\label{tab:us_ipt_hours}
\begin{tabular}{lccc}
\toprule
 & Full-time hours ($H^F$) & IPT hours ($H^{IP}$) & $\gamma^{IP}$ \\
\midrule
Mean  & 44.27 & 24.52 & 0.554 \\
SD    &  0.59 &  0.44 & 0.008 \\
Min   & 43.18 & 23.19 & 0.533 \\
Max   & 45.16 & 25.37 & 0.578 \\
$N$   & 38 & 38 & 38 \\
\bottomrule
\end{tabular}

\medskip
\begin{minipage}{\linewidth}
\footnotesize
\textit{Notes:} Same CPS ASEC sample and weighting as Table~\ref{tab:us_ipt_shares}, 1988--2025 ($N=38$).
Full-time and IPT groups are defined as in Table~\ref{tab:us_ipt_shares}.
$H^F_t$ and $H^{IP}_t$ are population-weighted means of \texttt{ahrsworkt} among full-time and IPT workers (valid actual hours only).
$\gamma^{IP}_t = H^{IP}_t/H^F_t$.
For years before 1988 in the quarterly figures, we set $\gamma^{IP}$ to the sample mean (0.56) and use the annual series from 1988 onward.
\end{minipage}
\end{table}

The average relative-hours ratio is 0.47 for total part-time workers and 0.56 for involuntary part-time workers. Although involuntary part-time workers supply somewhat more hours than the average part-time worker, the quantitative difference is relatively small. I nevertheless distinguish between the two measures for consistency with the corresponding employment definitions. In the Japanese application, however, separate hours data for involuntary part-time workers are unavailable. Therefore, I assume that the relative-hours ratio for involuntary part-time workers is equal to that of total part-time workers.

\textbf{Proposition 1:} Consider a point on the Beveridge curve with the share of full-time workers $\alpha$, recruiting cost $c$, Beveridge elasticity $\epsilon$, share of time $\gamma$ and the marginal rate of substitution between unemployment and employment $z$, tightness $\theta$ is inefficiently high if $\theta > \frac{\alpha+(1-\alpha)\cdot (\gamma + (1-\gamma) \cdot z)-z}{c \epsilon}$, inefficiently low if $\theta < \frac{\alpha+(1-\alpha)\cdot (\gamma + (1-\gamma) \cdot z)-z}{c \epsilon}$, efficient if \\

\begin{equation}\label{my_first_eqn}
\theta^{*} = \frac{\alpha+(1-\alpha)\cdot [\gamma + (1-\gamma) \cdot z]-z}{c \epsilon}
\end{equation}

The derivation treats the employment composition $\alpha$ and relative-hours parameter $\gamma$ as fixed at the point where efficiency is evaluated. The tangency condition therefore characterizes efficient labor-market tightness for a given employment composition and hours structure. In the empirical implementation, $\alpha_t$ and $\gamma_t$ are allowed to vary over time and enter the sufficient-statistics formula quarter by quarter.

\begin{proof}
Efficiency is achieved where the Beveridge curve is tangent to an isowelfare curve (\citeauthor{MS21b}, \citeyear{MS21b}). To locate the tangency point, we first need the slope of the isowelfare curve, which is minus the marginal rate of substitution between unemployment and vacancies in the welfare function. The slope of the Beveridge curve is $v'(u)$. Therefore,
\begin{align*}
-v'(u)
&= -\frac{\partial W / \partial u}{\partial W / \partial v} \\
&= \frac{-\partial W / \partial u - \partial W / \partial (n_{F}+n_{P})}{\partial W / \partial v} \\
&= \frac{-\partial W / \partial u + \alpha \cdot \partial W / \partial n_{F} + (1-\alpha) \cdot \partial W / \partial n_{P}}{\partial W / \partial v} \\
&= \frac{\frac{-\partial W / \partial u + \alpha \cdot \partial W / \partial n_{F} + (1-\alpha) \cdot \partial W / \partial n_{P}}{\partial W / \partial n_{F}}}{(\partial W / \partial v) / (\partial W / \partial n_{F})} \\
&= \frac{\alpha + (1-\alpha)\cdot [\gamma + (1-\gamma) z] - z}{c}.
\end{align*}
Thus,
\begin{equation}\label{eq:isowelfare_slope}
-v'(u)=\frac{\alpha + (1-\alpha)\cdot [\gamma + (1-\gamma) z] - z}{c}.
\end{equation}
The Beveridge elasticity satisfies $\epsilon \theta = -v'(u)$, where $\theta \equiv v/u$ denotes labor-market tightness. Thus, the efficient tightness is
\begin{equation}\label{eq:efficient_tightness}
\theta^{*}=\frac{\alpha + (1-\alpha)\cdot [\gamma + (1-\gamma) z] - z}{c\epsilon}.
\end{equation}
The Beveridge curve is isoelastic. Therefore, the efficient unemployment rate can be expressed as

\begin{equation}\label{eq:efficient_unemployment}
u^{*}
=
\left[
\frac{c\epsilon}
{\alpha+(1-\alpha)\cdot [\gamma + (1-\gamma) z]-z}
\cdot
\frac{v}{u^{-\epsilon}}
\right]^{\frac{1}{1+\epsilon}}.
\end{equation}
\end{proof}

If $\alpha=1$, all employed workers are full-time workers. The part-time margin is then shut down, and the model collapses to the baseline functional form in \citet{MS21b}.

\begin{figure}[H]
    \centering
    \includegraphics[width=0.7\textwidth]{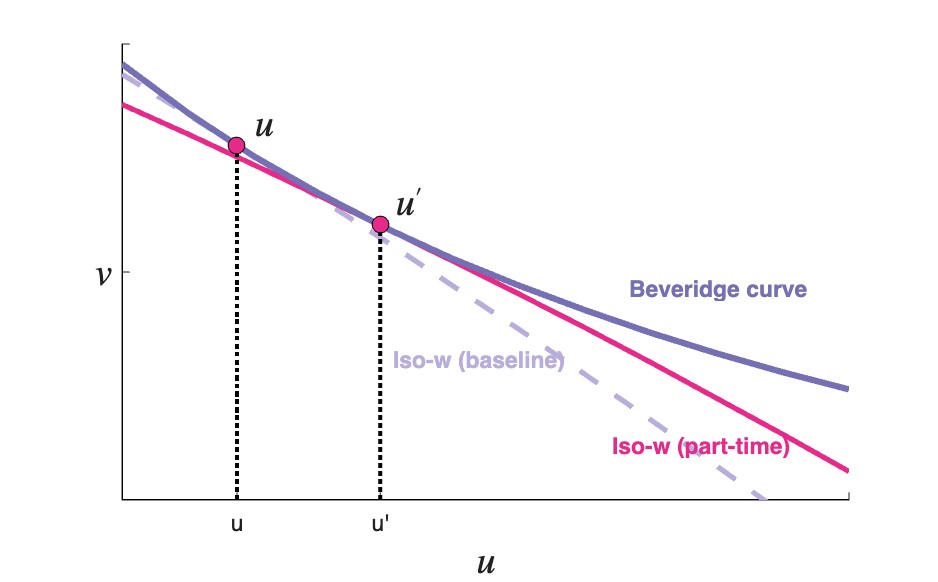}
    \caption{Efficient unemployment rate with and without part-time employment. \textit{Notes:} The dashed line shows the baseline isowelfare curve from \citet{MS21b}, while the solid line shows the isowelfare curve after incorporating part-time employment. Efficient unemployment is determined by the tangency between the Beveridge curve and the isowelfare curve. Accounting for part-time employment changes the slope of the isowelfare curve and shifts the efficient unemployment rate from u to u'.}   
    \label{fig:ustar}
\end{figure}

Figure~\ref{fig:ustar} illustrates the role of part-time employment in determining the efficient unemployment rate. Once part-time employment is introduced, the isowelfare curve becomes flatter. The reason is that part-time jobs contribute fewer effective hours than full-time jobs, since $\gamma<1$. Therefore, an additional employed worker generates less output when a larger share of employment is part-time. This reduces the social gain from pushing unemployment lower through additional vacancy creation.

As a result, the efficient allocation features lower labor-market tightness and a higher efficient unemployment rate. In the figure, the efficient point shifts slightly to the right along the Beveridge curve. Part-time employment can therefore make the observed unemployment rate appear lower by moving workers into employment, while still leaving the economy with lower effective labor input. This raises the efficient unemployment rate relative to the baseline model that treats all employment as full-time. This mechanism is central to the paper: accounting for part-time employment changes the benchmark against which labor-market slack should be measured.

\subsection{Estimation of remaining sufficient statistics}

Following \citet{MS21b}, I estimate the remaining sufficient statistics required to compute the efficient unemployment rate. These include the Beveridge elasticity, the social value of nonwork, and recruiting costs.

By regressing the log vacancy rate (using the vacancy proxy constructed by \citet{B10}) on the log unemployment rate (U.S. Bureau of Labor Statistics) from 1951Q1 to 2026Q1, I estimate the Beveridge elasticity in the United States. The statistical model has $m$ breaks and $m+1$ regimes:
\begin{equation}
\ln v(t) = \ln \hat{a}_j - \epsilon_j \ln u(t) + z(t), \qquad t = T_{j-1}+1,\ldots,T_j,
\label{eq:beveridge-regime}
\end{equation}
for $j=1,\ldots,m+1$. The observed independent variable is the logarithm of the unemployment rate, $\ln u(t)$, and the dependent variable is the logarithm of the vacancy rate, $\ln v(t)$. The $m$ break dates are denoted by $T_1,\ldots,T_m$; in this case, $T_0=0$ and $T_{m+1}=300$. The parameter $\epsilon_j$ is the Beveridge elasticity in regime $j$, and $\hat{a}_j$ is the intercept. These parameters, together with the break dates, are jointly estimated using the Bai--Perron algorithm of \citet{BP98,BP03}, which first determines the number of structural breaks $m$ and then estimates the parameters and break dates.

I follow the Bai--Perron setup of \citet{MS21b}, allowing for different variances of the errors across regimes and for autocorrelation in the errors. To obtain standard errors robust to autocorrelation and heteroskedasticity, the algorithm uses a quadratic kernel with automatic bandwidth selection based on an AR(1) approximation \citep{A1}. This setup also allows for different distributions of the independent and dependent variables across regimes. As required by the Bai--Perron procedure \citep{BP03}, I set the maximum number of breaks to three and the trimming parameter to 0.20, so that each regime contains at least 20\% of the sample (60 quarters).

The next step is to determine the number of structural breaks. To examine whether structural breaks are present, the algorithm runs supF tests of no structural break versus $m$ breaks for $m=1,2,3$. The tests reject the null hypothesis of no break at the 1\% significance level. I then use two information criteria---the Bayesian information criterion \citep{Yao88} and the modified Schwarz criterion \citep{LSZ97}---to select the number of breaks. Both criteria select three breaks.

\begin{figure}[p]

\subcaptionbox{Beveridge curve, 1951Q1--1965Q4}{
\includegraphics[width=0.48\textwidth]{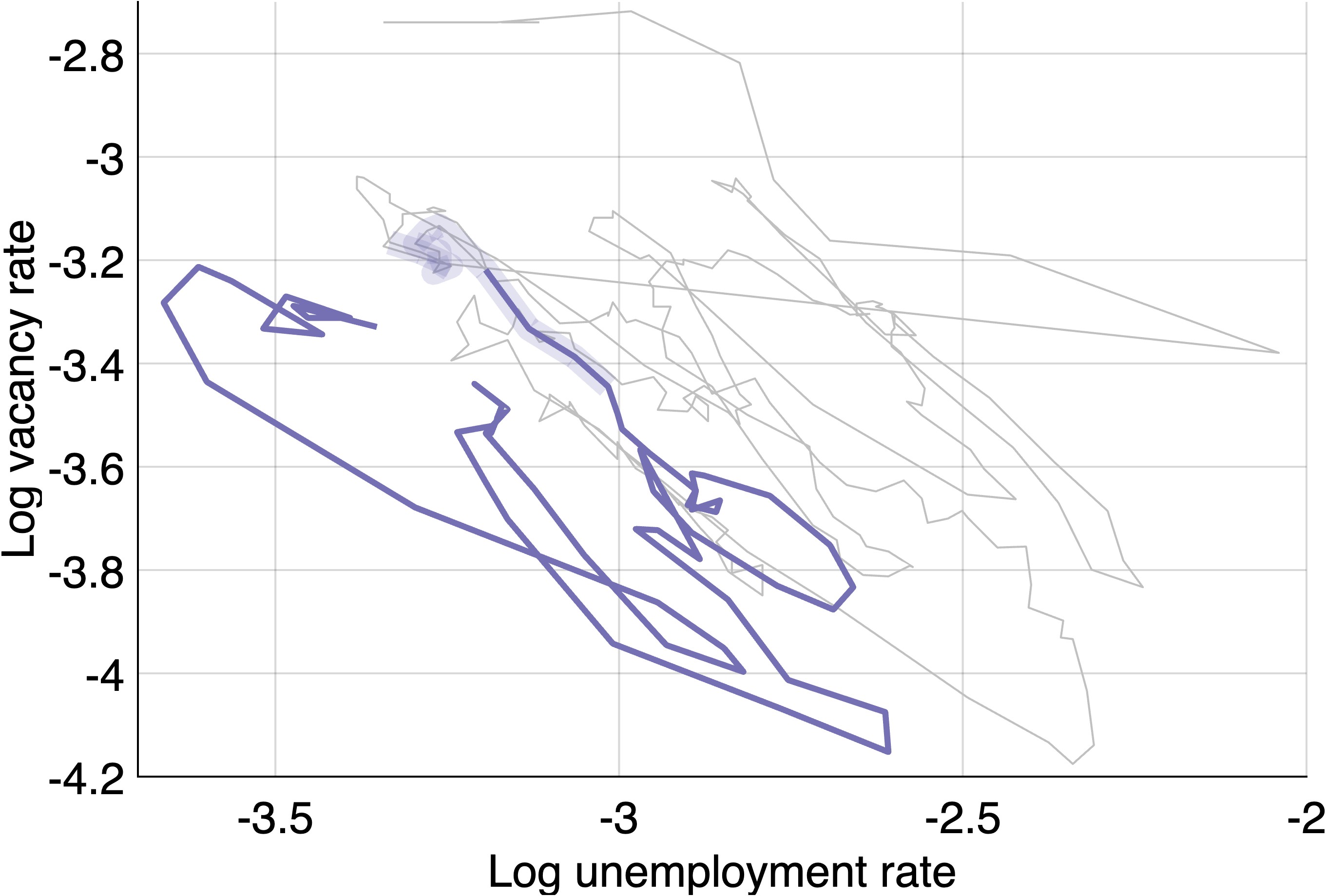}}
\hfill
\subcaptionbox{Beveridge curve, 1966Q1--1990Q2}{
\includegraphics[width=0.48\textwidth]{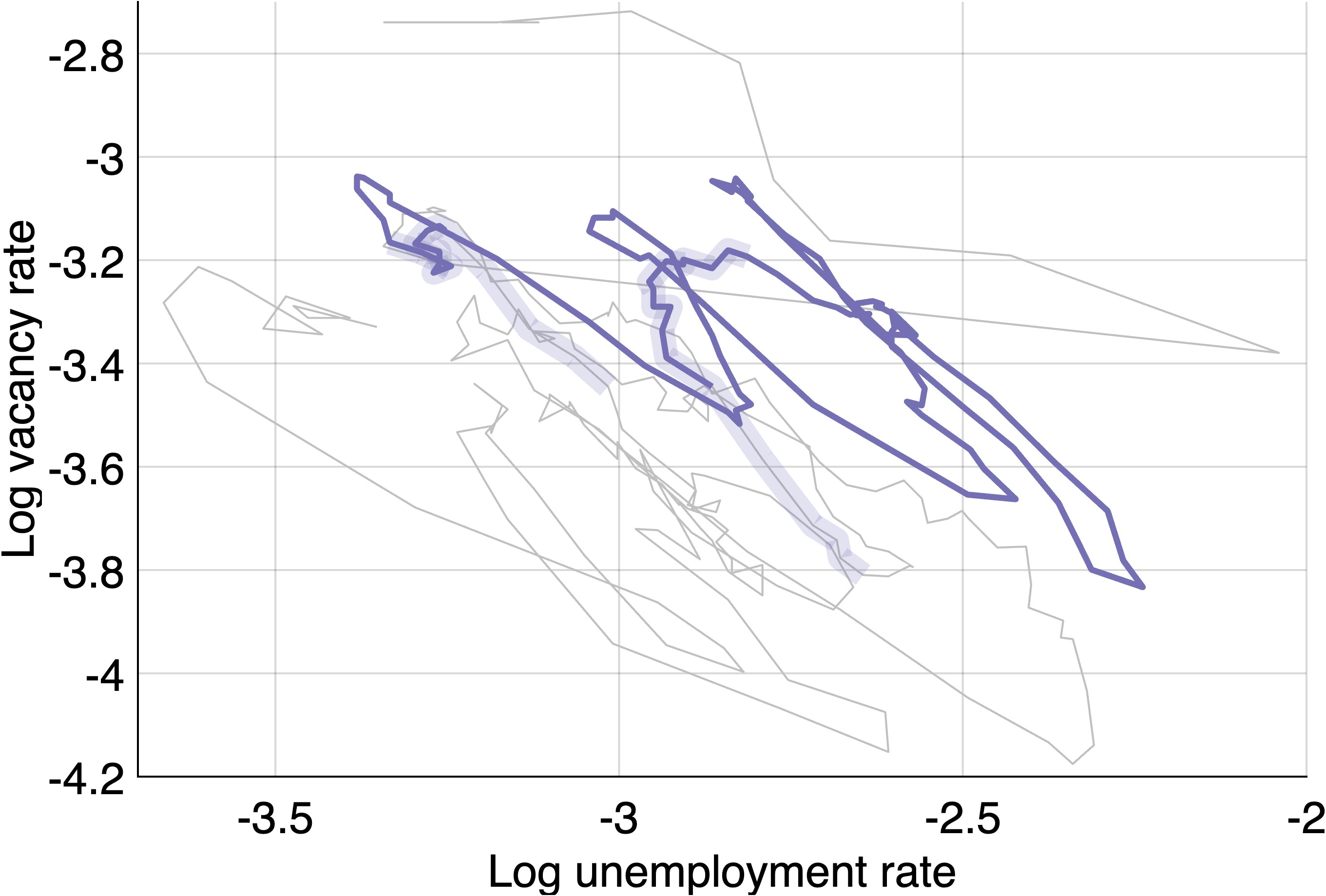}}

\vspace{0.3cm}

\subcaptionbox{Beveridge curve, 1990Q3--2011Q1}{
\includegraphics[width=0.48\textwidth]{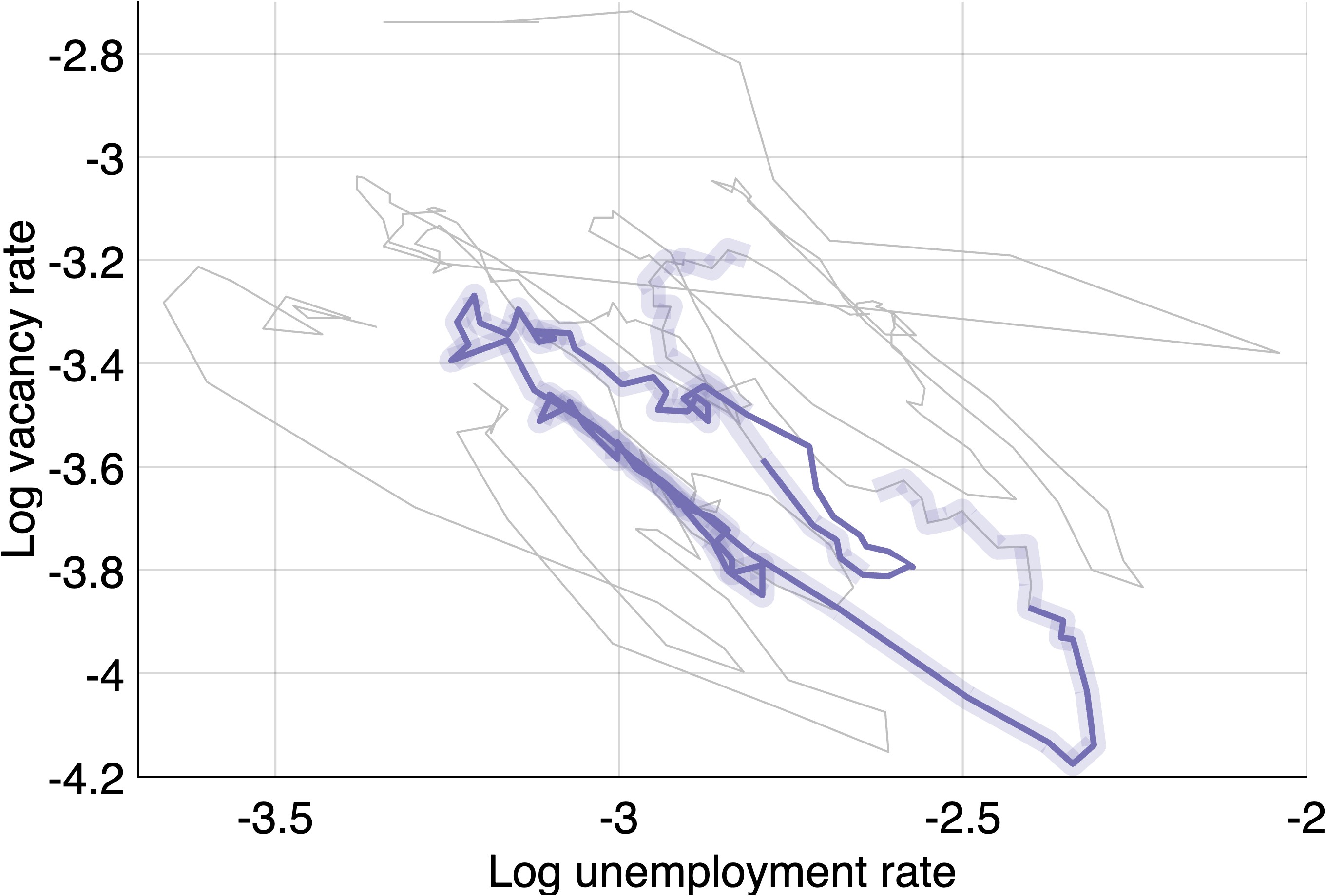}}
\hfill
\subcaptionbox{Beveridge curve, 2011Q2--2026Q1}{
\includegraphics[width=0.48\textwidth]{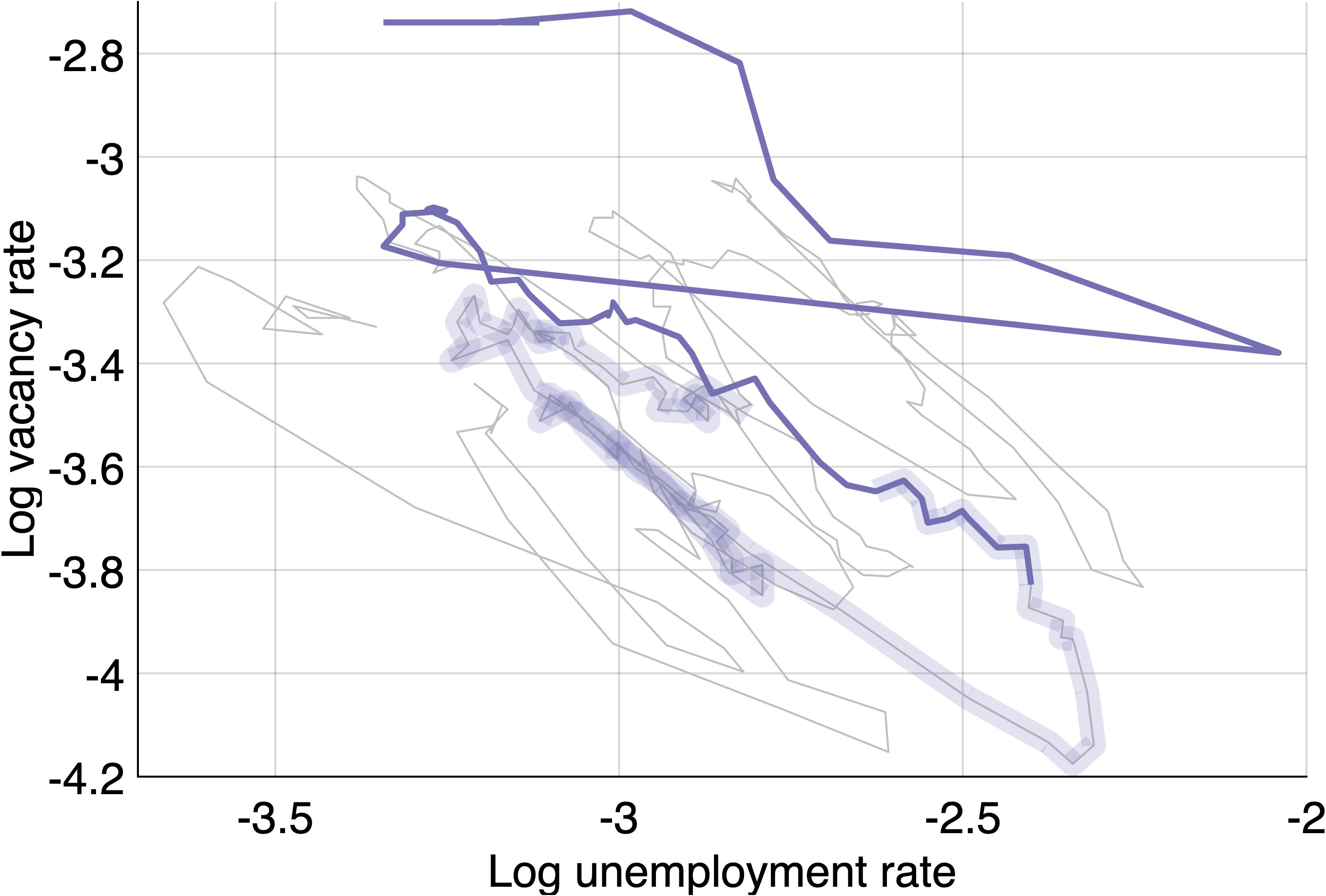}}

\caption{Beveridge curve estimation in the United States, 1951--2026. \textit{Notes:} The figure plots the Beveridge curve separately for four labor-market regimes identified using the structural break procedure of \citet{BP98,BP03}. The estimated regimes are 1951Q1--1965Q4, 1966Q1--1990Q2, 1990Q3--2011Q1, and 2011Q2--2026Q1.}

\label{f:graph1}

\end{figure}

The estimated break dates occur in 1965Q4, 1990Q3, and 2011Q1, yielding four Beveridge-curve regimes: 1951Q1--1965Q4, 1966Q1--1990Q2, 1990Q3--2011Q1, and 2011Q2--2026Q1. Table~\ref{tab:beveridge-elasticity} reports the resulting Beveridge elasticity estimates and corrected standard errors. The elasticity estimates range from 0.48 to 0.79 and average 0.68 over 1951--2026; the corrected standard errors range from 0.03 to 0.19. The overall fit is $R^2=0.77$. Figure~\ref{f:graph1} plots the estimated Beveridge curve within each regime.

The post-2020 period displays a distinct outward shift of the Beveridge curve, reflecting unusual labor-market disruptions associated with the COVID-19 pandemic. These disruptions include health concerns, changes in work preferences, and changes in the efficiency of job matching \citep{BFHS23}. Because the pandemic period represents a sharp departure from the historical Beveridge relationship, I present the main results separately for the pre-pandemic period (1951--2019) and the post-pandemic period (2020--2026). In the pre-pandemic sample, I estimate the Beveridge curve with the Bai--Perron procedure, setting the maximum number of breaks to five (with trimming parameter 0.15), as in \citet{MS21b}. In the post-pandemic subsample, the Bai--Perron break procedure is not reliable given the short sample, so I instead estimate the Beveridge elasticity using a simple OLS regression.

\begin{table}[H]
\centering
\caption{Beveridge elasticity estimates by regime, 1951Q1--2026Q1}
\label{tab:beveridge-elasticity}
\begin{tabular}{lcc}
\toprule
Regime & $\hat{\epsilon}_j$ & SE \\
\midrule
1951Q1--1965Q4 & 0.73 & 0.03 \\
1966Q1--1990Q2 & 0.48 & 0.07 \\
1990Q3--2011Q1 & 0.79 & 0.03 \\
2011Q2--2026Q1 & 0.71 & 0.19 \\
\bottomrule
\end{tabular}

\medskip
\begin{minipage}{0.95\linewidth}
\footnotesize
\textit{Notes:} Estimates are obtained from the Bai--Perron procedure with a maximum of three breaks and trimming parameter 0.20. Standard errors are corrected for autocorrelation and heteroskedasticity.
\end{minipage}
\end{table}

Second, I set the social value of nonwork following \citet{MS21b}. They measure this statistic from revealed-preference estimates in two studies. Using administrative data from the U.S. Army, \citet{BM18} use soldiers' reenlistment decisions as a natural experiment to estimate the cost of unemployment and find that home production and recreation during unemployment offset between 13\% and 35\% of earnings losses. In a separate field experiment, \citet{MP17} estimate that the value of nonwork time during unemployment is worth 58\% of predicted earnings. \citet{MS21b} translate these estimates into social values of nonwork by adjusting predicted earnings to the marginal product of labor and subtracting the value of public benefits received during unemployment (7\% of the marginal product of labor). This yields a plausible range of 0.03--0.49. \citet{MS21b} set the statistic to its midrange value, $z=0.26$, and I adopt the same calibration.

Third, I set the recruiting cost following \citet{MS21b}, who measure it from the 1997 National Employer Survey conducted by the Bureau of the Census, as in \citet{VR10}. The survey reports that establishments devote 3.2\% of labor costs to recruiting. Adjusting for the unemployment rate in 1997, \citet{MS21b} obtain a recruiting cost of $c=0.92$. Because there is no comprehensive time series of recruiting effort in the United States, they assume that the recruiting cost remains constant at its 1997 value, as is standard in matching models. I adopt the same value and assumption.

Together with the part-time employment statistics estimated above, these sufficient statistics allow me to compute efficient labor-market tightness, the efficient unemployment rate, and the unemployment gap under the extended framework.

\section{Application to the United States}

I first extend the baseline Beveridgean unemployment-gap results of \citet{MS21b} through 2026. Because the Beveridge curve displays unusual movements during and after the COVID-19 pandemic, I report the results separately for the pre-pandemic and post-pandemic periods.

I then incorporate part-time employment using two calibrations. The first uses total part-time employment, which includes both voluntary and involuntary part-time workers. Under this calibration, the full-time employment share is $\alpha^{P}=0.75$ and the relative-hours parameter is $\gamma^{P}=0.47$. The second uses involuntary part-time employment, defined as workers employed part time for economic reasons. Under this calibration, the corresponding values are $\alpha^{IP}=0.93$ and $\gamma^{IP}=0.56$. Comparing the two calibrations allows me to distinguish the aggregate-hours effect of part-time employment from the underemployment effect.

\subsection{Baseline results}

\begin{figure}[H]
    
\subcaptionbox{Efficient labor-market tightness, 1951--2019}{
\includegraphics[width=0.48\textwidth]{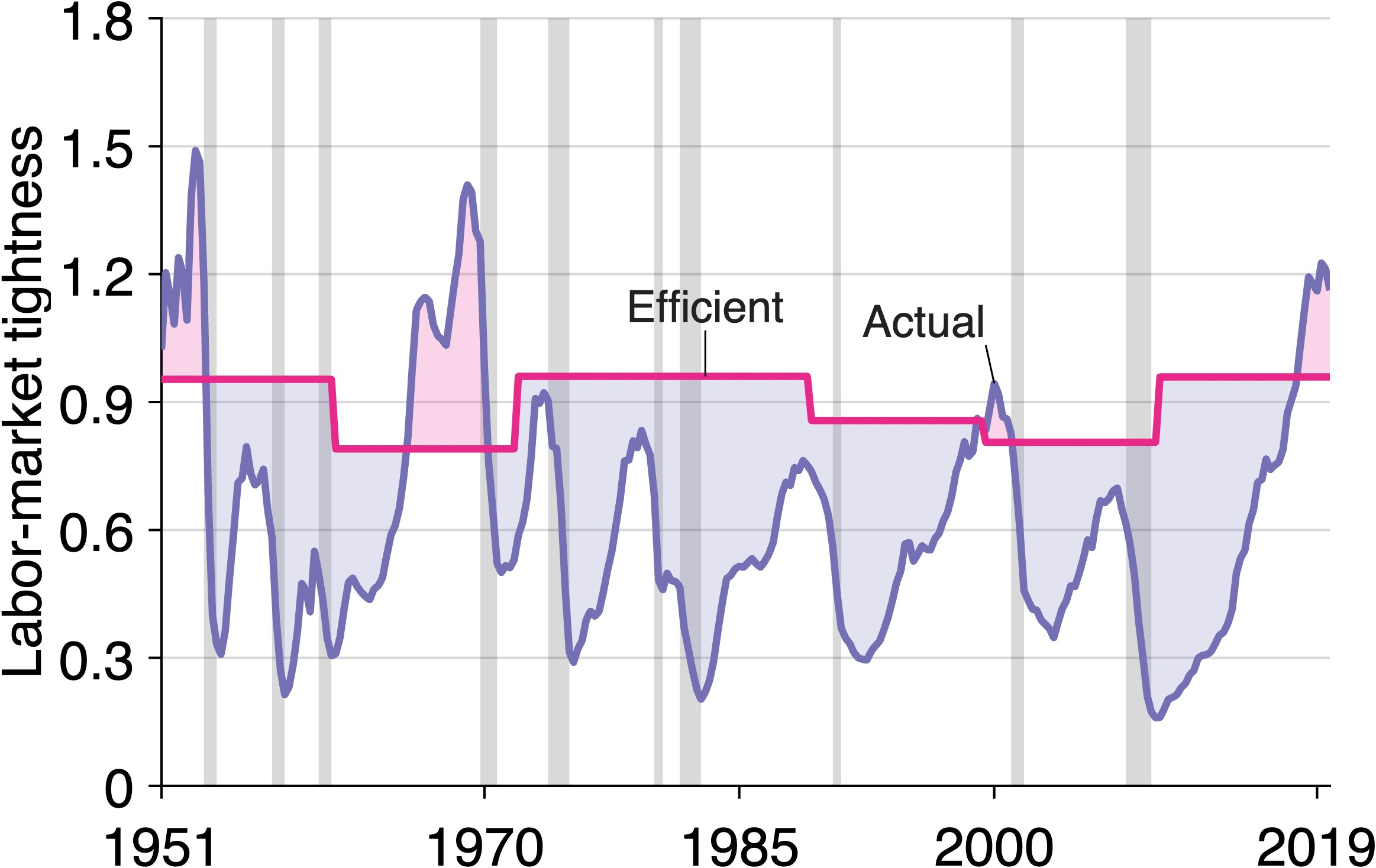}}
\hfill
\subcaptionbox{Efficient labor-market tightness, 2020--2026}{
\includegraphics[width=0.48\textwidth]{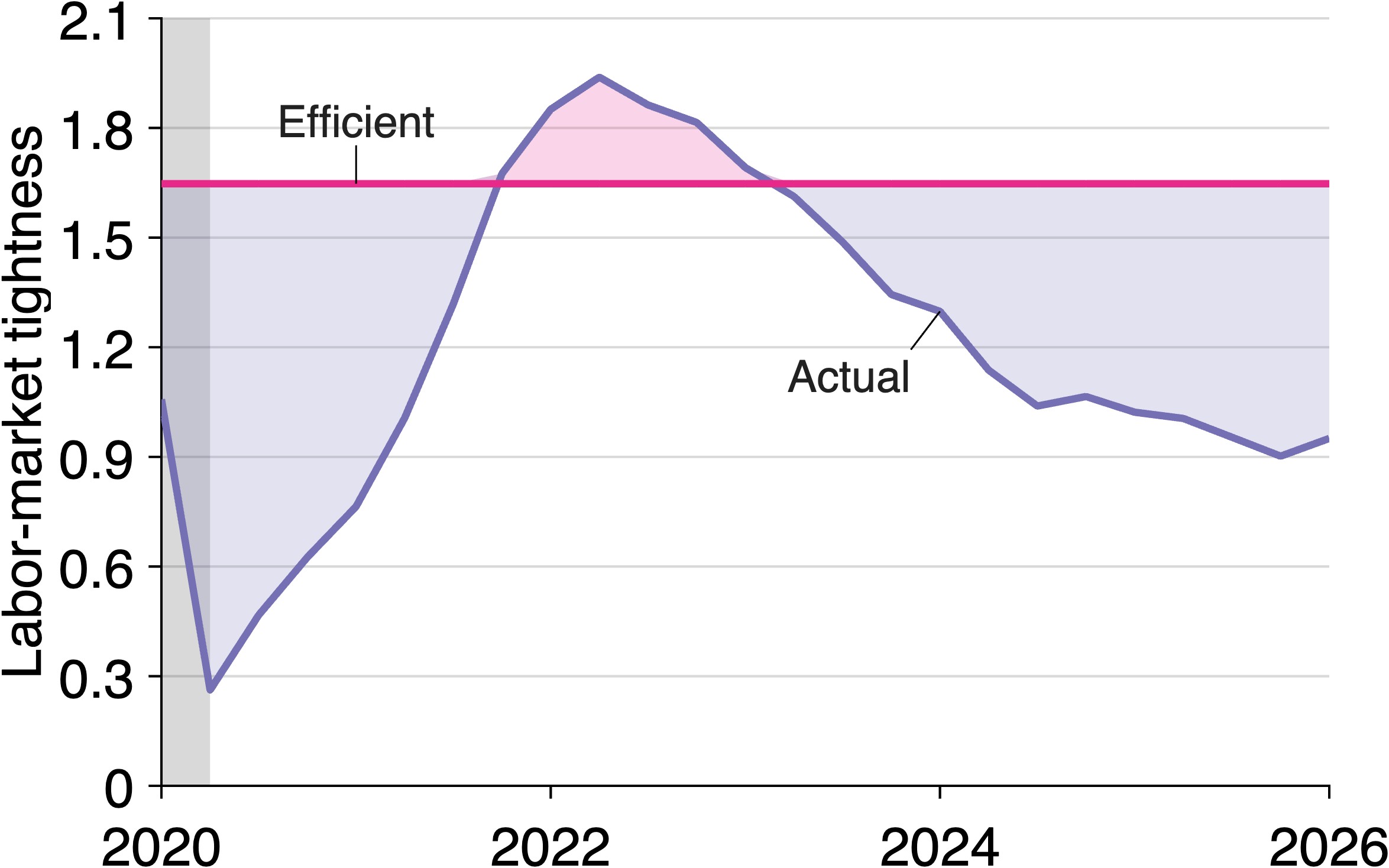}}

\vspace{0.2cm}

\subcaptionbox{Efficient unemployment rate, 1951--2019}{
\includegraphics[width=0.48\textwidth]{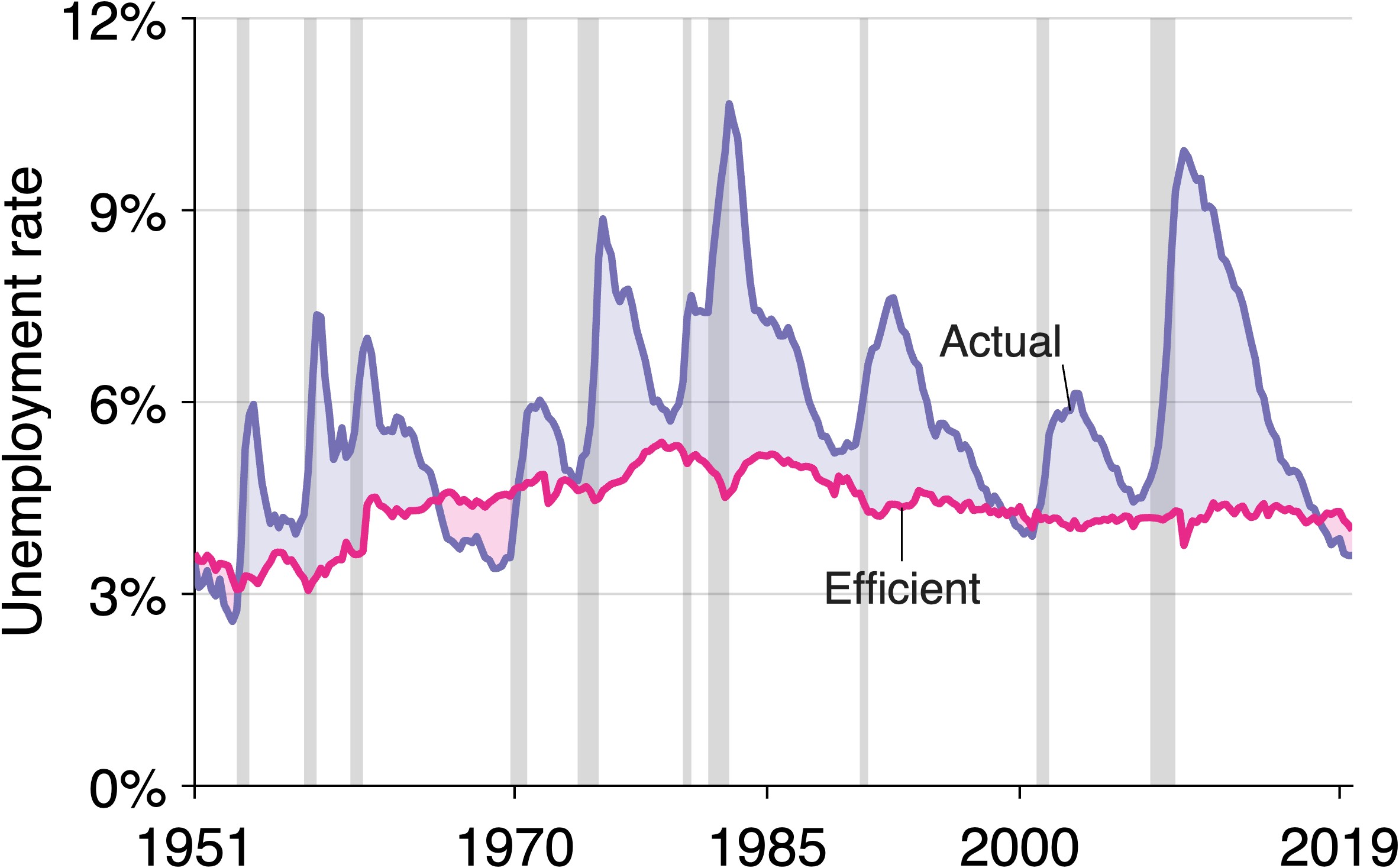}}
\hfill
\subcaptionbox{Efficient unemployment rate, 2020--2026}{
\includegraphics[width=0.48\textwidth]{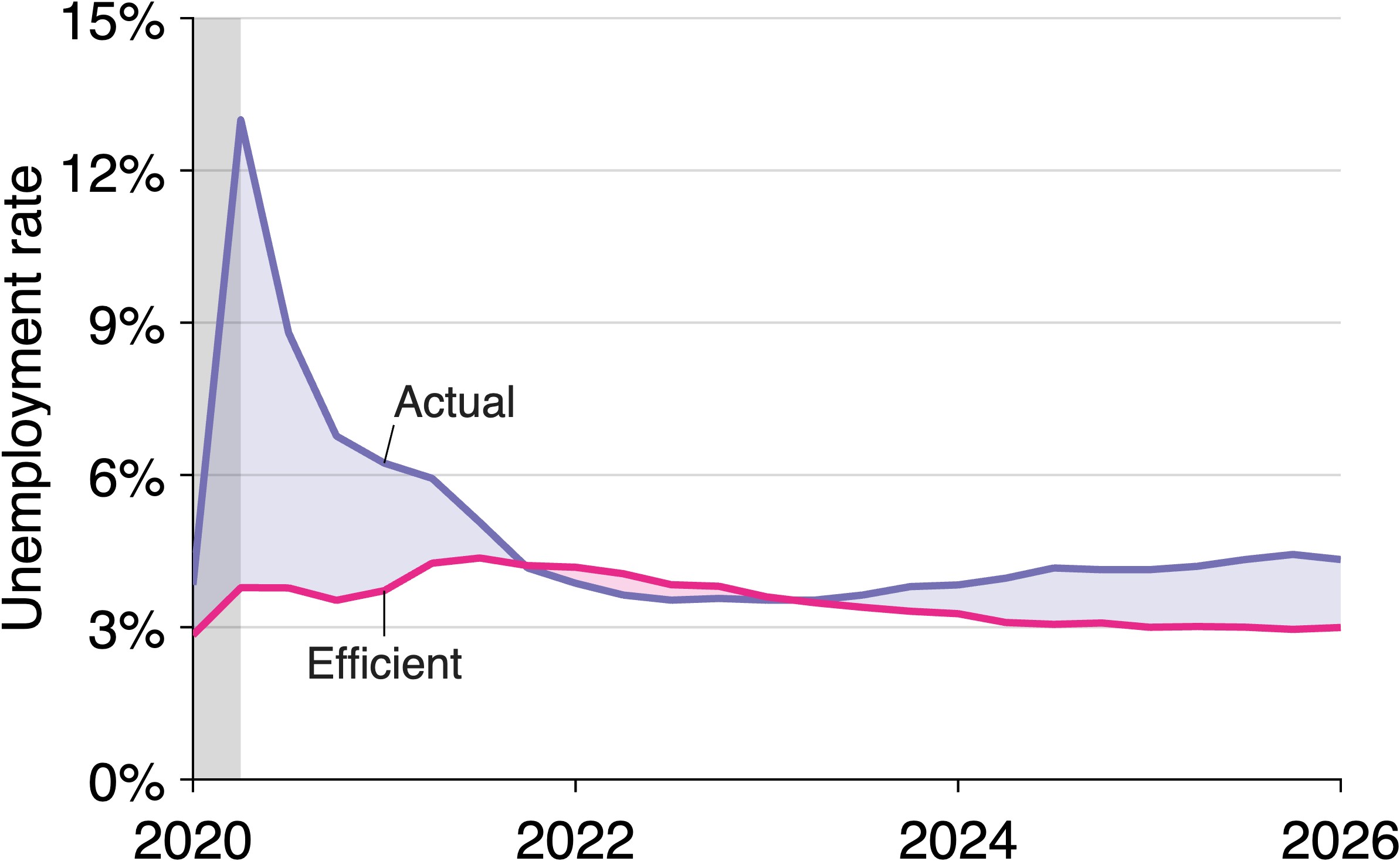}}

\vspace{0.2cm}

\subcaptionbox{Unemployment gap, 1951--2019}{
\includegraphics[width=0.48\textwidth]{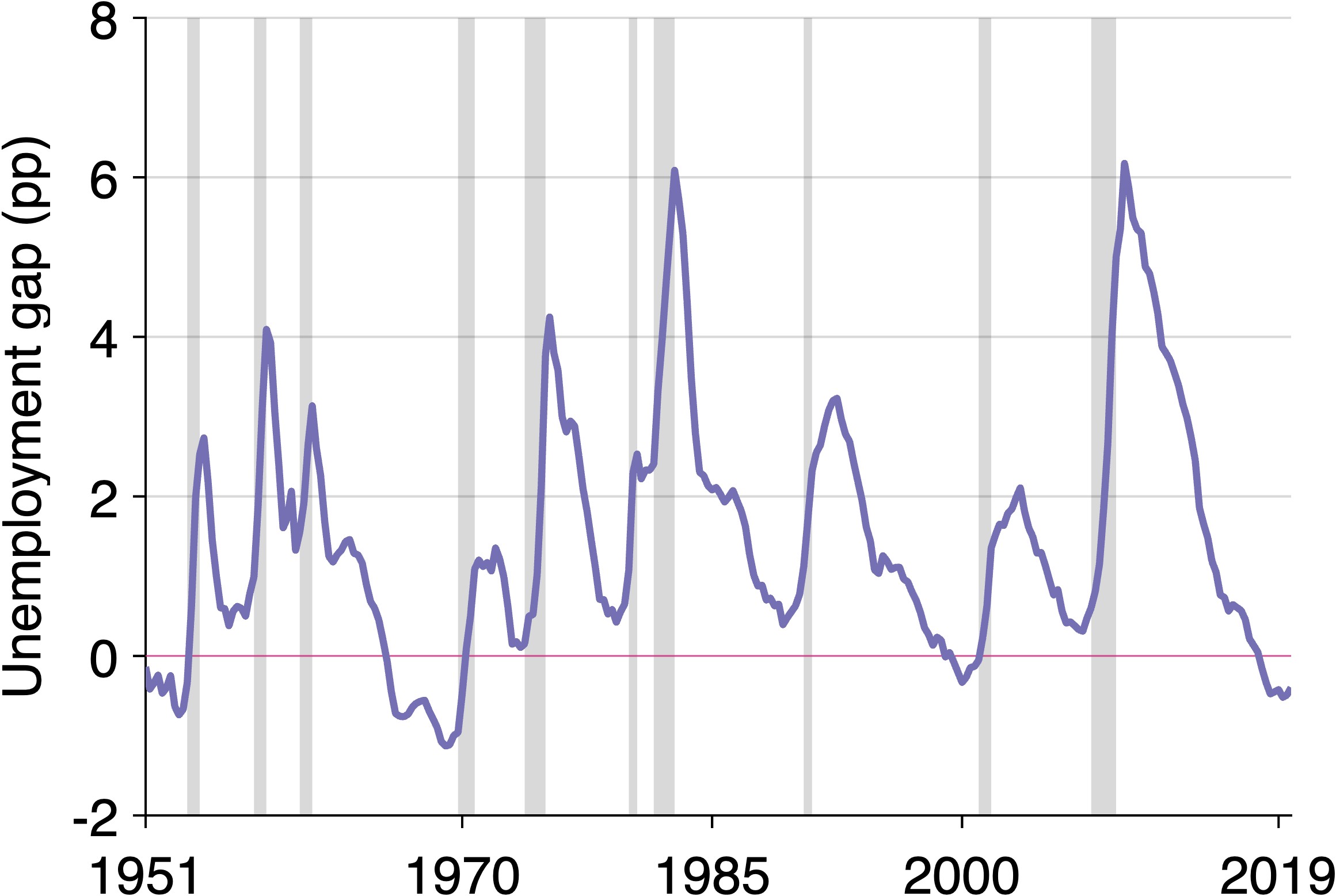}}
\hfill
\subcaptionbox{Unemployment gap, 2020--2026}{
\includegraphics[width=0.48\textwidth]{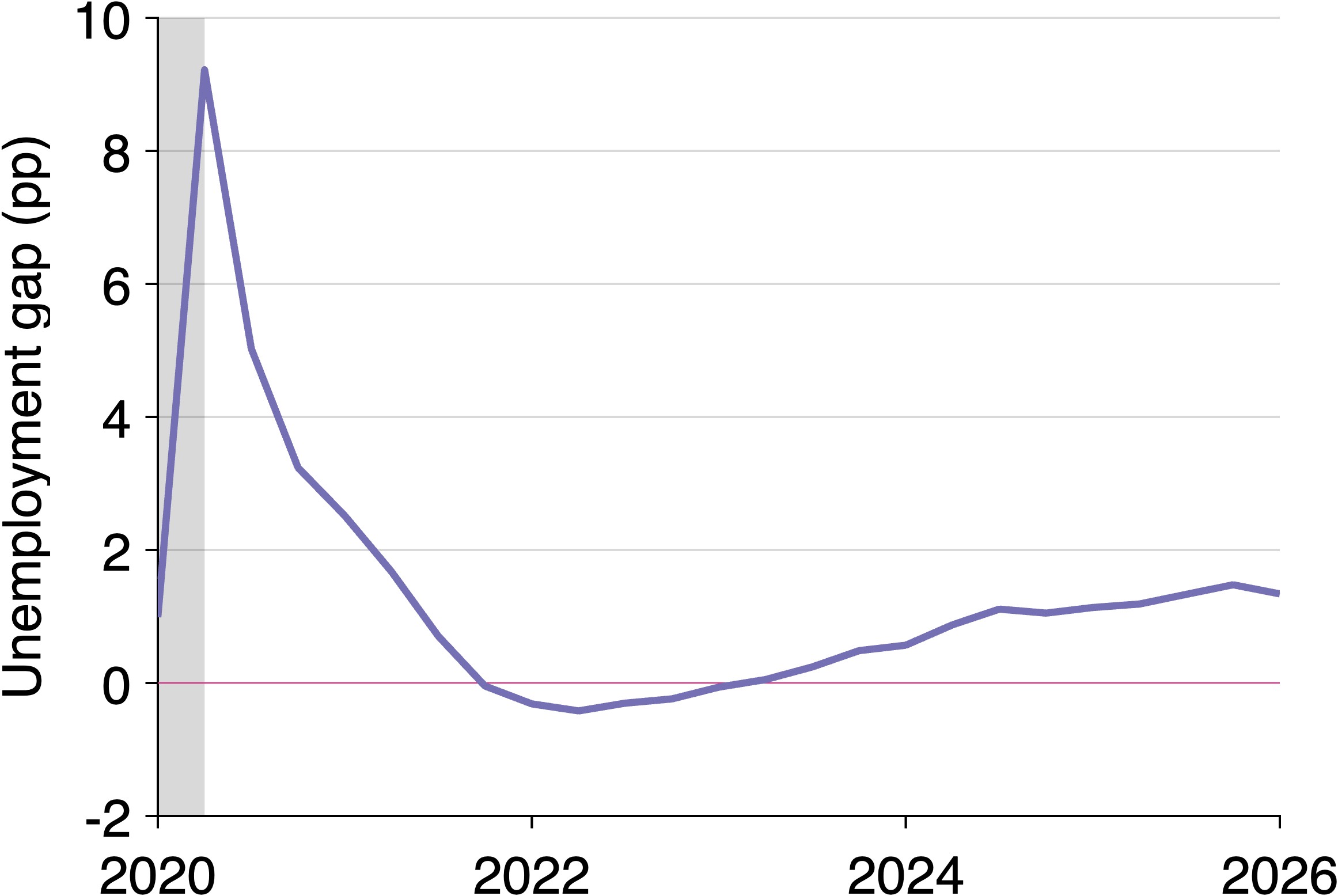}}

\caption{Efficient labor-market tightness, efficient unemployment, and unemployment gap in the United States under the full-time baseline, 1951--2026. \textit{Notes:} The figure reports the baseline results following \citet{MS21b}, in which employment is treated as homogeneous and all employed workers are effectively treated as full-time workers. The sample is split between the pre-pandemic period, 1951--2019, and the post-pandemic period, 2020--2026. These results provide the benchmark against which the part-time employment extensions are compared.}
\label{f:graph_fulltime_baseline}

\end{figure}

Figure~\ref{f:graph_fulltime_baseline} reports the baseline efficient-unemployment results for the United States through 2026. The figure separates the pre-pandemic and post-pandemic periods because the Beveridge curve displays unusual movements during and after the COVID-19 pandemic.

Panels A and B plot actual and efficient labor-market tightness. Before the pandemic, actual tightness rose during several well-known episodes, including the Korean War period, the late-1960s peak of the Vietnam War, and the dot-com boom of 1999--2000. After the pandemic, actual tightness increased sharply in 2021--2022 before easing in 2023. Because the post-pandemic sample is short, the Bai--Perron structural-break procedure is not applicable to this subsample. I therefore estimate the post-pandemic Beveridge elasticity using a simple OLS regression, which yields an elasticity of 1.65. This estimate serves as the post-pandemic benchmark.

Panels C and D report the efficient unemployment rate. Before the pandemic, the efficient unemployment rate averages 4.3\% and ranges from 3.0\% to 5.4\%. After the pandemic, the efficient unemployment rate averages 3.5\% and ranges from 2.9\% to 4.4\%, while the actual unemployment rate averages 4.8\%. Thus, under the baseline framework that treats employment as homogeneous and does not account for part-time employment, the United States generally exhibits a positive unemployment gap. This suggests that labor-market slack remains present even before incorporating part-time employment into the framework.

\subsection{Incorporation of part-time employment}
\subsubsection{Total part-time employment}
\begin{figure}[H]
\centering

\subcaptionbox{Efficient labor-market tightness with part-time employment, 1951--2019}{
\circledgraphfour{0.48\textwidth}{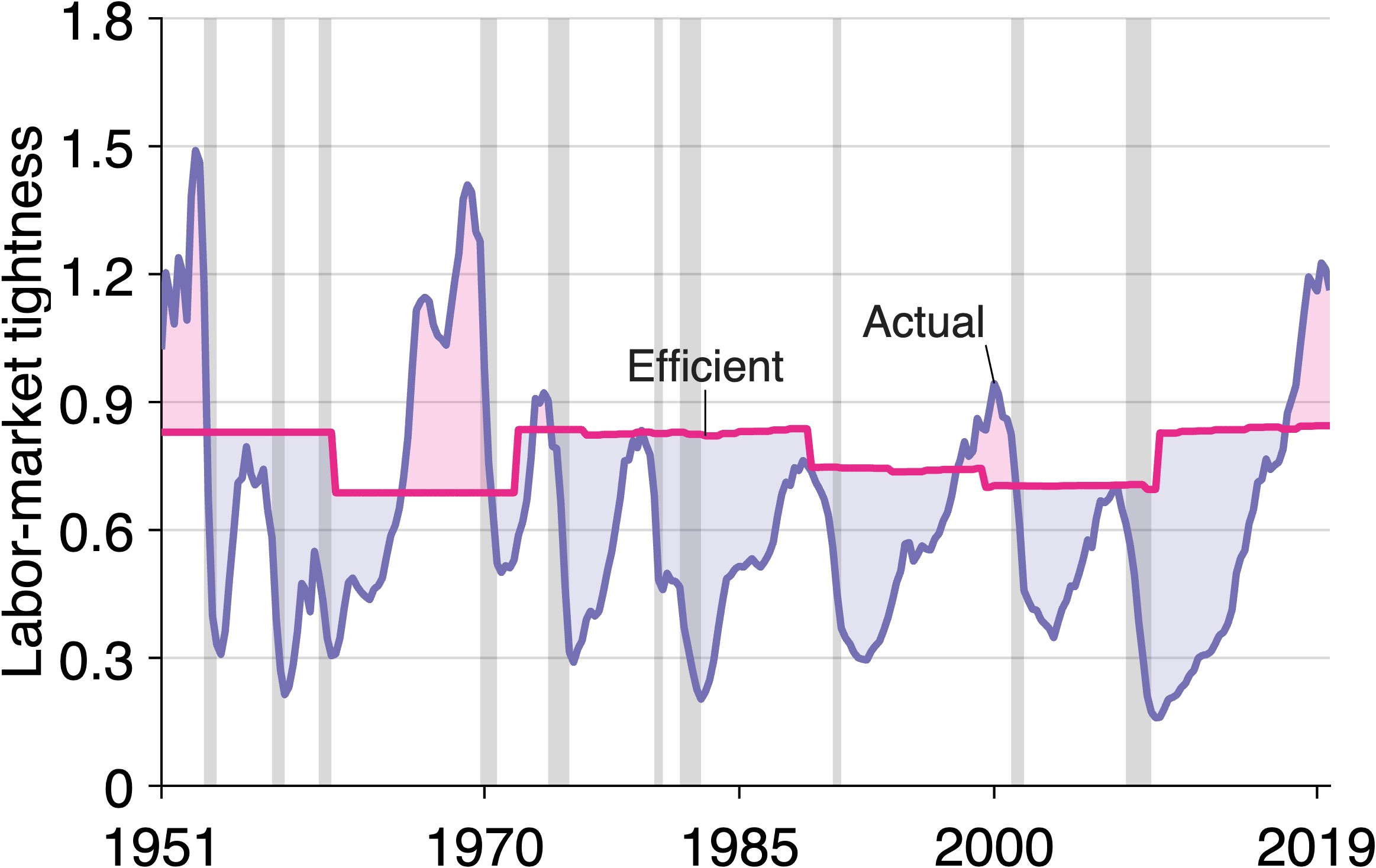}{3,2.5}{0.2}{3.6,2.3}{0.2}{4.5,2.1}{0.2}{6.3,2}}
\hfill
\subcaptionbox{Efficient labor-market tightness with part-time employment, 2020--2026}{
\includegraphics[width=0.48\textwidth]{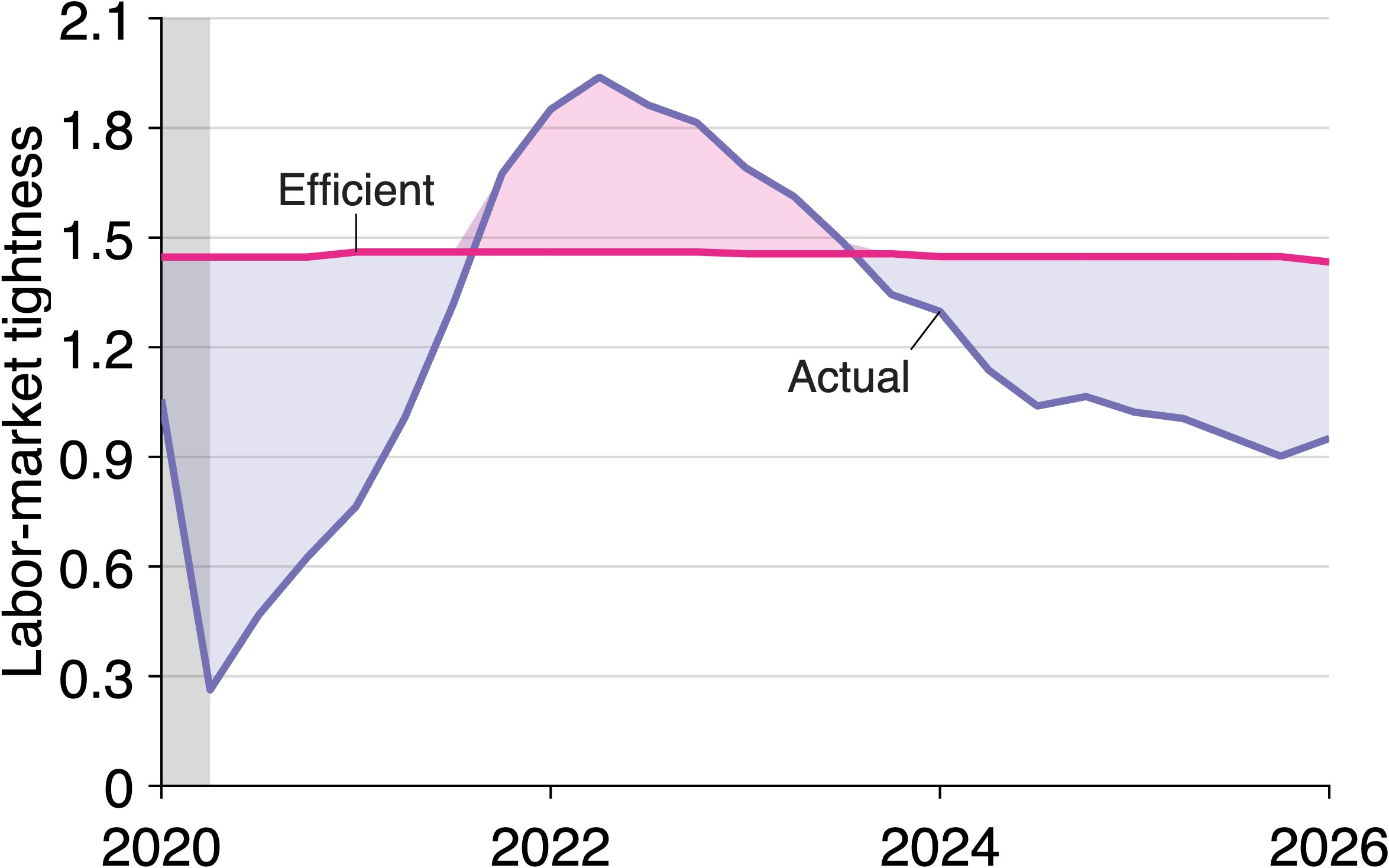}}

\vspace{0.2cm}

\subcaptionbox{Efficient unemployment rate with part-time employment, 1951--2019}{
\includegraphics[width=0.48\textwidth]{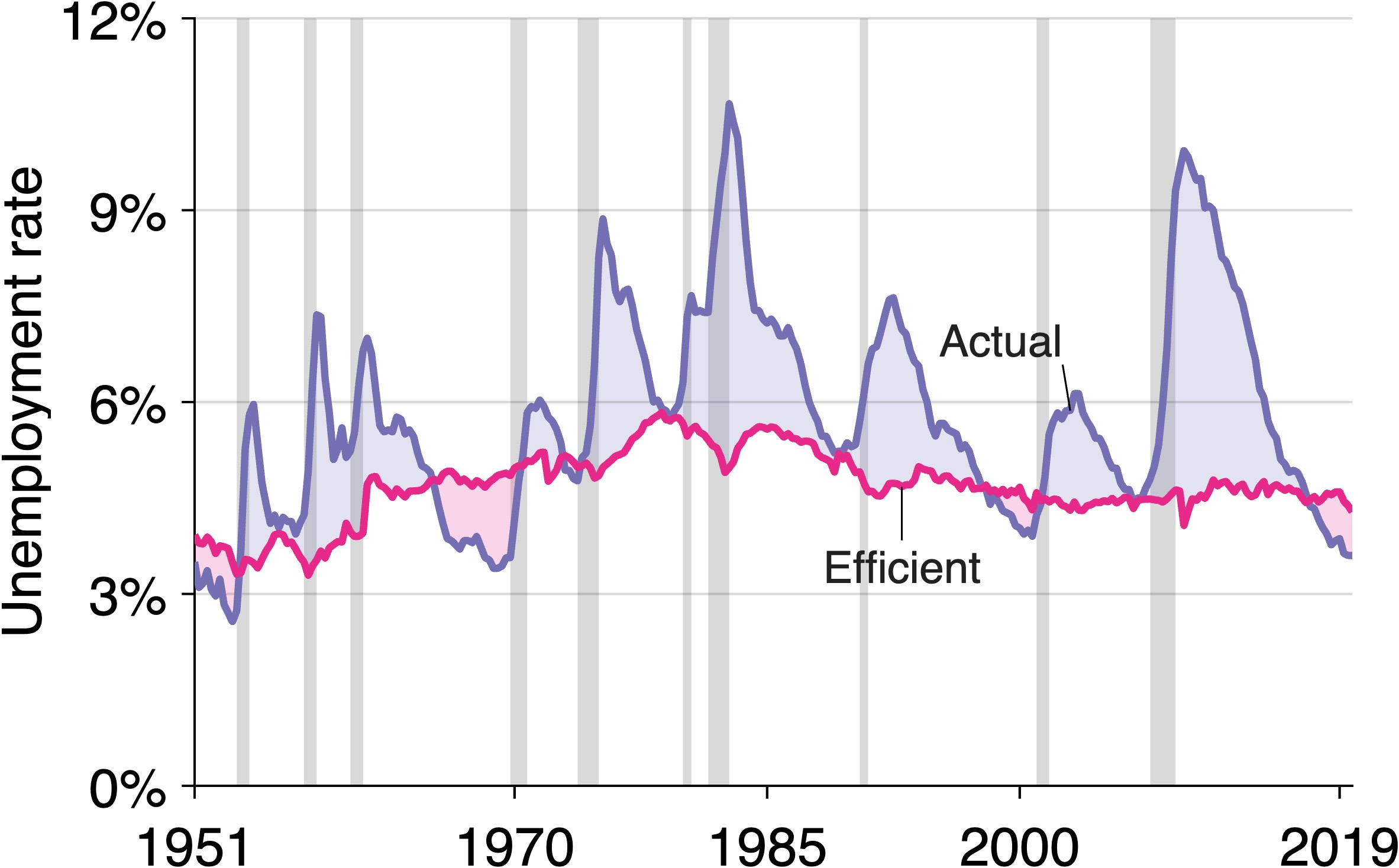}}
\hfill
\subcaptionbox{Efficient unemployment rate with part-time employment, 2020--2026}{
\includegraphics[width=0.48\textwidth]{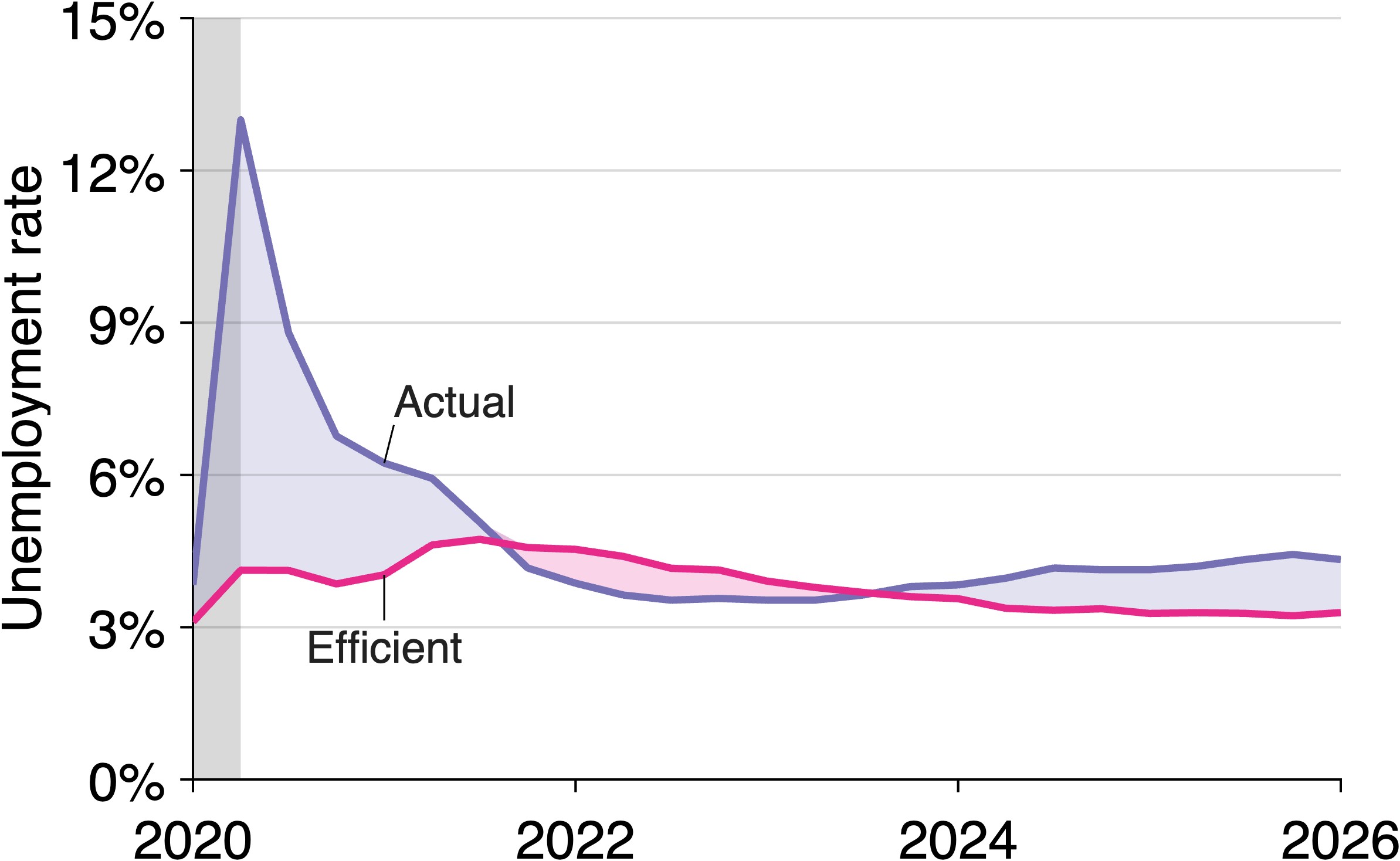}}

\vspace{0.2cm}

\subcaptionbox{Unemployment gap, 1951--2019}{
\includegraphics[width=0.48\textwidth]{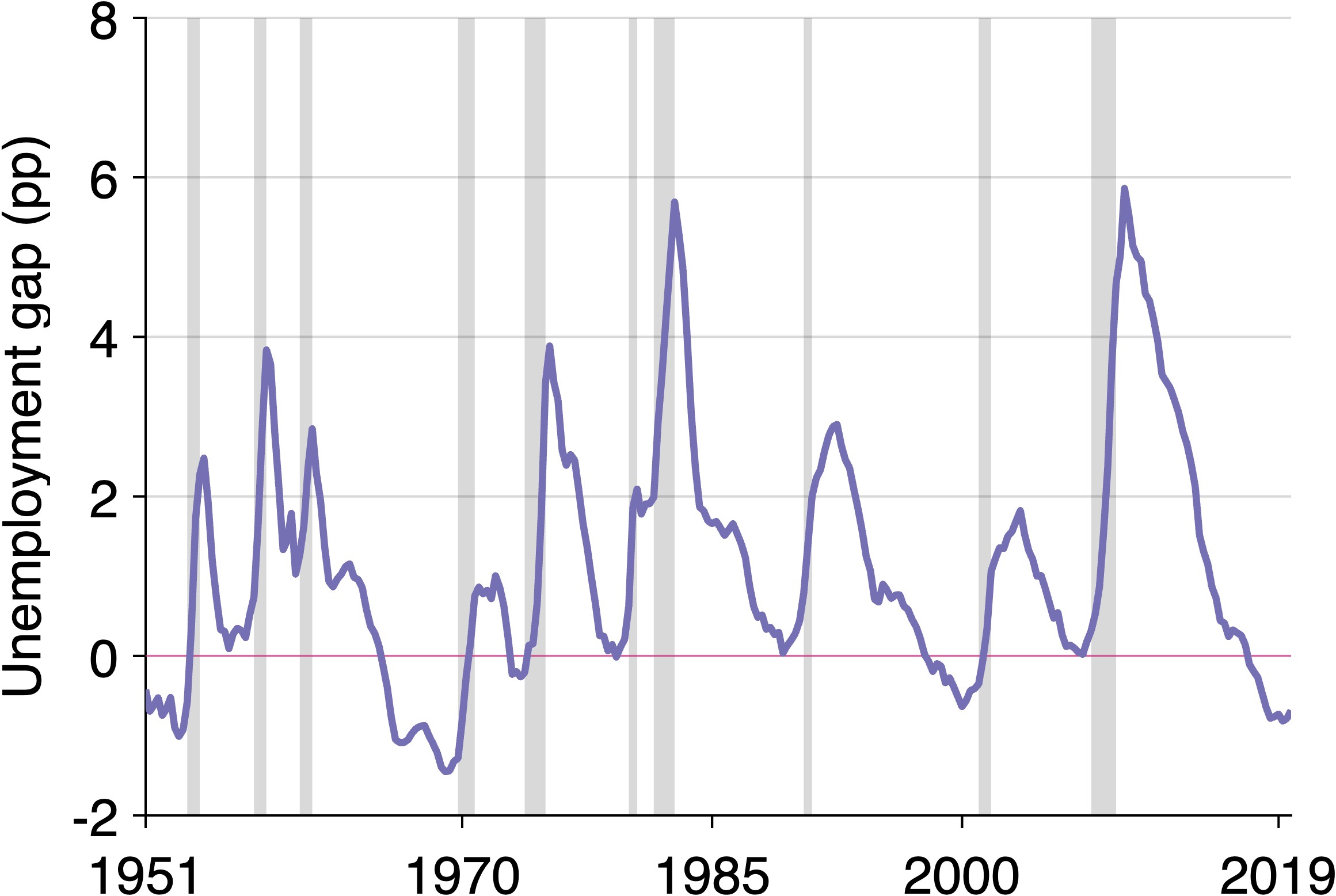}}
\hfill
\subcaptionbox{Unemployment gap, 2020--2026}{
\includegraphics[width=0.48\textwidth]{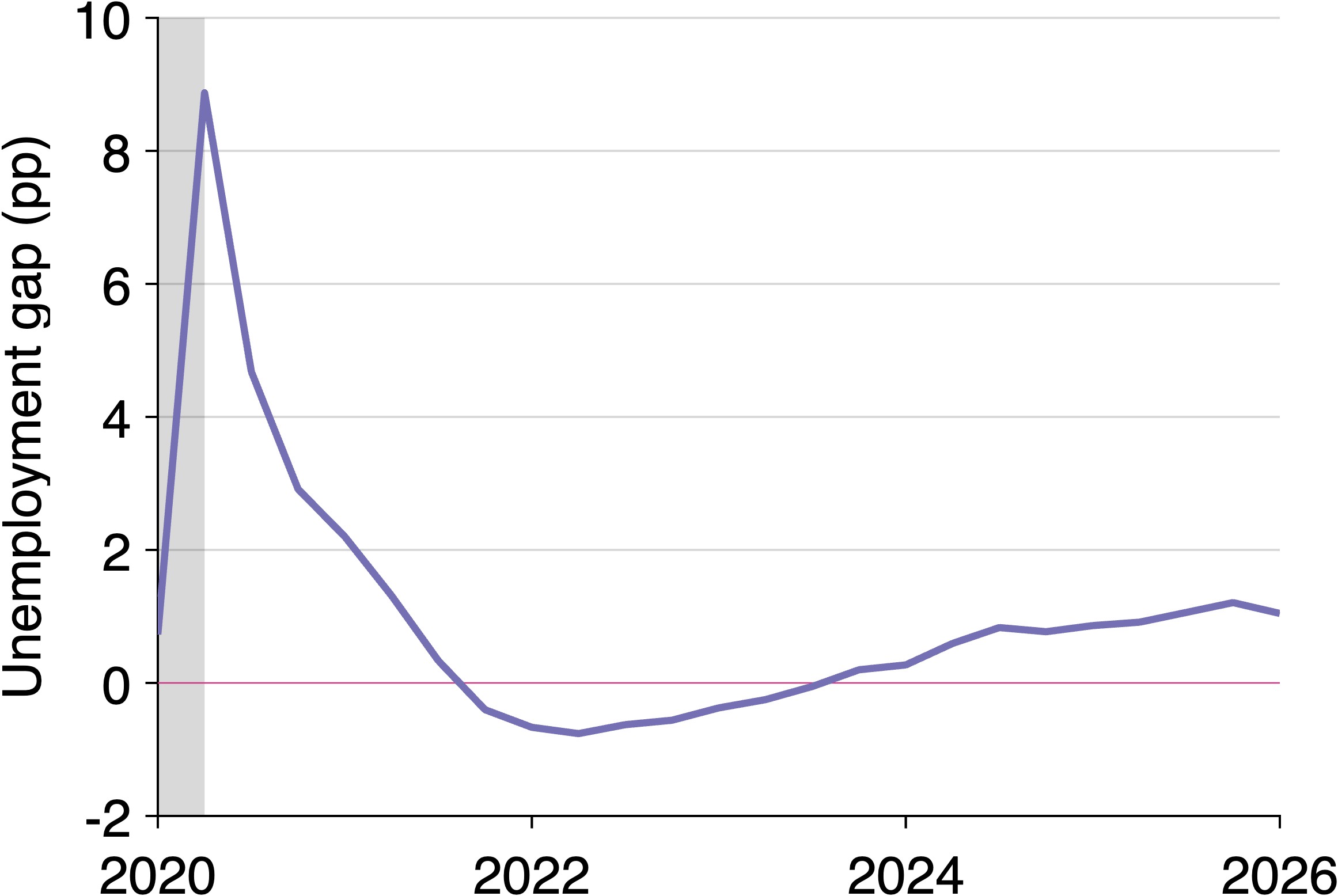}}

\caption{Efficient labor-market tightness, efficient unemployment, and unemployment gap with part-time employment in the United States, 1951--2026. \textit{Notes:} The figure reports efficient labor-market tightness, the efficient unemployment rate, and the unemployment gap under the extended framework with part-time employment. The calibration sets the full-time employment share to $\alpha^{P}=0.75$ and the relative-hours parameter to $\gamma^P=0.47$. The sample is split between the pre-pandemic period, 1951--2019, and the post-pandemic period, 2020--2026.}
\label{f:us_pt}

\end{figure}

I now incorporate total part-time employment using the sufficient statistics estimated in Section~4. Under this extension, the calibration sets the full-time employment share to $\alpha^{P}=0.75$ and the relative-hours parameter to $\gamma^P=0.47$. The resulting measures remain broadly consistent with the full-time baseline, but several meaningful differences emerge. These differences are most visible in Panel A of Figure~\ref{f:us_pt}, where the circled episodes show actual labor-market tightness approaching or exceeding efficient labor-market tightness before several major downturns.

In particular, the extended framework identifies episodes in which actual
labor-market tightness approaches or reaches the efficient allocation before several
major recessions: the 1973--75 recession associated with the first oil shock, the early-1980s recessions associated with the Volcker disinflation period, the 1990--91
recession associated with the oil-price shock and credit tightening, and the
2007--09 Great Recession associated with the global financial crisis. In these
episodes, accounting for part-time employment lowers the efficient-tightness
benchmark, so that actual tightness converges toward---and in some cases
effectively meets---the efficient allocation. Under the baseline framework that
treats all employment as full-time, the same episodes instead appear inefficiently
slack. The part-time extension therefore suggests that labor-market conditions were closer to the efficient allocation immediately before these downturns than implied by the full-time baseline.

Before COVID, actual labor-market tightness averages 0.62, while efficient tightness
under the part-time calibration averages 0.78. After COVID, actual tightness averages
1.21, while efficient tightness averages 1.43. Thus, the extended framework continues
to indicate a generally slack labor market on average, but it also reveals several
episodes in which actual tightness approaches the efficient level.

The corresponding efficient unemployment rate also shifts under the part-time calibration. Before COVID, actual unemployment averages 5.77\%, while efficient unemployment averages 4.68\%, with a range from 3.30\% to 5.81\%. After COVID, actual unemployment averages 4.82\%, compared with an efficient unemployment rate of 3.86\%, which ranges from 3.13\% to 4.80\%. These estimates imply a positive unemployment gap on average, but the gap narrows substantially during the highlighted pre-recession episodes.

\begin{figure}[H]
\centering

\subcaptionbox{Efficient labor-market tightness comparison, 1951--2019}{
\includegraphics[width=0.48\textwidth]{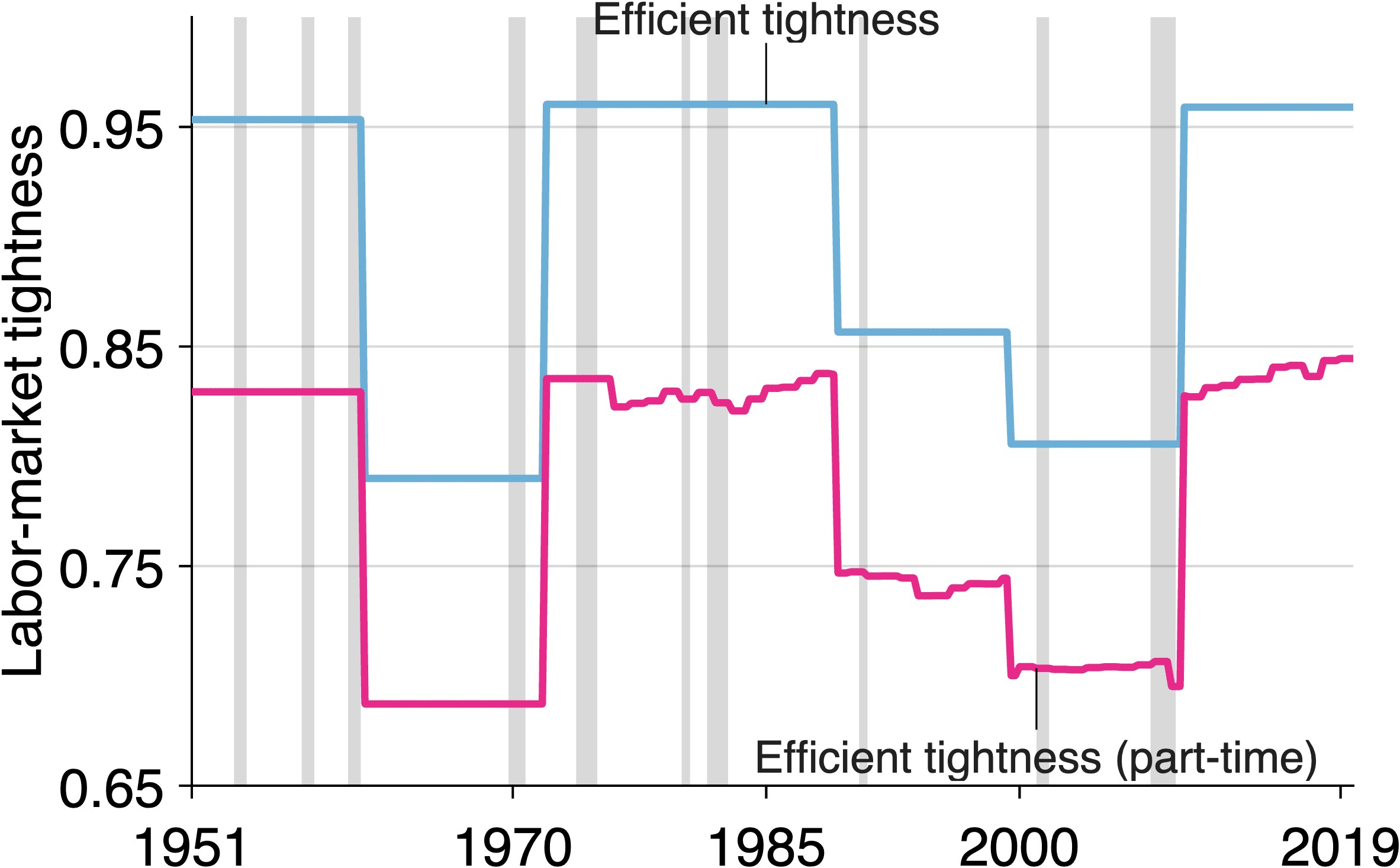}}
\hfill
\subcaptionbox{Efficient labor-market tightness comparison, 2020--2026}{
\includegraphics[width=0.48\textwidth]{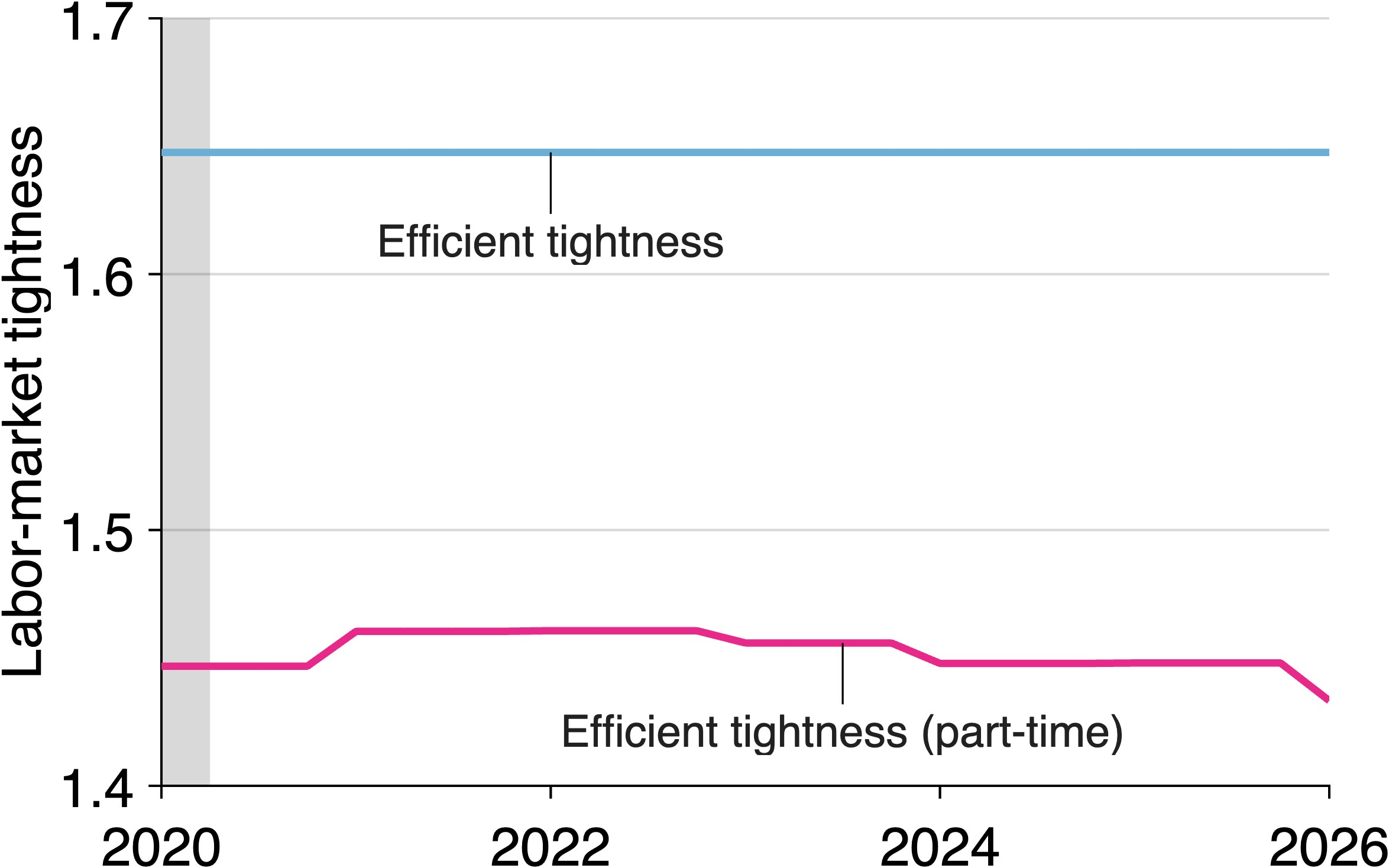}}

\vspace{0.1cm}

\subcaptionbox{Efficient unemployment rate comparison, 1951--2019}{
\includegraphics[width=0.48\textwidth]{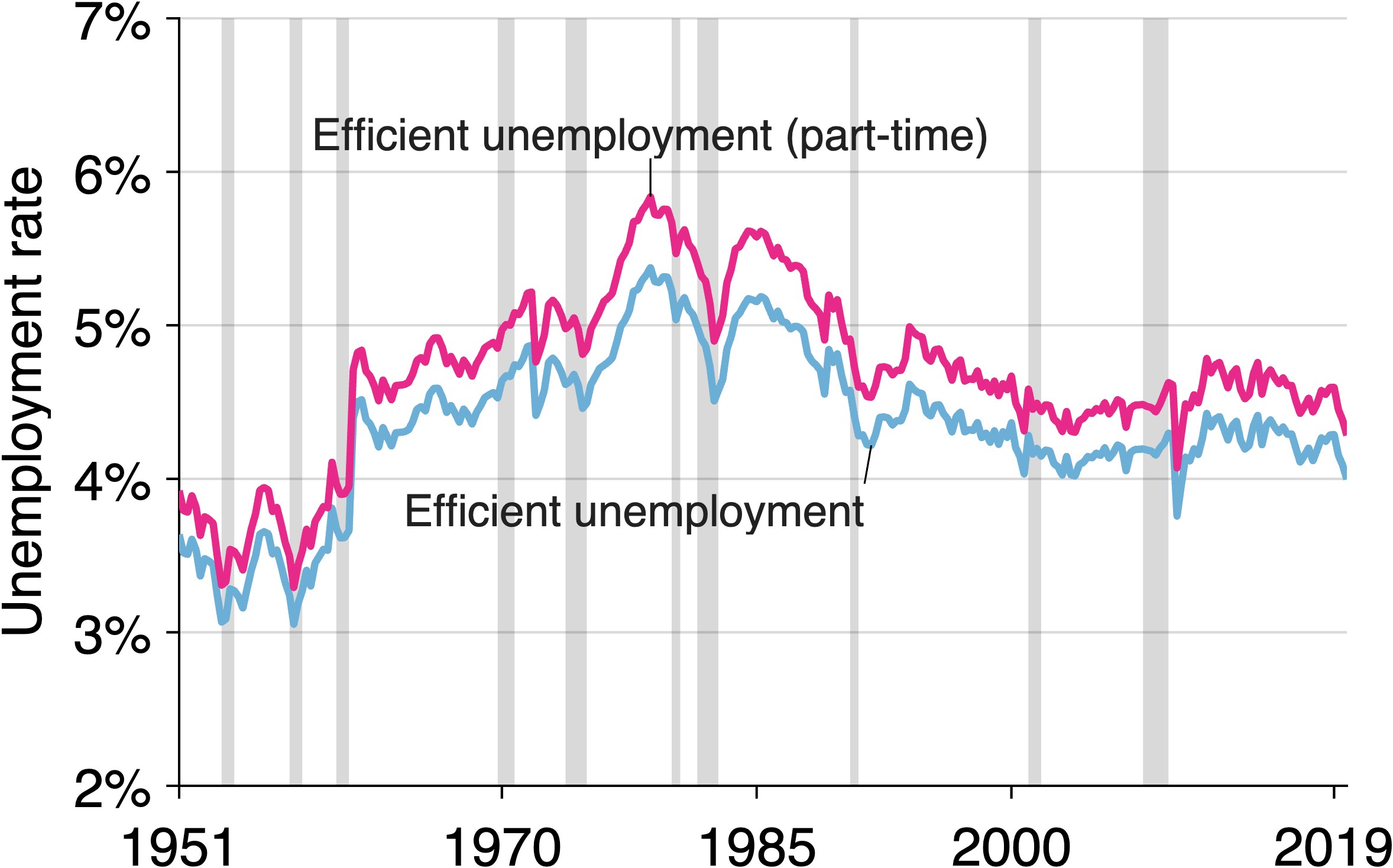}}
\hfill
\subcaptionbox{Efficient unemployment rate comparison, 2020--2026}{
\includegraphics[width=0.48\textwidth]{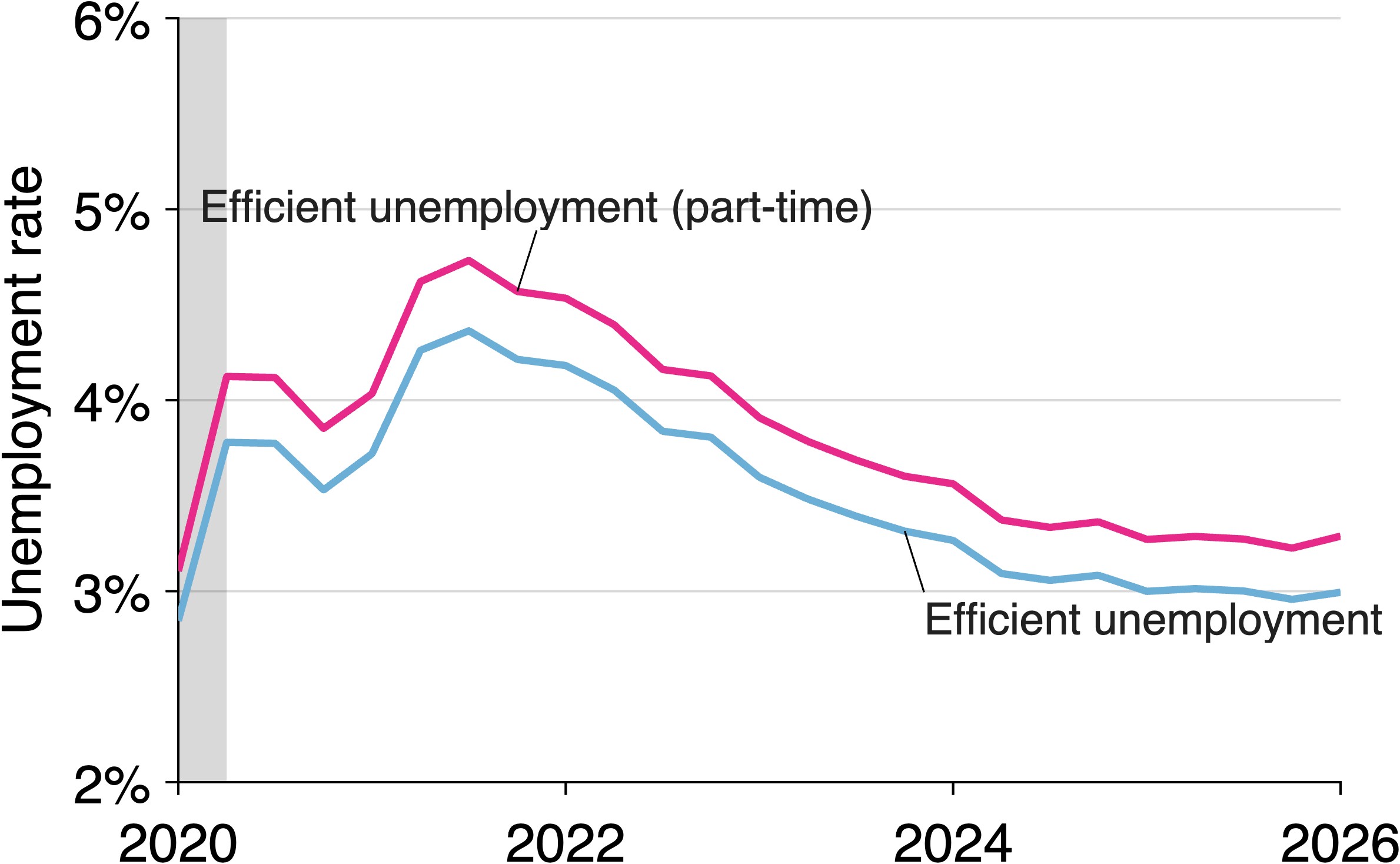}}

\vspace{0.1cm}

\subcaptionbox{Unemployment gap comparison, 1951--2019}{
\includegraphics[width=0.48\textwidth]{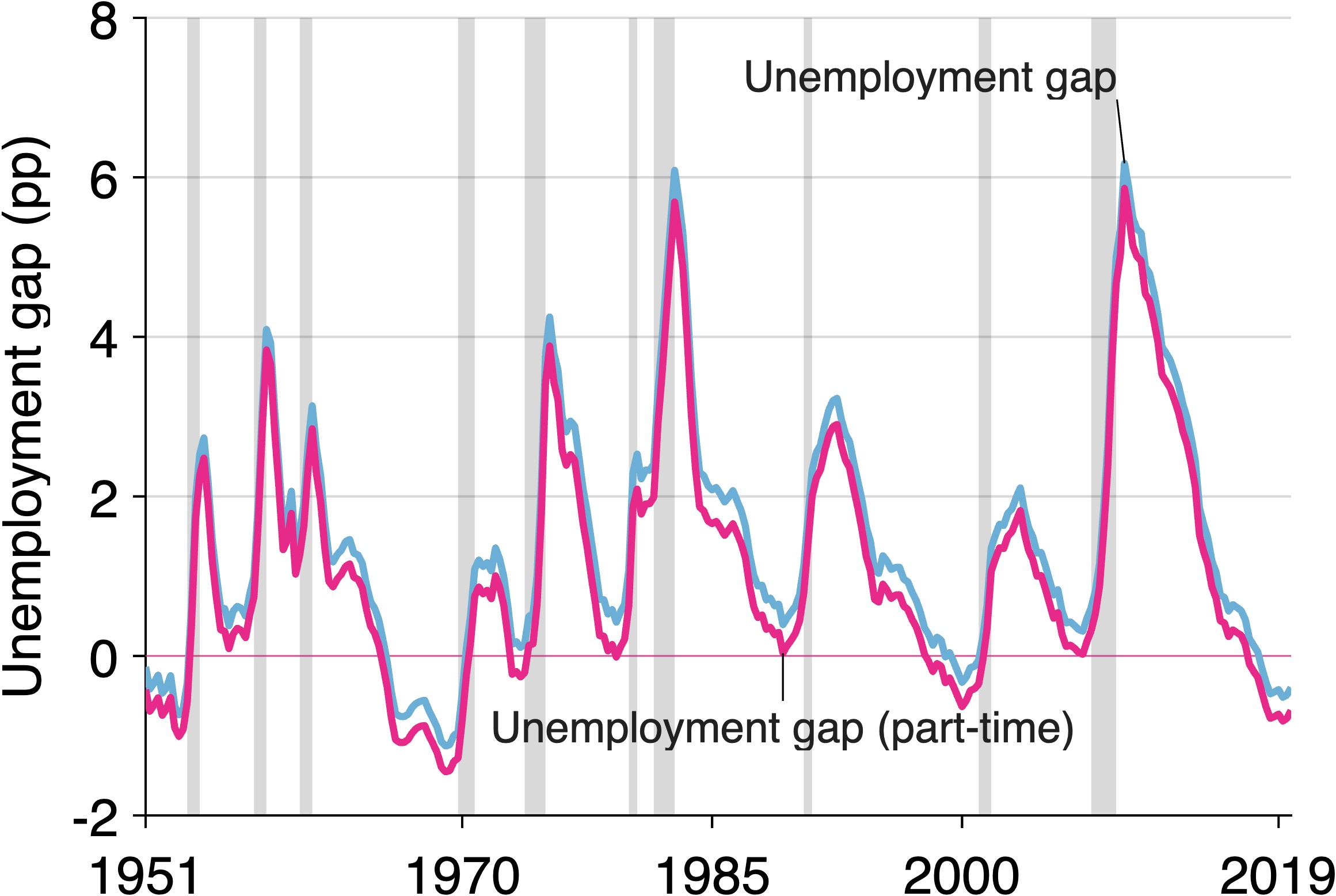}}
\hfill
\subcaptionbox{Unemployment gap comparison, 2020--2026}{
\includegraphics[width=0.48\textwidth]{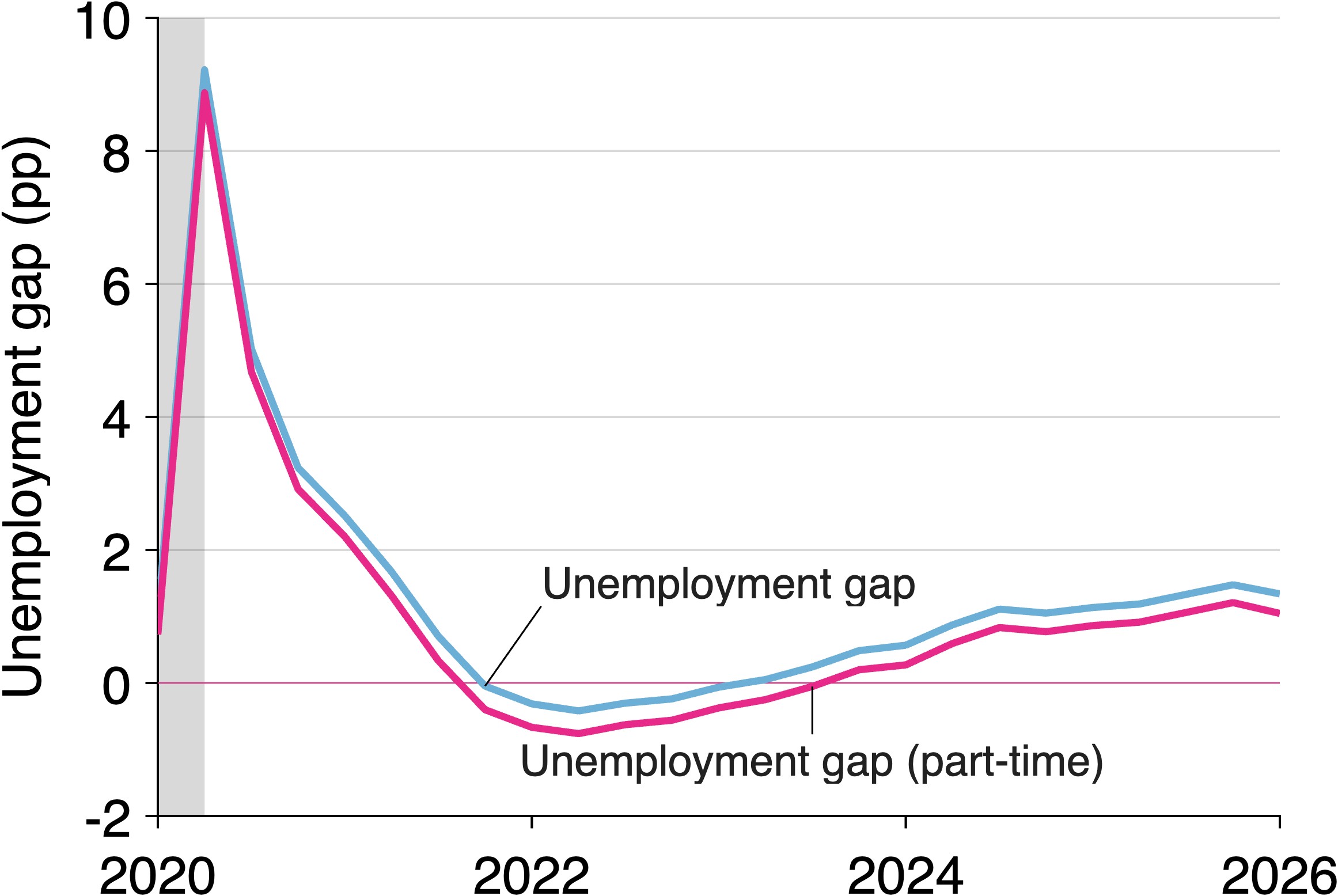}}

\caption{Comparison of efficient labor-market tightness, efficient unemployment, and unemployment gaps with and without total part-time employment in the United States. \textit{Notes:} The figure compares the baseline framework of \citet{MS21b} and the extended framework that incorporates total part-time employment.}
\label{f:graph3}

\end{figure}

Figure~\ref{f:graph3} makes this comparison explicit by plotting the baseline results against the part-time-employment extension. Although the quantitative differences are modest, they are systematic. Incorporating part-time employment shifts the efficient benchmark by lowering efficient labor-market tightness and raising efficient unemployment. Consequently, the distance between actual and efficient labor-market conditions narrows during several pre-recession episodes, including those preceding the 1973–75, 1980, 1990–91, and 2007–09 recessions. This pattern suggests that labor-market data often contain information about the buildup of cyclical pressures before downturns. The finding is consistent with \citet{MS_recession_started}, who show that labor-market indicators provide useful information for assessing recession risk and identifying turning points in the business cycle.

\subsubsection{Incorporation of involuntary part-time} 

I next focus on involuntary part-time employment, defined as employment part time for economic reasons. This exercise is not a full decomposition of employment because voluntary part-time workers are excluded from the construction of $\alpha^{IP}$. Instead, it isolates the underemployment margin by asking how the efficient-unemployment benchmark changes when the part-time margin is restricted to workers who would prefer additional hours.

Under this definition, involuntary part-time employment represents a much smaller share of employment than total part-time employment. The calibration therefore sets the full-time employment share to $\alpha^{IP}=0.93$ and the relative-hours parameter to $\gamma^{IP}=0.56$.

\begin{figure}[H]
\centering

\subcaptionbox{Efficient labor-market tightness with involuntary part-time, 1951--2019}{
\includegraphics[width=0.48\textwidth]{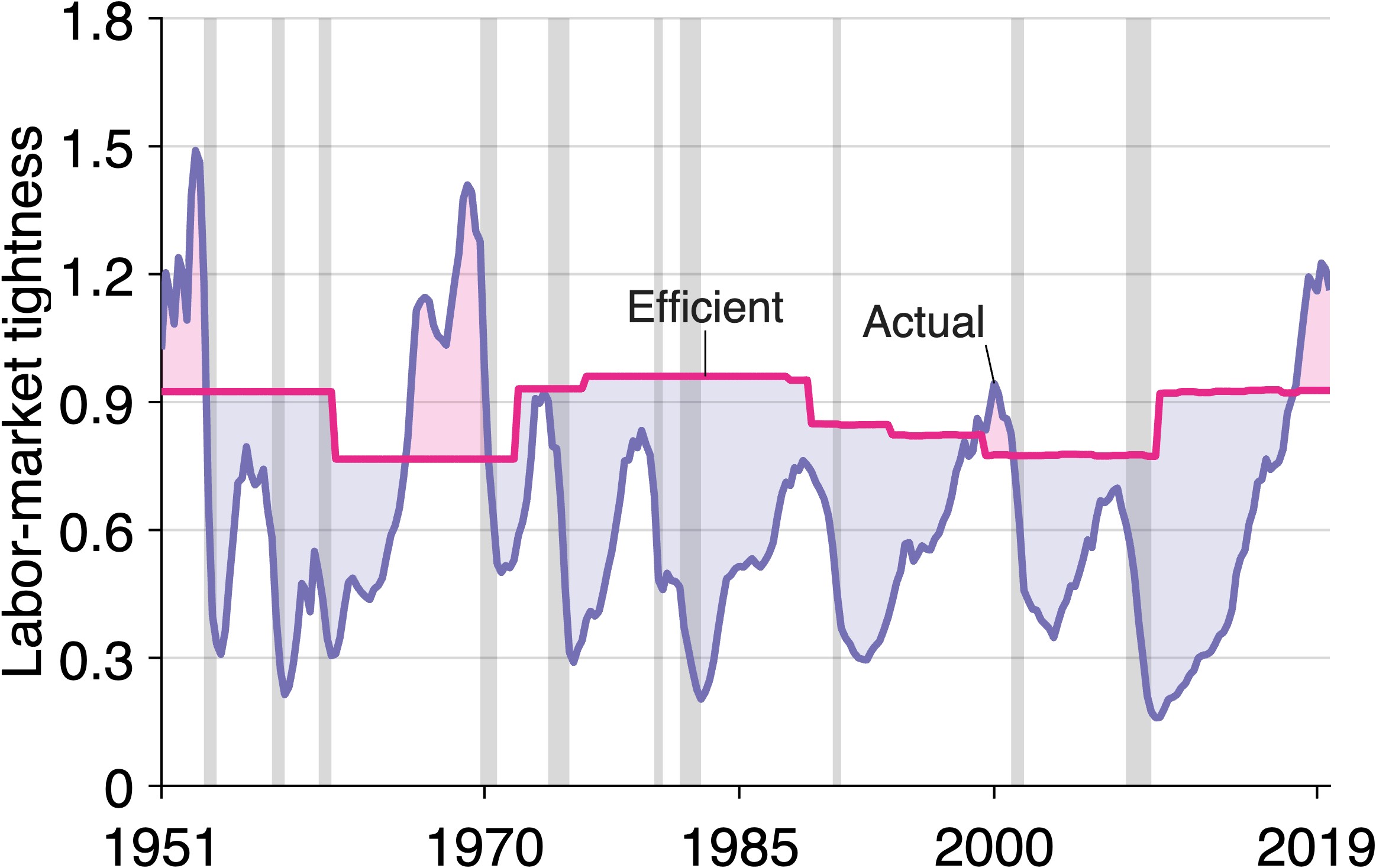}}
\hfill
\subcaptionbox{Efficient labor-market tightness with involuntary part-time, 2020--2026}{
\includegraphics[width=0.48\textwidth]{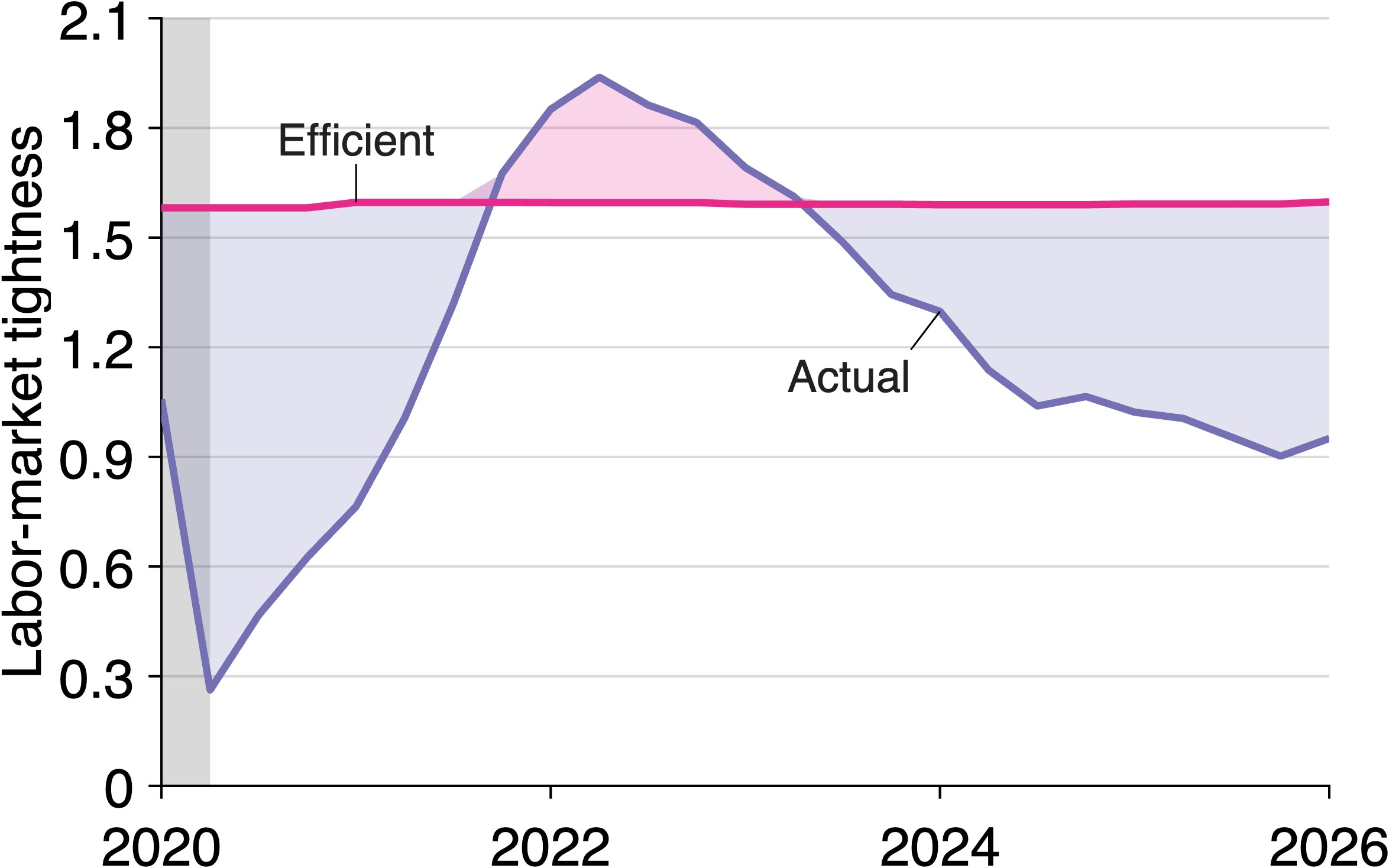}}

\vspace{0.2cm}

\subcaptionbox{Efficient unemployment rate with involuntary part-time, 1951--2019}{
\includegraphics[width=0.48\textwidth]{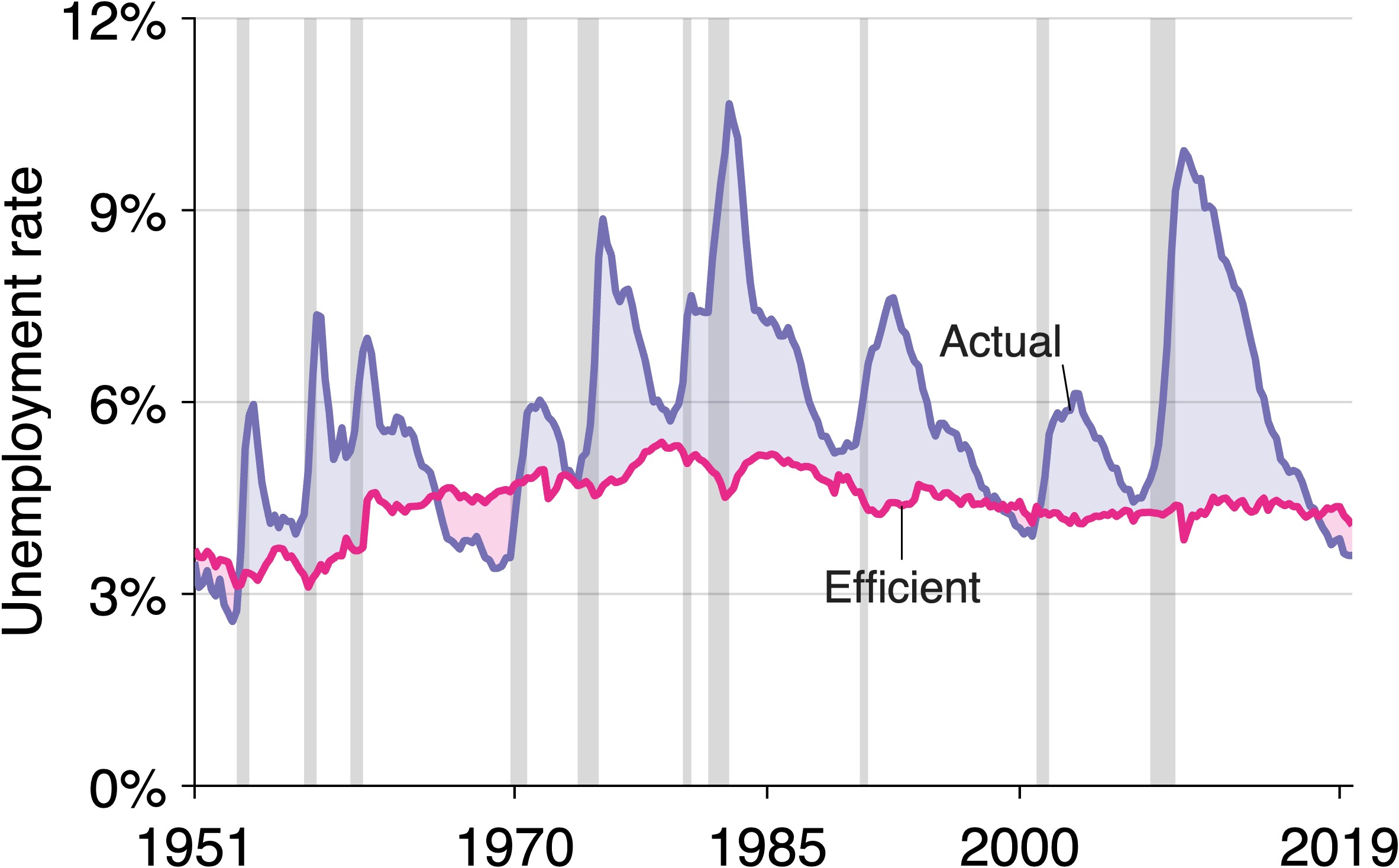}}
\hfill
\subcaptionbox{Efficient unemployment rate with involuntary part-time, 2020--2026}{
\includegraphics[width=0.48\textwidth]{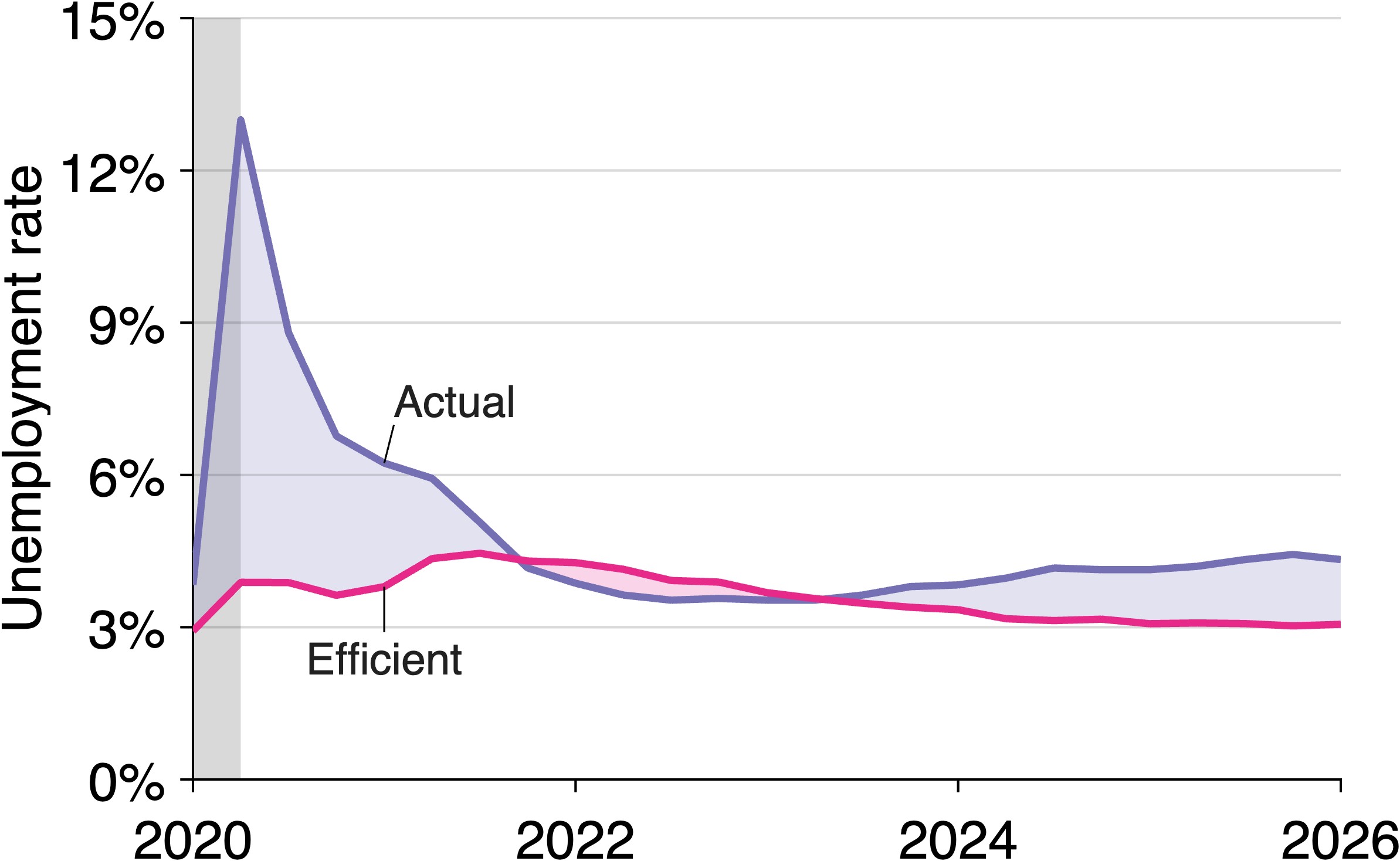}}

\vspace{0.2cm}

\subcaptionbox{Unemployment gap with involuntary part-time, 1951--2019}{
\includegraphics[width=0.48\textwidth]{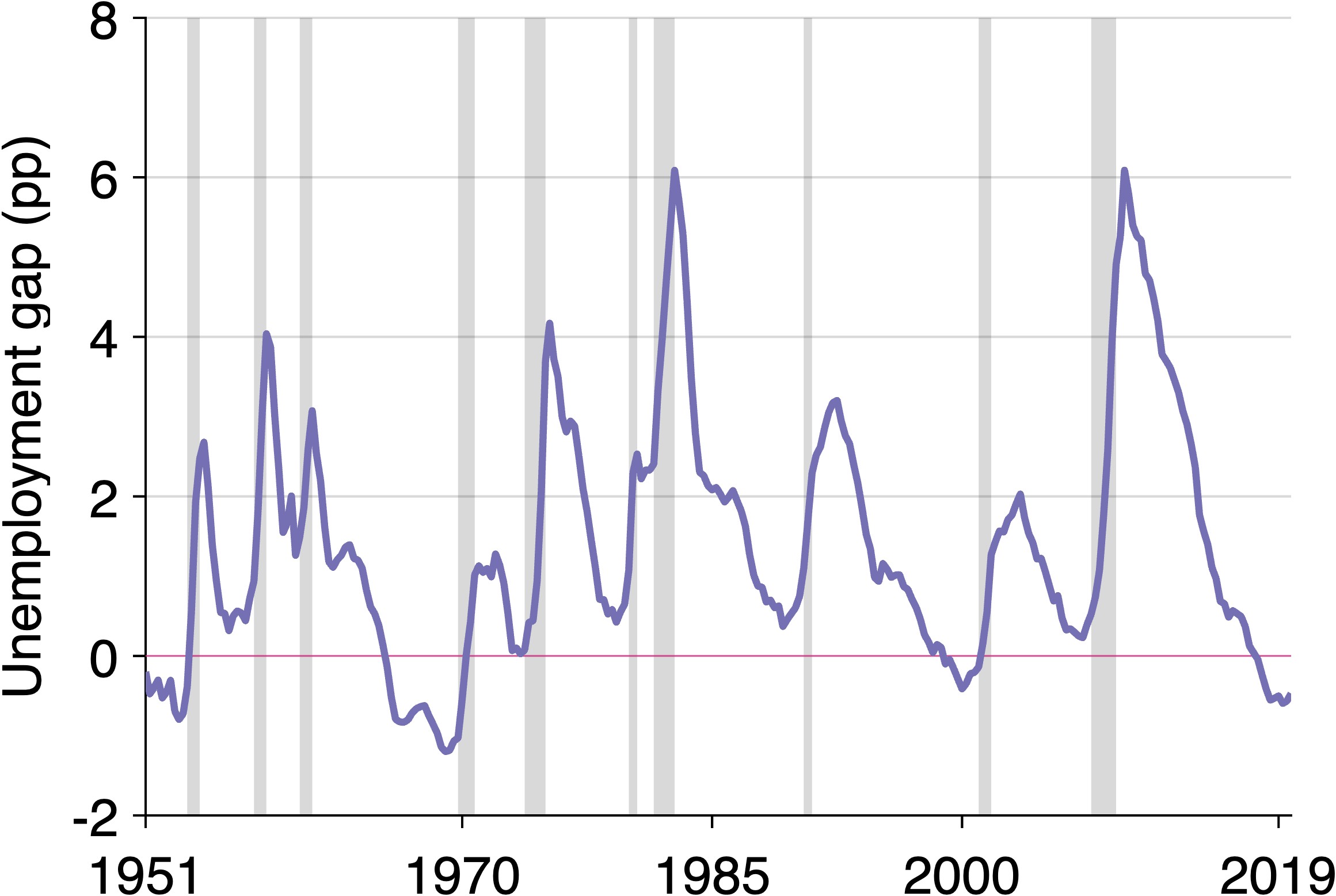}}
\hfill
\subcaptionbox{Unemployment gap with involuntary part-time, 2020--2026}{
\includegraphics[width=0.48\textwidth]{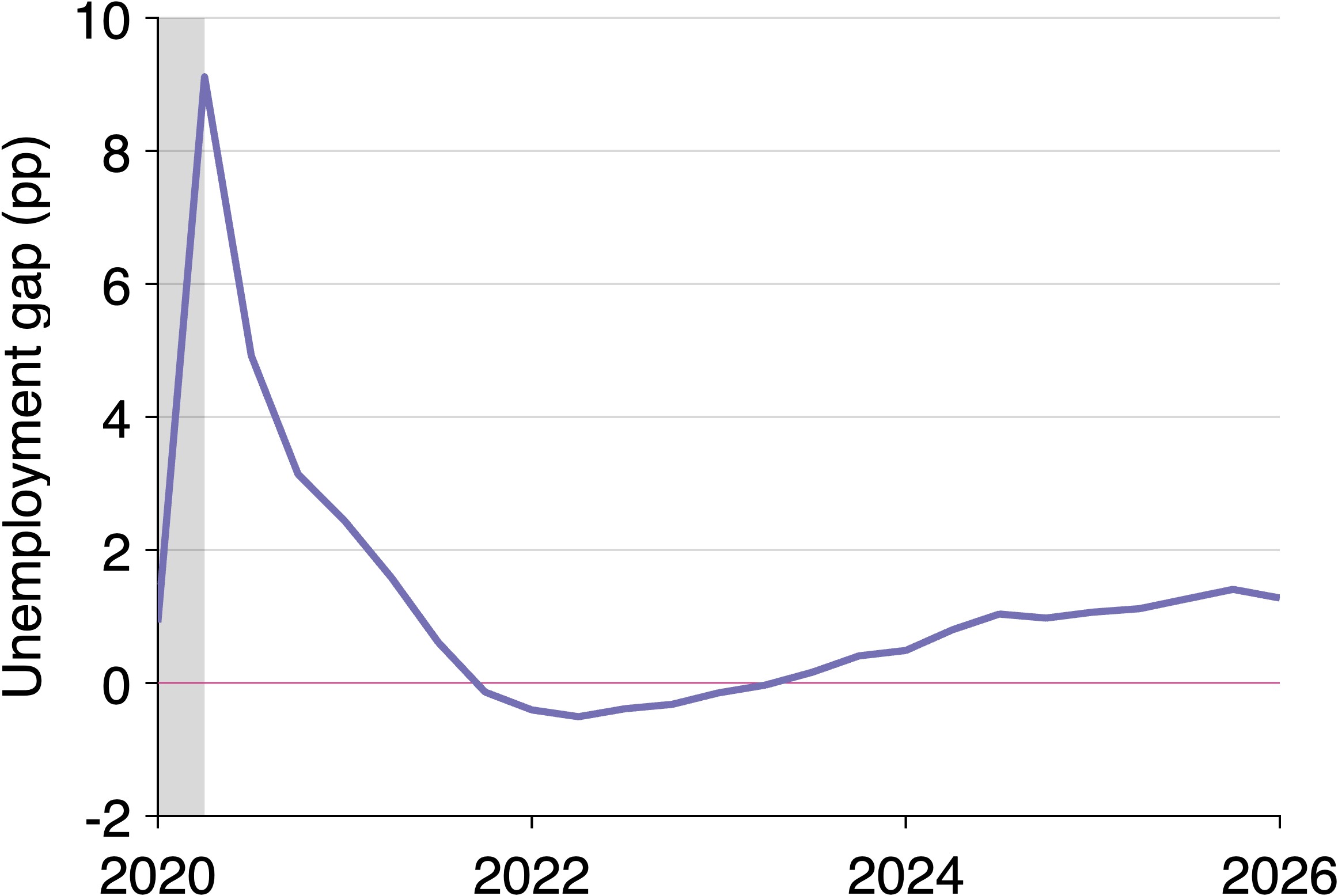}}

\caption{Efficient labor-market tightness, efficient unemployment, and unemployment gap in the United States. \textit{Notes:} The figure reports efficient labor-market tightness, the efficient unemployment rate, and the unemployment gap under the extended framework with involuntary part-time employment. The sample is split between the pre-pandemic period, 1951--2019, and the post-pandemic period, 2020--2026.}
\label{f:us_ipt}

\end{figure}

Figure~\ref{f:us_ipt} reports efficient labor-market tightness, efficient unemployment, and the unemployment gap under the involuntary part-time extension. Before COVID, actual labor-market tightness averages 0.620, while efficient tightness averages 0.867. After COVID, actual tightness averages 1.206, compared with an efficient level of 1.597. Efficient tightness ranges from 0.766 to 0.951 before COVID. In the post-pandemic sample, efficient tightness fluctuates only slightly around 1.597 because the Beveridge elasticity is estimated over a single regime and the remaining sufficient statistics exhibit limited variation.

The corresponding efficient unemployment rate averages 4.42 percent before COVID and 3.58 percent after COVID, compared with actual unemployment rates of 5.77 percent and 4.82 percent, respectively. Before COVID, efficient unemployment ranges from 3.10 percent to 5.47 percent, while after COVID it ranges from 2.91 percent to 4.46 percent. These results suggest that incorporating involuntary part-time employment shifts the efficient benchmark but does not substantially alter the qualitative behavior of the unemployment gap.

\begin{figure}[H]
\centering

\subcaptionbox{Efficient labor-market tightness comparison, 1951--2019}{
\includegraphics[width=0.48\textwidth]{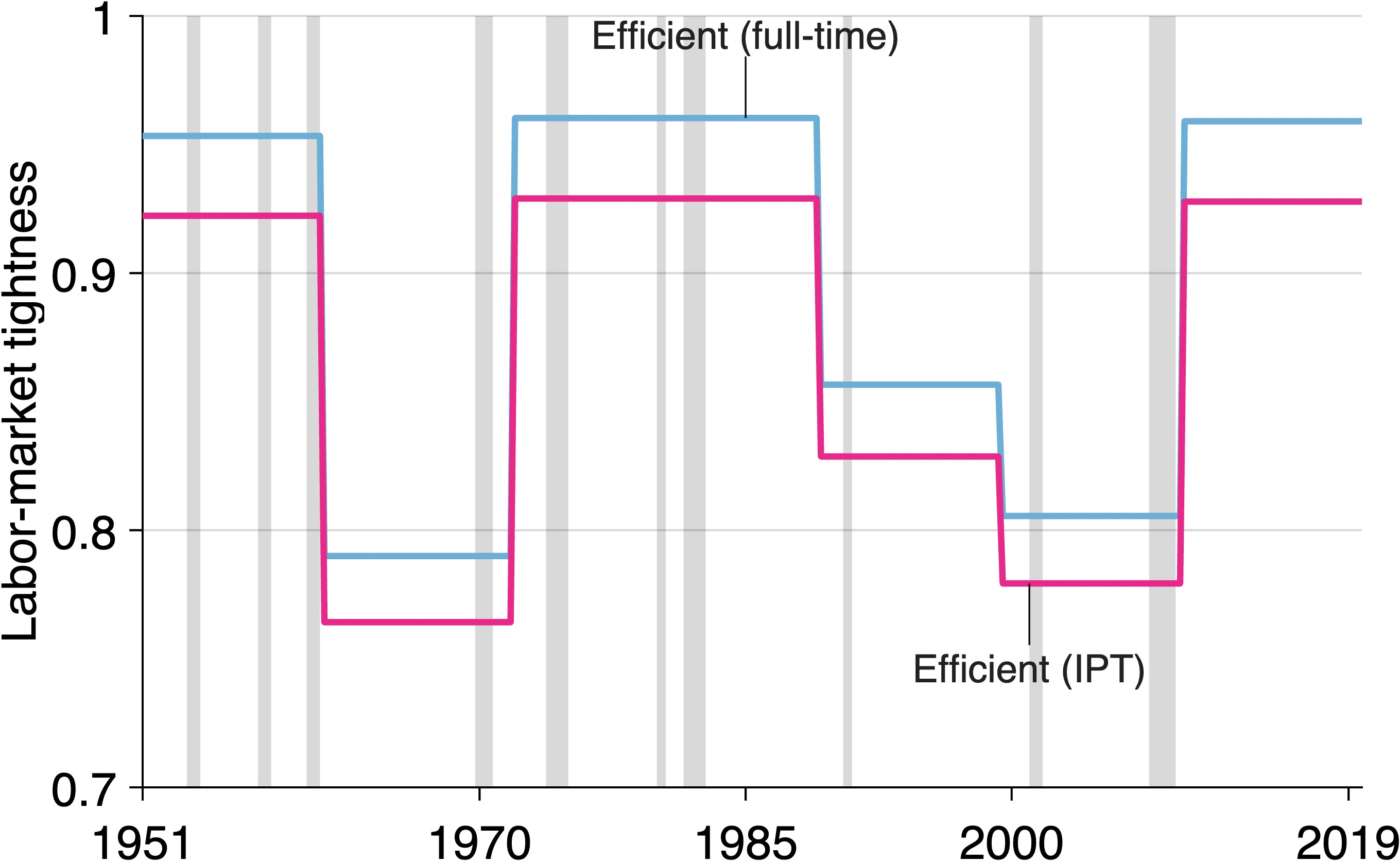}}
\hfill
\subcaptionbox{Efficient labor-market tightness comparison, 2020--2026}{
\includegraphics[width=0.48\textwidth]{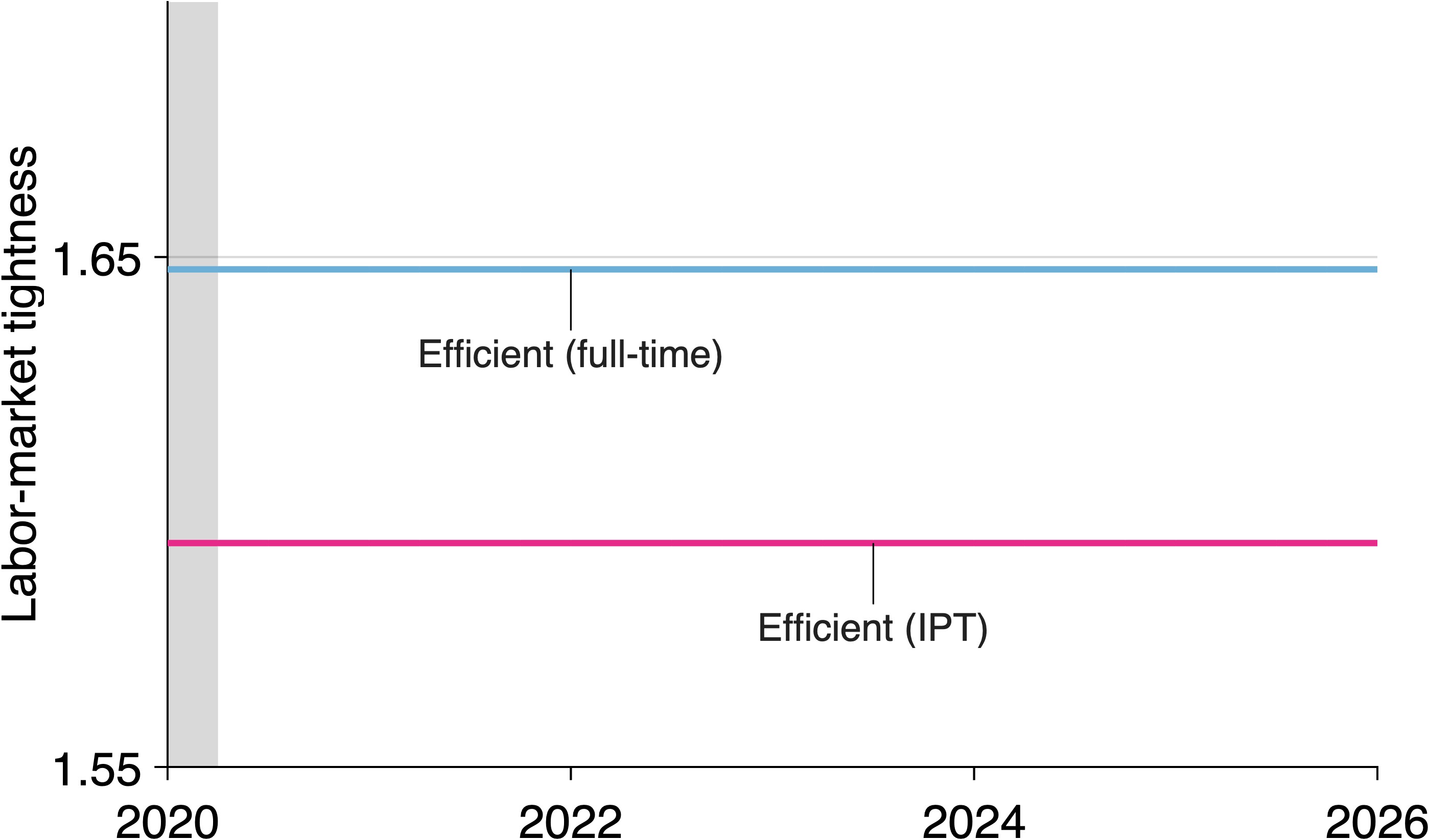}}

\vspace{0.1cm}

\subcaptionbox{Efficient unemployment rate comparison, 1951--2019}{
\includegraphics[width=0.48\textwidth]{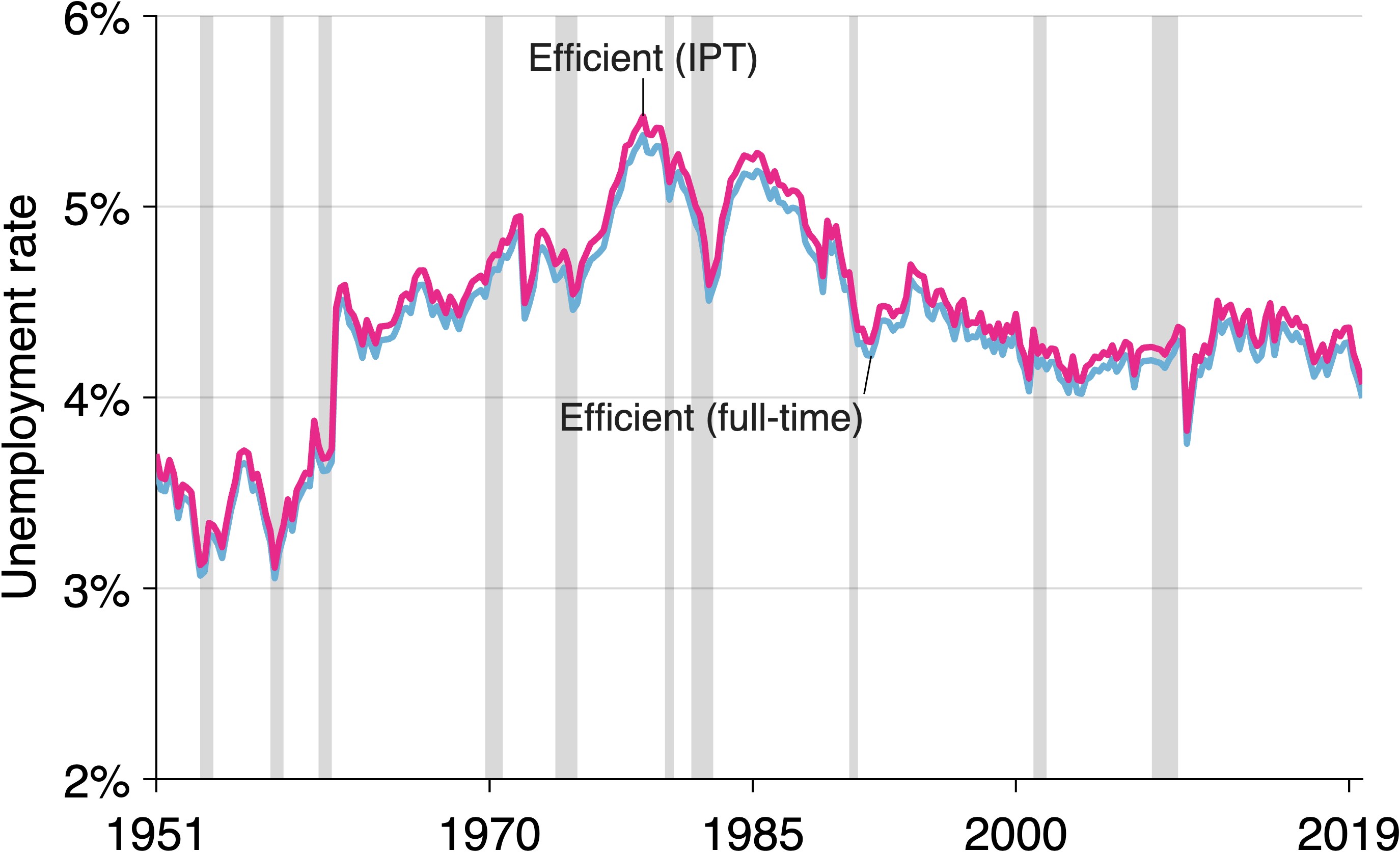}}
\hfill
\subcaptionbox{Efficient unemployment rate comparison, 2020--2026}{
\includegraphics[width=0.48\textwidth]{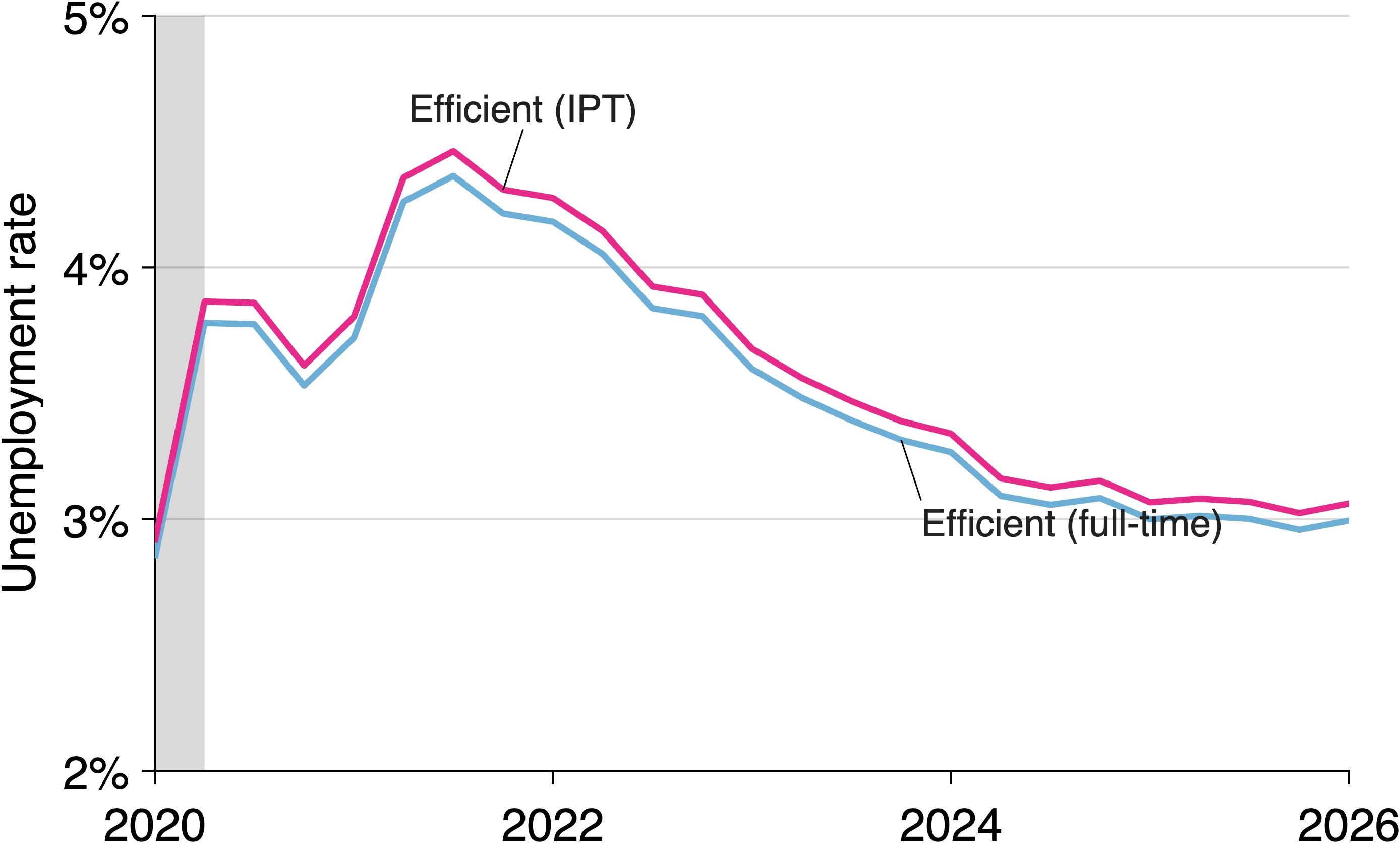}}

\vspace{0.1cm}

\subcaptionbox{Unemployment gap comparison, 1951--2019}{
\includegraphics[width=0.48\textwidth]{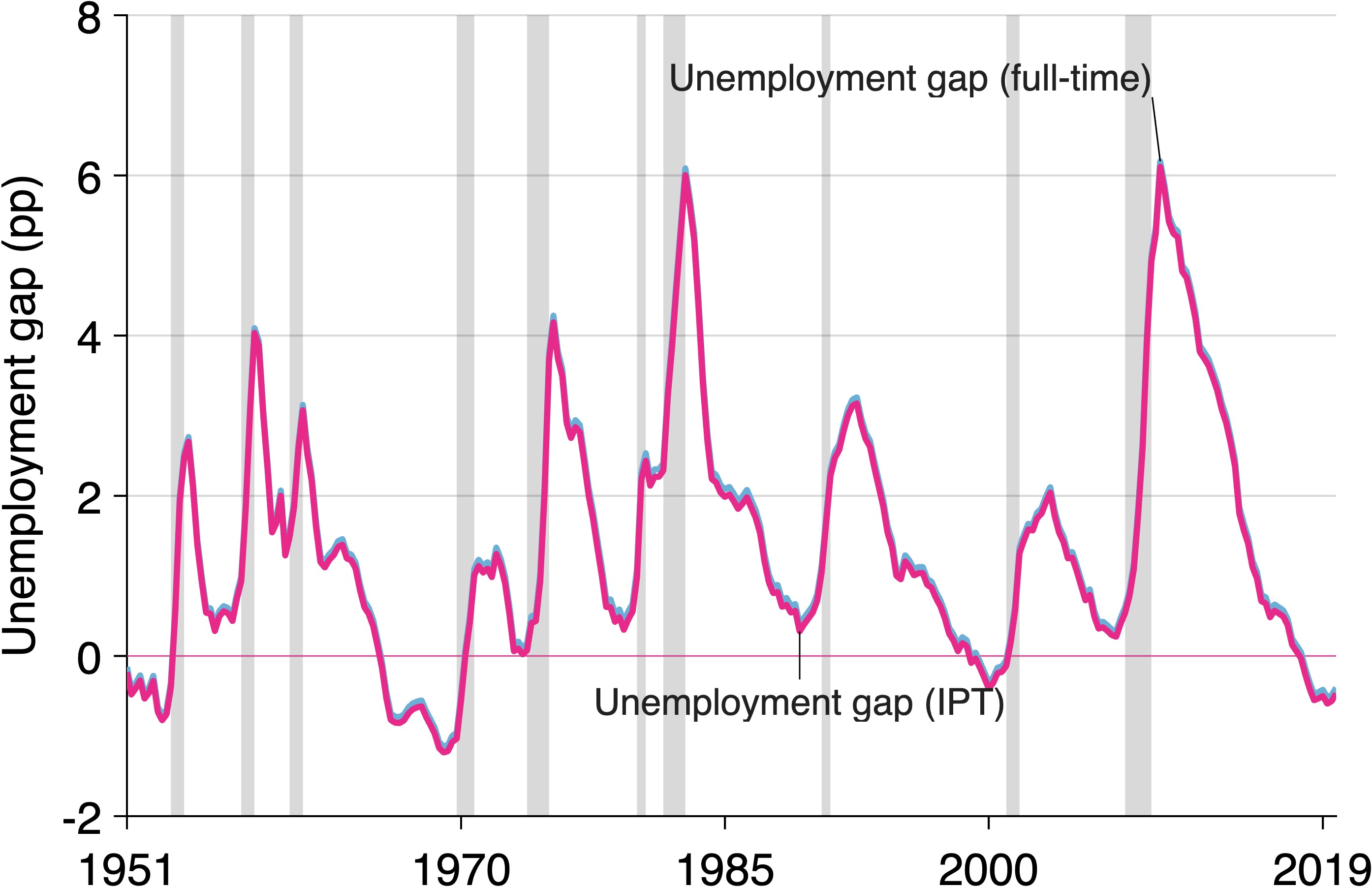}}
\hfill
\subcaptionbox{Unemployment gap comparison, 2020--2026}{
\includegraphics[width=0.48\textwidth]{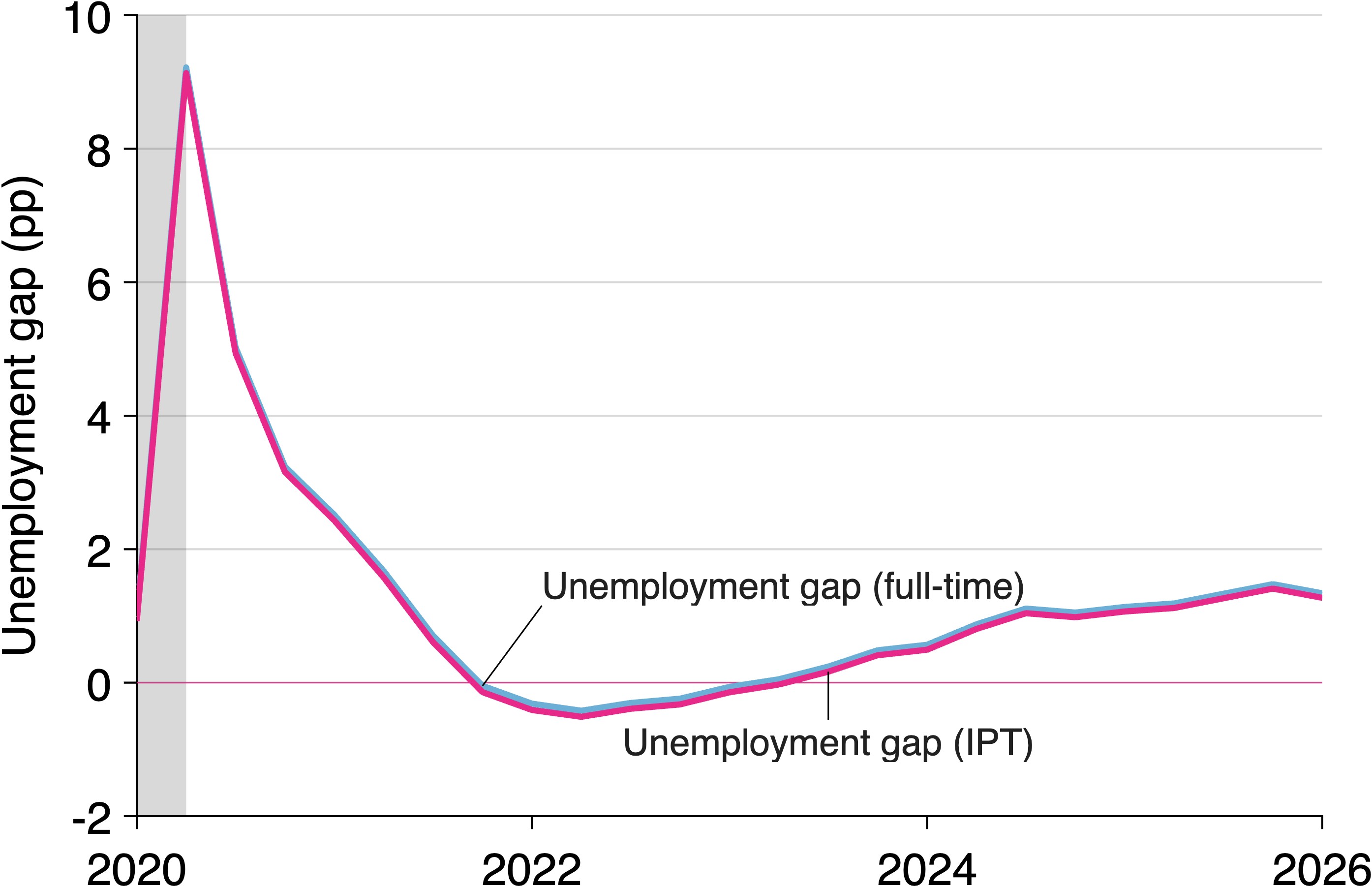}}

\caption{Comparison of efficient labor-market tightness, efficient unemployment, and unemployment gaps with and without involuntary part-time employment in the United States. \textit{Notes:} The figure compares the baseline framework of \citet{MS21b} and the extended framework that incorporates involuntary part-time employment.}
\label{f:us_ipt_compare}

\end{figure}

Figure~\ref{f:us_ipt_compare} compares these results with the full-time baseline. The involuntary part-time extension produces estimates that are close to the baseline because workers employed part time for economic reasons account for only a small fraction of employment in the United States. Consequently, incorporating involuntary part-time employment has only a modest effect on measured labor-market efficiency. Nevertheless, this group remains economically important because it captures workers who would prefer to work more hours but are unable to do so, making it a useful indicator of labor-market slack.

\subsection{Beveridgean unemployment gap in the United States, 1951--2019, 2020--2026}

Using the estimated Beveridge elasticity, social value of nonwork, recruiting cost, employment shares, and relative working hours, I compute efficient unemployment for the United States under both total part-time and involuntary part-time employment definitions. I also estimate the framework over the full 1951--2026 sample with five structural breaks. Because the Beveridge curve becomes unusually flat during the COVID-19 period, full-sample estimates provide limited information about underlying labor-market efficiency. I therefore focus on separate pre-COVID and post-COVID specifications as the main analysis, while reporting the full-sample results in Appendix A for completeness.

Under the total part-time calibration, efficient labor-market tightness averages 0.78 before COVID and 1.43 after COVID, compared with actual tightness of 0.62 and 1.21. The corresponding efficient unemployment rate averages 4.68\% before COVID and 3.86\% after COVID, compared with actual unemployment rates of 5.77\% and 4.82\%. The unemployment gap remains strongly countercyclical and widens sharply in recessions.

On average, the U.S. labor market remains slack relative to the efficient allocation, both before and after the pandemic. This average pattern, however, does not rule out episodic tightness. Actual tightness approaches or exceeds the efficient benchmark in several pre-recession episodes, suggesting that the part-time extension changes the interpretation of labor-market conditions at business-cycle peaks.

Accounting for part-time employment lowers efficient labor-market tightness and raises efficient unemployment because part-time workers contribute fewer effective hours than full-time workers. Thus, some convergence between actual and efficient labor-market conditions is a direct consequence of incorporating differences in labor input. The relevant question is whether this adjustment is quantitatively important. The results suggest that it is. Relative to the full-time baseline results, the unemployment gap narrows substantially in several episodes once differences in hours worked are taken into account, indicating that headcount employment alone may overstate labor-market slack.

As a result, the extended framework identifies several episodes in which actual labor-market tightness approaches or exceeds the efficient benchmark, most notably before the 1973--75 recession associated with the First Oil Crisis and again around the early-1980s downturns, before the 1990--91 recession, and before the 2007--09 Global Financial Crisis. These episodes appear considerably closer to the efficient allocation than under the full-time benchmark, highlighting the importance of the hours margin when assessing labor-market conditions.

Under the involuntary part-time calibration, efficient tightness averages 0.87 before COVID and 1.60 after COVID, and efficient unemployment averages 4.42\% and 3.58\%. Because involuntary part-time employment accounts for only a small share of total employment in the United States, these estimates remain close to the full-time baseline of \citet{MS21b}.

Overall, accounting for part-time employment mainly affects the efficient benchmark through effective labor input: part-time workers supply fewer market hours than full-time workers, so identical measured unemployment can imply different welfare depending on employment composition. This adjustment matters more under the total part-time definition than under the involuntary part-time definition, but the cyclical behavior of the unemployment gap remains broadly similar to the baseline framework.

\subsection{Robustness of U.S. results}

To further assess the robustness of the findings, I follow the procedure in \citet{MS21b}. Figures~\ref{f:alter} and~\ref{f:alter1} report alternative calibrations of the sufficient statistics for the 1951--2019 and 2020--2026 samples, respectively. Figures~\ref{f:inverse} and~\ref{f:inverse1} report the corresponding inverse-optimum sufficient statistics. Summary statistics for all panels appear in Tables~\ref{tab:fig8a_summary}--\ref{tab:fig8c_summary} and Tables~\ref{tab:fig9a_summary}--\ref{tab:fig9c_summary} in Appendix A.

\subsubsection{Alternative calibrations}

Under the extended framework with part-time employment, Proposition~1 implies that labor-market tightness $\theta \equiv v/u$ is efficient when
\begin{equation}\label{eq:theta_efficient}
\theta^*
=
\frac{\alpha + (1-\alpha)\left[\gamma + (1-\gamma)z\right] - z}
{c\,\epsilon}.
\end{equation}
Imposing the isoelastic Beveridge curve $v = \bar{\alpha}\,u^{-\epsilon}$ and solving for the unemployment rate yields the efficient unemployment rate
\begin{equation}\label{eq:ustar_robust}
u^{*}
=
\left[
\frac{c\,\epsilon}
{\alpha + (1-\alpha)\left[\gamma + (1-\gamma)z\right] - z}
\cdot
\frac{v}{u^{-\epsilon}}
\right]^{\frac{1}{1+\epsilon}}.
\end{equation}

Equation~\eqref{eq:ustar_robust} is evaluated quarter by quarter using observed unemployment $u$ and vacancy $v$, the branch-specific Beveridge elasticity $\epsilon$ estimated with Bai--Perron break dates, baseline calibrations $z=0.26$ and $c=0.92$, and the part-time parameters $\alpha=0.75$ (whole part-time series) and time-varying $\gamma$ from U.S.\ working-hours data.

\begin{figure}[H]
\centering

\subcaptionbox{Beveridge elasticity: $\epsilon$ in the 95\% confidence interval}{
\includegraphics[width=0.48\textwidth]{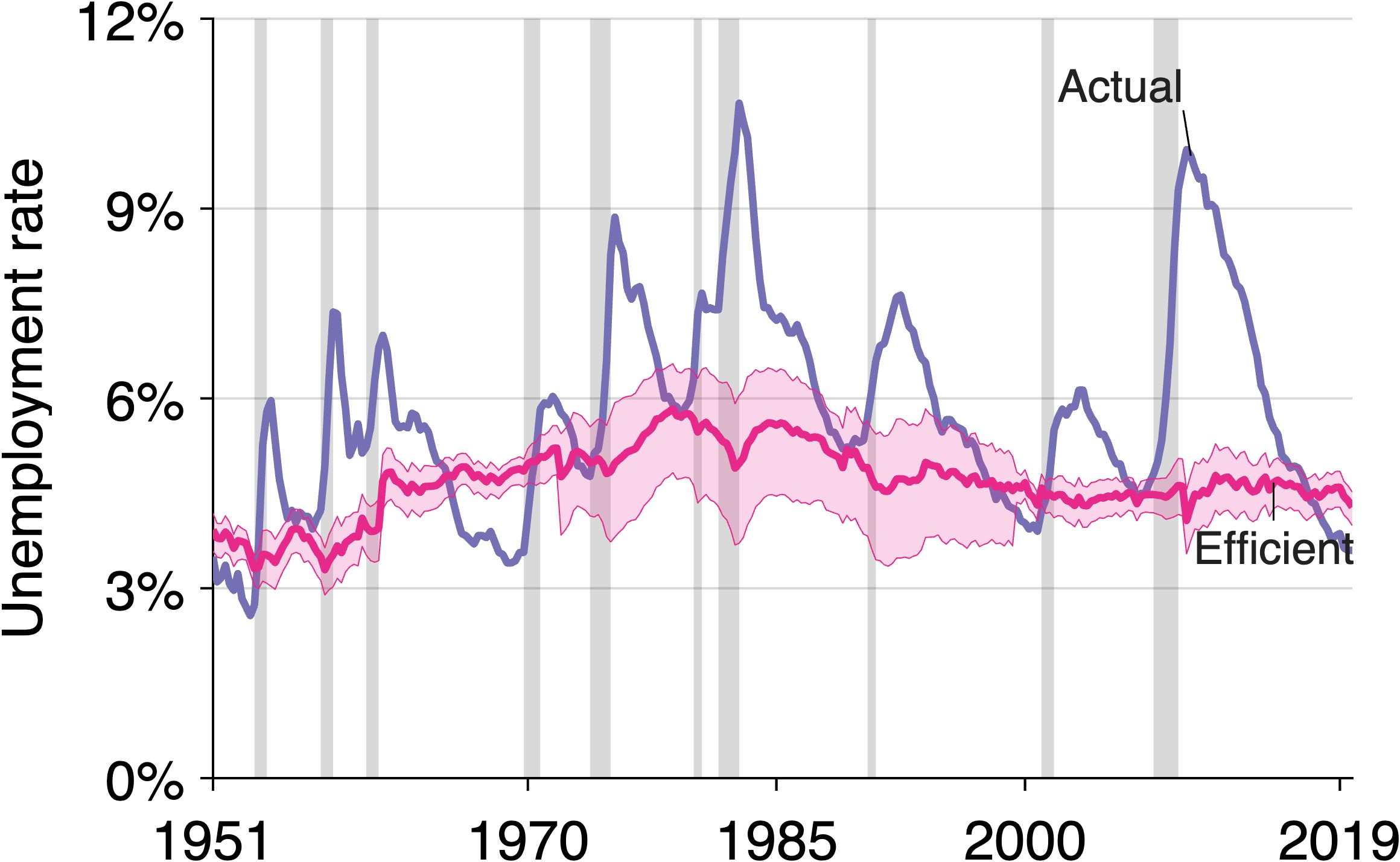}}
\hfill
\subcaptionbox{Social value of nonwork: $0.03<z<0.49$}{
\includegraphics[width=0.48\textwidth]{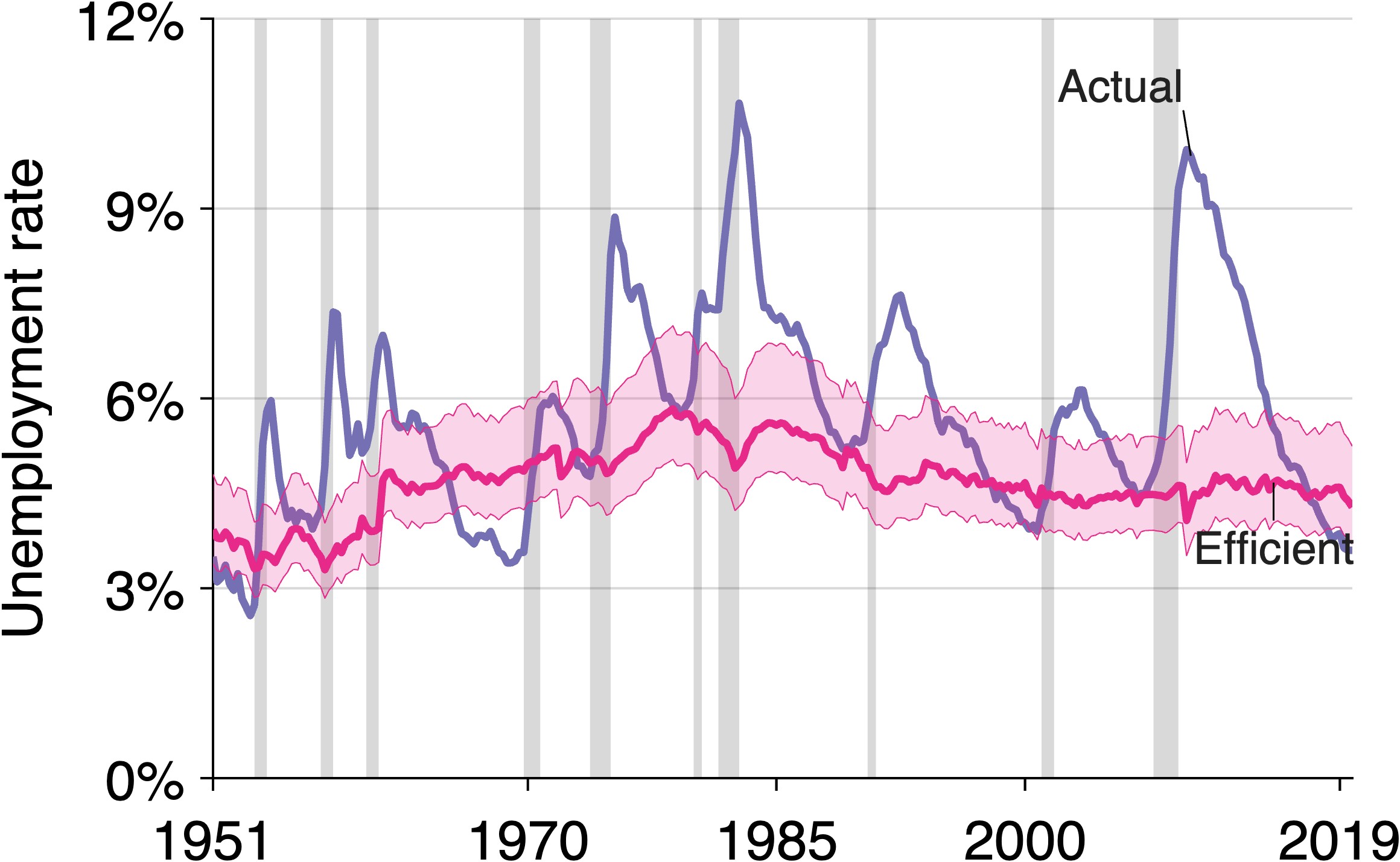}}

\vspace{0.3cm}

\subcaptionbox{Recruiting cost: $0.61<c<1.23$}{
\includegraphics[width=0.48\textwidth]{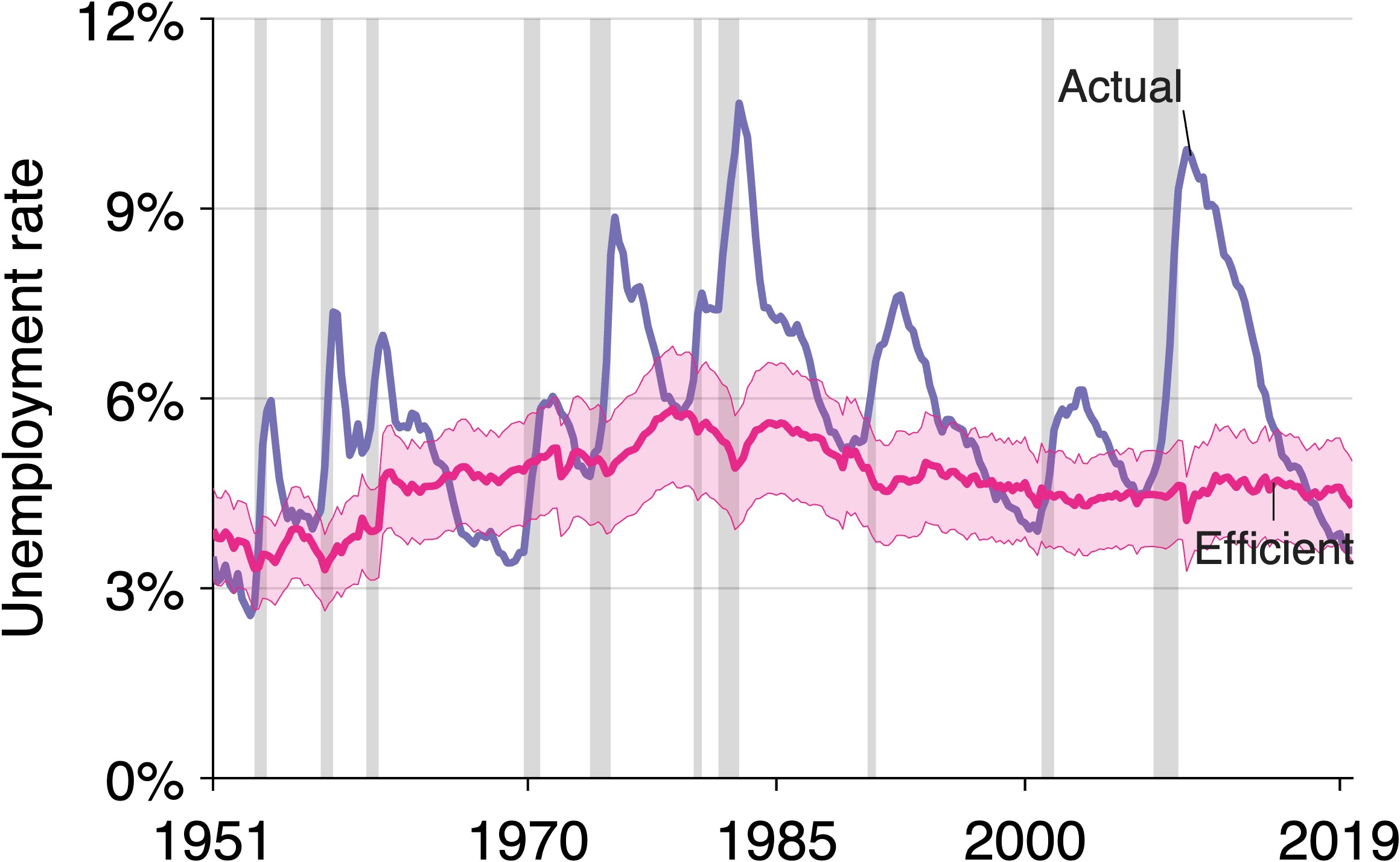}}

\caption{Efficient unemployment rates for alternative sufficient statistics in the United States, 1951--2019. \textit{Notes:} The figure reports the efficient unemployment rate implied by the model under alternative values of the Beveridge elasticity, the social value of nonwork, and recruiting costs. Each panel varies one sufficient statistic while holding the remaining sufficient statistics at their baseline values.}
\label{f:alter}

\end{figure}

\begin{figure}[H]
\centering

\subcaptionbox{Beveridge elasticity: $\epsilon$ in the 95\% confidence interval}{
\includegraphics[width=0.48\textwidth]{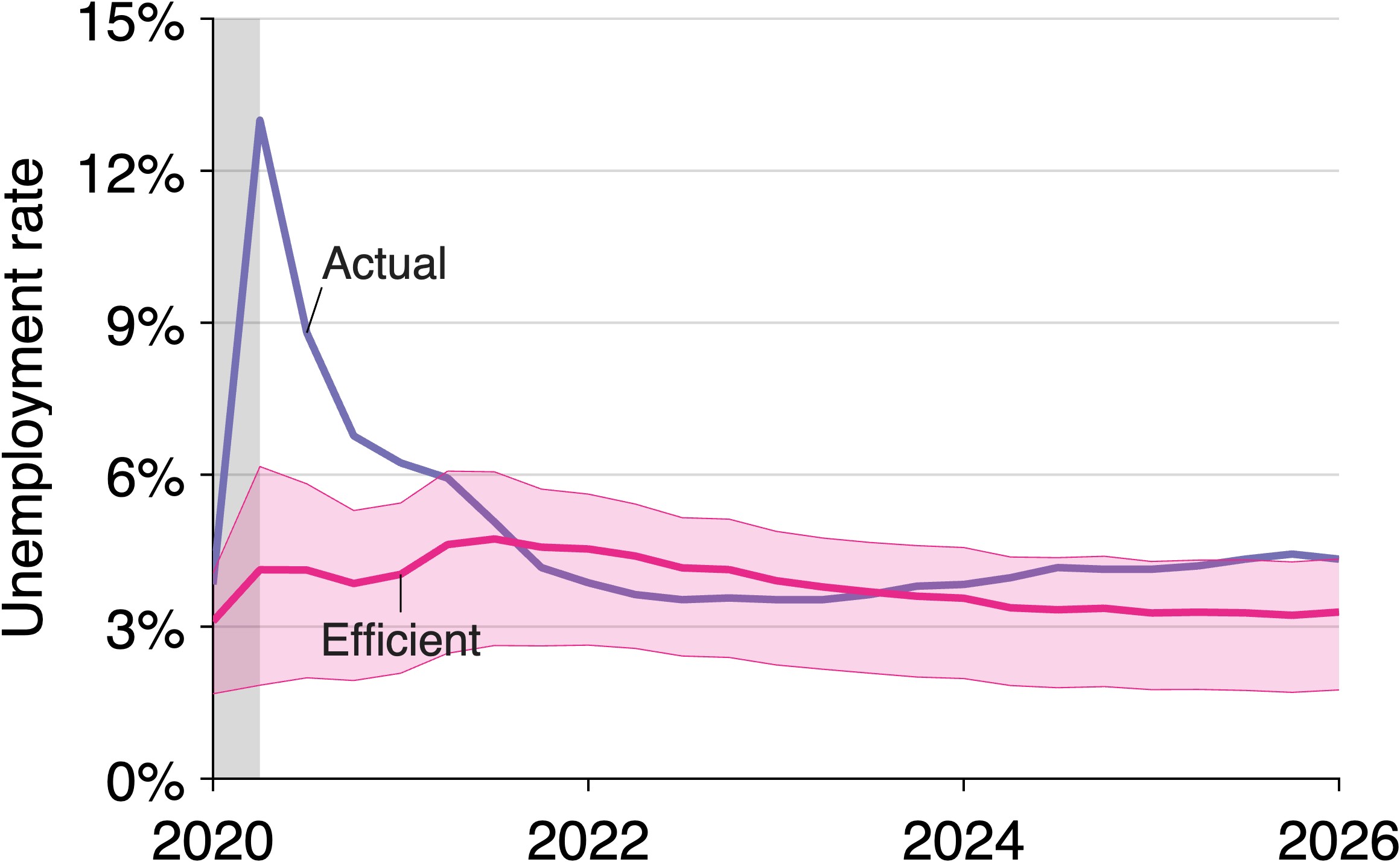}}
\hfill
\subcaptionbox{Social value of nonwork: $0.03<z<0.49$}{
\includegraphics[width=0.48\textwidth]{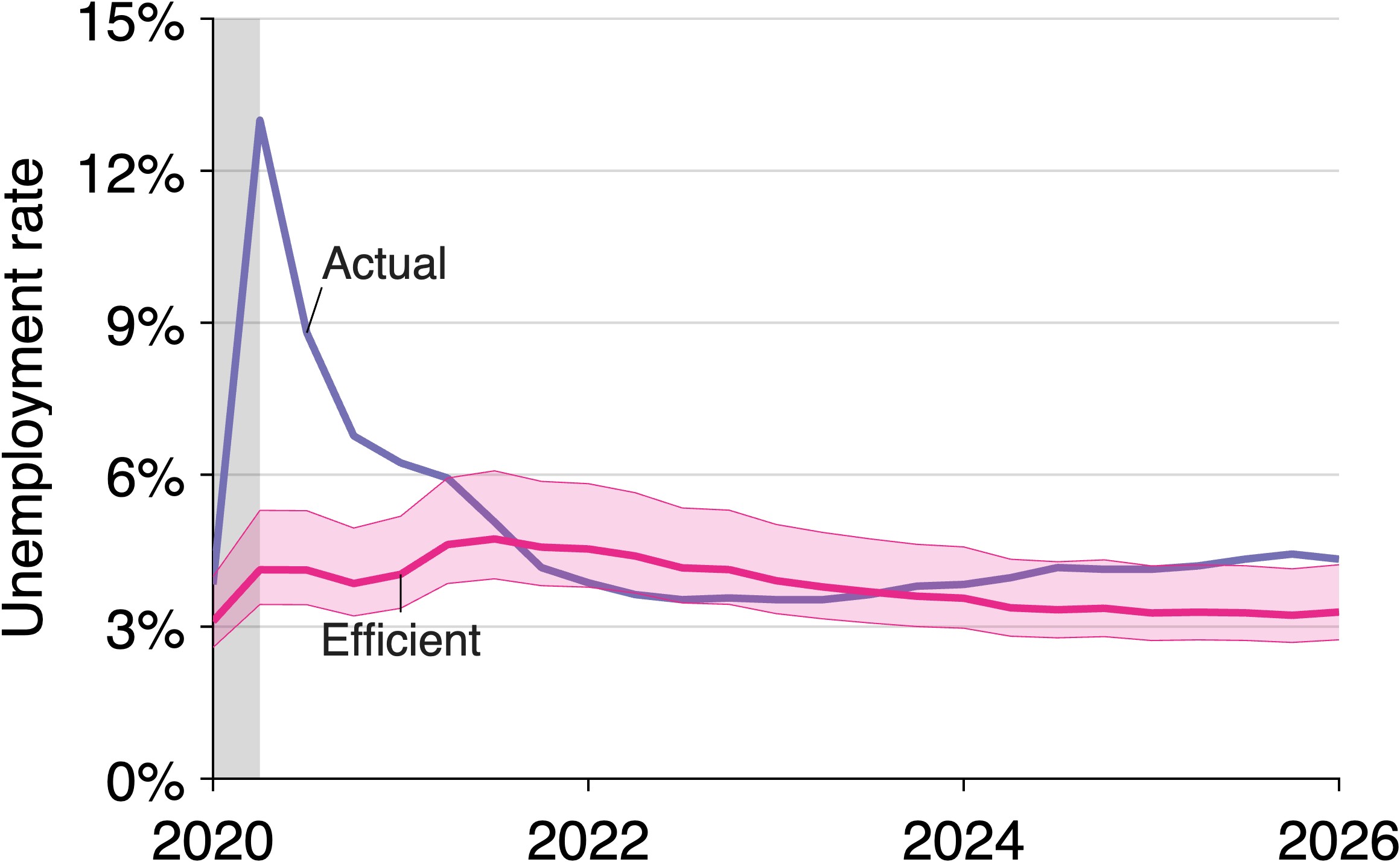}}

\vspace{0.3cm}

\subcaptionbox{Recruiting cost: $0.61<c<1.23$}{
\includegraphics[width=0.48\textwidth]{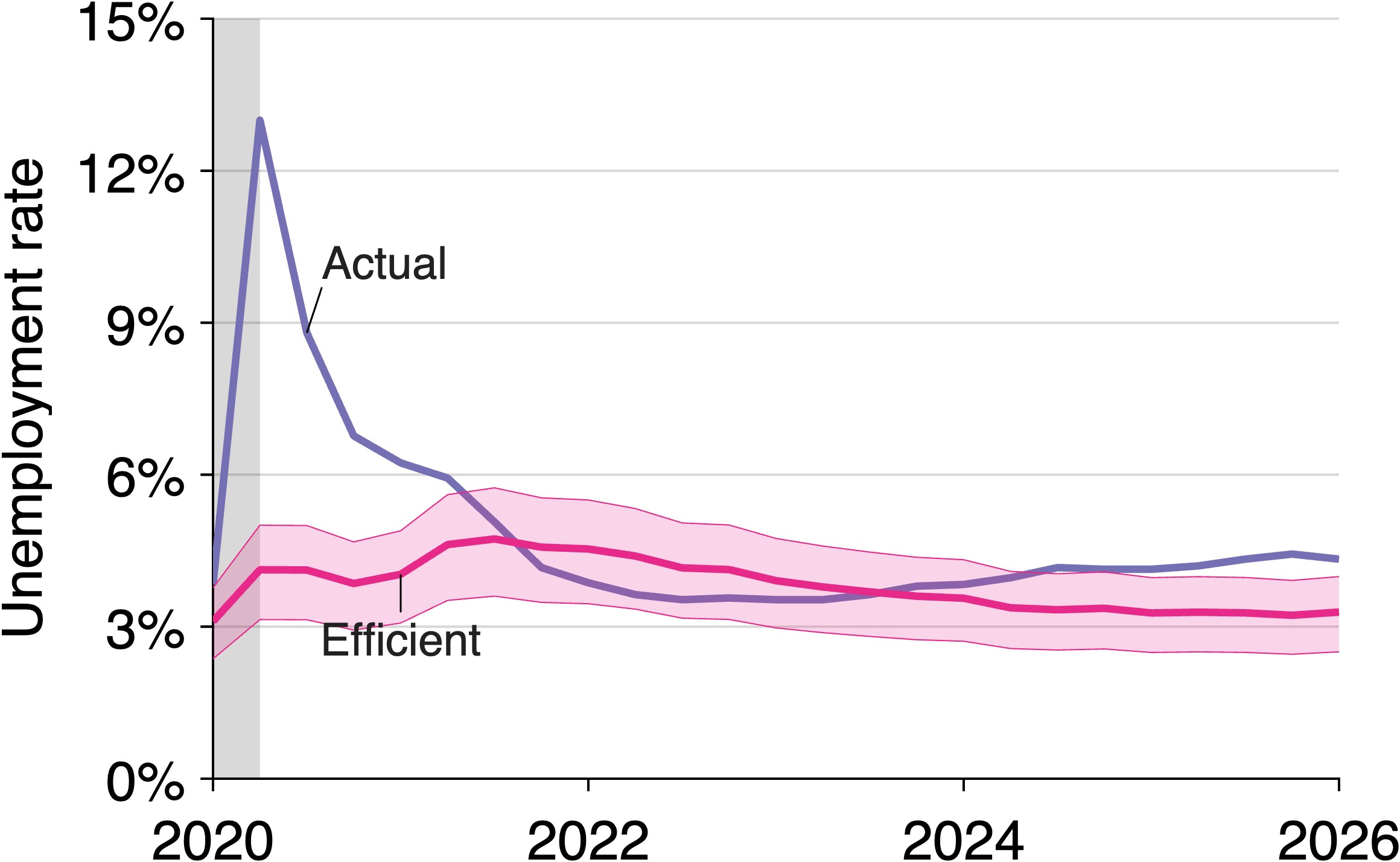}}

\caption{Efficient unemployment rates for alternative sufficient statistics in the United States, 2020--2026. \textit{Notes:} The figure reports the efficient unemployment rate implied by the model under alternative values of the Beveridge elasticity, the social value of nonwork, and recruiting costs. Each panel varies one sufficient statistic while holding the remaining sufficient statistics at their baseline values.}
\label{f:alter1}

\end{figure}

For each panel of Figures~\ref{f:alter} and~\ref{f:alter1}, I substitute alternative values of one sufficient statistic into \eqref{eq:ustar_robust} while holding the others fixed at baseline:
\begin{itemize}
    \item \textbf{Panel A (Beveridge elasticity).} I set $\epsilon$ to its branch-specific point estimate and to the lower and upper ends of its 95\% confidence interval, computed from the corrected standard errors in the Bai--Perron output.
    \item \textbf{Panel B (social value of nonwork).} I set $z$ to its baseline value $0.26$ and to the endpoints of the calibration range $[0.03,\,0.49]$.
    \item \textbf{Panel C (recruiting cost).} I set $c$ to its baseline value $0.92$ and to the endpoints of the calibration range $[\tfrac{2}{3}\times 0.92,\,\tfrac{4}{3}\times 0.92]$.
\end{itemize}
In each panel, the thick pink line plots $u^*$ under the baseline calibration of the displayed statistic; the thin pink lines and shaded pink area trace $u^*$ at the lower and upper bounds; and the purple line plots the actual unemployment rate.

The efficient unemployment rate is stable across alternative calibrations of the Beveridge elasticity, the social value of nonwork, and recruiting costs. Consistent with \citet{MS21b}, calibration uncertainty has limited effects before COVID. Over 1951--2019, varying the Beveridge elasticity within its 95\% confidence interval shifts the efficient unemployment rate by about $\pm 0.5$ percentage point on average, with a maximum deviation of 1.3 percentage points (Table~\ref{tab:fig8a_summary}). Similar patterns hold for the social value of nonwork and recruiting costs. Alternative calibrations typically move the efficient rate by less than 1 percentage point on average, and the maximum deviations remain below 1.3 percentage points (Tables~\ref{tab:fig8b_summary} and~\ref{tab:fig8c_summary}). Overall, the main unemployment-gap patterns are not driven by a particular choice of Beveridge elasticity, social value of nonwork, or recruiting cost.

The 2020--2026 sample exhibits somewhat greater sensitivity, reflecting the unusually flat Beveridge curve during and immediately after the pandemic. Under alternative calibrations, the efficient unemployment rate differs from baseline by about 0.7--1.0 percentage point on average in panel~A, with a maximum deviation of 1.4 percentage points in 2020Q2. Panels~B and~C show similarly modest post-COVID changes, with maximum deviations between 1.1 and 1.3 percentage points. Overall, the efficient unemployment rate remains quantitatively stable across a wide range of parameter values.

\subsubsection{Inverse-optimum sufficient statistics}

\begin{figure}[H]
\centering

\subcaptionbox{Inverse-optimum Beveridge elasticity}{
\includegraphics[width=0.48\textwidth]{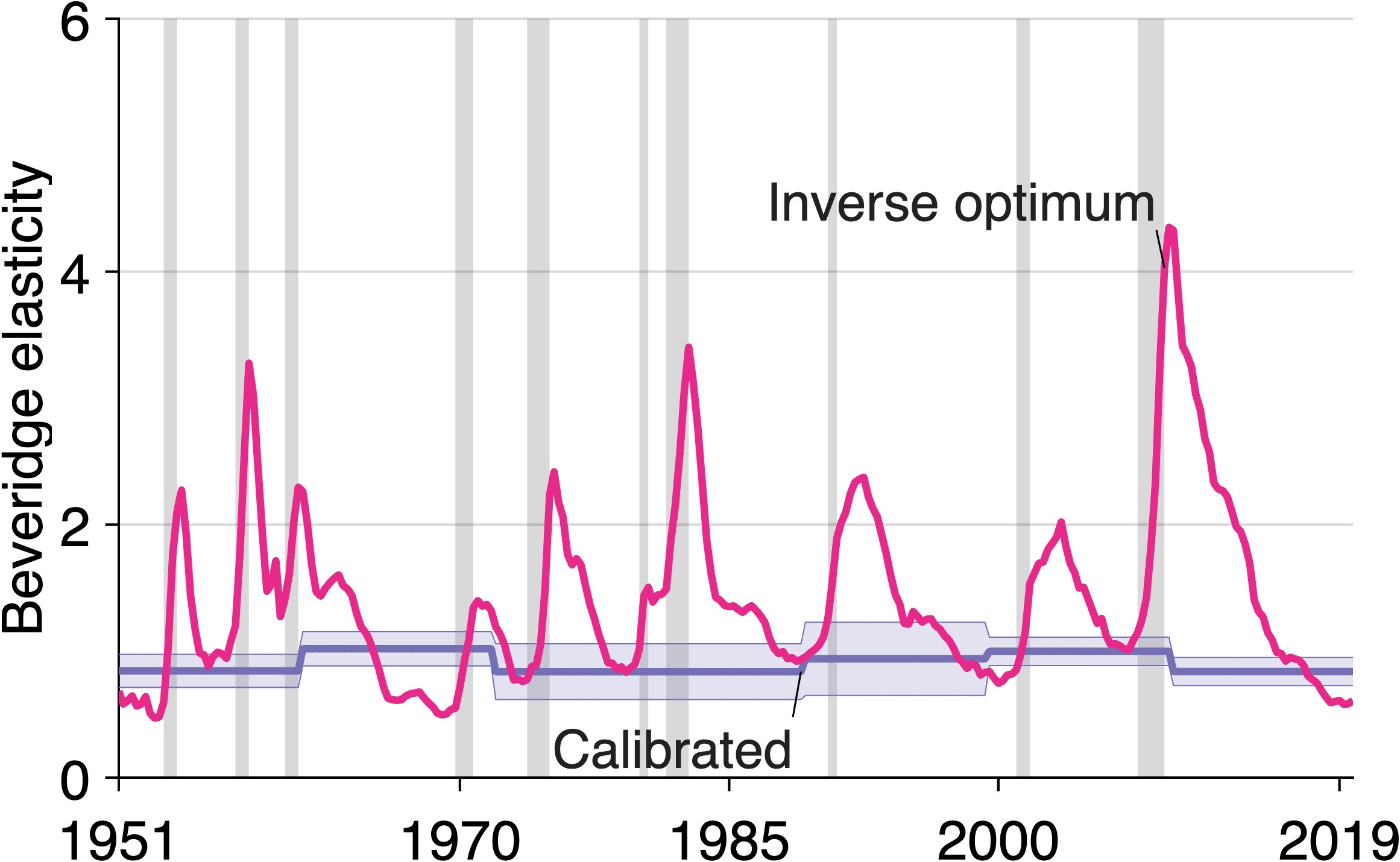}}
\hfill
\subcaptionbox{Inverse-optimum social value of nonwork}{
\includegraphics[width=0.48\textwidth]{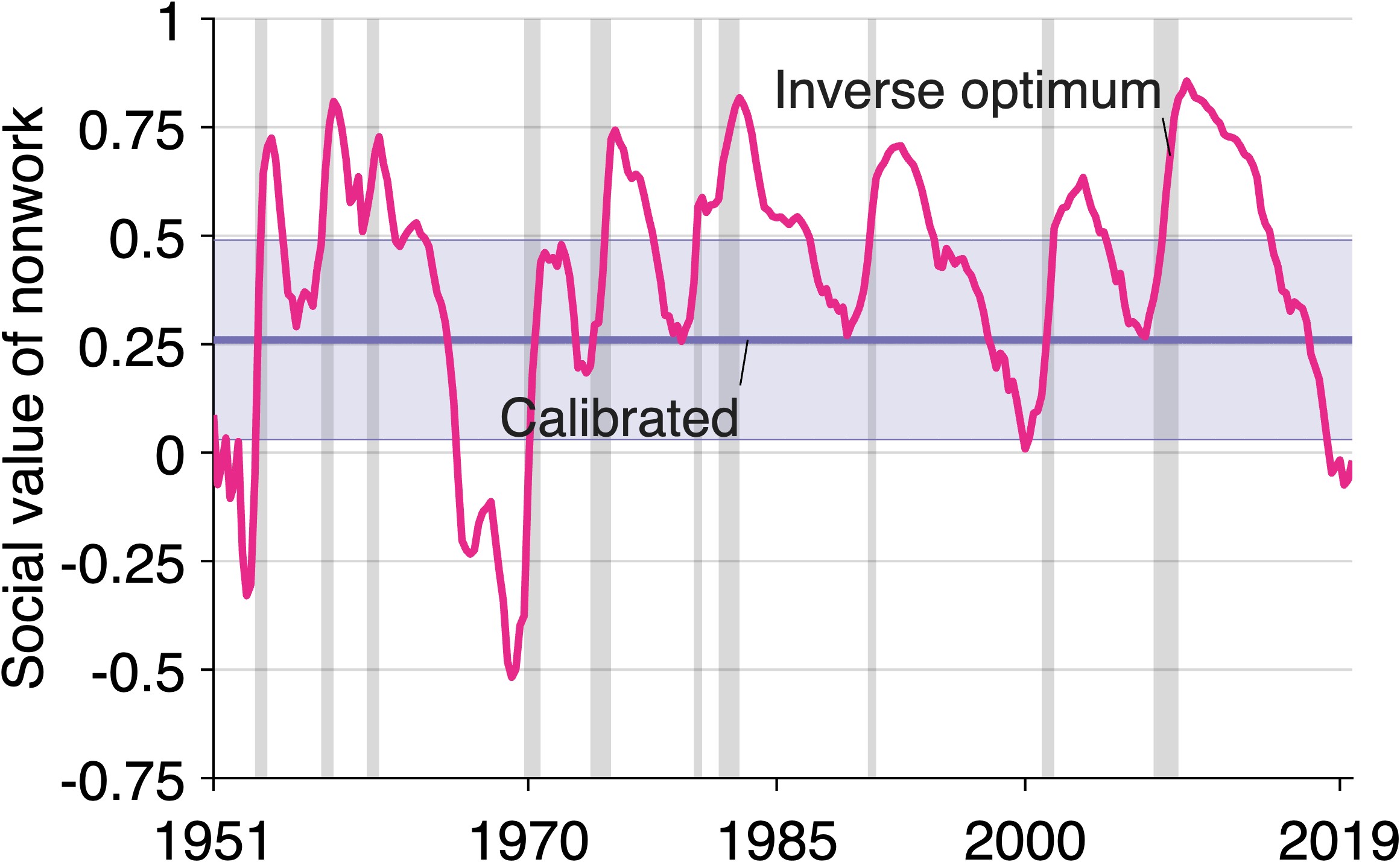}}

\vspace{0.3cm}

\subcaptionbox{Inverse-optimum recruiting cost}{
\includegraphics[width=0.48\textwidth]{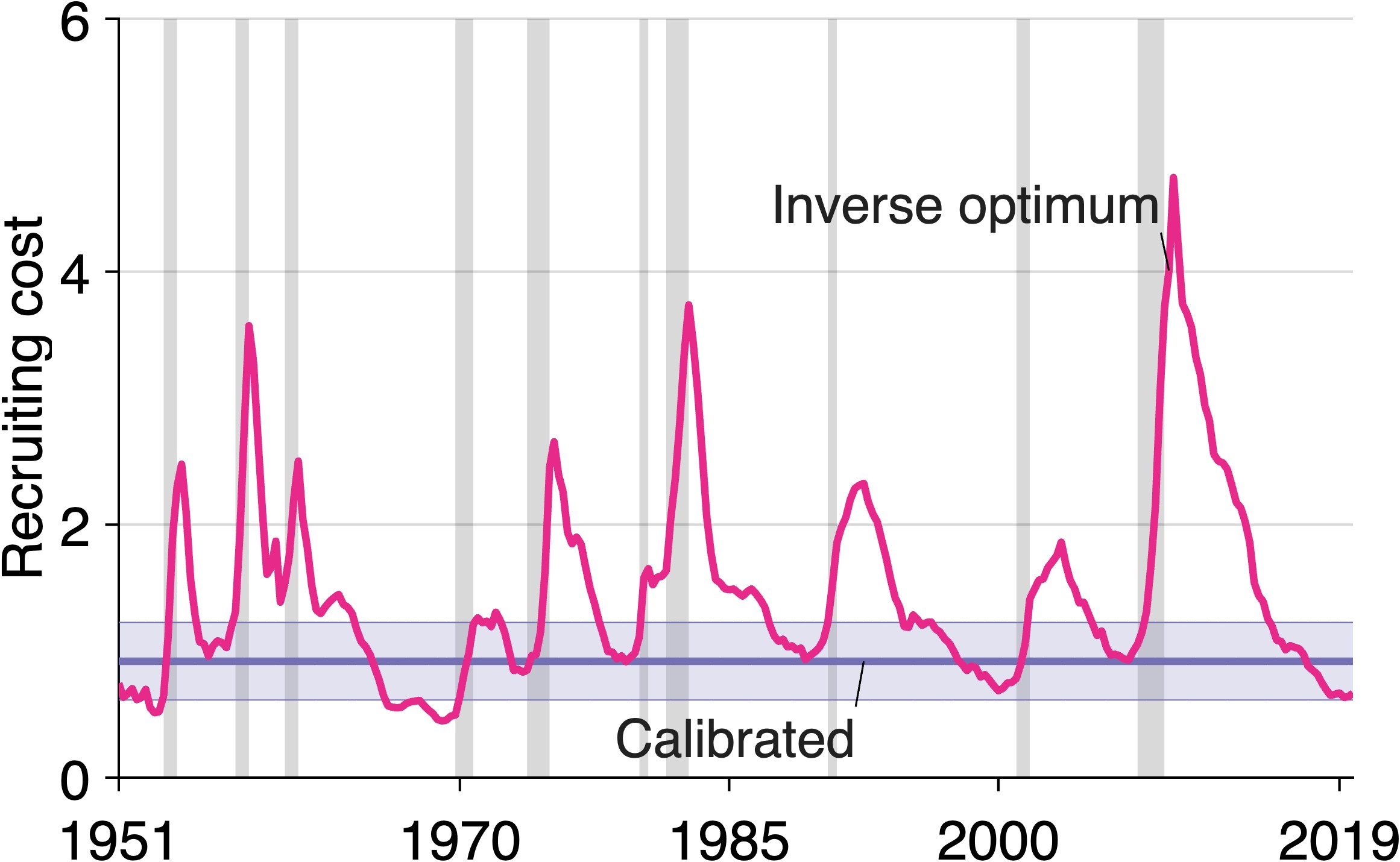}}

\caption{Inverse-optimum sufficient statistics in the United States, 1951--2019. \textit{Notes:} The figure reports the inverse-optimum Beveridge elasticity, social value of nonwork, and recruiting cost implied by the efficient-unemployment framework. Each panel recovers the value of a sufficient statistic consistent with the observed unemployment rate and labor-market tightness.}
\label{f:inverse}

\end{figure}

\begin{figure}[H]
\centering

\subcaptionbox{Inverse-optimum Beveridge elasticity}{
\includegraphics[width=0.48\textwidth]{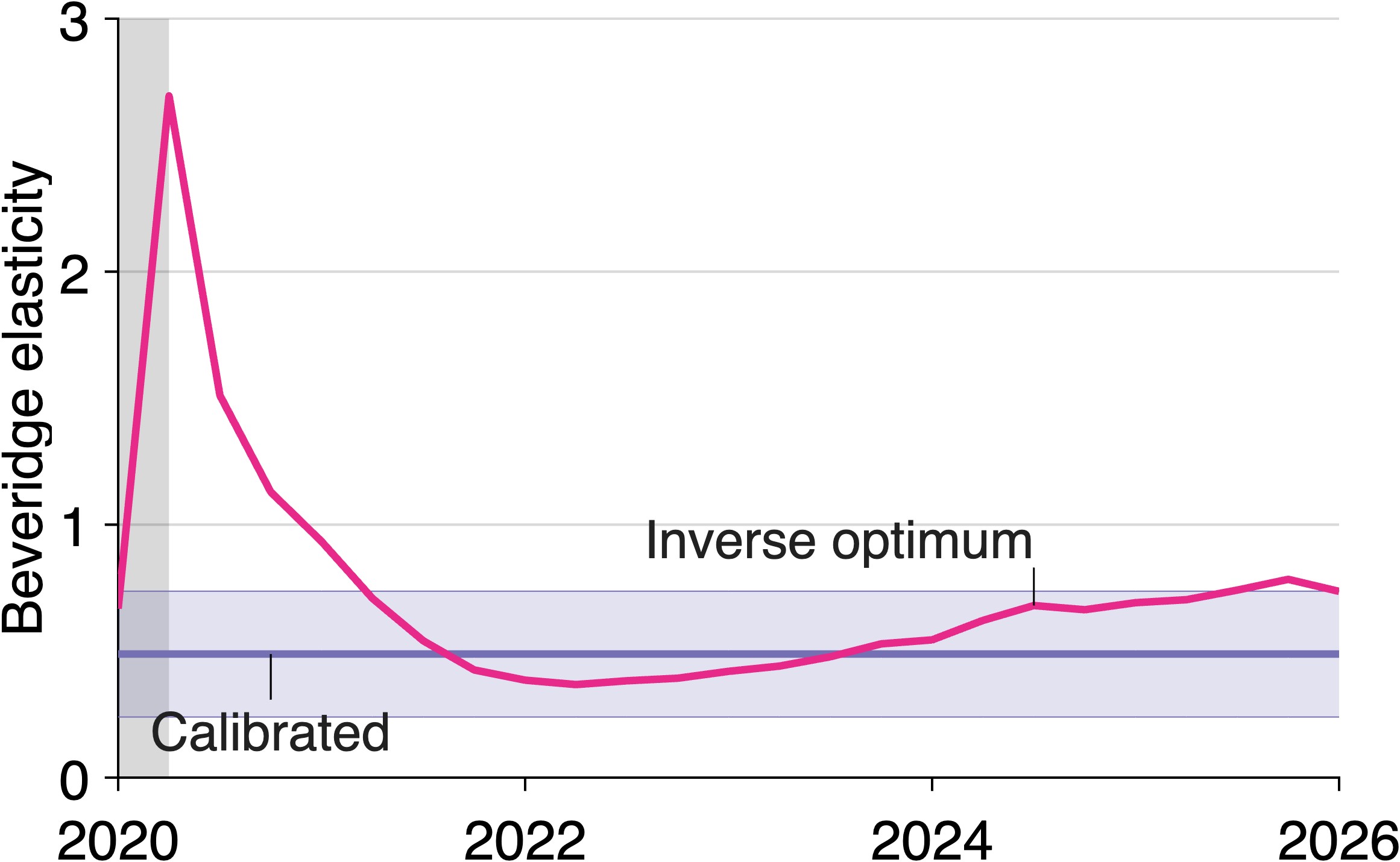}}
\hfill
\subcaptionbox{Inverse-optimum social value of nonwork}{
\includegraphics[width=0.48\textwidth]{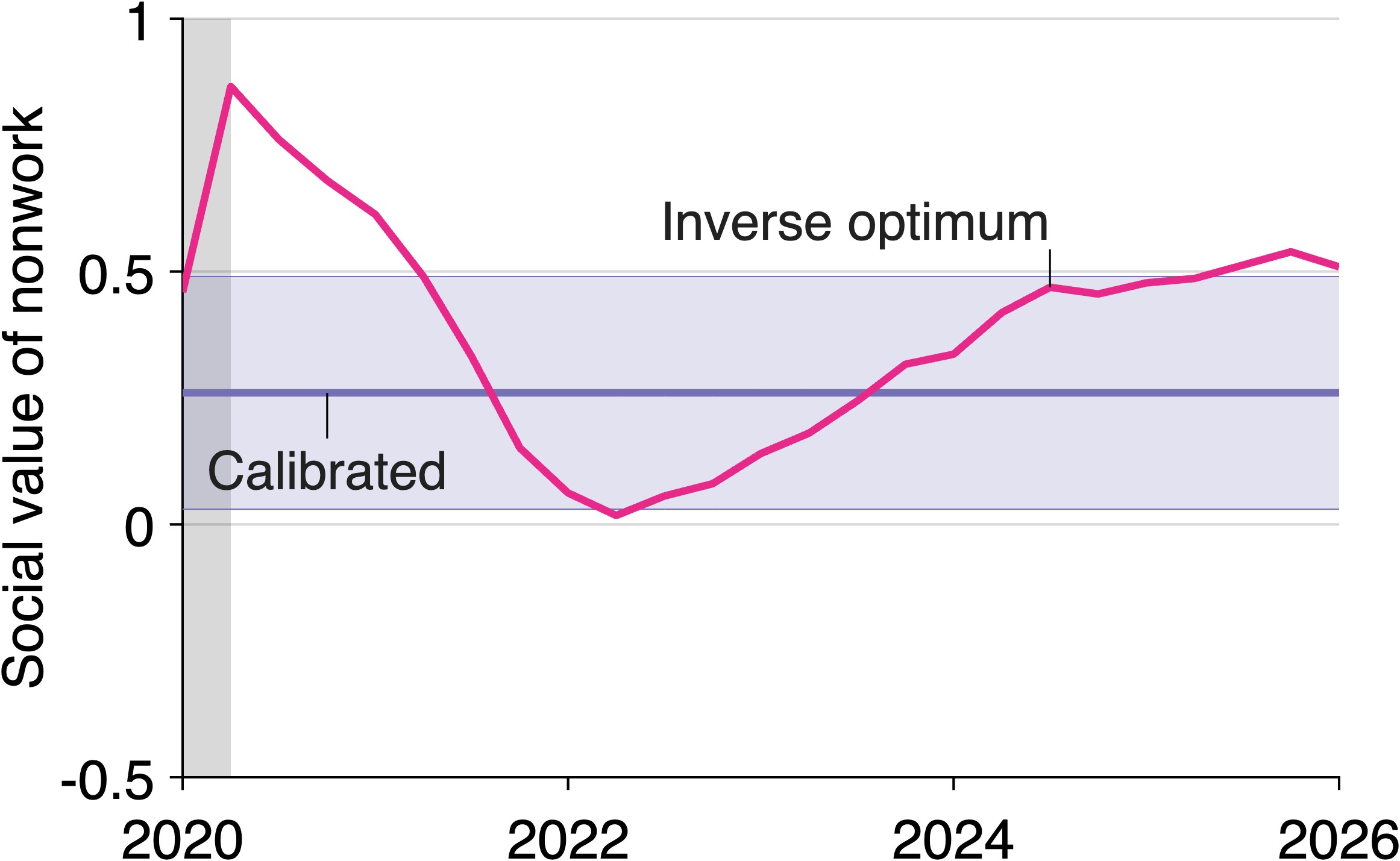}}

\vspace{0.3cm}

\subcaptionbox{Inverse-optimum recruiting cost}{
\includegraphics[width=0.48\textwidth]{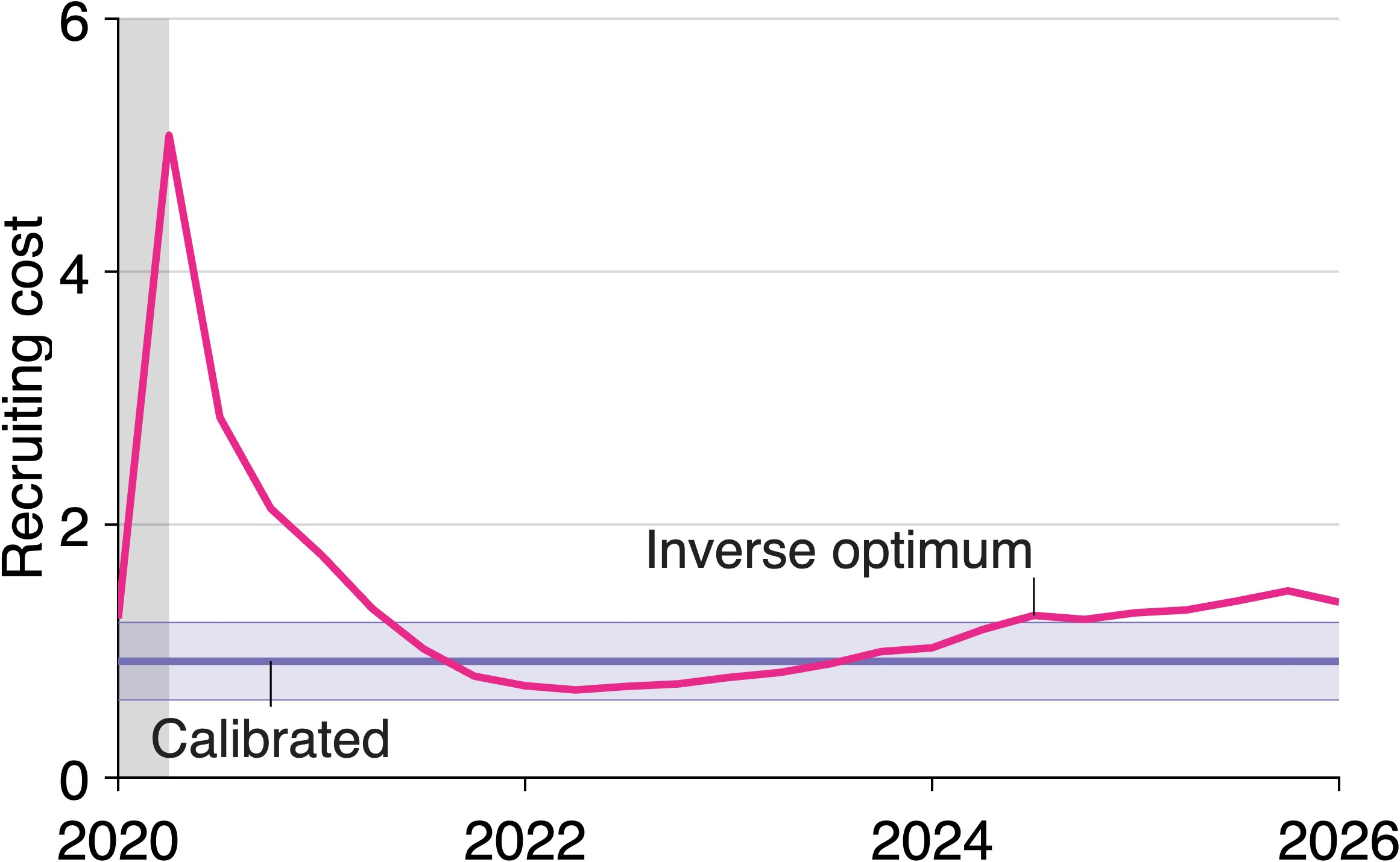}}

\caption{Inverse-optimum sufficient statistics in the United States, 2020--2026. \textit{Notes:} The figure reports the inverse-optimum Beveridge elasticity, social value of nonwork, and recruiting cost implied by the efficient-unemployment framework. Each panel recovers the value of a sufficient statistic consistent with the observed unemployment rate and labor-market tightness.}
\label{f:inverse1}

\end{figure}

The inverse-optimum exercise asks which value of each sufficient statistic would make \emph{observed} labor-market tightness efficient, holding the other sufficient statistics at their baseline calibrations. From \eqref{eq:theta_efficient}, actual tightness is efficient when the Beveridge elasticity satisfies
\begin{equation}\label{eq:eps_star}
\epsilon^{*}
=
\frac{\alpha + (1-\alpha)\left[\gamma + (1-\gamma)z\right] - z}
{c\,\theta}.
\end{equation}
Similarly, actual tightness is efficient when the social value of nonwork satisfies
\begin{equation}\label{eq:z_star}
z^{*}
=
\frac{\alpha + (1-\alpha)\gamma - c\,\epsilon\,\theta}
{1-(1-\alpha)(1-\gamma)},
\end{equation}
and when the recruiting cost satisfies
\begin{equation}\label{eq:c_star}
c^{*}
=
\frac{\alpha + (1-\alpha)\left[\gamma + (1-\gamma)z\right] - z}
{\epsilon\,\theta}.
\end{equation}
In each panel of Figures~\ref{f:inverse} and~\ref{f:inverse1}, the pink line plots the corresponding inverse-optimum series ($\epsilon^*$, $z^*$, or $c^*$); the thick purple line plots the calibrated value; and the thin purple lines and shaded purple area mark the 95\% confidence interval (panel~A) or calibration range (panels~B and~C).

Over 1951--2019, the inverse-optimum sufficient statistics frequently depart from their calibrations, especially in recessions, reproducing the patterns in \citet{MS21b}. The inverse-optimum Beveridge elasticity averages 1.42, compared with a calibrated value of about 0.91 and lies above its 95\% confidence interval in 61\% of quarters (Table~\ref{tab:fig9a_summary}). The inverse-optimum social value of nonwork averages 0.41, exceeds its calibration range in 46\% of quarters, and turns negative in 12\% of quarters (Table~\ref{tab:fig9b_summary}). The inverse-optimum recruiting cost averages 1.46, well above the calibrated $c=0.92$, and exceeds its calibration range in 51\% of quarters (Table~\ref{tab:fig9c_summary}). These departures indicate that, under the calibrated sufficient statistics, observed labor-market tightness was often lower than efficient---consistent with inefficient slack, especially in downturns.

Over 2020--2026, the inverse-optimum values align more closely with their calibrations. The inverse-optimum Beveridge elasticity averages 0.64 and lies within its 95\% confidence interval in 76\% of quarters; the inverse-optimum social value of nonwork averages 0.28 and lies within $[0.03,\,0.49]$ in 84\% of quarters; and the inverse-optimum recruiting cost averages 1.20 and lies within its calibration range in 76\% of quarters. Large deviations are concentrated in 2020Q2, when unemployment spiked and the Beveridge curve was unusually flat.

The efficient unemployment rate is stable across a wide range of plausible calibrations, consistent with \citet{MS21b}. The inverse-optimum exercise further shows that pre-COVID labor-market outcomes can be reconciled with efficiency only under often-implausible parameter values. These robustness exercises support the main U.S. finding: accounting for part-time employment does not overturn the conclusion that the labor market was generally inefficiently slack, especially during recessions. The post-COVID evidence is weaker because the unusually flat Beveridge curve reduces the informativeness of the sufficient-statistics approach, but it remains broadly consistent with this conclusion.

\section{Application to Japan}

Part-time employment is a central and growing feature of the Japanese labor market, as documented in Figures~\ref{fig:oecd_jpn} and~\ref{fig:japan_pt_ft}. From 2002 to 2025, full-time employment remained relatively stable while part-time employment increased substantially. Table~\ref{tab:pt_share} indicates that Japan had, on average, 48.96 million full-time workers and 13.75 million part-time workers during this period. Part-time workers accounted for 21.9 percent of total employment among full-time and part-time workers.

However, Figure~\ref{fig:japan_ipt_pt} and Tables~\ref{tab:pt_share}--\ref{tab:invpt_share} show that the rise in part-time employment is not primarily driven by involuntary part-time work. Involuntary part-time employment accounts for only 5.9 percent of full-time and involuntary part-time employment on average, and only 22.9 percent of total part-time employment. Moreover, the involuntary part-time share has declined over time. This distinction is important because total part-time employment captures the broader hours-margin mechanism, while involuntary part-time employment captures hidden underemployment.

I therefore apply the extended Beveridgean unemployment framework to Japan using total part-time employment. This specification captures the mechanism emphasized in the model. Part-time workers count as employed in headcount terms, but they supply fewer effective hours than full-time workers. I then consider a more restrictive specification based on involuntary part-time employment. This second exercise asks whether the results are driven by hidden underemployment rather than by the broader expansion of lower-hours employment.

\begin{figure}[H]
\centering

\subcaptionbox{Beveridge curve, 1970Q1--1983Q2}
{\includegraphics[width=0.48\textwidth]{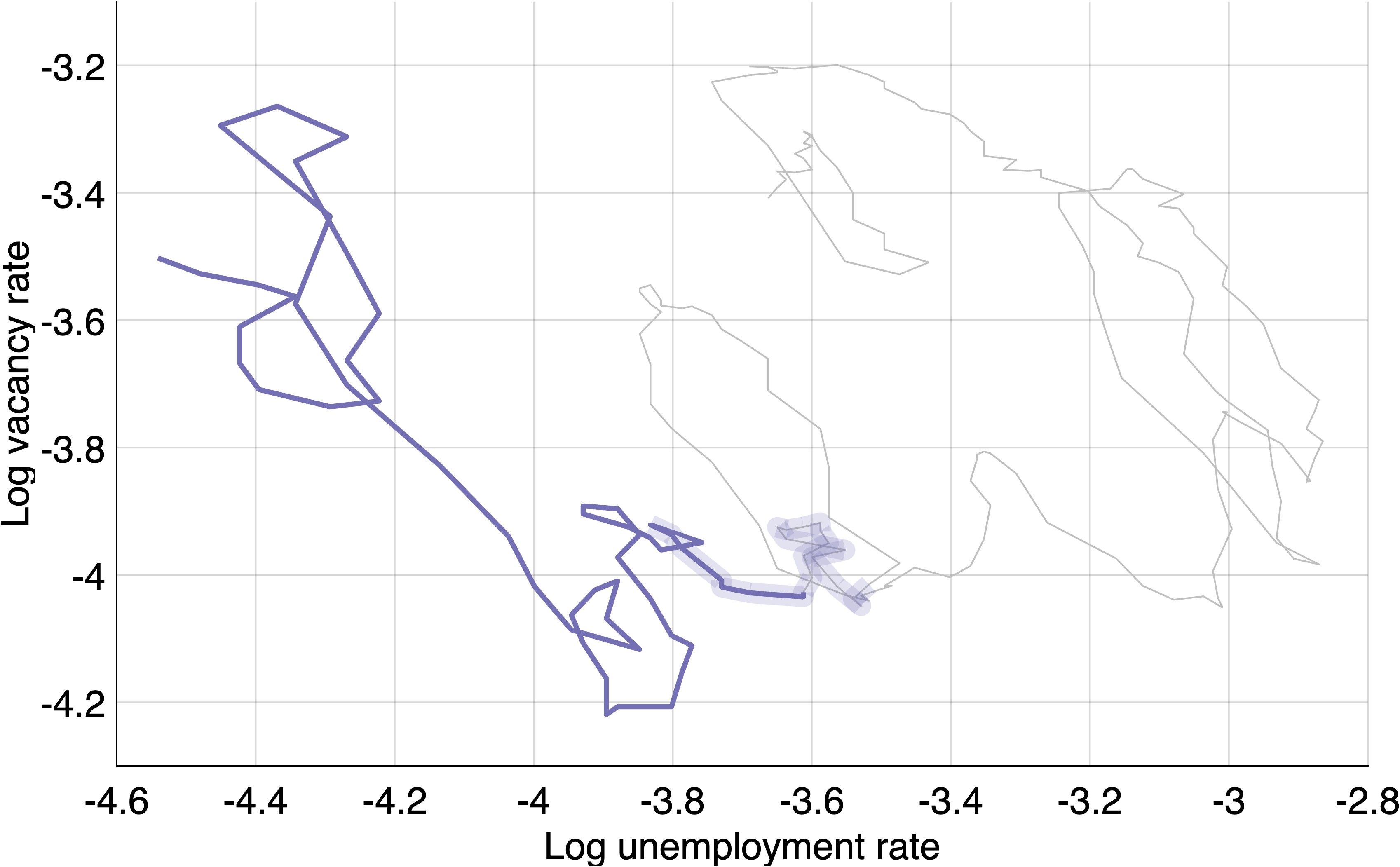}}\hfill
\subcaptionbox{Beveridge curve, 1983Q3--2000Q1}
{\includegraphics[width=0.48\textwidth]{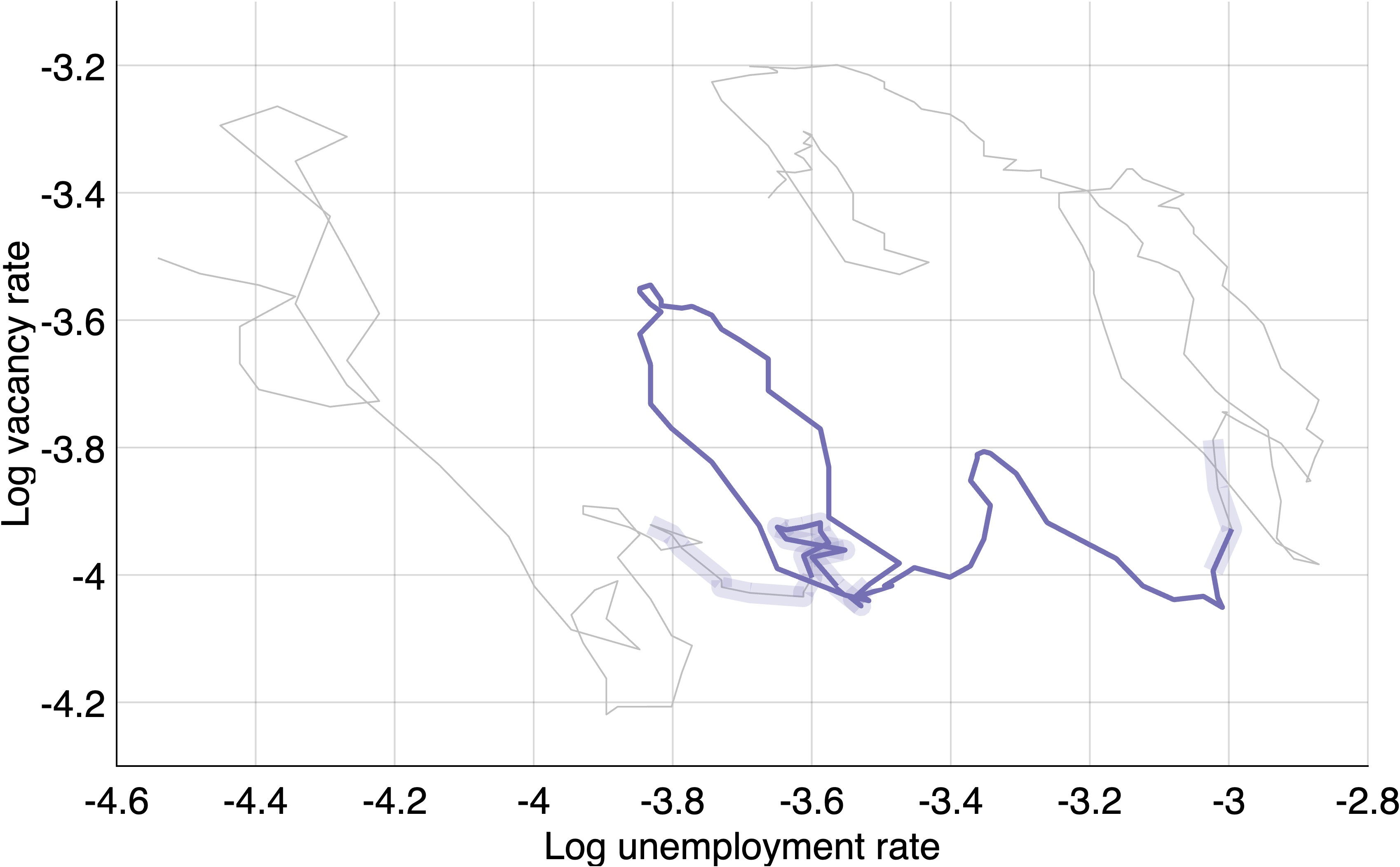}}\hfill
\subcaptionbox{Beveridge curve, 2000Q2--2025Q3}
{\includegraphics[width=0.48\textwidth]{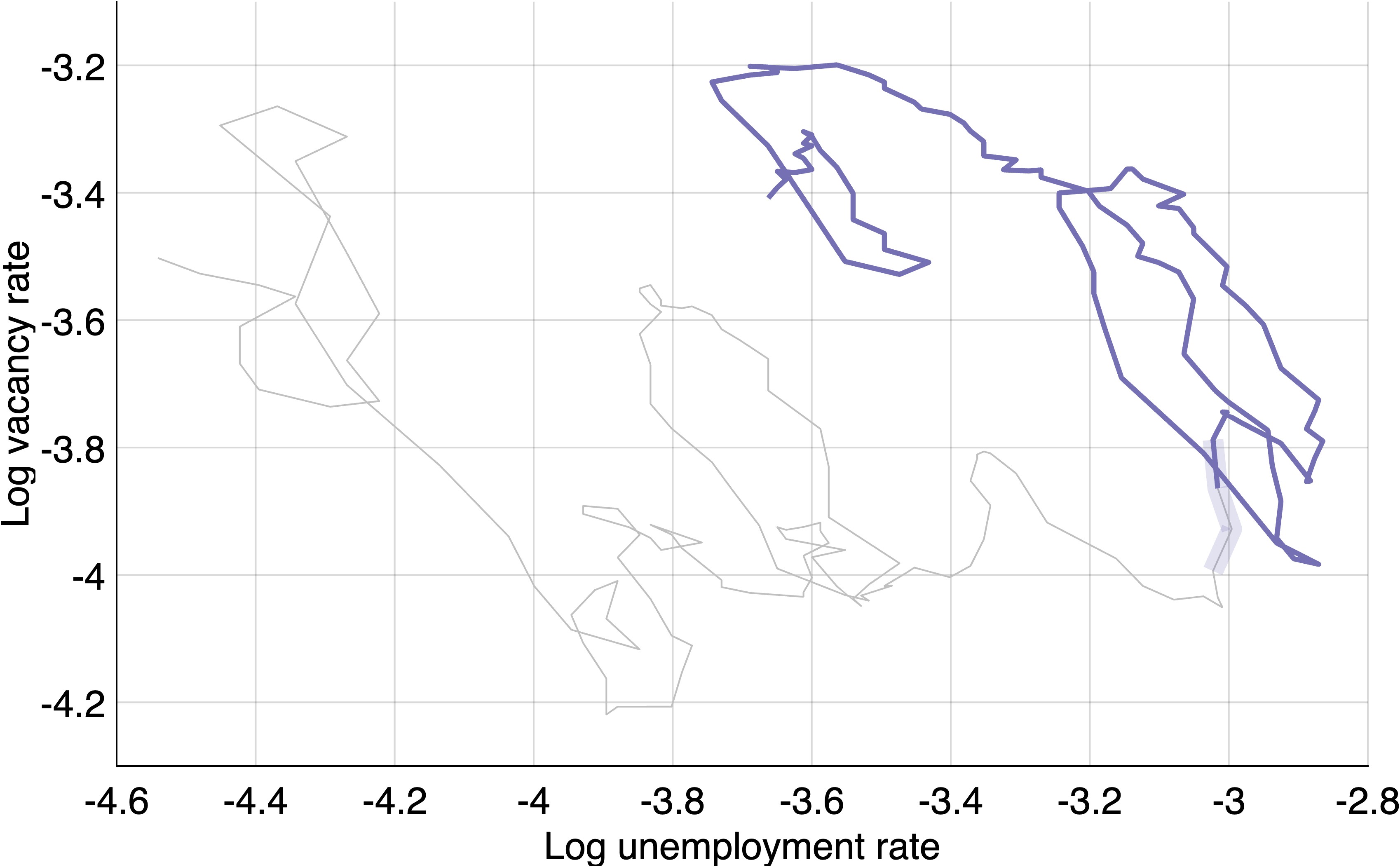}}

\caption{Beveridge curve estimation in Japan, 1970--2025. \textit{Notes}: The number of structural breaks and their dates are estimated using the Bai--Perron (1998, 2003) algorithm.}
\label{f:bc_japan}
\end{figure}

To implement the measurement for Japan, I first estimate the Beveridge elasticity by regressing the log vacancy rate on the log unemployment rate. The unemployment rate is obtained from the OECD database, and the vacancy rate is obtained from Statistics Bureau of Japan. The estimation sample runs from 1970Q1 to 2025Q3 (223 quarterly observations). As in the U.S. application, I allow for structural breaks in the Beveridge relationship using the Bai--Perron algorithm of \citet{BP98,BP03}, following the setup in \citet{MS21b}. I allow for different variances of the errors across regimes, for autocorrelation in the errors, and for different distributions of the independent and dependent variables across regimes. Standard errors are corrected for autocorrelation and heteroskedasticity using a quadratic kernel with automatic bandwidth selection based on an AR(1) approximation \citep{A1}. I set the maximum number of breaks to two and the trimming parameter to 0.20, so that each regime contains at least 20\% of the sample.

The supF tests reject the null hypothesis of no structural break at the 1\% significance level. Both the Bayesian information criterion \citep{Yao88} and the modified Schwarz criterion \citep{LSZ97} select two breaks. The estimated break dates occur in 1983Q2 and 2000Q1, yielding three Beveridge-curve regimes: 1970Q1--1983Q2, 1983Q3--2000Q1, and 2000Q2--2025Q3. The resulting Beveridge elasticity estimates are 0.86, 0.46, and 0.63, with corrected standard errors of 0.15, 0.10, and 0.08, respectively. The overall fit is $R^2=0.80$. Figure~\ref{f:bc_japan} plots the estimated Beveridge curve within each regime. The figure shows a clear outward shift in the Japanese Beveridge curve over time, indicating that the same unemployment rate has been associated with higher vacancy rates in later subperiods.

For the remaining sufficient statistics, I set $z=0.26$ and $c=0.92$ following \citet{MS21b}. I construct $\alpha$ from the OECD part-time employment share and set $\gamma=0.56$ using hours data from the Statistics Bureau of Japan, Labour Force Survey Detailed Tabulation. The Japanese hours data imply average weekly hours of 24.19 for part-time workers and 43.59 for full-time workers, so $\gamma=H^P/H^F\approx 0.56$. Using the U.S. values of $z$ and $c$ keeps the cross-country comparison transparent and isolates the role of Japan's employment composition and relative hours. However, these parameters may differ across countries because recruiting institutions, employment protection, job-search behavior, and the value of nonmarket time are not identical in Japan and the United States. The Japan results should therefore be interpreted as measuring how incorporating part-time employment changes the efficient-unemployment benchmark while holding the baseline preference and recruiting-cost parameters fixed. Accordingly, the levels of efficient unemployment should not be interpreted as directly comparable across the United States and Japan. The relevant comparison is within each country, between the full-time benchmark and the part-time calibrations. To assess the sensitivity of this choice, I later vary $z$ and $c$ over plausible ranges and compute inverse-optimum values of the sufficient statistics.

The Japanese calibration combines OECD employment counts with hours data from the Labour Force Survey Detailed Tabulation. This introduces a potential definitional difference because part-time employment is not always measured using the same 35-hour threshold used in the U.S. CPS classification. I therefore interpret the Japan results as using the OECD/LFS classification of part-time employment rather than imposing a U.S.-specific hours cutoff. The relative-hours parameter $\gamma$ partially addresses this issue by measuring the actual hours gap between part-time and full-time workers in the Japanese data. Nevertheless, differences in national definitions of part-time employment should be kept in mind when comparing the magnitude of the part-time adjustment across countries.

\begin{figure}[H]
\centering
\includegraphics[width=0.7\textwidth]{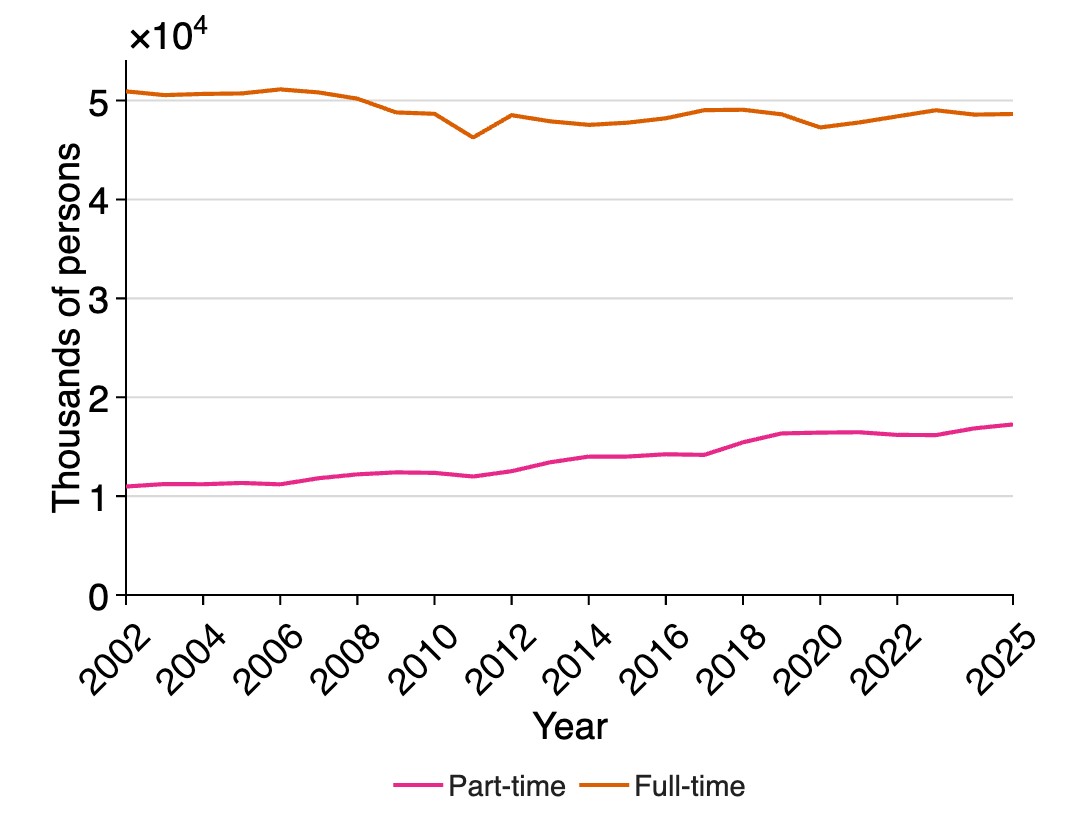}

\caption{Full-time and part-time employment in Japan, 2002--2025. \textit{Notes}: The figure reports annual full-time and part-time employment in Japan. Full-time employment remained relatively stable over the sample period, fluctuating around 49 million workers. In contrast, part-time employment increased steadily from approximately 11 million workers in 2002 to more than 17 million workers in 2025, reflecting the growing importance of part-time work in the Japanese labor market.  \textit{Source}: OECD Employment Database.}
\label{fig:japan_pt_ft}
\end{figure}

\begin{figure}[H]
\centering
\includegraphics[width=0.7\textwidth]{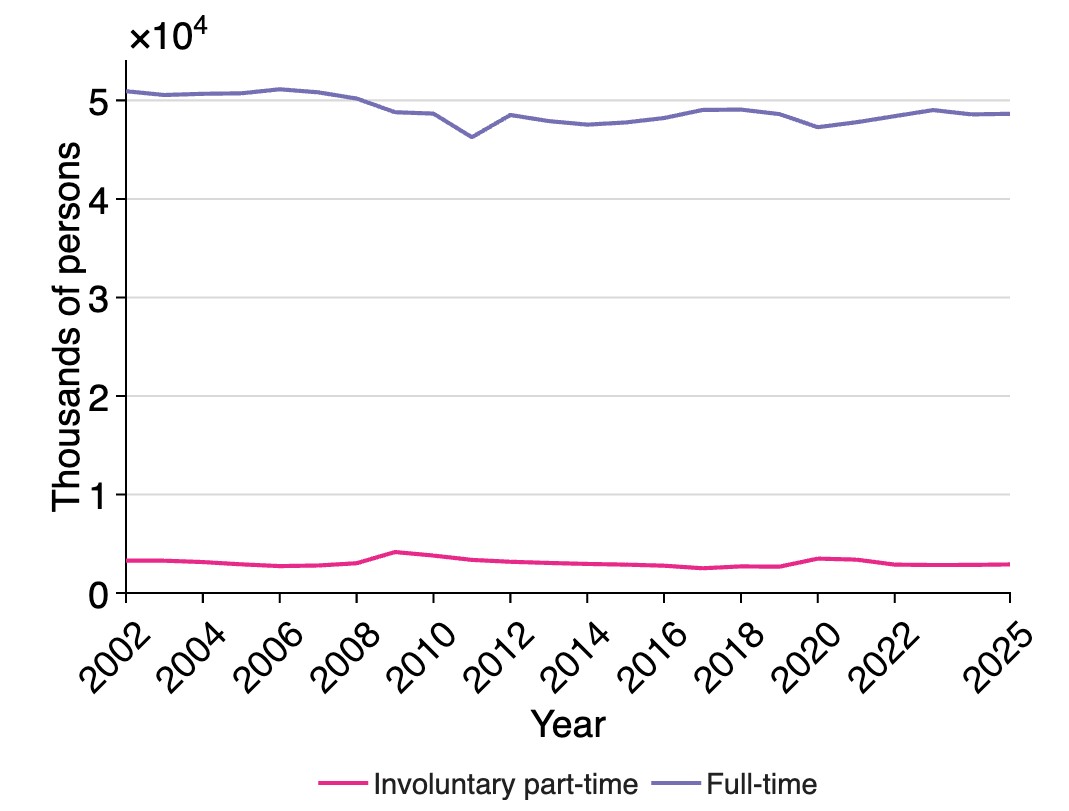}

\caption{Involuntary part-time employment and full time in Japan, 2002--2025. \textit{Notes}: The figure reports total part-time employment, involuntary part-time employment, and the involuntary part-time share in Japan. Although total part-time employment increased steadily over the sample period, involuntary part-time employment remained much smaller and the involuntary share declined over time. On average, involuntary part-time workers account for about 20--25 percent of total part-time employment. \textit{Source}: OECD Employment Database.}
\label{fig:japan_ipt_share}
\end{figure}

\begin{figure}[H]
\centering
\includegraphics[width=0.7\textwidth]{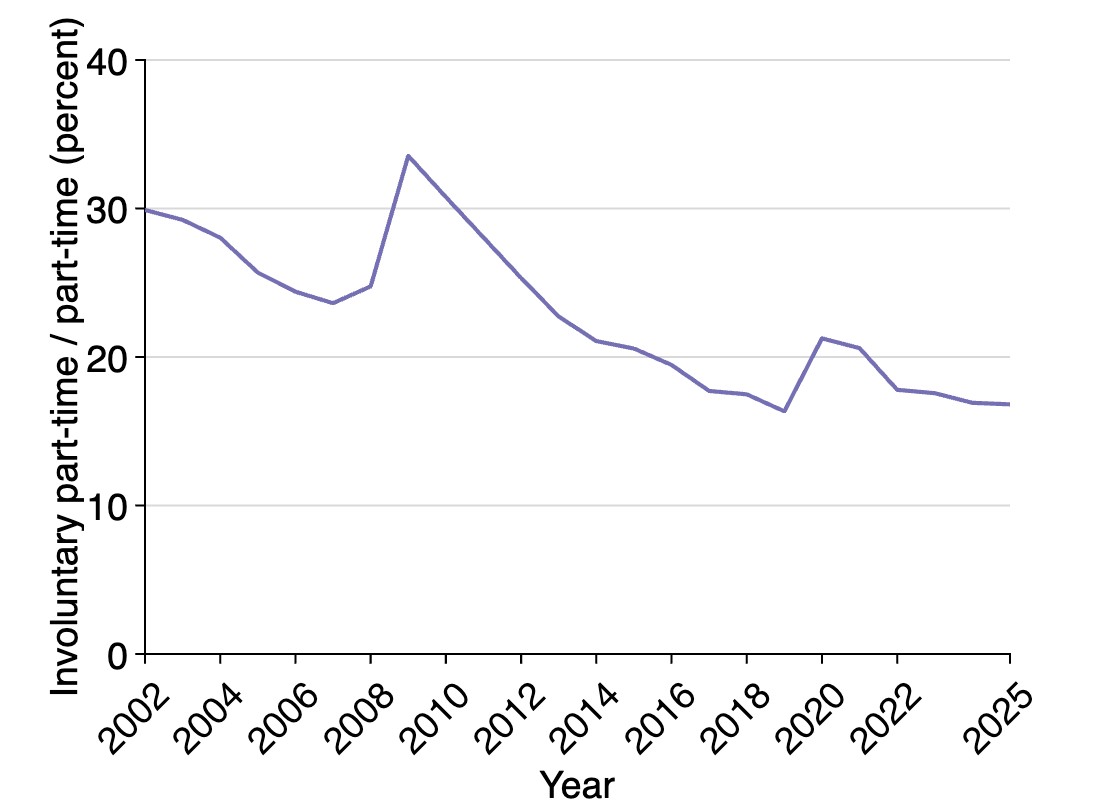}

\caption{Involuntary part-time employment as a share of total part-time employment in Japan, 2002--2025. \textit{Notes:} The figure reports the ratio of involuntary part-time employment to total part-time employment. Although part-time employment expanded substantially over the sample period, the involuntary share declined from roughly 30 percent to less than 20 percent. \textit{Source:} OECD Employment Database.}
\label{fig:japan_ipt_pt}
\end{figure}

As shown in Figure~\ref{fig:japan_pt_ft}, full-time employment remained relatively stable over 2002--2025, whereas part-time employment increased steadily. At the same time, the nature of part-time employment also changed. Figure~\ref{fig:japan_ipt_share} shows that although total part-time employment expanded substantially, the share of involuntary part-time workers declined from roughly 30 percent to less than 20 percent over the same period.

\begin{table}[H]
\centering
\caption{Summary Statistics of Full-Time and Part-Time Employment}
\label{tab:pt_share}
\begin{tabular}{ccccc}
\hline
Statistic & Full-time ($n_F$) & Part-time ($n_P$) & $\alpha^{P}$ & $1-\alpha^{P}$ \\
\hline
Mean & 48.96 million & 13.75 million & 0.781 & 0.219 \\
SD   & 1.31 million  & 2.15 million  & 0.971 & 0.029 \\
Min  & 46.26 million & 10.97 million & 0.823 & 0.177 \\
Max  & 51.13 million & 17.25 million & 0.738 & 0.262 \\
N    & 24            & 24            & 24    & 24    \\
\hline
\end{tabular}

\vspace{0.2cm}
\footnotesize
\textit{Notes}: $n_F$ denotes full-time employment, $n_P$ denotes total part-time employment, and $n_P/(n_F+n_P)$ is the share of part-time employment in total employment. The sample contains 24 annual observations from 2002 to 2025. \textit{Source}: OECD Database.
\end{table}

\begin{table}[H]
\centering
\caption{Summary Statistics of Full-Time and Involuntary Part-Time Employment}
\label{tab:ipt_summary}
\begin{tabular}{ccccc}
\hline
Statistic & Full-time ($n_F$) & Involuntary part-time ($n_{IP}$) & $\alpha^{IP}$ & $1-\alpha^{IP}$ \\
\hline
Mean & 48.96 million & 3.06 million & 0.941 & 0.059 \\
SD   & 1.32 million  & 0.38 million & 0.007 & 0.007 \\
Min  & 46.26 million & 2.51 million & 0.921 & 0.049 \\
Max  & 51.13 million & 4.16 million & 0.951 & 0.079 \\
N    & 24            & 24           & 24    & 24    \\
\hline
\end{tabular}

\vspace{0.2cm}
\footnotesize
\textit{Notes}: $n_F$ denotes full-time employment, $n_{IP}$ denotes involuntary part-time employment, and $\alpha = n_F/(n_F+n_{IP})$. The sample contains 24 annual observations. \textit{Source}: OECD database. 
\end{table}

\begin{table}[H]
\centering
\caption{Summary Statistics of Part-Time and Involuntary Part-Time Employment}
\label{tab:invpt_share}
\begin{tabular}{cccc}
\hline
Statistic & Part-time ($n_P$) & Involuntary part-time ($n_{IP}$) & $n_{IP}/n_P$ \\
\hline
Mean & 13.76 million & 3.06 million & 0.229 \\
SD   & 2.15 million  & 0.38 million & 0.051 \\
Min  & 10.97 million & 2.51 million & 0.163 \\
Max  & 17.25 million & 4.16 million & 0.335 \\
N    & 24            & 24           & 24    \\
\hline
\end{tabular}

\vspace{0.2cm}
\footnotesize
\textit{Notes}: $n_P$ denotes total part-time employment, $n_{IP}$ denotes involuntary part-time employment, and $n_{IP}/n_P$ is the share of part-time workers who report working part-time for economic reasons. The sample contains 24 annual observations. \textit{Source}: OECD database. 
\end{table}

\begin{figure}[H]
\centering
\includegraphics[width=0.75\textwidth]{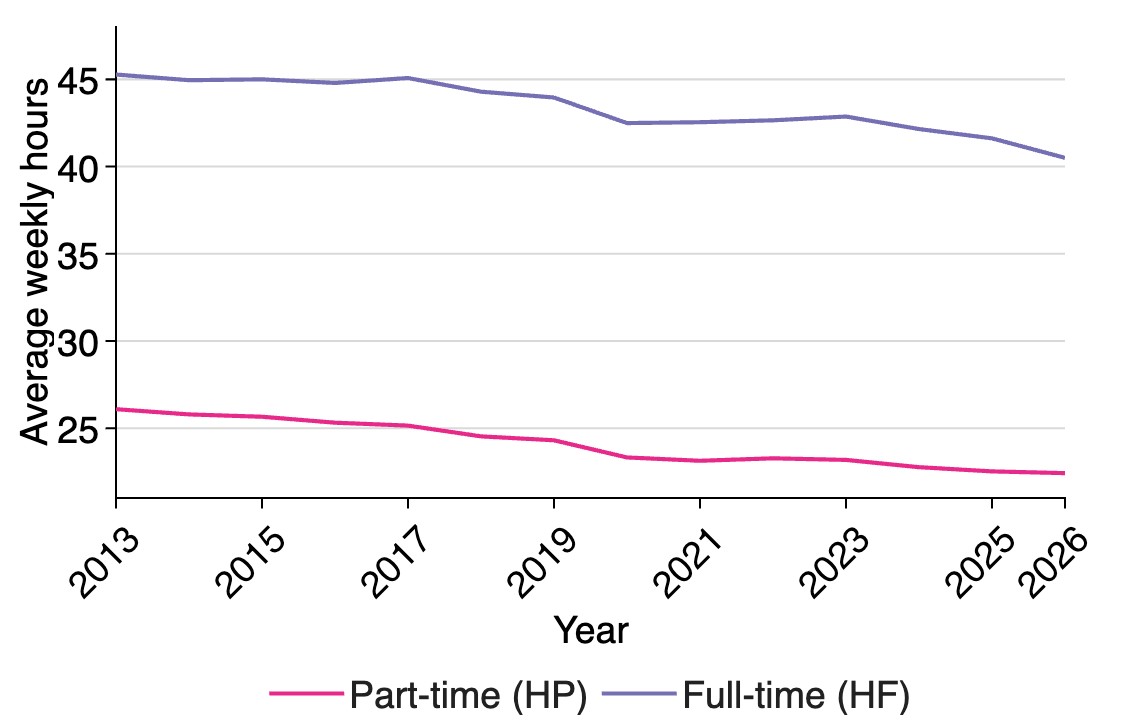}

\caption{Average weekly hours for part-time and full-time workers in Japan, 2013--2026. \textit{Notes:} The figure reports average weekly hours for part-time workers ($H_P$) and full-time workers ($H_F$). Part-time workers consistently work far fewer hours than full-time workers. Over the sample period, average weekly hours decline for both groups: full-time hours fall from about 45.5 to 40.5 hours per week, while part-time hours fall from about 26 to 22.5 hours per week, with a sharper decline for full-time workers around 2020. On average, $H_P = 24.2$ hours and $H_F = 43.6$ hours, implying $\gamma = H_P/H_F \approx 0.56$. \textit{Source:} Statistics Bureau of Japan, \emph{Labour Force Survey Detailed Tabulation}.}
\label{fig:japan_hours}
\end{figure}

\begin{table}[H]
\centering
\caption{Summary Statistics of Weekly Hours and Relative Hours in Japan}
\label{tab:hours_summary}
\begin{tabular}{llll}
\hline
Statistic & $H_P$ & $H_F$ & $\gamma$ \\
\hline
Mean  & 24.19 & 43.59 & 0.555 \\
SD    & 1.30  & 1.90  & 0.017 \\
Min   & 21.80 & 37.40 & 0.522 \\
Max   & 26.70 & 46.30 & 0.600 \\
N     & 160   & 160   & 160   \\
\hline
\end{tabular}

\vspace{0.2cm}
\footnotesize
\textit{Notes}: $H_P$ denotes average weekly hours worked by part-time employees, $H_F$ denotes average weekly hours worked by full-time employees, and $\gamma = H_P/H_F$ is the relative-hours parameter used in the calibration. The sample consists of monthly observations from January 2013 to April 2026. \textit{Source}: Statistics Bureau of Japan, \emph{Labour Force Survey Detailed Tabulation}.
\end{table}

The calibration of the part-time specification uses two additional inputs. First, I calibrate the relative-hours parameter, $\gamma$, using average weekly hours worked by full-time and part-time workers. Figure~\ref{fig:japan_hours} shows that part-time workers consistently work substantially fewer hours than full-time workers in Japan. Table~\ref{tab:hours_summary} reports the corresponding summary statistics. From January 2013 to April 2026, part-time workers worked 24.19 hours per week on average, while full-time workers worked 43.59 hours. I therefore set
\[
\gamma=\frac{H_P}{H_F}=0.56,
\]
so that a part-time worker supplies about 56 percent as many weekly hours as a full-time worker.

Second, I calibrate the employment-composition parameter using the full-time and part-time employment shares. Table~\ref{tab:pt_share} shows that, from 2002 to 2025, Japan had on average 48.96 million full-time workers and 13.75 million part-time workers. Part-time workers accounted for 21.9 percent of full-time and part-time employment. I therefore set the pre-2002 full-time share to
\[
\alpha = 1-0.219 \approx 0.78.
\]
For 2002--2025, I use the annual OECD time series of the full-time and part-time employment shares directly. This allows the part-time share to vary over time in the post-2002 period, while using the average post-2002 composition as the calibration for the earlier period. Although OECD total part-time employment data are available before 2002, I start the main Japan employment-share calibration in 2002 to align the total part-time and involuntary part-time series. Since the involuntary part-time series is not available before 2002, using 2002 as the common starting point keeps the two calibrations comparable. For the pre-2002 period, I therefore use the 2002--2025 average full-time share, $\alpha=0.78$, in the total part-time calibration.

This choice may overstate the historical importance of part-time employment if Japan's part-time share was lower before 2002. In that case, the pre-2002 part-time calibration would overstate the upward shift in efficient unemployment and the downward shift in efficient labor-market tightness before 2002. 

As a robustness check, I recompute the pre-2002 results using a higher full-time share, $\alpha=0.90$, which corresponds to a smaller historical part-time share. Appendix Figure~\ref{f:japan_alpha090_robustness} reports the robustness check using $\alpha=0.90$ before 2002. The alternative calibration produces a smaller pre-2002 part-time adjustment, as expected, but the post-2002 paths are unchanged because both calibrations use observed OECD part-time employment shares from 2002 onward. Importantly, the main episodes emphasized below---the period around the Global Financial Crisis, the aftermath of the 2011 earthquake, and the post-pandemic period---occur after 2002 and therefore use observed OECD part-time employment shares rather than the backward extrapolation.

The extended Beveridgean unemployment framework changes the interpretation of employment growth in Japan. A rise in employment mechanically lowers the unemployment rate, but it does not necessarily imply a proportional increase in effective labor input when a growing share of employment consists of part-time jobs. Since part-time workers supply fewer hours than full-time workers, part-time employment can reduce measured unemployment while still leaving effective labor utilization lower than suggested by headcount employment.

\begin{figure}[H]
\centering

\subcaptionbox{Labor-market tightness without part-time (full-time only), Japan}
{\includegraphics[width=0.45\textwidth]{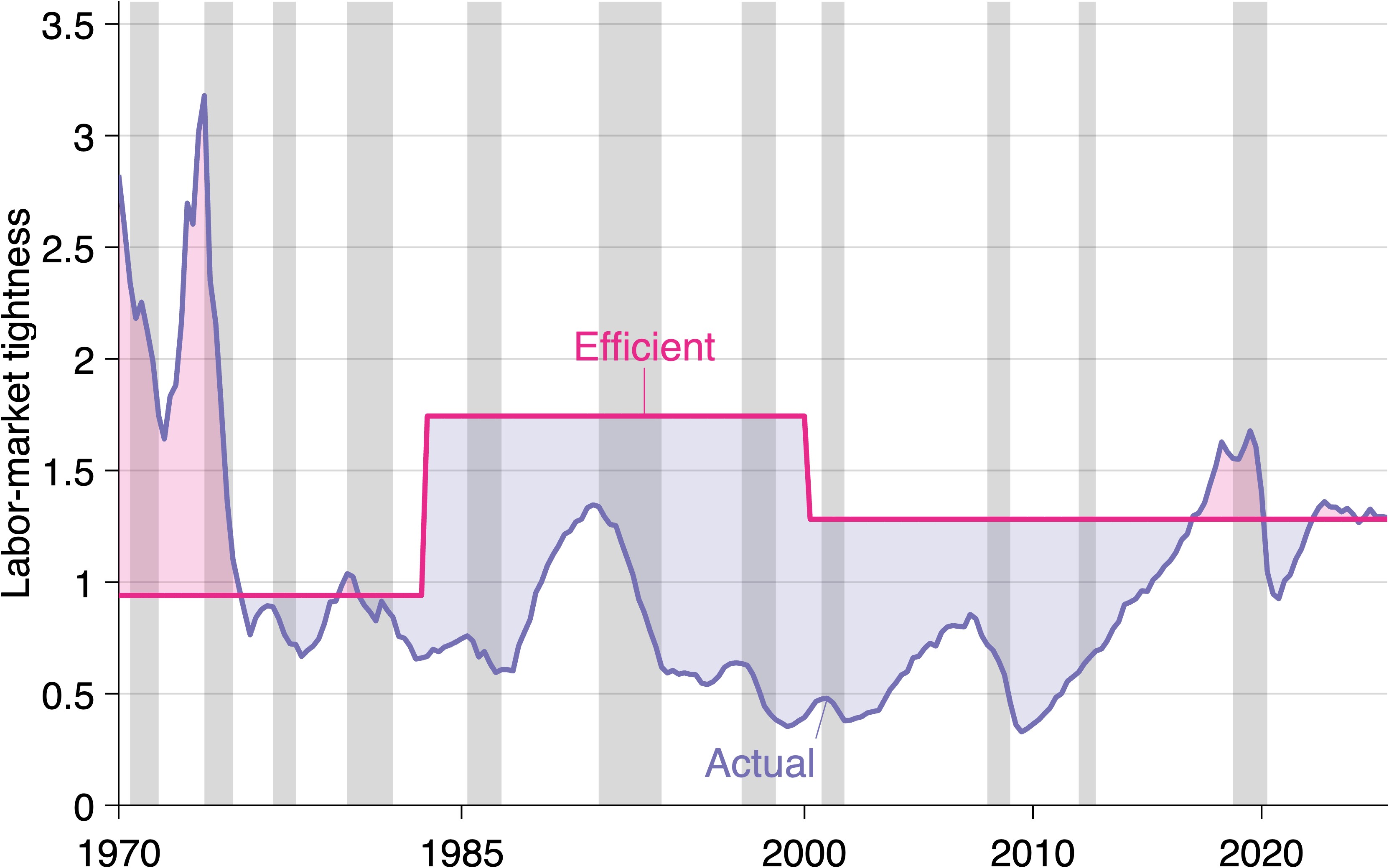}}\hfill
\subcaptionbox{Labor-market tightness with total part-time, Japan}
{\includegraphics[width=0.45\textwidth]{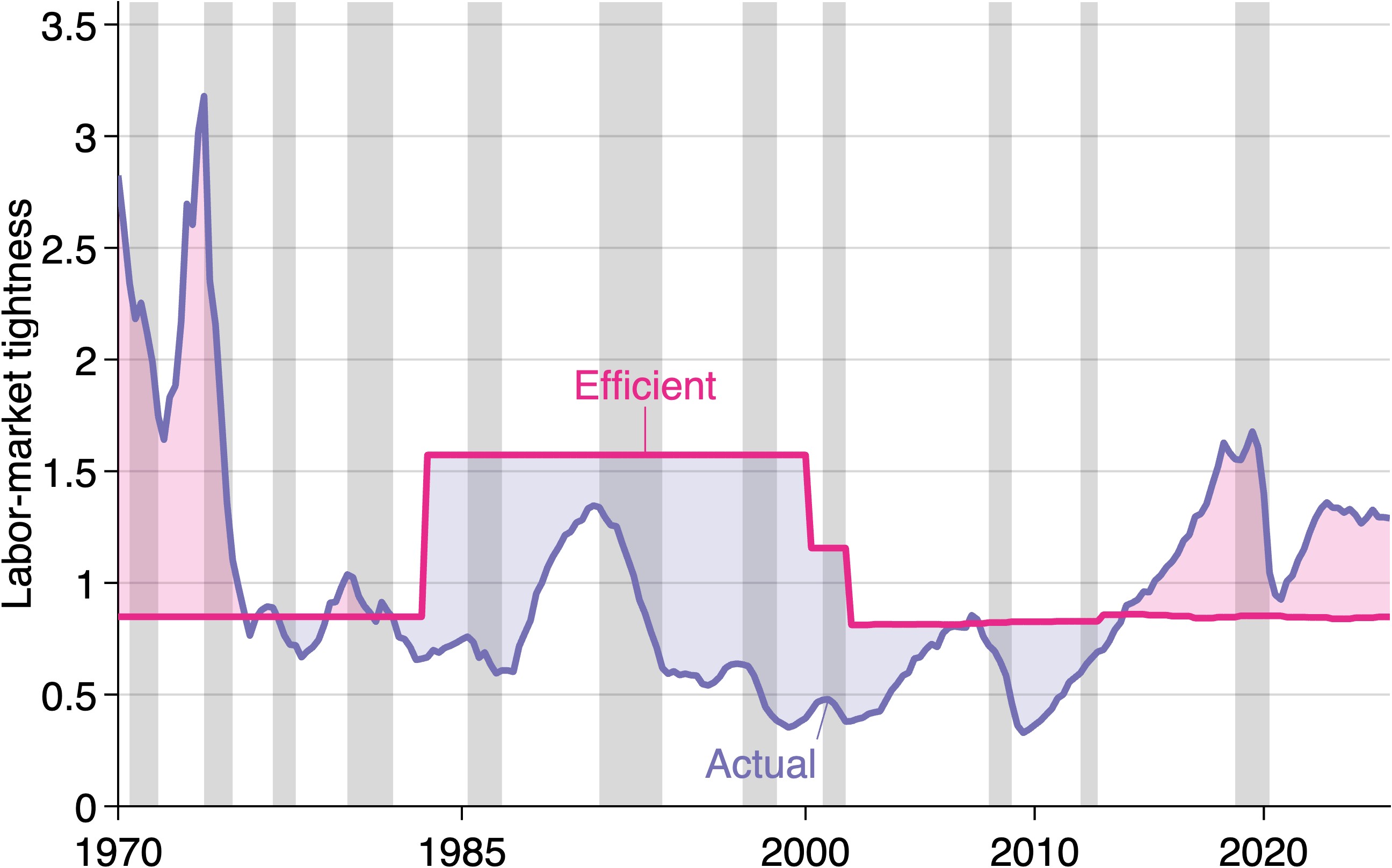}}\fspace

\subcaptionbox{Efficient unemployment rate without part-time (full-time only), Japan}
{\includegraphics[width=0.45\textwidth]{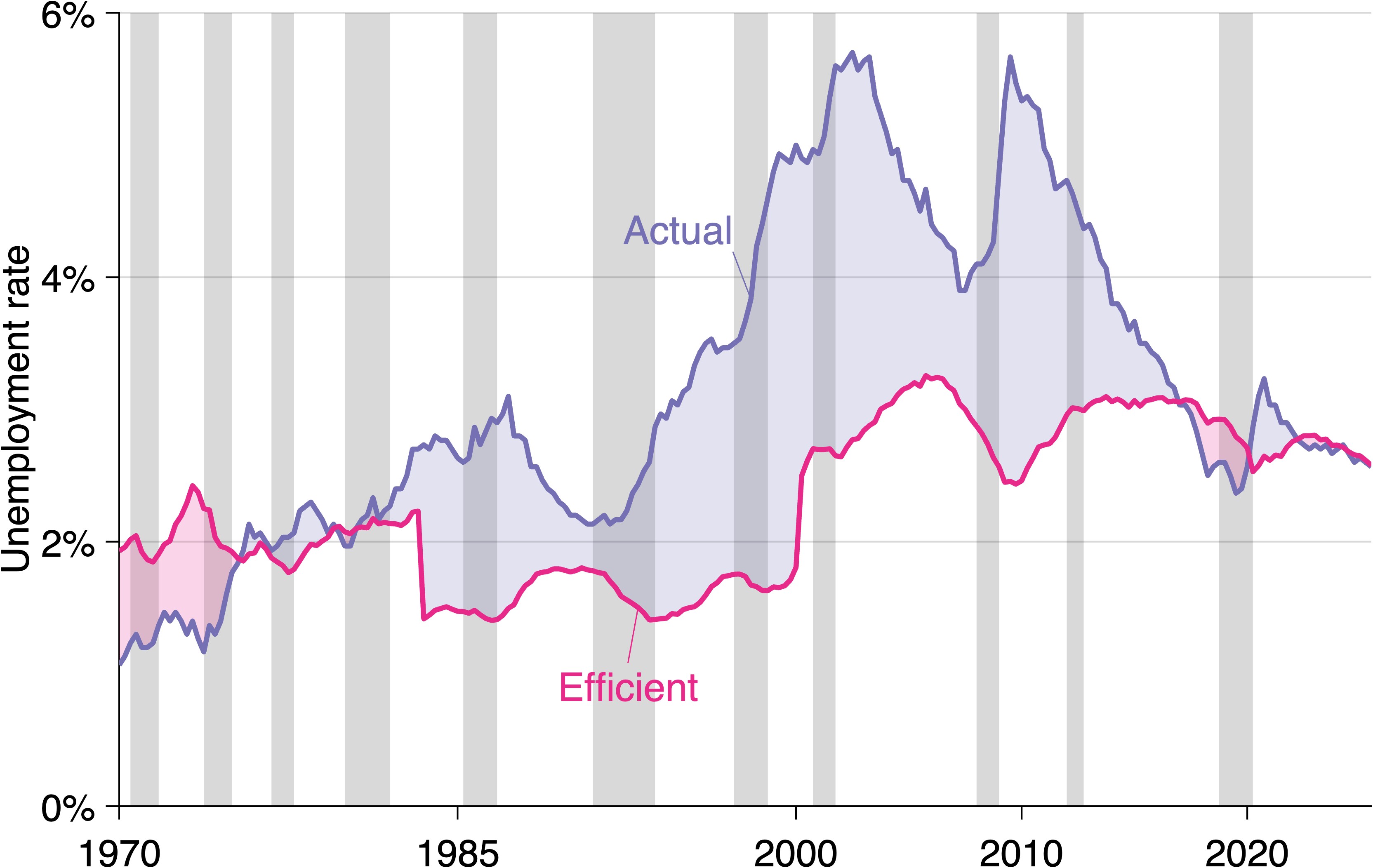}}\hfill
\subcaptionbox{Efficient unemployment rate with total part-time, Japan}
{\includegraphics[width=0.45\textwidth]{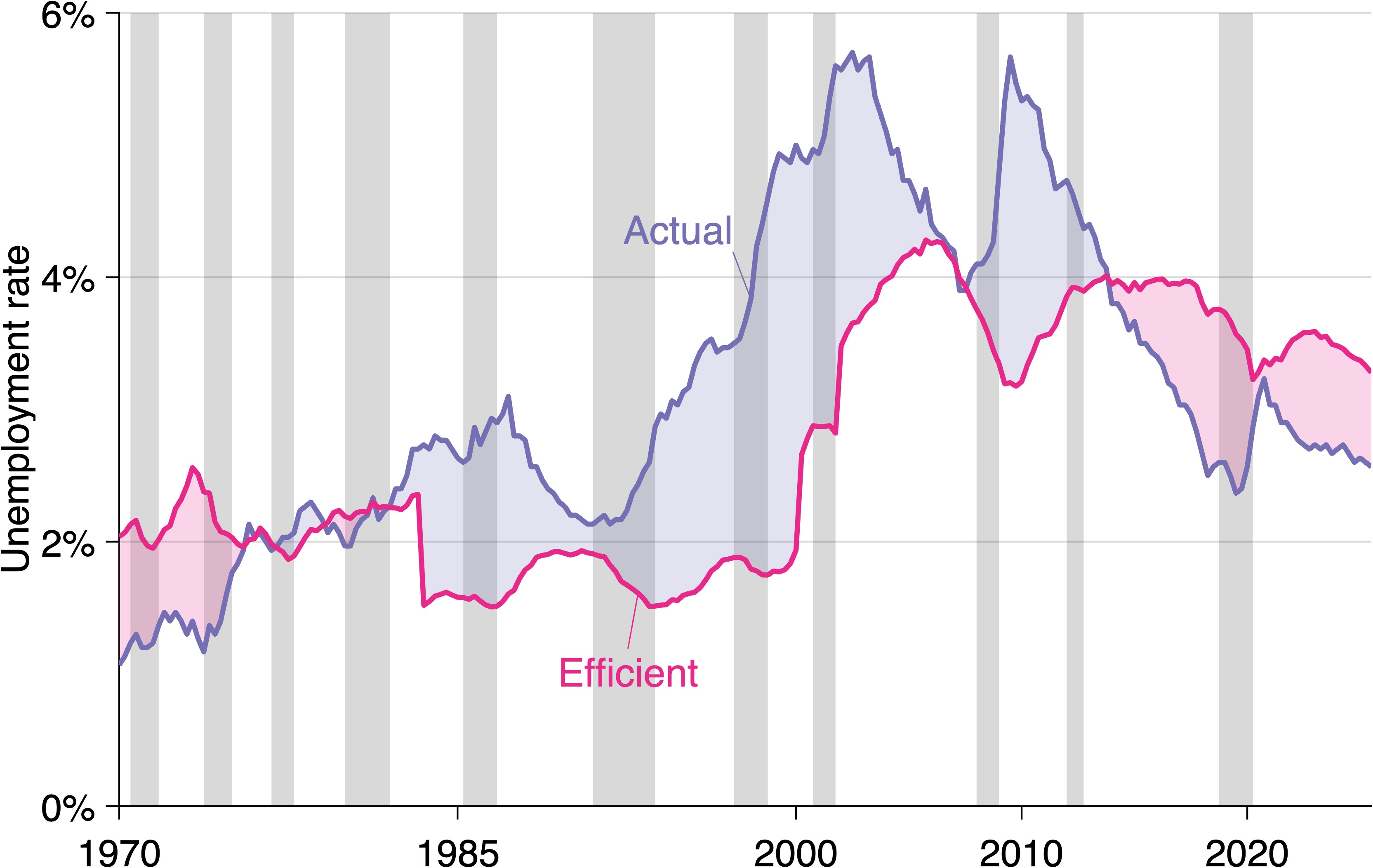}}\fspace

\subcaptionbox{Unemployment gap without part-time (full-time only), Japan}
{\includegraphics[width=0.45\textwidth]{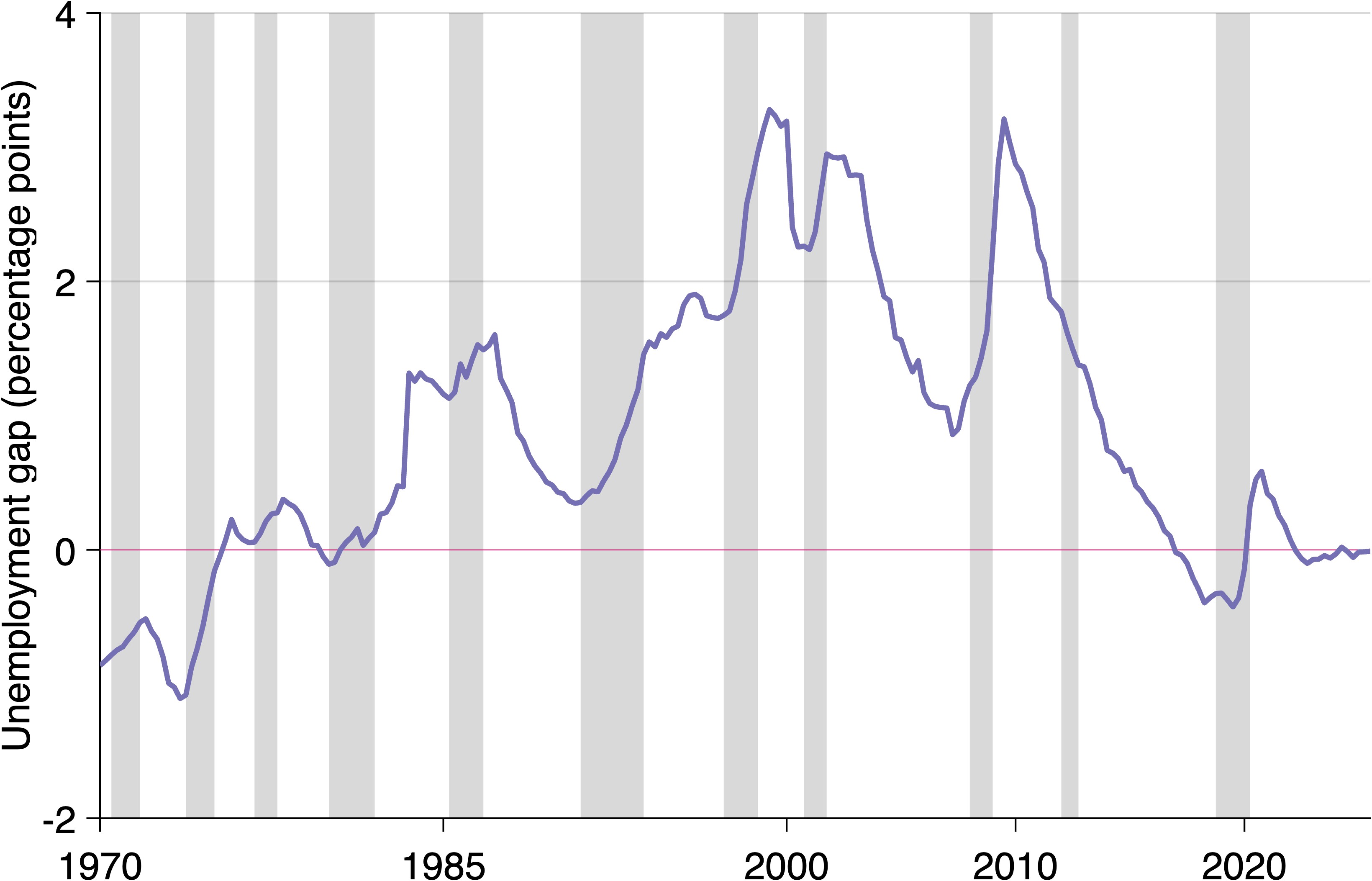}}
\hfill
\subcaptionbox{Unemployment gap with total part-time, Japan}
{\includegraphics[width=0.45\textwidth]{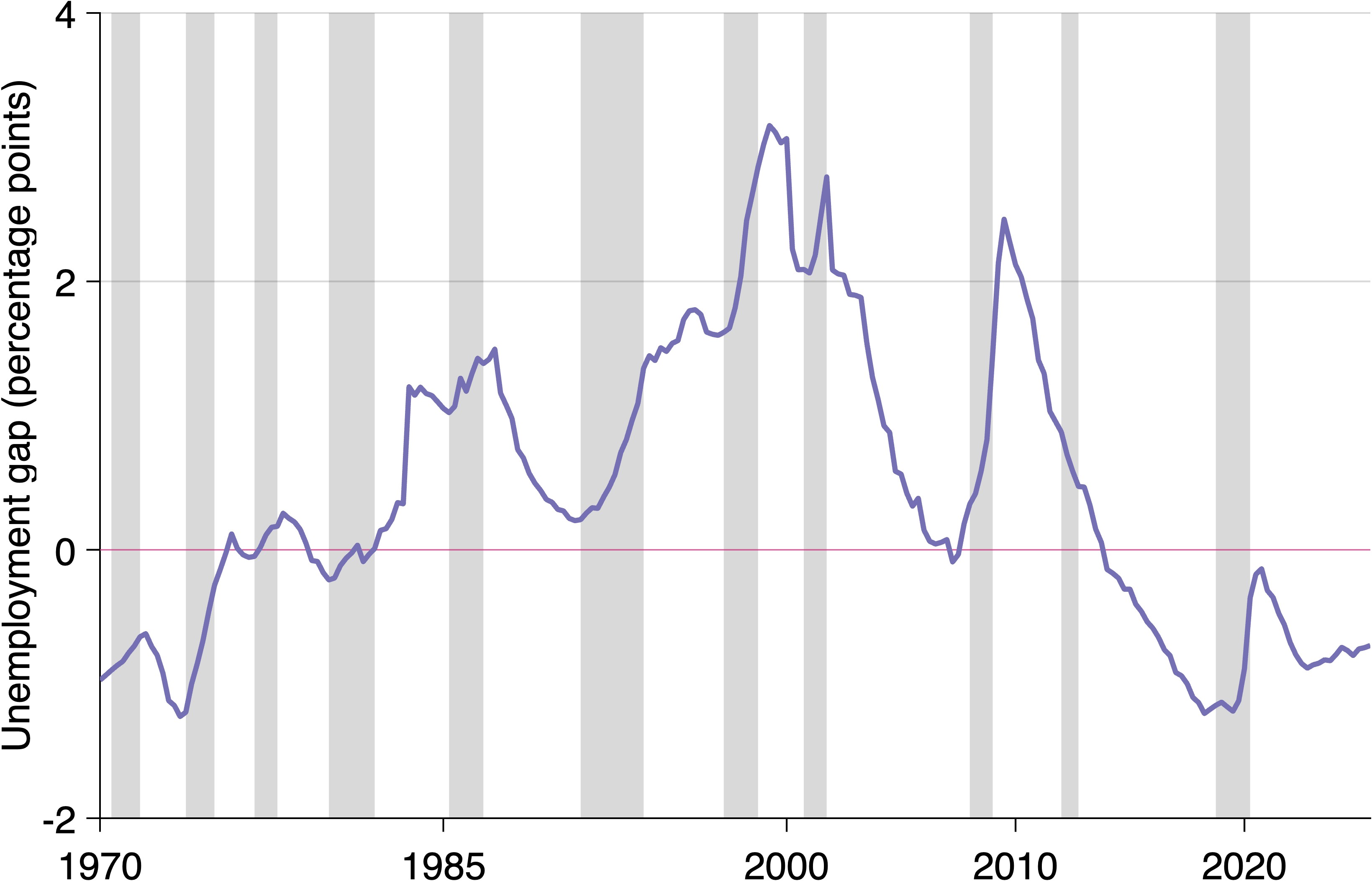}}\fspace

\caption{Efficient labor-market tightness, efficient unemployment rate, and unemployment gap in Japan, 1970--2025. Left column: full-time benchmark ($\alpha = 1$). Right column: total part-time employment incorporation ($\alpha = 0.78$ through 2001, part-time share from OECD thereafter). Shaded areas are recessions. Panels A--D compare efficient and actual series; panel E--F show the unemployment gap.}
\label{f:japan_efficiency}
\end{figure}

Figure~\ref{f:japan_efficiency} compares the efficient-unemployment results under the full-time benchmark and the part-time calibration. Under the full-time benchmark, average efficient labor-market tightness is 1.34 and the average efficient unemployment rate is 2.28 percent. After incorporating part-time employment, efficient tightness falls to 1.07 and the efficient unemployment rate rises to 2.71 percent. This shift reflects the lower effective labor input supplied by part-time workers. Because reducing unemployment increasingly adds lower-hours employment, the welfare gain from additional vacancy creation is smaller, which lowers efficient tightness and raises the efficient unemployment benchmark.

The unemployment gap also changes. Under the full-time benchmark, the average unemployment gap is 0.88 percentage points. After incorporating part-time employment, the average gap falls to 0.45 percentage points. These results suggest that ignoring part-time employment overstates efficient labor-market tightness and understates the efficient unemployment rate. In headcount terms, Japan may appear more slack than it is relative to the part-time-adjusted benchmark. Once the hours margin is taken into account, the efficient unemployment rate shifts upward and the measured unemployment gap becomes smaller.

\begin{figure}[H]
\centering

\subcaptionbox{Efficient unemployment rate, $\alpha = 1$}
{\includegraphics[width=0.48\textwidth]{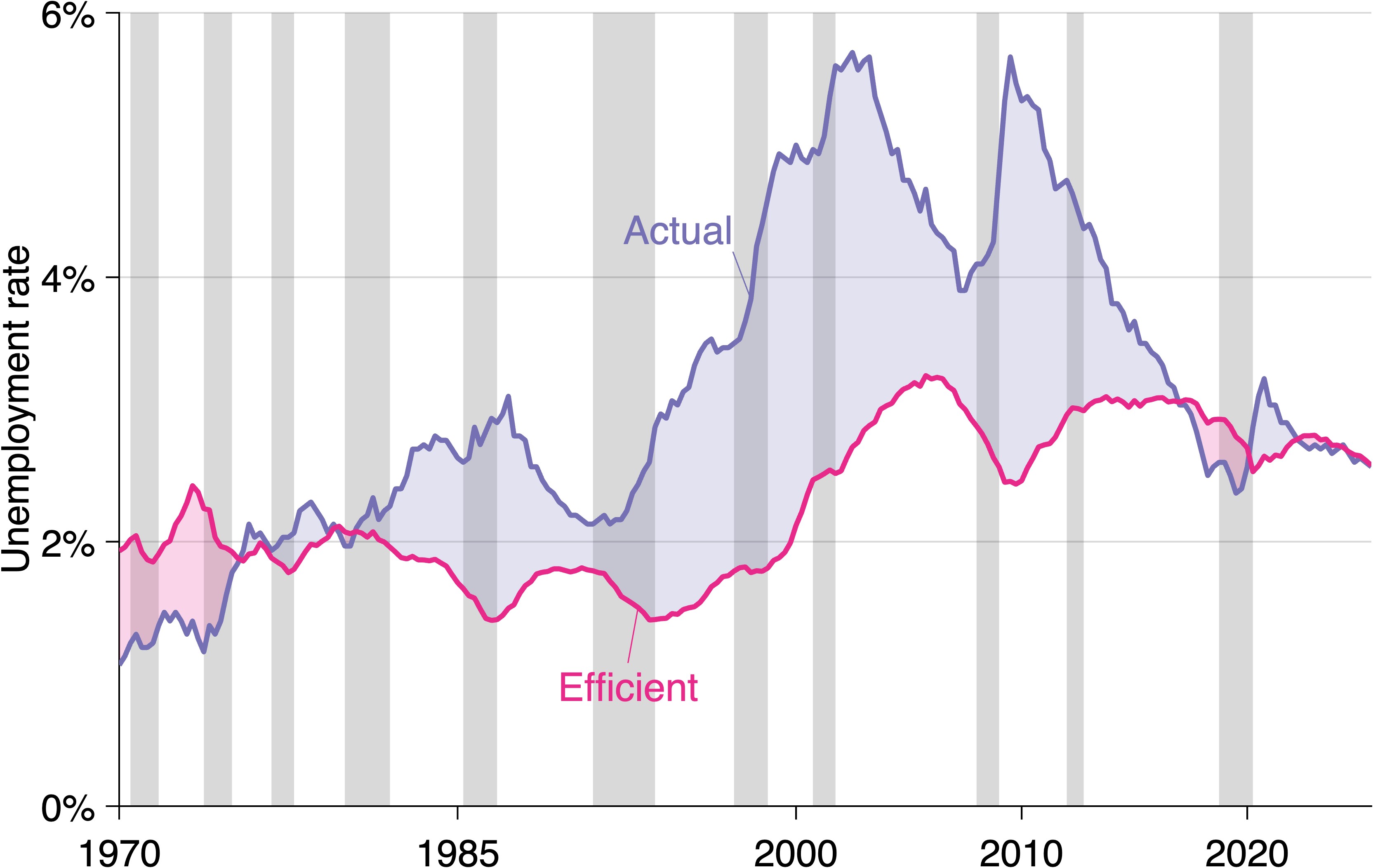}}\hfill
\subcaptionbox{Efficient unemployment rate, $\alpha = 0.78$}
{\includegraphics[width=0.48\textwidth]
{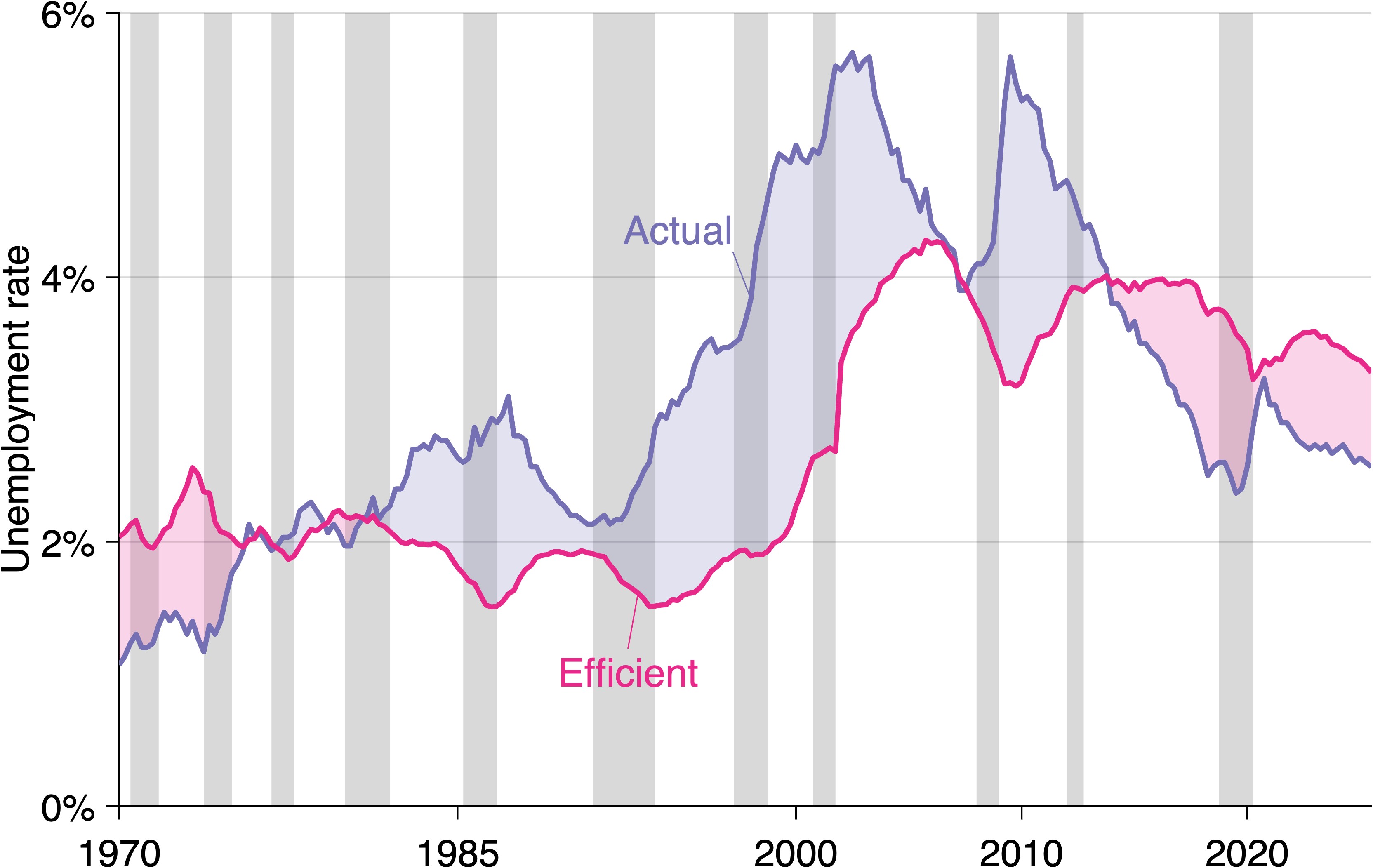}}\hfill
\vspace{0.1cm}
\subcaptionbox{Efficient unemployment rate, $\alpha = 0.94$}
{\includegraphics[width=0.48\textwidth]{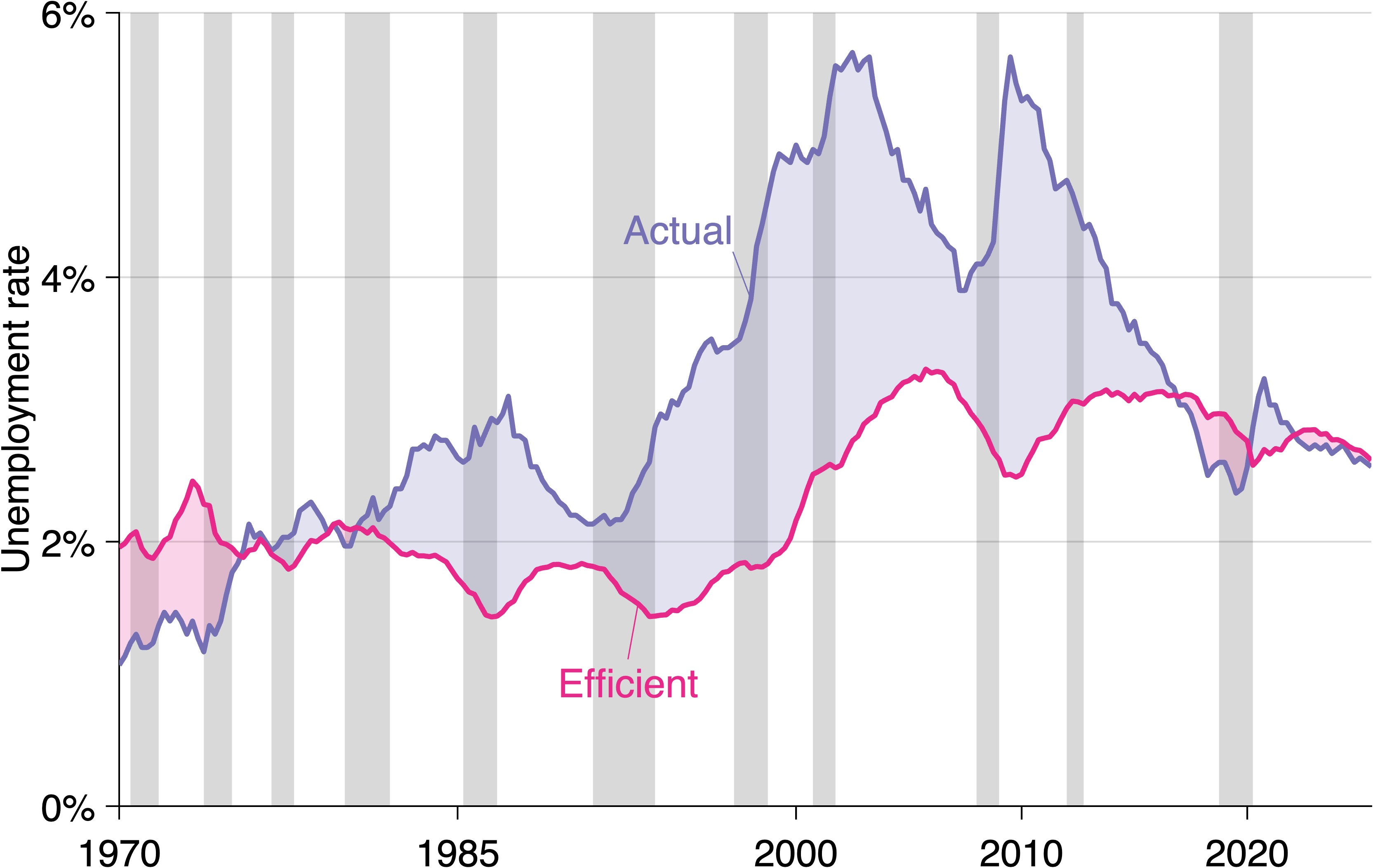}}

\caption{Efficient unemployment in Japan using smoothed Beveridge elasticities estimated over Bai--Perron regimes. The figure compares the full-time benchmark, with $\alpha=1$, to part-time calibrations. For total part-time employment, $\alpha=0.78$ through 2001 and follows the OECD part-time employment share from 2002 onward. The $\alpha=0.94$ case uses the corresponding involuntary part-time employment path.}
\label{f:japan_epsilon}
\end{figure}

The efficient unemployment series in Figure~\ref{f:japan_efficiency} inherits several discrete movements from the Bai--Perron estimation because the Beveridge elasticity is allowed to change across regimes. These movements are mechanical consequences of the piecewise elasticity estimates rather than abrupt changes in labor-market fundamentals. To verify that the main conclusion is not driven by these discontinuities, Figure~\ref{f:japan_epsilon} recomputes the efficient unemployment rate after smoothing the Beveridge elasticity around the estimated break dates. The smoothed series removes the sharp jumps while leaving the substantive comparison across the full-time, total part-time, and involuntary part-time calibrations largely unchanged. This confirms that the differences across calibrations are not driven by the discontinuous treatment of structural breaks.

I next examine whether this result reflects hidden underemployment or the broader hours margin associated with part-time work.

\subsection{2008 Financial Crisis}

Before the global financial crisis, the actual unemployment rate and the efficient unemployment rate move close together under the part-time calibration. This suggests that the Japanese labor market was near its efficient allocation during this period once differences in hours worked are taken into account. Without accounting for part-time employment, the same episode appears more slack because the full-time baseline treats all employed workers as supplying the same effective labor input. The part-time calibration raises the efficient unemployment benchmark by recognizing that a larger share of employment consists of lower-hours jobs. As a result, the observed unemployment rate lies closer to the efficient rate than under the full-time benchmark.

In this sense, the results suggest that labor-market conditions contain information about cyclical turning points. The part-time extension does not merely alter the level of efficient unemployment; it also changes the interpretation of several pre-recession episodes, indicating that employment composition may be relevant for recession monitoring and business-cycle assessment.

\subsection{2011 Earthquake Aftermath}
After the 2011 earthquake, the part-time calibration indicates a clear tightening of the Japanese labor market. Actual unemployment falls below efficient unemployment, implying that the labor market became tight relative to the efficient benchmark. This tightening is less visible under the full-time benchmark, which treats all employed workers as supplying the same effective labor input. Once part-time employment is incorporated, the lower hours supplied by part-time workers raise the efficient-unemployment benchmark, making the post-2011 tightening more apparent.

This pattern is consistent with the labor-market disruption following the earthquake. The shock generated job losses and regional mismatch, while reconstruction-related vacancies also increased. The part-time calibration therefore captures a feature of the period that is much less visible under the full-time benchmark. More broadly, the post-2011 period marks the beginning of a longer phase in which Japan’s labor market appears increasingly tight, a pattern that continues into the post-pandemic period.

\subsection{Pandemic Period}

Around the COVID-19 pandemic, the labor market initially moved closer to the efficient benchmark. Part of this convergence reflects a contraction in labor supply and labor-force participation. In the post-pandemic period, however, the Japanese labor market tightened again. Actual unemployment remains below the efficient unemployment rate, indicating that labor-market tightness increased relative to the efficient benchmark.

This tightening becomes much more apparent once part-time employment is incorporated into the analysis. Even though Japan appears inefficiently slack on average over the full sample, the periods surrounding the global financial crisis, the aftermath of the 2011 earthquake, and the post-pandemic recovery all exhibit episodes of labor-market tightness that persist into recent years. In several of these episodes, the full-time benchmark would continue to classify the labor market as relatively slack. Accounting for part-time employment therefore provides a more informative assessment of labor-market conditions.

\begin{figure}[H]
\centering

\subcaptionbox{Labor-market tightness: part-time  ($\alpha=0.78$) vs full-time ($\alpha=1$)}
{\includegraphics[width=0.48\textwidth]{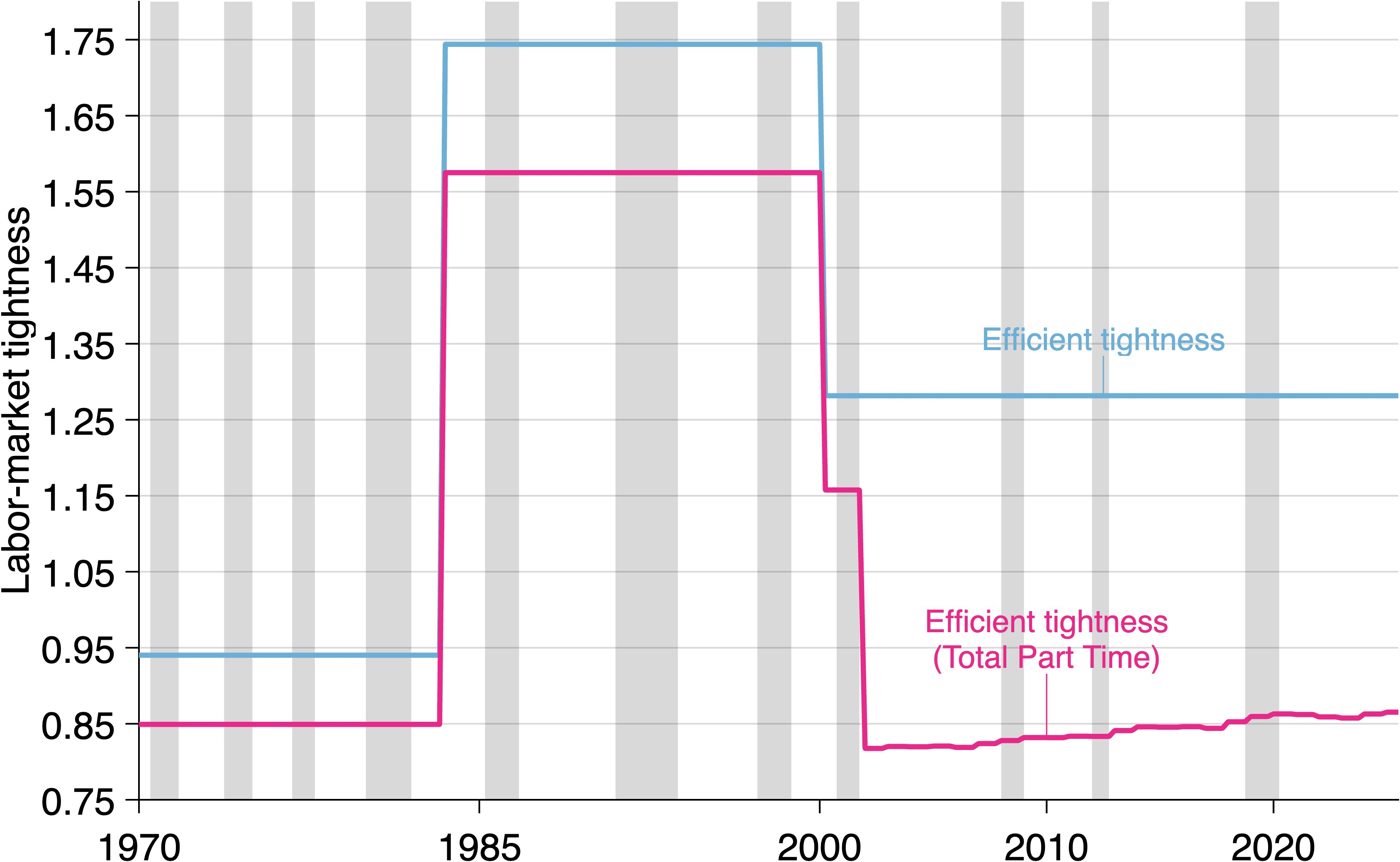}}\hfill
\subcaptionbox{Labor-market tightness: IPT ($\alpha=0.94$) vs full-time ($\alpha=1$)}
{\includegraphics[width=0.48\textwidth]{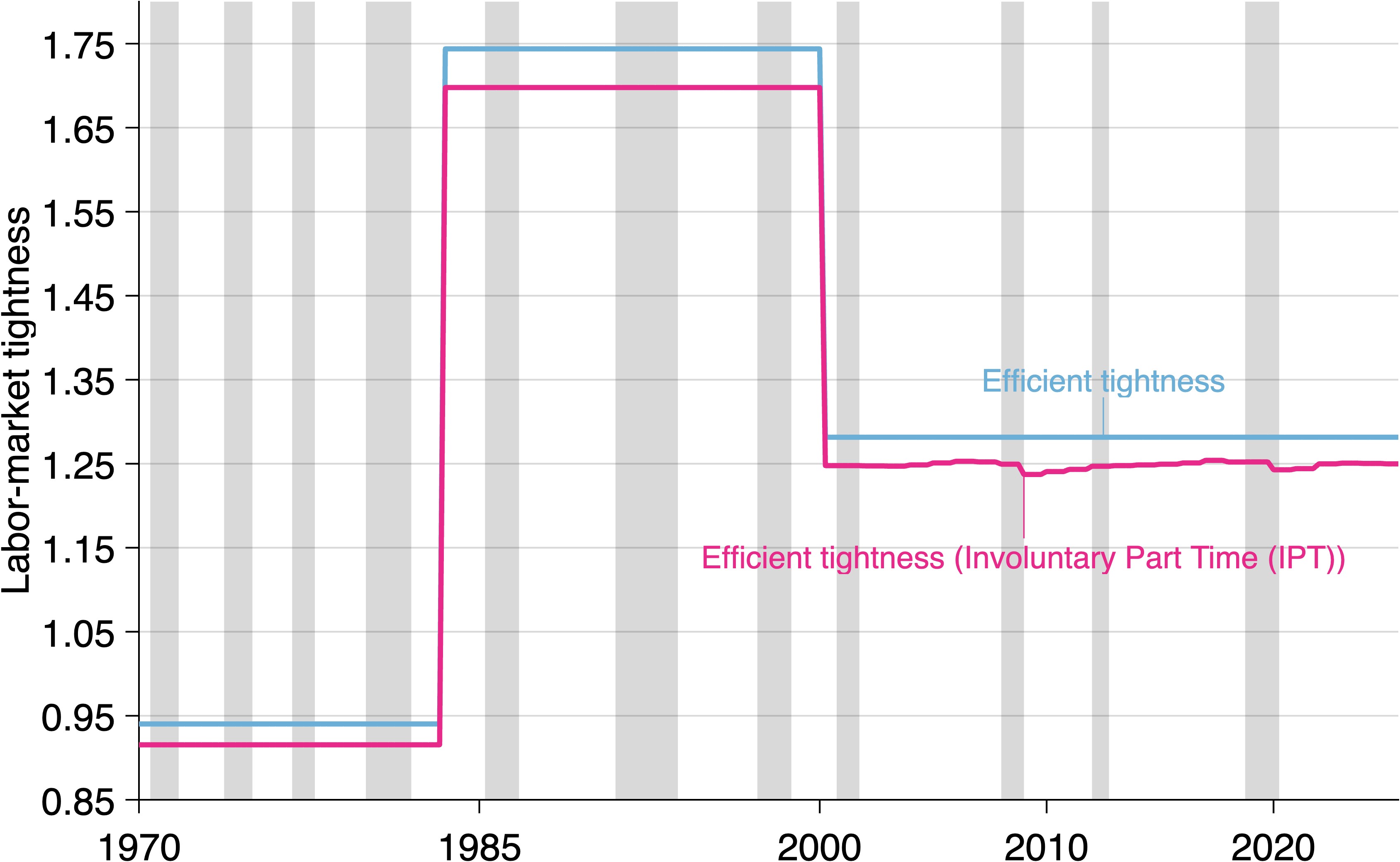}}

\vspace{0.25cm}

\subcaptionbox{Efficient unemployment rate: part-time  ($\alpha=0.78$) vs full-time ($\alpha=1$)}
{\includegraphics[width=0.48\textwidth]{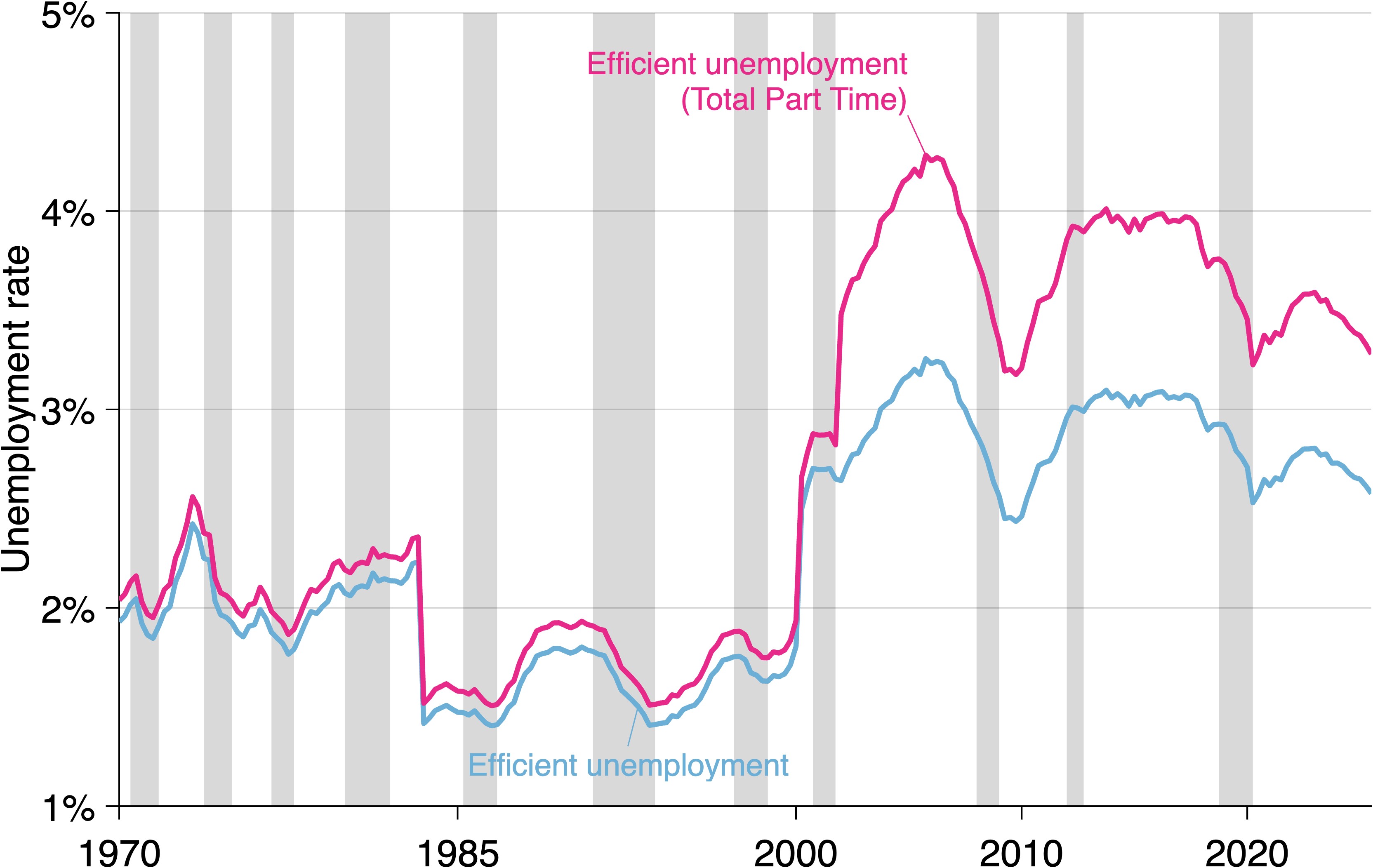}}\hfill
\subcaptionbox{Efficient unemployment rate: IPT($\alpha=0.94$) vs full-time ($\alpha=1$)}
{\includegraphics[width=0.48\textwidth]{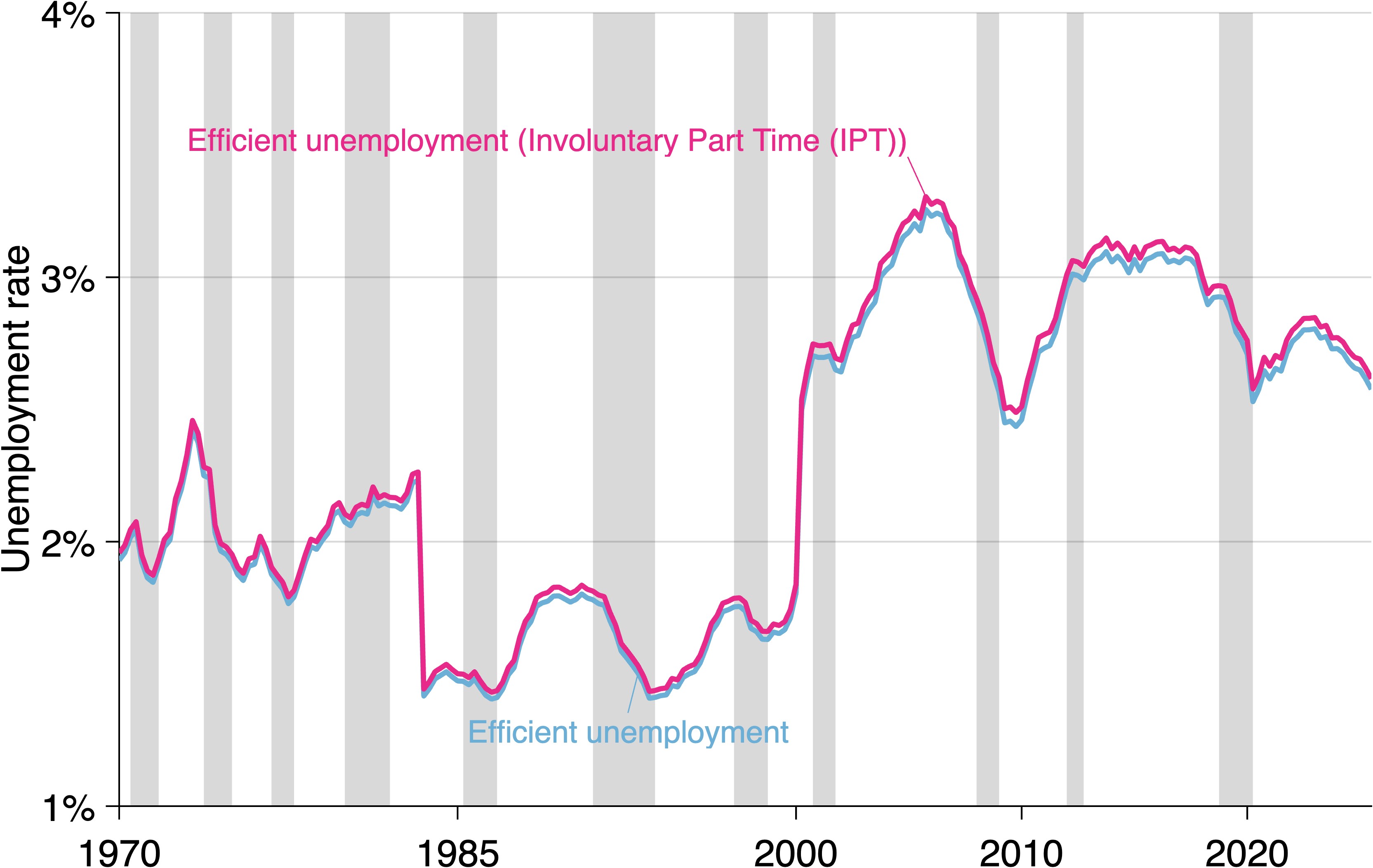}}

\vspace{0.25cm}

\subcaptionbox{Unemployment gap: part-time ($\alpha=0.78$) vs full-time ($\alpha=1$)}
{\includegraphics[width=0.48\textwidth]{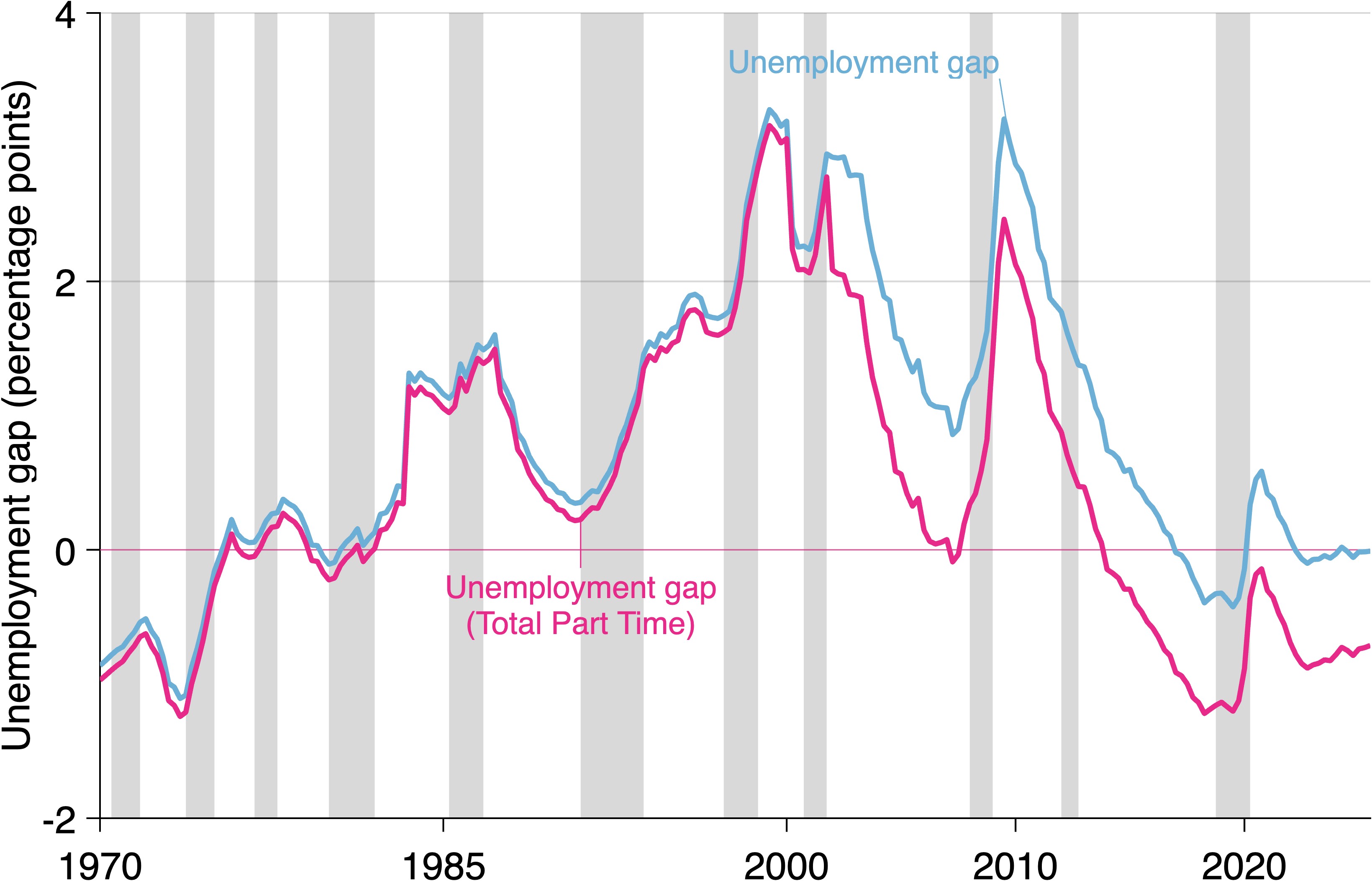}}\hfill
\subcaptionbox{Unemployment gap: IPT ($\alpha=0.94$) vs full-time ($\alpha=1$)}
{\includegraphics[width=0.48\textwidth]{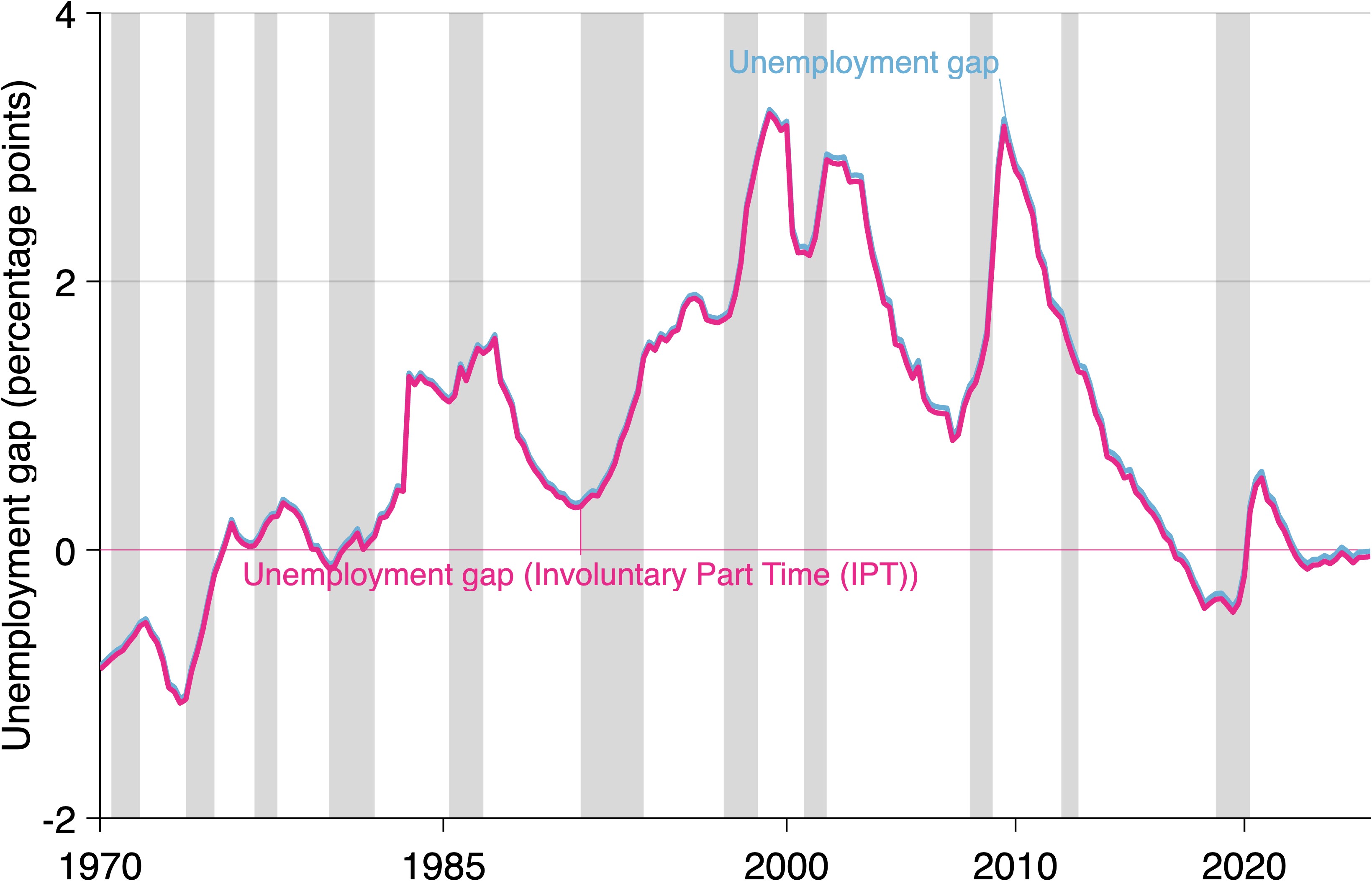}}

\caption{Labor-market tightness, efficient unemployment rate, and unemployment gap in Japan, 1970--2025. Each panel compares a part-time specification to the full-time benchmark ($\alpha=1$): Left column:  $\alpha=0.78$ through 2001 (OECD part-time share from 2002 thereafter); right column: $\alpha=0.94$ through 2001 (involuntary part-time share from OECD thereafter). }
\label{f:japan_comparison}
\end{figure}

\subsection{Involuntary part-time}

Because hours for involuntary part-time workers are not available for Japan, I set $\gamma^{IP}=\gamma$ in this exercise. The comparison between total part-time employment and involuntary part-time employment therefore mainly reflects differences in employment composition, captured by $\alpha$, rather than differences in relative hours. The involuntary part-time results should be interpreted as isolating the underemployment share of part-time employment under the same relative-hours assumption used for total part-time workers. To distinguish between the labor-utilization effect of total part-time employment and the underemployment effect of involuntary part-time employment, Figure~\ref{f:japan_comparison} compares two alternative specifications. The left column incorporates total part-time employment, using $\alpha^{P}=0.78$ before 2002 and the OECD part-time employment share from 2002 onward. The right column incorporates involuntary part-time employment, using $\alpha^{IP}=0.94$ before 2002 and the OECD involuntary part-time employment share from 2002 onward.

The results differ substantially across the two specifications. When total part-time employment is incorporated, efficient labor-market tightness falls noticeably, the efficient unemployment rate rises, and the unemployment gap becomes considerably smaller relative to the full-time benchmark. In contrast, incorporating IPT produces only minor changes in all three measures. The estimated paths of efficient labor-market tightness, efficient unemployment, and the unemployment gap remain very close to those obtained under the full-time specification.

The difference in magnitude reflects the relative size of the two groups. As reported in Tables~\ref{tab:pt_share}, \ref{tab:ipt_summary}, and \ref{tab:invpt_share}, part-time workers account for 21.9 percent of full-time and part-time employment on average, whereas involuntary part-time workers account for only 5.9 percent. Moreover, involuntary part-time workers represent only 22.9 percent of total part-time employment. Consequently, most part-time employment in Japan is not classified as involuntary.

These findings suggest that the quantitative importance of part-time employment in Japan stems primarily from the hours margin rather than from hidden underemployment. The large effect of total part-time employment indicates that employment growth increasingly occurs through lower-hours jobs. By contrast, the relatively small effect of IPT implies that hidden labor-market slack plays a much more limited role. Therefore, the rise of part-time employment in Japan appears to reflect a structural shift in labor utilization rather than a large increase in workers who are unable to obtain full-time employment.

This pattern also helps explain why Japan can have a large and growing part-time sector without a correspondingly large increase in hidden slack. Much of the rise in part-time employment appears to reflect demographic and labor-supply changes, including greater participation by women and older workers in shorter-hour jobs, rather than workers being unable to obtain full-time employment.

\subsection{Inverse-optimum sufficient statistics}

\begin{figure}[H]
\centering

\subcaptionbox{Beveridge elasticity}{
\includegraphics[width=0.48\textwidth]{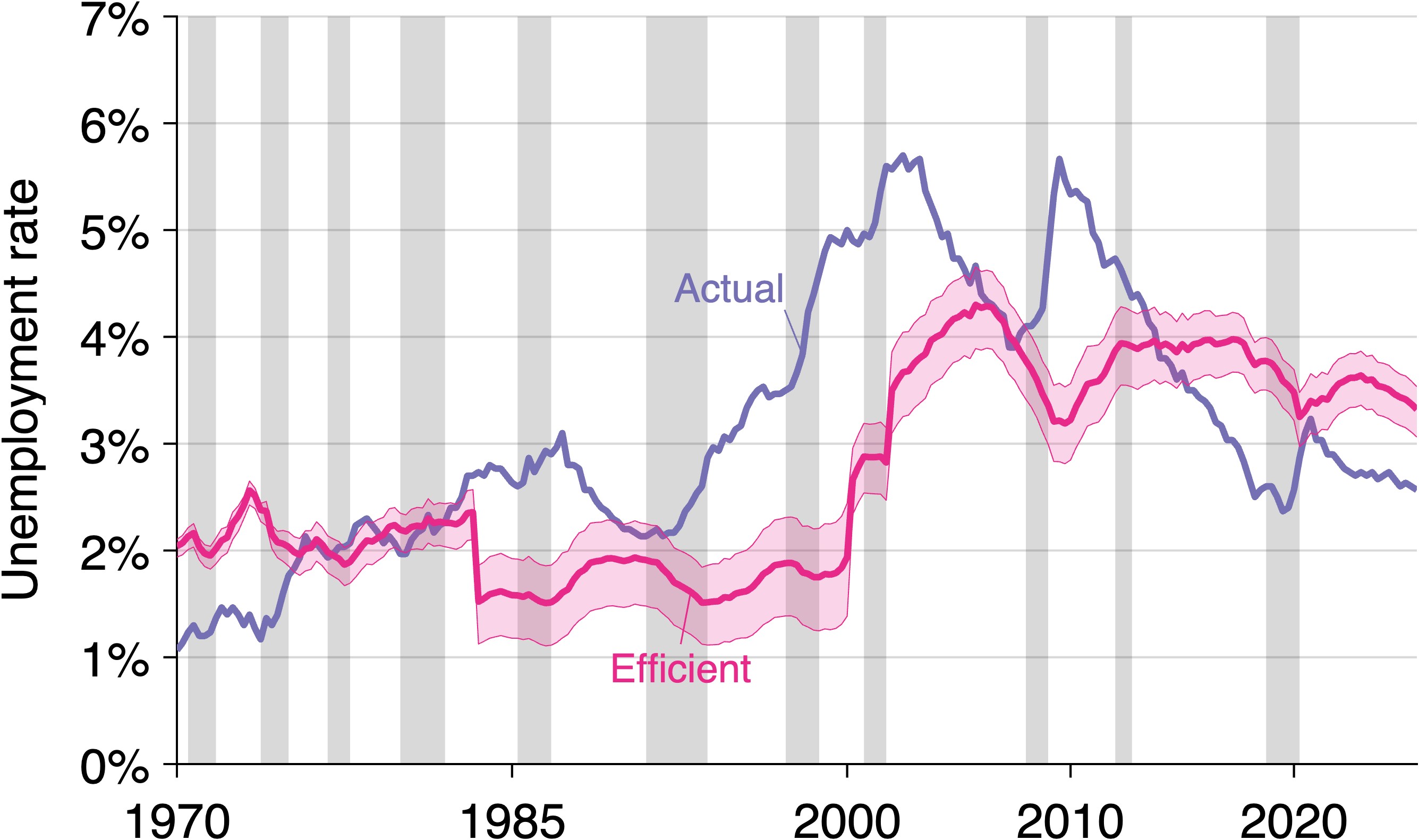}}
\hfill
\subcaptionbox{Social value of nonwork}{
\includegraphics[width=0.48\textwidth]{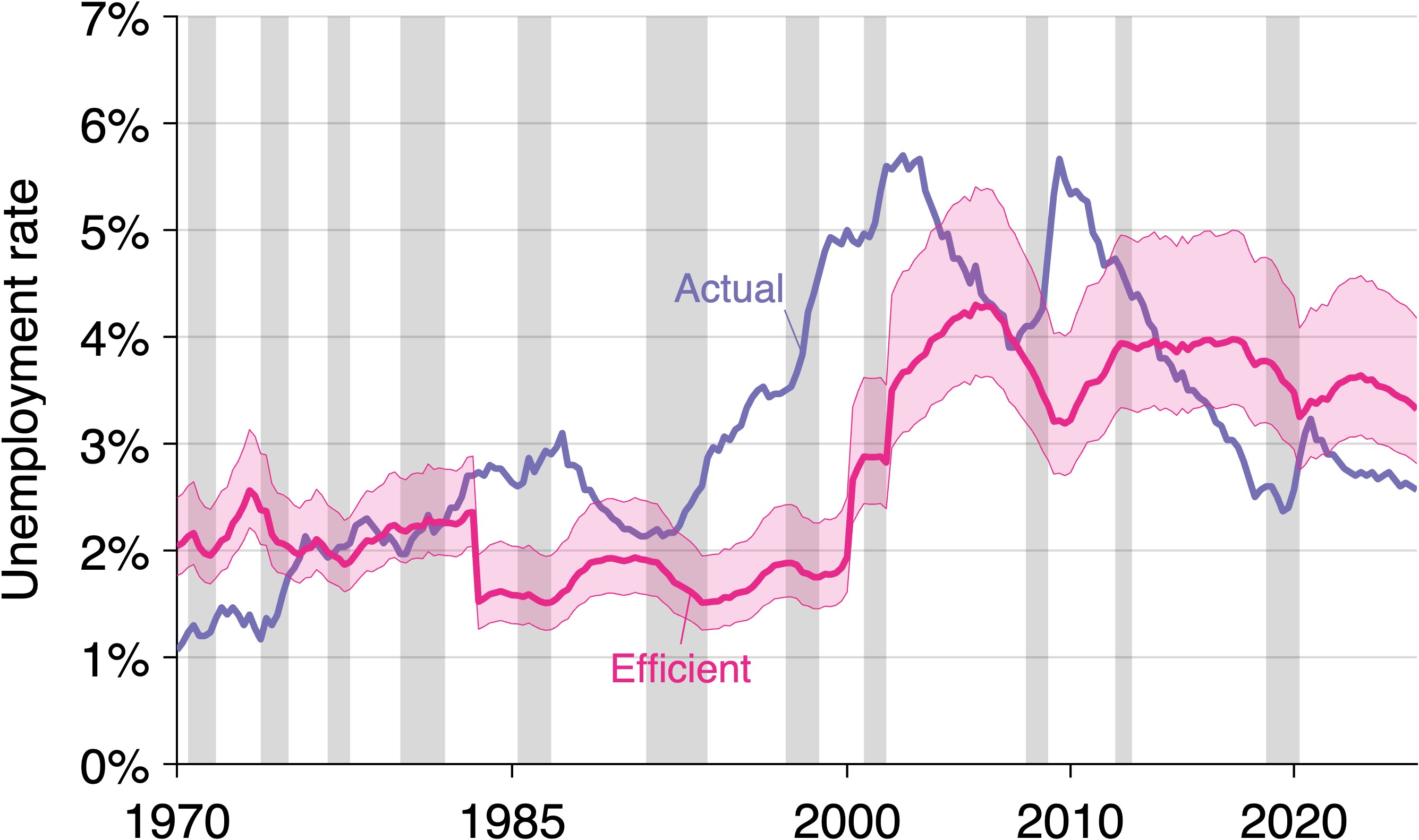}}

\vspace{0.3cm}

\subcaptionbox{Recruiting cost}{
\includegraphics[width=0.48\textwidth]{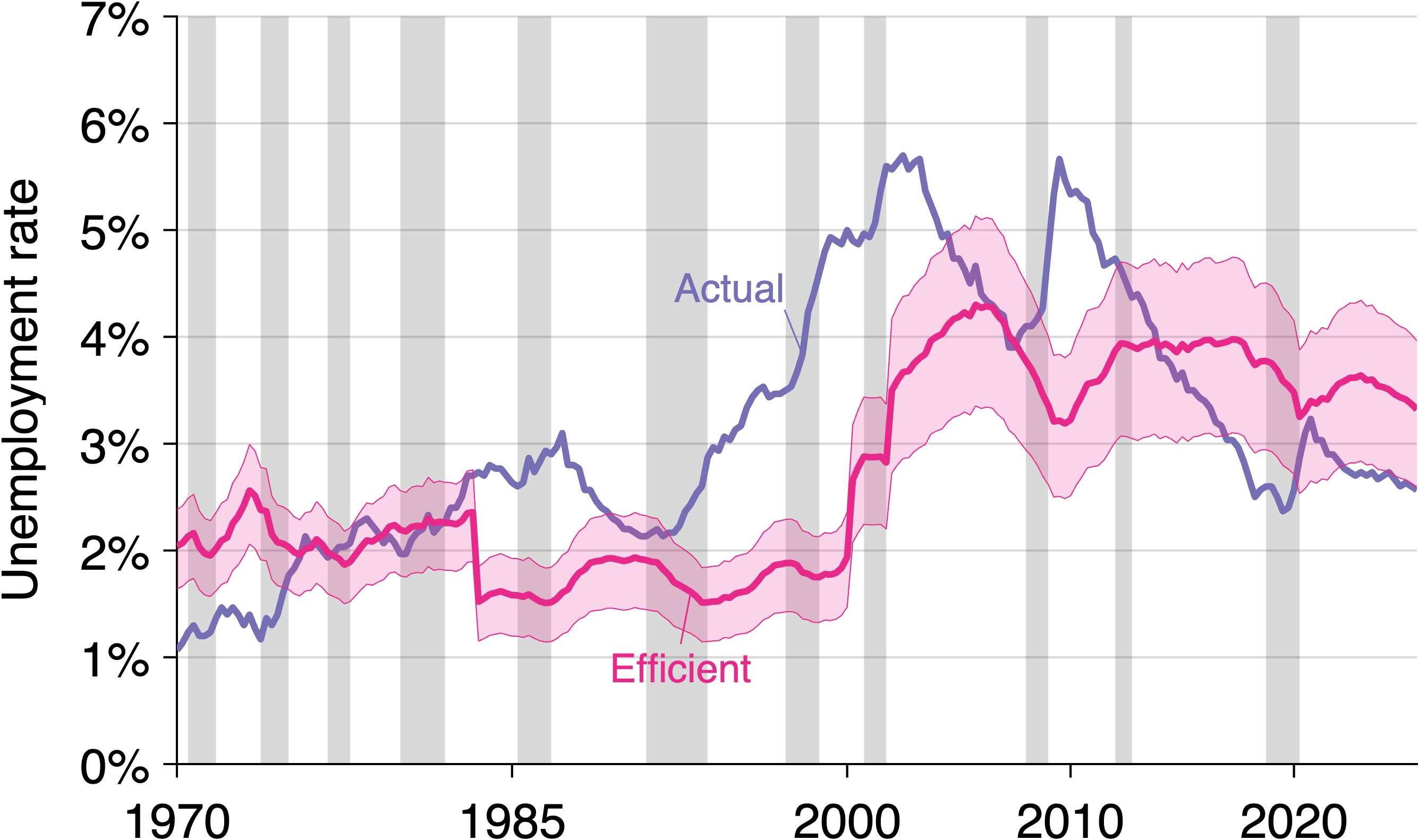}}

\caption{Efficient unemployment rates for alternative sufficient statistics in Japan, 1970--2025.}
\label{f:alter_jp}
\end{figure}

There remains uncertainty about the precise values of the sufficient statistics. Yet, Figure~\ref{f:alter_jp} shows that the efficient unemployment rate is remarkably stable across plausible calibrations of the Beveridge elasticity, the social value of nonwork, and recruiting costs. When the Beveridge elasticity is at the bottom (top) end of its 95\% confidence interval, the efficient unemployment rate tracks the baseline series but is on average 0.33 percentage point lower (0.28 percentage point higher); at no point does it deviate by more than 0.55 percentage point from the baseline calibration, and the full envelope spans only 1.1\% to 4.6\% (Table~\ref{tab:fig8a_summary}). The corresponding average shifts for alternative calibrations of the social value of nonwork are 0.41 and 0.69 percentage point, with a maximum deviation of 1.1 percentage points and an envelope of 1.3\%--5.4\% (Table~\ref{tab:fig8b_summary}); for recruiting costs, they are 0.59 and 0.52 percentage point, with a maximum deviation of 0.9 percentage point and an envelope of 1.1\%--5.1\% (Table~\ref{tab:fig8c_summary}). Although alternative parameter values shift the level of efficient unemployment, they do not materially alter the qualitative assessment of labor-market conditions. 

\begin{figure}[H]
\centering

\subcaptionbox{Inverse-optimum Beveridge elasticity}{
\includegraphics[width=0.48\textwidth]{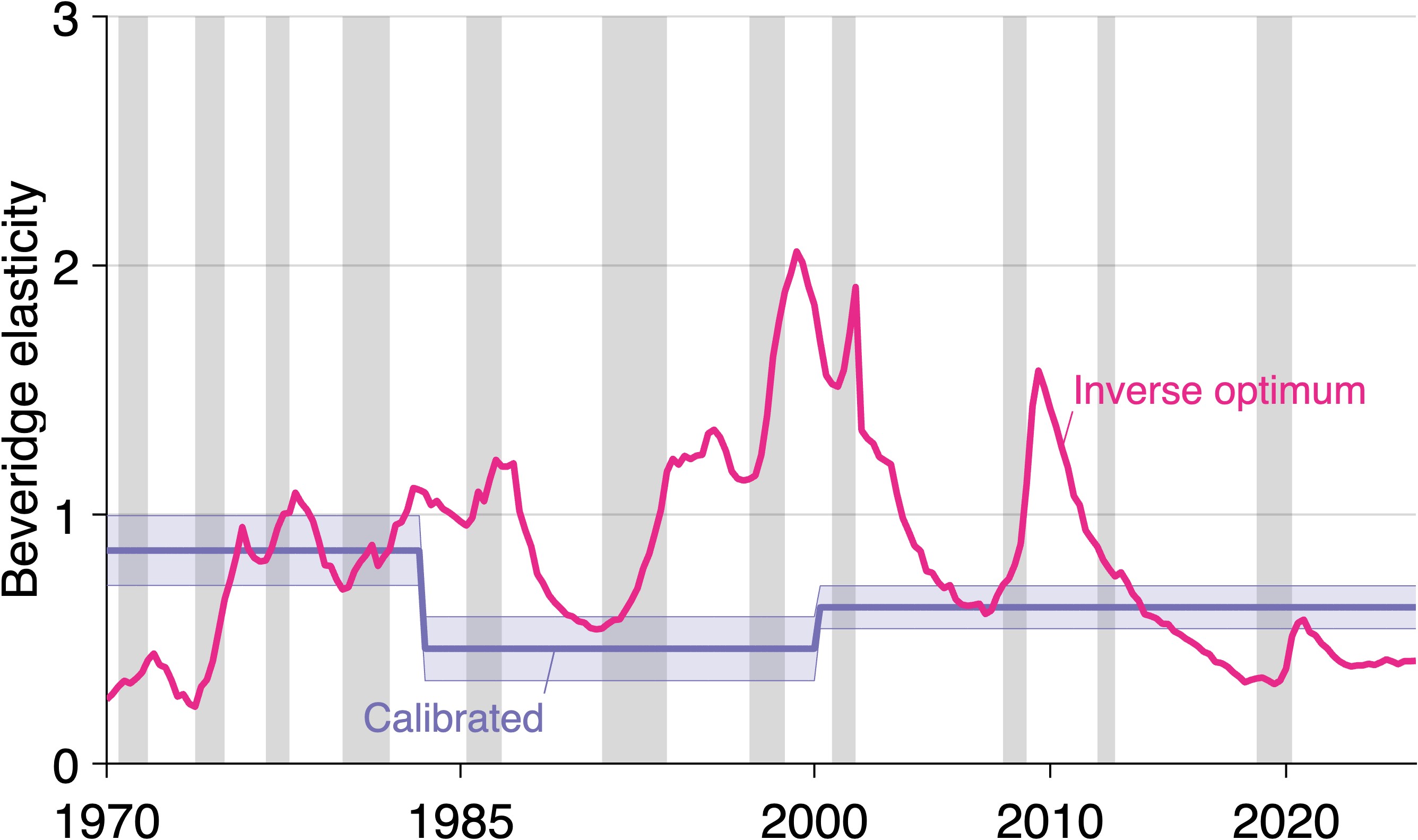}}
\hfill
\subcaptionbox{Inverse-optimum social value of nonwork}{
\includegraphics[width=0.48\textwidth]{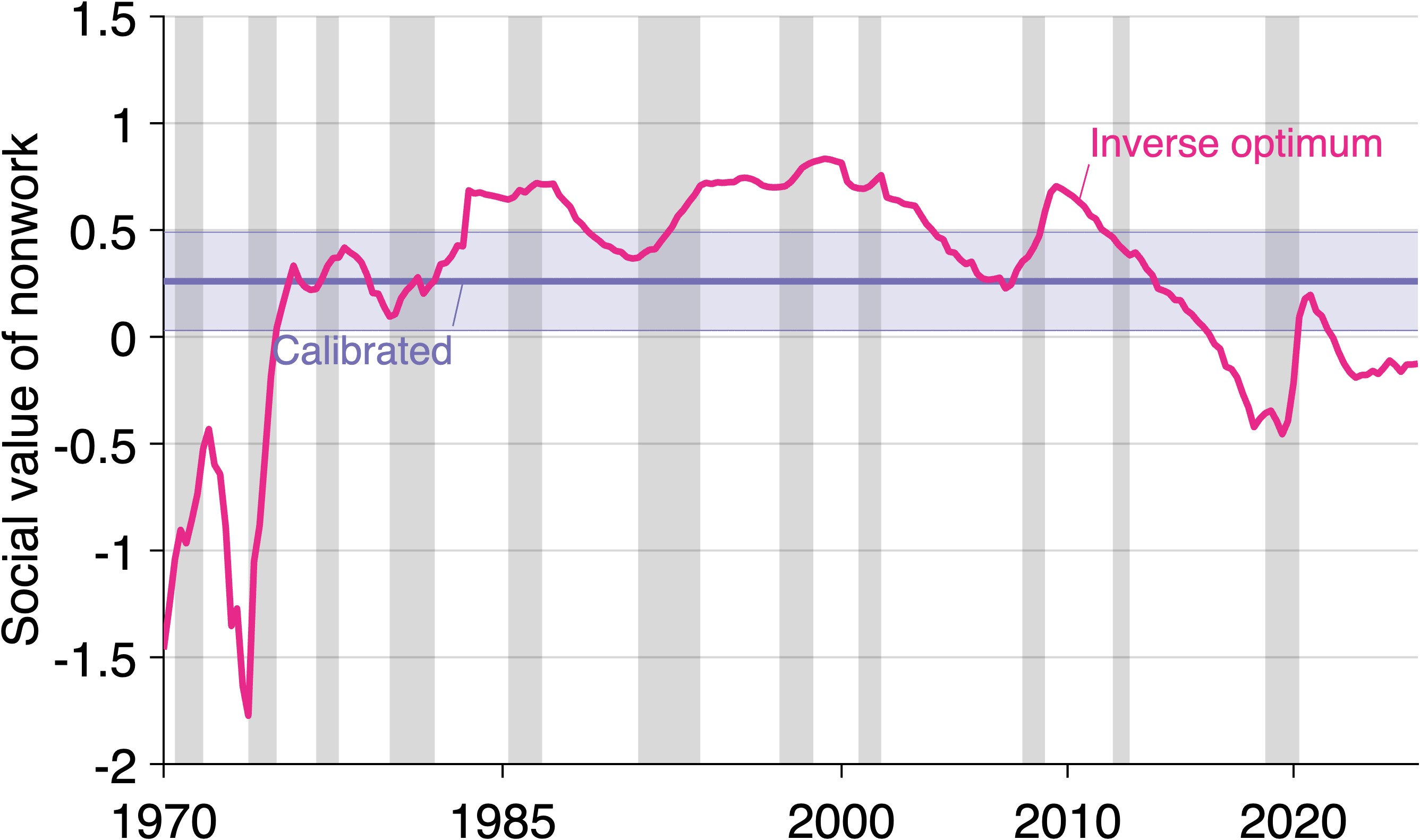}}

\vspace{0.3cm}

\subcaptionbox{Inverse-optimum recruiting cost}{
\includegraphics[width=0.48\textwidth]{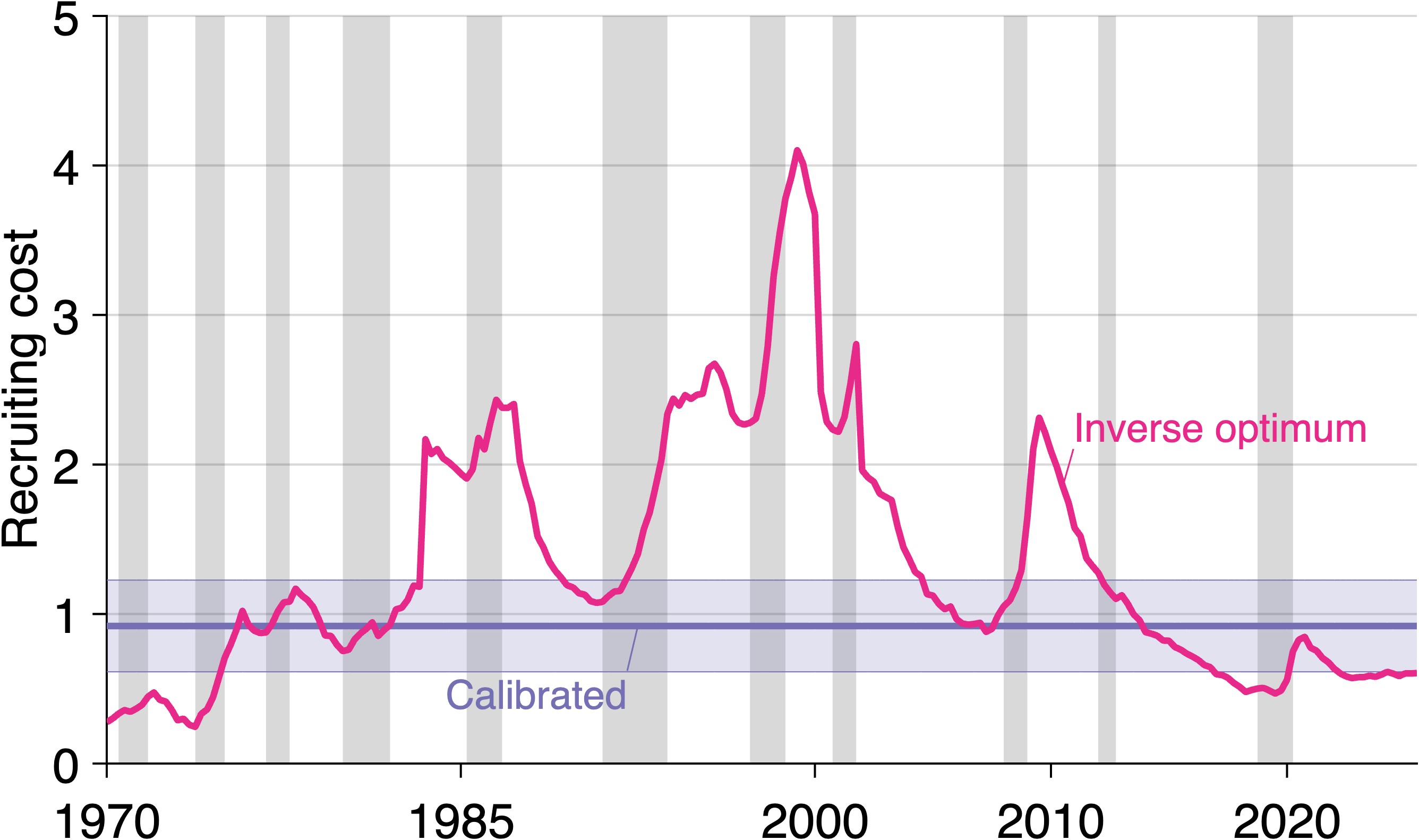}}

\caption{Inverse-optimum sufficient statistics in Japan, 1970--2025.}
\label{f:inverse_jp}
\end{figure}

Figure~\ref{f:inverse_jp} provides further evidence of robustness. The inverse-optimum Beveridge elasticity, social value of nonwork, and recruiting cost frequently depart from their calibrated values, particularly during the Lost Decades and other periods of prolonged labor-market weakness. On average, the inverse-optimum Beveridge elasticity is 0.84, compared with a calibrated value of 0.63; it falls within the 95\% confidence interval in only 24\% of quarters and lies above the upper bound in 50\% (Table~\ref{tab:fig9a_summary}). The inverse-optimum social value of nonwork averages 0.27 and falls within the calibration range $[0.03,\,0.49]$ in 42\% of quarters; it turns negative in tight-labor episodes, reaching $-1.77$ in 1973Q4, and exceeds the upper bound in 36\% of quarters (Table~\ref{tab:fig9b_summary}). The inverse-optimum recruiting cost averages 1.34, well above the calibrated $c=0.92$, and lies within its calibration range in 46\% of quarters while exceeding the upper bound in 40\% (Table~\ref{tab:fig9c_summary}). In many episodes, reconciling observed labor-market outcomes with efficiency would require a Beveridge elasticity substantially above its estimated value, an implausibly high social value of nonwork, or recruiting costs far exceeding their calibrated level. 

Taken together, the results indicate that Japan's labor market was frequently inefficiently slack during the Lost Decades, but moved closer to efficiency after 2011 and appears increasingly tight in the post-pandemic period.


\section{Conclusion}\label{s:ccl}

This paper extends the efficient-unemployment framework of \citet{MS21b} by allowing
employed workers to differ in hours supplied. The original framework treats all
employed workers as homogeneous. I extend the sufficient-statistics formula through
two additional objects, the employment-composition parameter $\alpha$ and the
relative-hours parameter $\gamma$. When all workers are full-time, so that $\alpha = 1$, the extended formula collapses to the original Michaillat--Saez expression, providing a consistency check on the extension. The same welfare function is used throughout; the total and involuntary part-time calibrations differ only in how employment is partitioned when measuring $\alpha$ and $\gamma$.

Applying the extended framework to the United States from 1951 to 2026 shows that incorporating involuntary part-time employment produces only limited changes in efficient unemployment and the unemployment gap. The resulting series remain close to the baseline estimates, reflecting the relatively small role of involuntary part-time employment in the U.S. labor market. Incorporating total part-time employment has more important implications. On average, the unemployment gap remains positive and strongly countercyclical, so the broad assessment of labor-market slack is unchanged. At the same time, the total part-time calibration reveals a different cyclical pattern around major downturns. Before the 1973--75, early-1980s, 1990--91, and 2007--09 recessions, actual labor-market tightness approaches or reaches the efficient allocation once differences in hours worked are taken into account. Under the full-time benchmark, these same episodes instead appear inefficiently slack. Accounting for part-time employment therefore changes the interpretation of labor-market conditions around several recession turning points, highlighting the importance of the hours margin when assessing cyclical conditions.

The Japanese evidence points to the same mechanism but with quantitatively larger effects. As in the United States, incorporating involuntary part-time employment produces only limited changes in efficient unemployment and unemployment gaps. By contrast, incorporating total part-time employment has substantial effects on the efficient benchmark. The average efficient unemployment rate rises, efficient labor-market tightness falls, and the unemployment gap becomes substantially smaller than under the full-time benchmark. Although Japan remains inefficiently slack on average over the full sample, accounting for part-time employment changes the interpretation of several important episodes. Before the global financial crisis, the Japanese labor market appears close to efficient under the part-time calibration. After the 2011 earthquake, the part-time calibration reveals a tighter labor market, and during the post-pandemic period it again indicates increasing tightness. Thus, while the U.S. results mainly alter the interpretation of conditions around recession turning points, the Japanese results reveal extended periods of labor-market tightness that are much less visible under the full-time benchmark.

Additional exercises using involuntary part-time employment indicate that these results are not primarily driven by underemployment. Involuntary part-time workers account for only a small share of total part-time employment in both countries, and a particularly small and declining share in Japan. Restricting the part-time margin to this group therefore yields estimates that remain close to the full-time benchmark.

This exercise should be interpreted cautiously. Because hours worked by involuntary part-time workers are unavailable for Japan, I set $\gamma^{IP}=\gamma$. The difference between the total and involuntary part-time results therefore reflects mainly the employment-share statistic $\alpha$, rather than a separately measured hours margin.

Overall, the results suggest that accounting for employment heterogeneity can materially affect the measurement of labor-market efficiency. In both countries, incorporating total part-time employment raises efficient unemployment and lowers efficient labor-market tightness because part-time workers supply fewer effective hours than full-time workers. The effect is modest for involuntary part-time employment, indicating that underemployment alone is not the primary mechanism. Instead, the main channel operates through differences in labor input across workers. More generally, the findings highlight the importance of measuring labor input not only by the number of workers employed but also by the amount of work they perform.

\bibliography{bibliography}

@article{BB19,
	author = {David N.F. Bell and David G. Blanchflower},
	doi = {10.1016/j.jebo.2019.03.018},
	journal = {Journal of Economic Behavior \& Organization},
	pages = {180--196},
	title = {The well-being of the overemployed and the underemployed and the rise in depression in the UK},
	volume = {161},
	year = {2019}}

@article{P,
	author = {Christopher A. Pissarides},
	doi = {10.7551/mitpress/9780262013635.003.0023},
	journal = {Understanding Inflation and the Implications for Monetary Policy: A Phillips Curve Retrospective. MIT Press, Cambridge},
    pages = {235--242},
	title = {Comments on "A New Method for Estimating Time Variation in the NAIRU" by William T. Dickens},
	year = {2009}}

@article{vbl20,
  author  = {Valletta, Robert G. and Bengali, Leila and van der List, Catherine},
  title   = {Cyclical and Market Determinants of Involuntary Part-Time Employment},
  journal = {Journal of Labor Economics},
  volume  = {38},
  number  = {1},
  pages   = {67--93},
  year    = {2020},
  doi     = {10.1086/704496}
}

@article{BM18,
	author = {Mark Borgschulte and Paco Martorell},
	doi = {10.1257/app.20160257},
	journal = {American Economic Journal: Applied Economics},
	number = {3},
	pages = {101--127},
	title = {Paying to Avoid Recession: Using Reenlistment to Estimate the Cost of Unemployment},
	volume = {10},
	year = {2018}}

@article{MS21b,
	author = {Pascal Michaillat and Emmanuel Saez},
	doi = {10.1016/j.pubecp.2021.100009},
	journal = {Journal of Public Economics Plus},
	pages = {100009},
	title = {Beveridgean Unemployment Gap},
	volume = {2},
	year = {2021}}

@article{c9,
	author = {Raj Chetty},
	doi = {10.1146/annurev.economics.050708.142910},
	journal = {Annual Review of Economics},
	pages = {451–87},
	title = {Sufficient Statistics for Welfare Analysis: A Bridge Between Structural and Reduced-Form Methods},
	volume = {1},
	year = {2009}}

@article{MP17,
	author = {Alexandre Mas and Amanda Pallais},
	doi = {10.3386/w23906},
	journal = {NBER Working Paper 23906},
	title = {Labor Supply and the Value of Non-Work Time: Experimental Estimates from the Field},
	year = {2017}}

@article{BP98,
	author = {Jushan Bai and Pierre Perron},
	doi = {10.2307/2998540},
	journal = {Econometrica},
    number = {1},
	pages = {47-78},
	title = {Estimating and Testing Linear Models with Multiple Structural Changes},
	volume = {66},
	year = {1998}}

@article{A1 ,
	author = {Donald W. K. Andrews},
	doi = {10.2307/2938229},
	journal = {Econometrica},
    number = {3},
	pages = {817-858},
	title = {Heteroskedasticity and Autocorrelation Consistent Covariance Matrix Estimation},
	volume = {59},
	year = {1991}}

@article{BP03,
	author = {Jushan Bai and Pierre Perron},
	doi = {10.1002/jae.659},
	journal = {Journal of Applied Econometrics},
    number = {1},
	pages = {1-22},
	title = {Computation and analysis of multiple structural change models},
	volume = {18},
	year = {2003}
}

@article{Yao88,
  author  = {Yao, Yi-Ching},
  title   = {Estimating the Number of Change-Points via Schwarz' Criterion},
  journal = {Statistics \& Probability Letters},
  volume  = {6},
  number  = {3},
  pages   = {181--189},
  year    = {1988}
}

@article{LSZ97,
	author = {Jian Liu and Shiying Wu and James V. Zidek},
	doi = {10.1016/0167-7152(88)90118-6},
	journal = {Statistica Sinica},
        number = {7},
	pages = {497-525},
	title = {On Segmented Multivariate Regression},
	year = {1997}}

@article{BFHS23,
	author = {Gadi Barlevy and R. Jason Faberman and Bart Hobijn and Ayşegül Şahin},
	doi = {10.3386/w31783},
	journal = {NBER Working Paper 31783},
	title = {The Shifting Reasons for Beveridge-Curve Shifts},
	year = {2023}}

@article{B10,
	author = {Regis Barnichon},
	doi = {10.1016/j.econlet.2010.08.029},
	journal = {Economics Letters},
    number = {3},
	pages = {175--178},
	title = {Building a composite help-wanted index: Dataset},
    volume = {109},
	year = {2010}}

@article{BLS_parttime,
  author  = {{U.S. Bureau of Labor Statistics}},
  title   = {Part-Time Workers},
  journal = {U.S. Bureau of Labor Statistics},
  year    = {2024},
  volume  = { },
  number  = { },
  pages   = { },
  note    = {Persons who usually work fewer than 35 hours per week are classified as part-time workers. \url{https://www.bls.gov/cps/definitions.htm}}
}

@article{BLS_PTER,
  author  = {{U.S. Bureau of Labor Statistics}},
  title   = {Persons at Work Part Time for Economic Reasons},
  journal = {U.S. Bureau of Labor Statistics},
  year    = {2024},
  volume  = { },
  number  = { },
  pages   = { },
  note    = {Persons working part time because of slack work, business conditions, or inability to find full-time work. \url{https://www.bls.gov/cps/definitions.htm}}
}

@article{BLS_underutilization,
  author  = {{U.S. Bureau of Labor Statistics}},
  title   = {Alternative Measures of Labor Underutilization for States},
  journal = {U.S. Bureau of Labor Statistics},
  year    = {2026},
  volume  = { },
  number  = { },
  pages   = { },
  note    = {Accessed June 12, 2026. \url{https://www.bls.gov/lau/stalt.htm}}
}

@article{MS_recession_started,
  title={Has the recession started?},
  author={Michaillat, Pascal and Saez, Emmanuel},
  journal={Oxford Bulletin of Economics and Statistics},
  volume={87},
  number={6},
  pages={1047--1058},
  year={2025},
  publisher={Wiley Online Library}
}

@techreport{VR10,
  author       = {Villena-Rold{\'a}n, Benjam{\'i}n},
  title        = {Aggregate Implications of Employer Search and Recruiting Selection},
  institution  = {Centro de Econom{\'i}a Aplicada, Universidad de Chile},
  type         = {Documentos de Trabajo},
  number        = {271},
  year          = {2010}
}

\newpage

\appendix

\section{Notation relative to Michaillat and Saez (2021)}
\label{app:notation}

This paper follows the Beveridgean unemployment-gap framework of \citet{MS21b}, but uses slightly different notation. In \citet{MS21b}, the social value of nonwork is denoted by $\zeta$ and recruiting cost by $\kappa$. In this paper, I denote the same objects by $z$ and $c$:
\[
z \equiv \zeta,
\qquad
c \equiv \kappa.
\]

In \citet{MS21b}, the tangency condition is stated as follows:

\begin{quote}
\textbf{Proposition 1.} In a Beveridge diagram, efficiency is achieved at the point where the Beveridge curve is tangent to an isowelfare curve. Hence, the efficient unemployment rate is implicitly defined by
\[
v'(u)
=
-\frac{1-\zeta}{\kappa},
\]
where $\zeta<1$ is the social value of nonwork and $\kappa>0$ is the recruiting cost.
\end{quote}

Using the notation of this paper, this condition becomes
\begin{equation}
v'(u)
=
-\frac{1-z}{c}.
\label{eq:ms_prop1_ownnotation}
\end{equation}

With an isoelastic Beveridge curve, \citet{MS21b} then obtain the following efficient-tightness condition:

\begin{quote}
\textbf{Proposition 2.} Consider a point on the Beveridge curve with tightness $\theta$, Beveridge elasticity $\epsilon$, social value of nonwork $\zeta$, and recruiting cost $\kappa$. Tightness is inefficiently high if
\[
\theta>\frac{1-\zeta}{\kappa\epsilon},
\]
inefficiently low if
\[
\theta<\frac{1-\zeta}{\kappa\epsilon},
\]
and efficient if
\[
\theta=\frac{1-\zeta}{\kappa\epsilon}.
\]
\end{quote}

Under the notation used in this paper, the full-time benchmark is
\begin{equation}
\theta^{*}
=
\frac{1-z}{c\epsilon}.
\label{eq:ms_prop2_ownnotation}
\end{equation}

The part-time extension replaces the full-time employment term $1$ with effective employment input,
\[
\alpha+(1-\alpha)\left[\gamma+(1-\gamma)z\right],
\]
where $\alpha$ is the full-time employment share and $\gamma$ is the relative-hours parameter. The efficient-tightness condition with part-time employment is therefore
\begin{equation}
\theta^{*}
=
\frac{
\alpha+(1-\alpha)\left[\gamma+(1-\gamma)z\right]-z
}
{c\epsilon}.
\label{eq:pt_eff_tightness_appendix}
\end{equation}

When $\alpha=1$, equation~\eqref{eq:pt_eff_tightness_appendix} collapses to the full-time benchmark in equation~\eqref{eq:ms_prop2_ownnotation}.

\section{Additional results}
\subsection{Additional results for the United States}

\begin{figure}[H]
\centering
\includegraphics[width=0.9\textwidth]{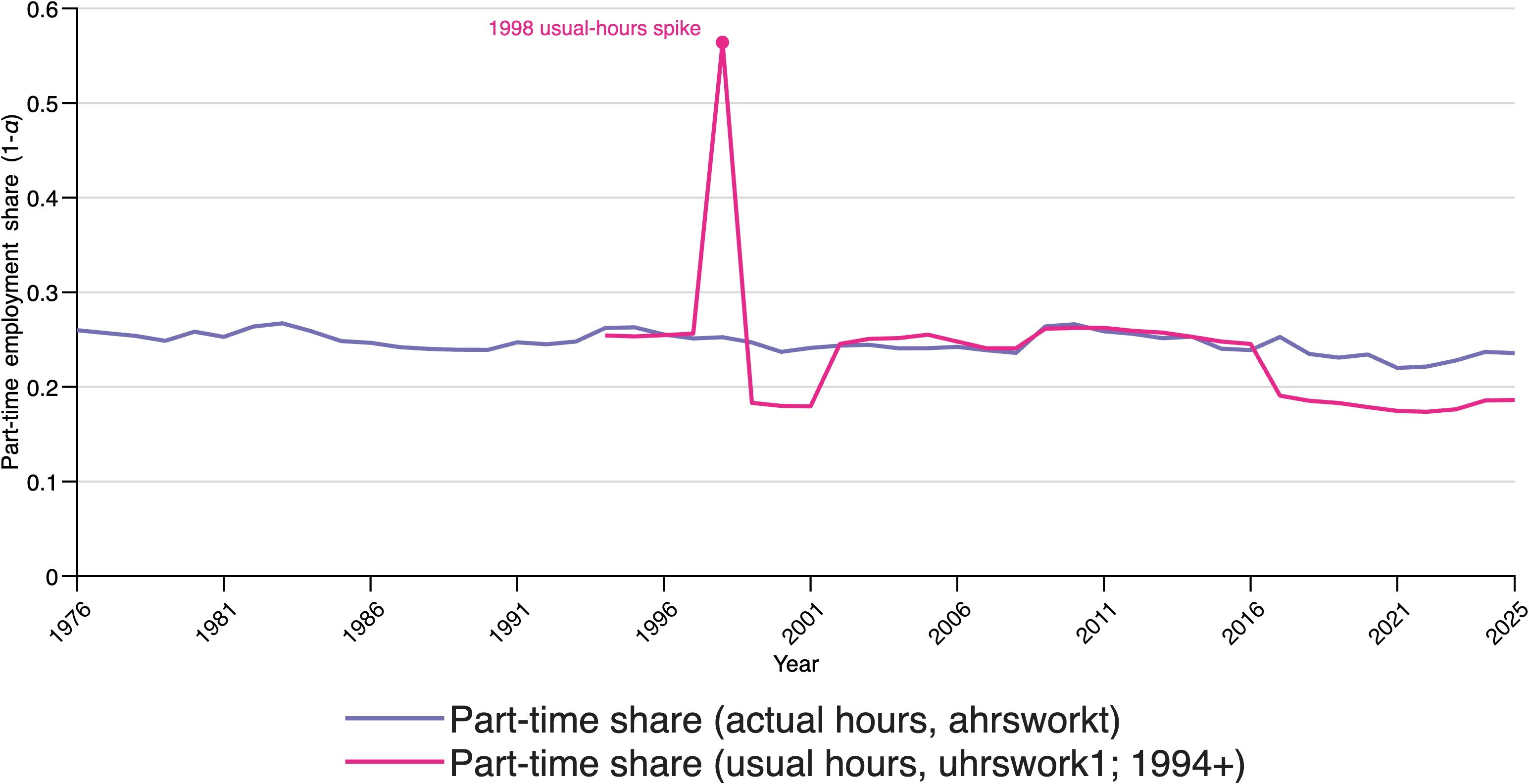}
\caption{Part-time employment shares based on actual and usual hours in the CPS ASEC. 
\textit{Notes:} The figure compares part-time employment shares constructed using actual hours worked and usual hours worked. Workers are classified as part-time if their reported hours are fewer than 35 per week. The usual-hours measure produces a large spike in 1998 that is absent from the actual-hours measure.}
\label{f:actual_usual_pt}
\end{figure}

\begin{figure}[H]
\centering

\subcaptionbox{Efficient unemployment rate, 1951--2019}{
\includegraphics[width=0.48\textwidth]{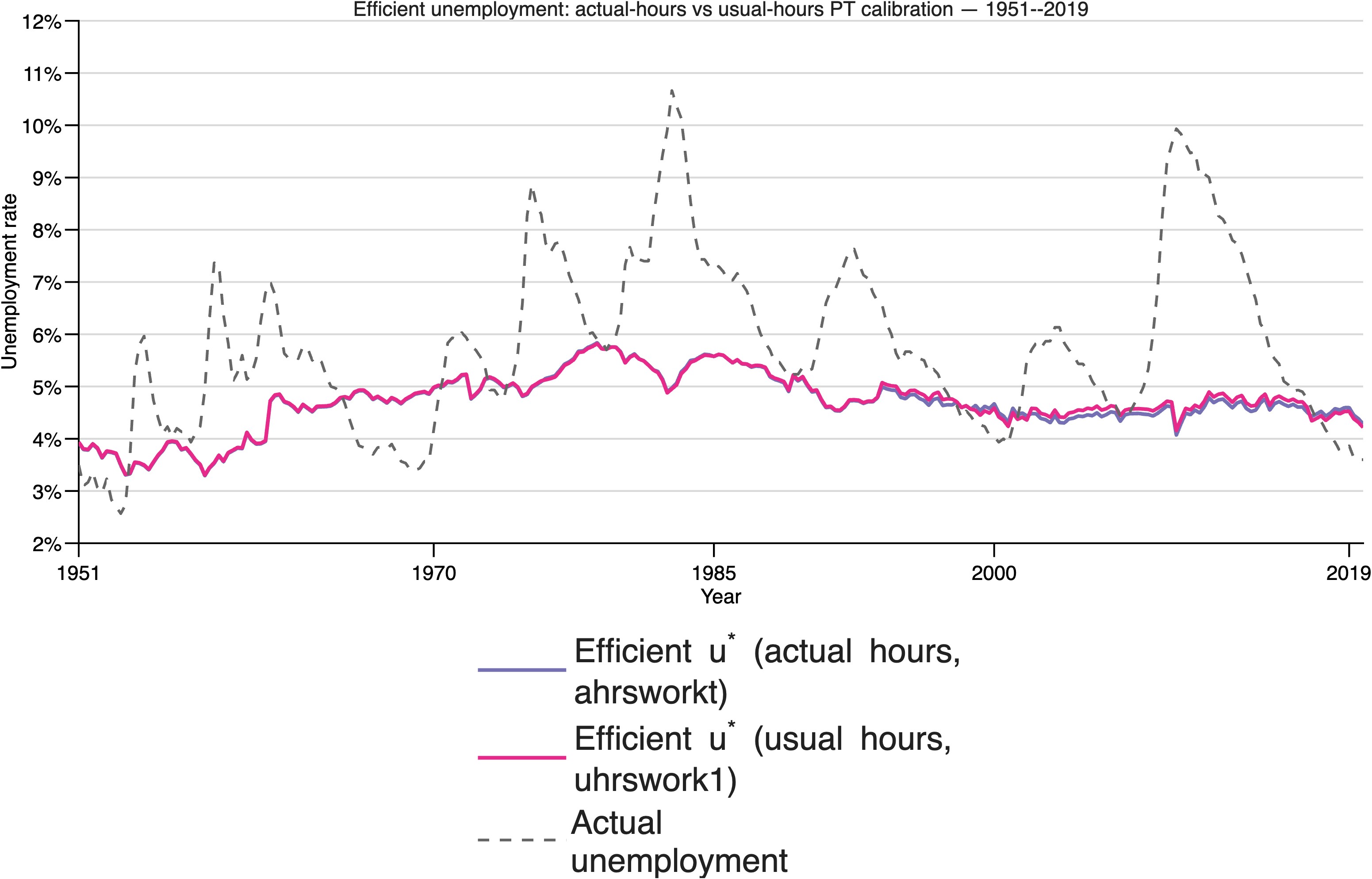}}
\hfill
\subcaptionbox{Efficient unemployment rate, 2020--2026}{
\includegraphics[width=0.48\textwidth]{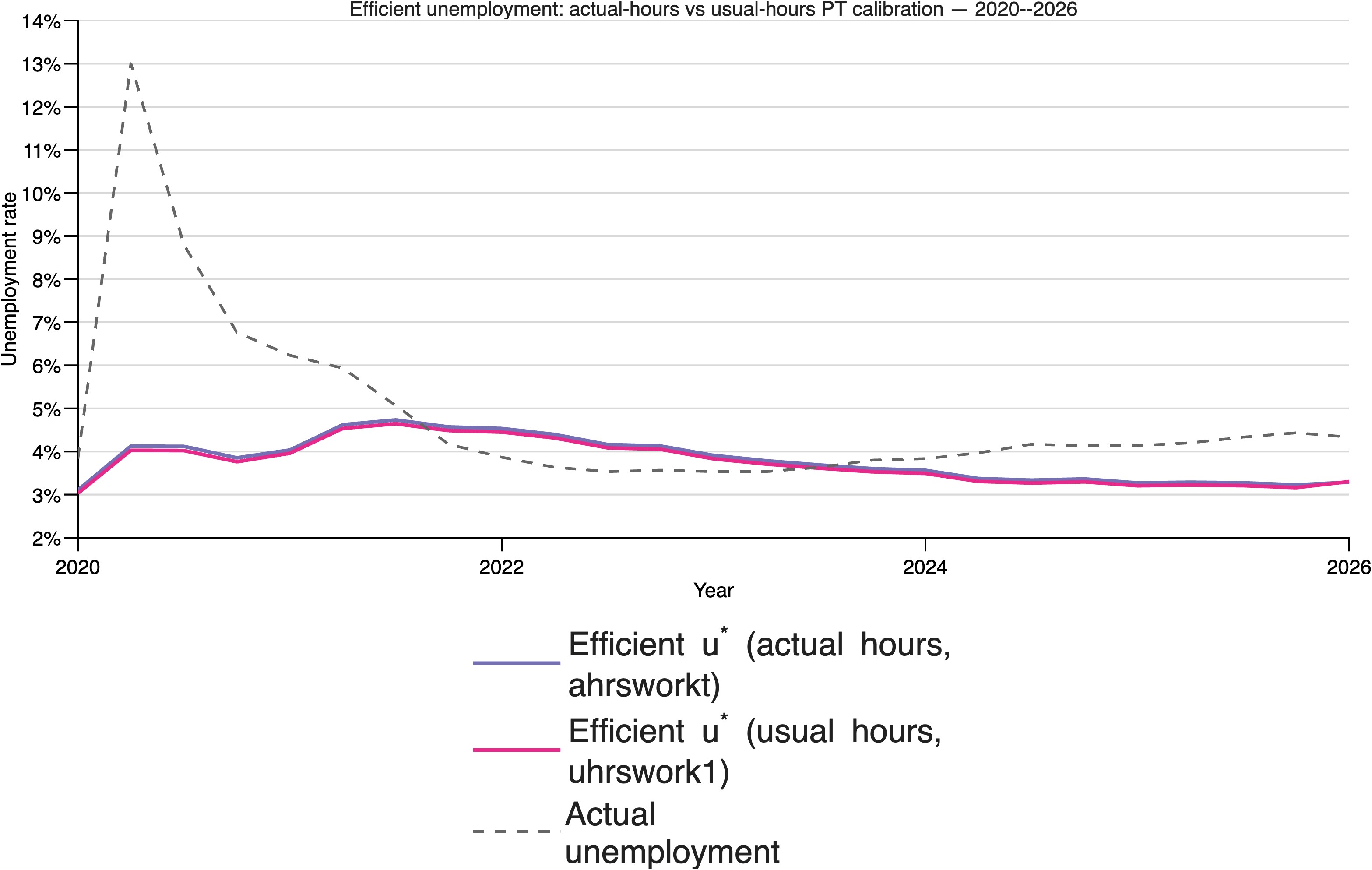}}

\vspace{0.3cm}

\subcaptionbox{Unemployment gap, 1951--2019}{
\includegraphics[width=0.48\textwidth]{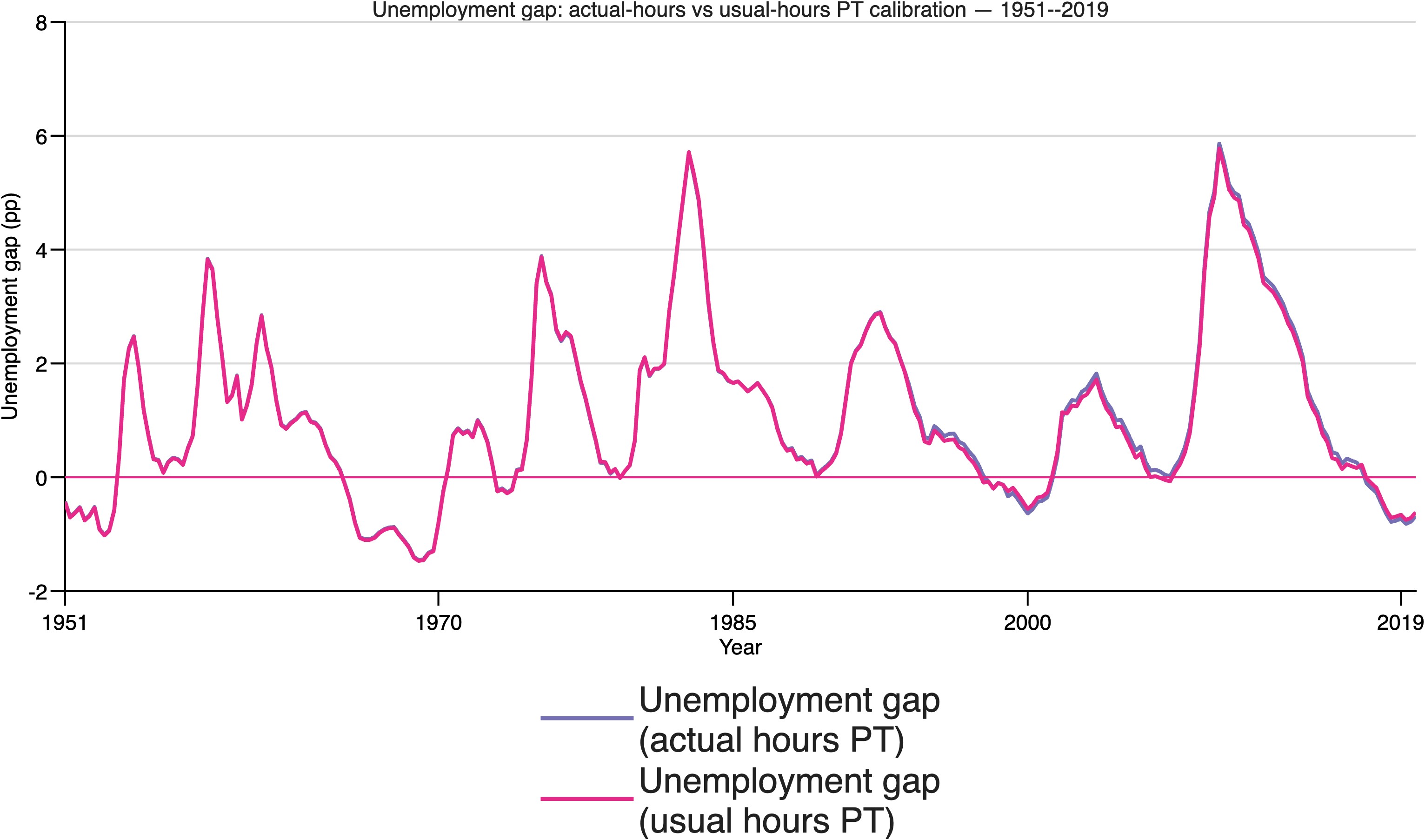}}
\hfill
\subcaptionbox{Unemployment gap, 2020--2026}{
\includegraphics[width=0.48\textwidth]{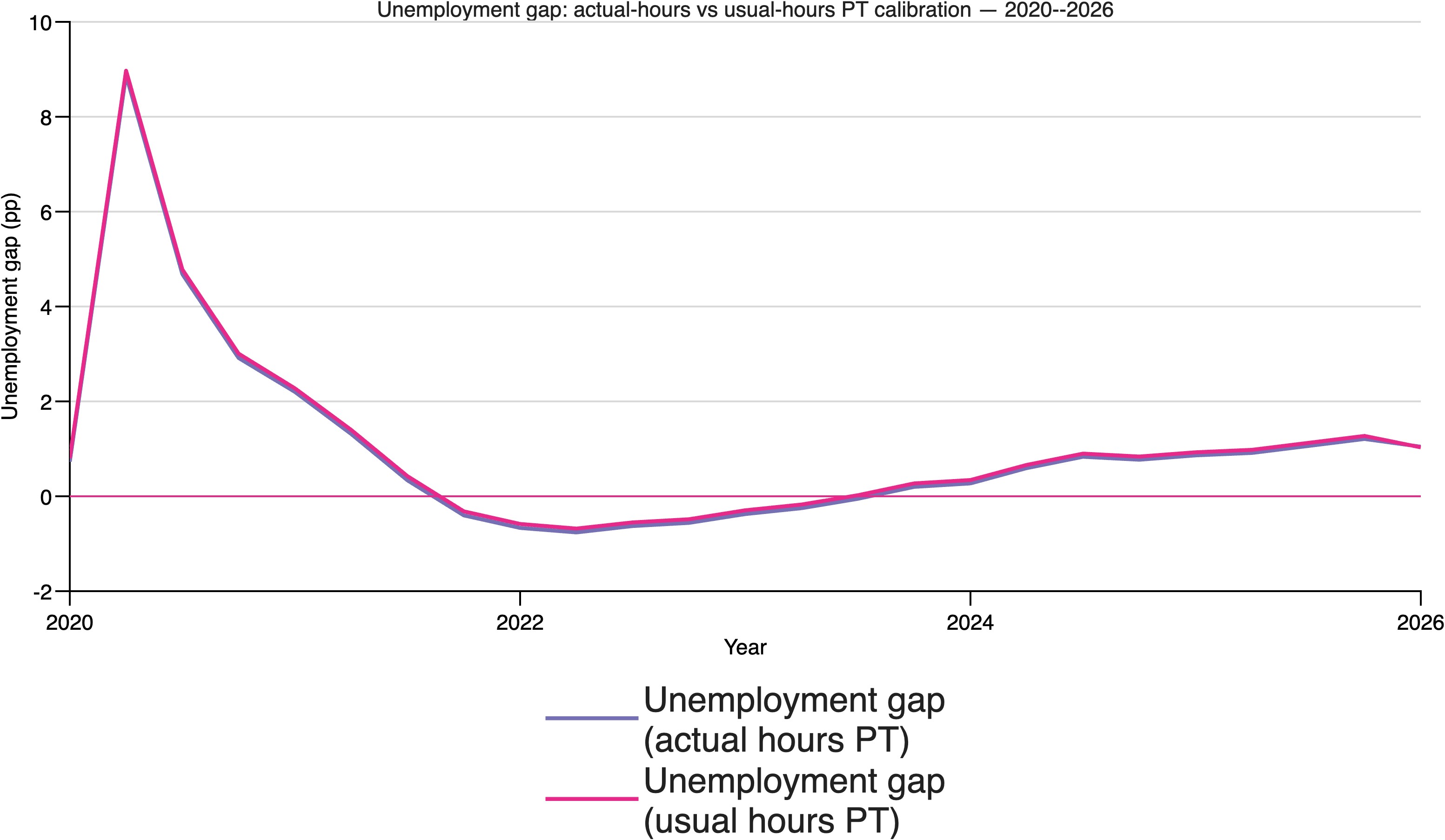}}

\caption{Robustness to actual-hours and usual-hours part-time classifications in the United States. 
\textit{Notes:} The figure compares efficient unemployment rates and unemployment gaps constructed using part-time employment classifications based on actual hours worked and usual hours worked. Workers are classified as part-time if reported hours are fewer than 35 per week. The actual-hours measure uses \texttt{ahrsworkt}, while the usual-hours measure uses \texttt{uhrswork1}.}
\label{f:actual_usual_robustness}
\end{figure}

\begin{figure}[H]
\centering

\subcaptionbox{Efficient labor-market tightness with total part-time employment, 1951--2026}{
\includegraphics[width=0.48\textwidth]{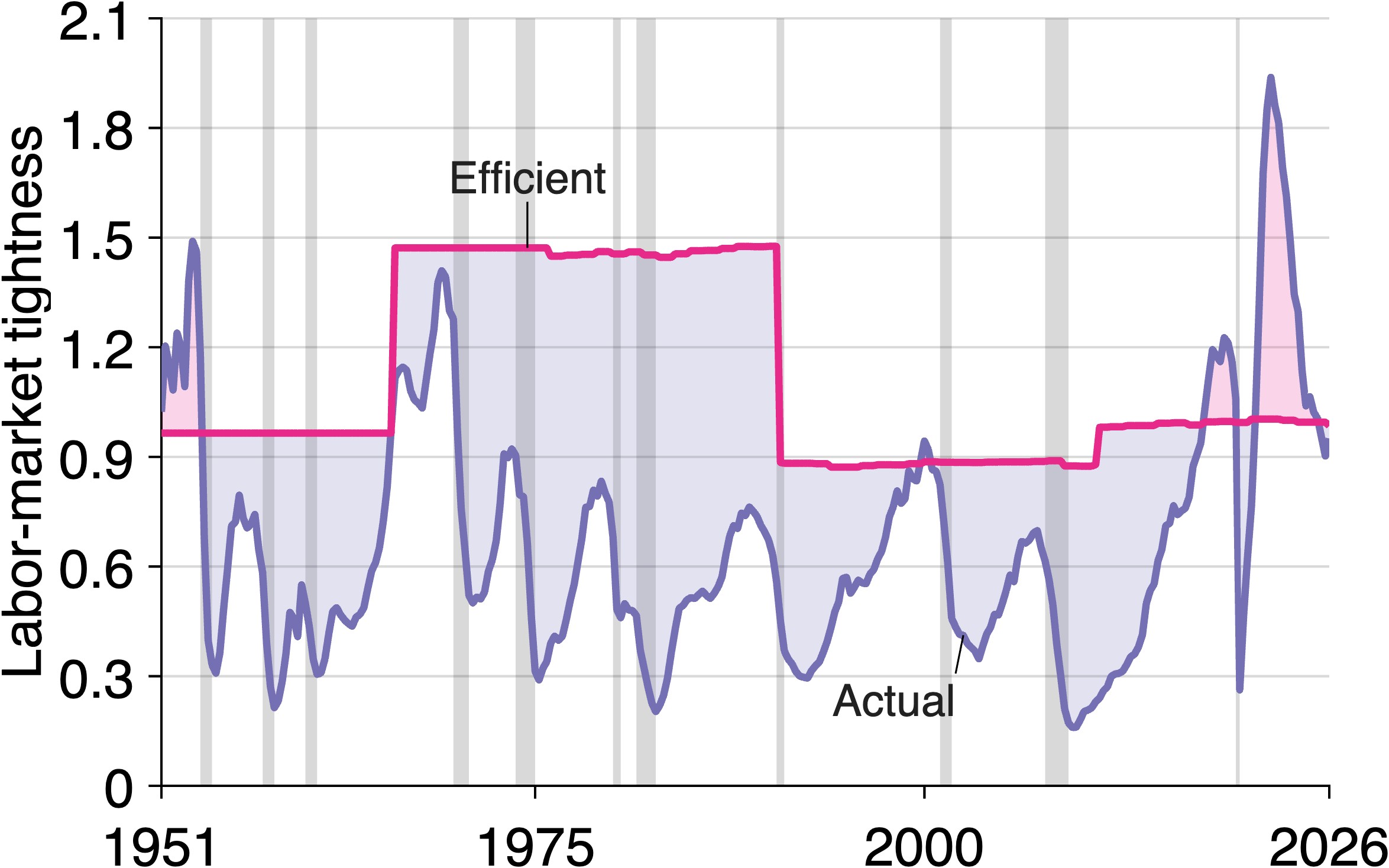}}
\hfill
\subcaptionbox{Efficient unemployment rate with total part-time employment, 1951--2026}{
\includegraphics[width=0.48\textwidth]{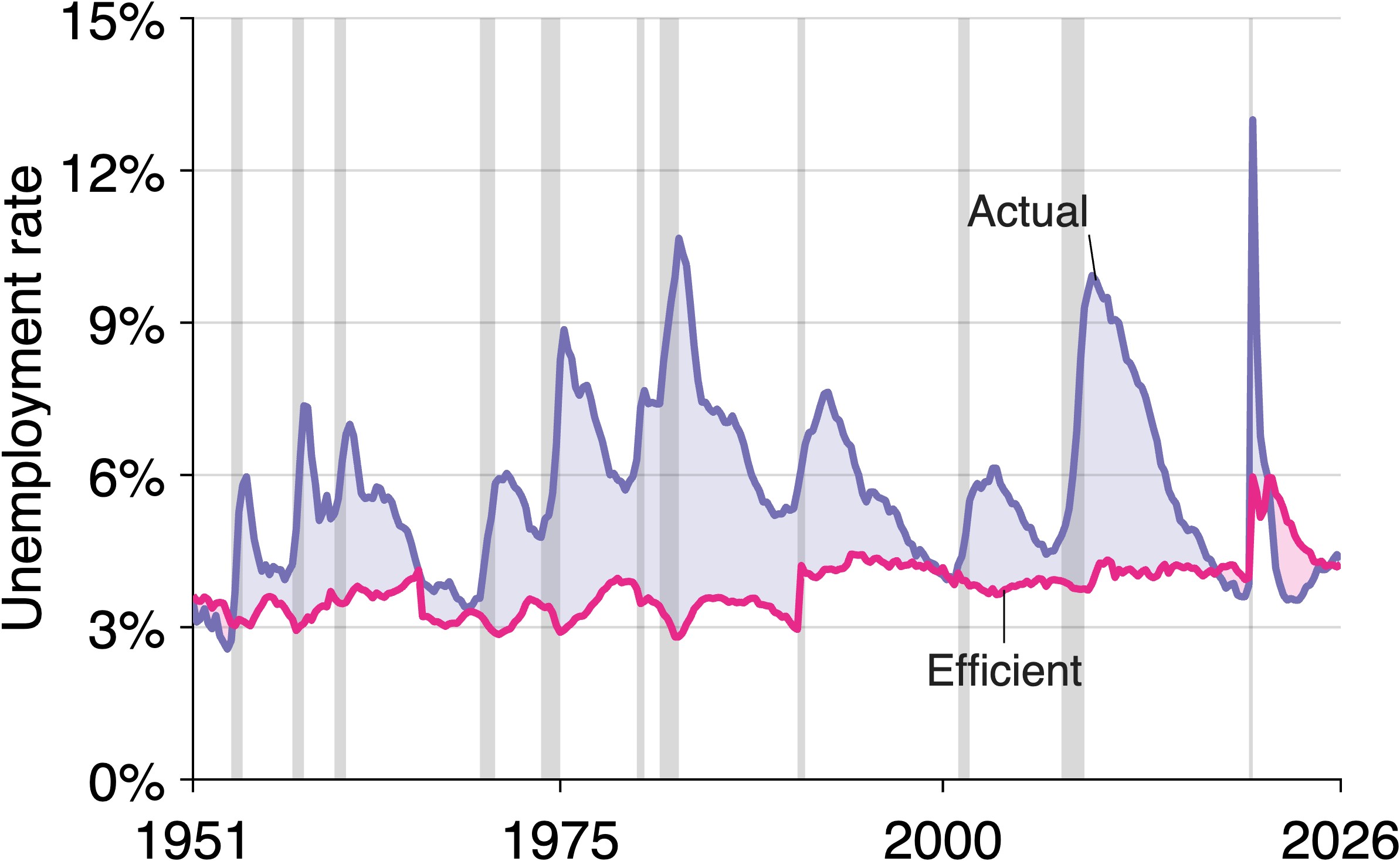}}

\vspace{0.3cm}

\subcaptionbox{Unemployment gap with total part-time employment, 1951--2026}{
\includegraphics[width=0.48\textwidth]{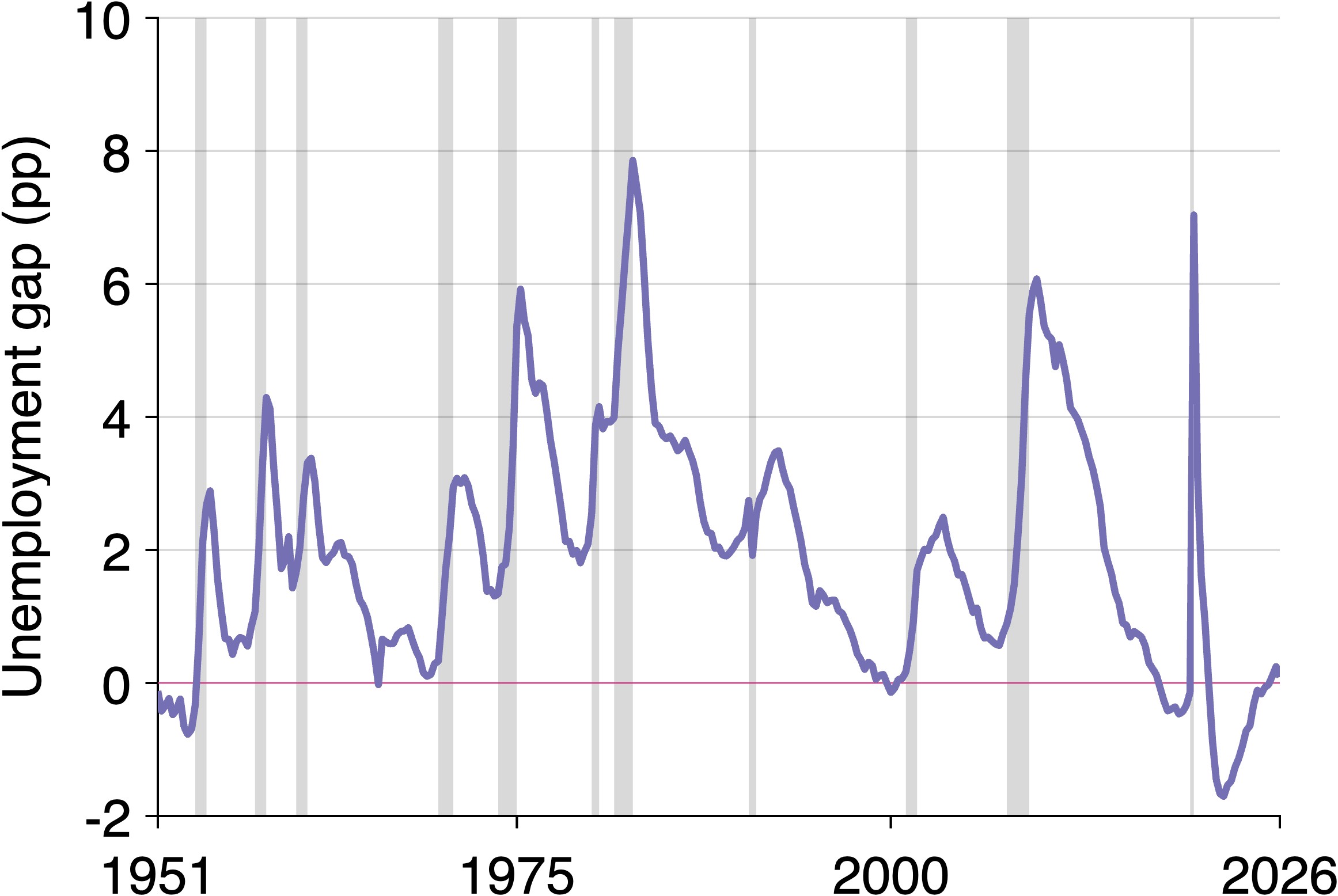}}

\caption{Efficient labor-market tightness, efficient unemployment, and unemployment gap in the United States, 1951--2026. \textit{Notes:} The figure reports efficient labor-market tightness, the efficient unemployment rate, and the unemployment gap under the extended framework that incorporates total part-time employment. The unemployment gap is measured as the difference between the actual unemployment rate and the efficient unemployment rate implied by the model.}
\label{f:graph2}

\end{figure}

\begin{figure}[H]
\centering

\subcaptionbox{Efficient labor-market tightness comparison, 1951--2026}
{\includegraphics[width=0.48\textwidth]{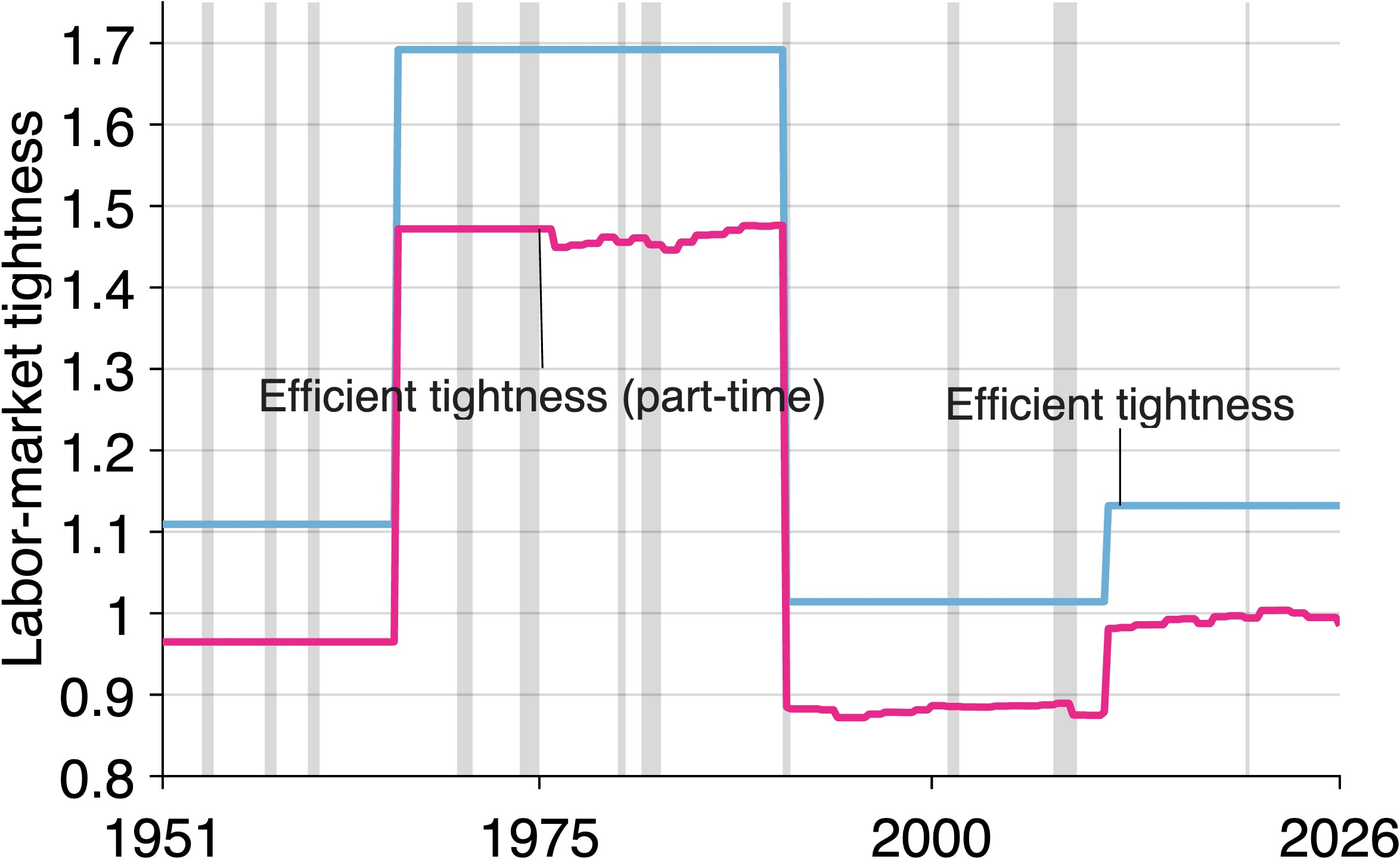}}
\hfill
\subcaptionbox{Efficient unemployment rate comparison, 1951--2026}
{\includegraphics[width=0.48\textwidth]{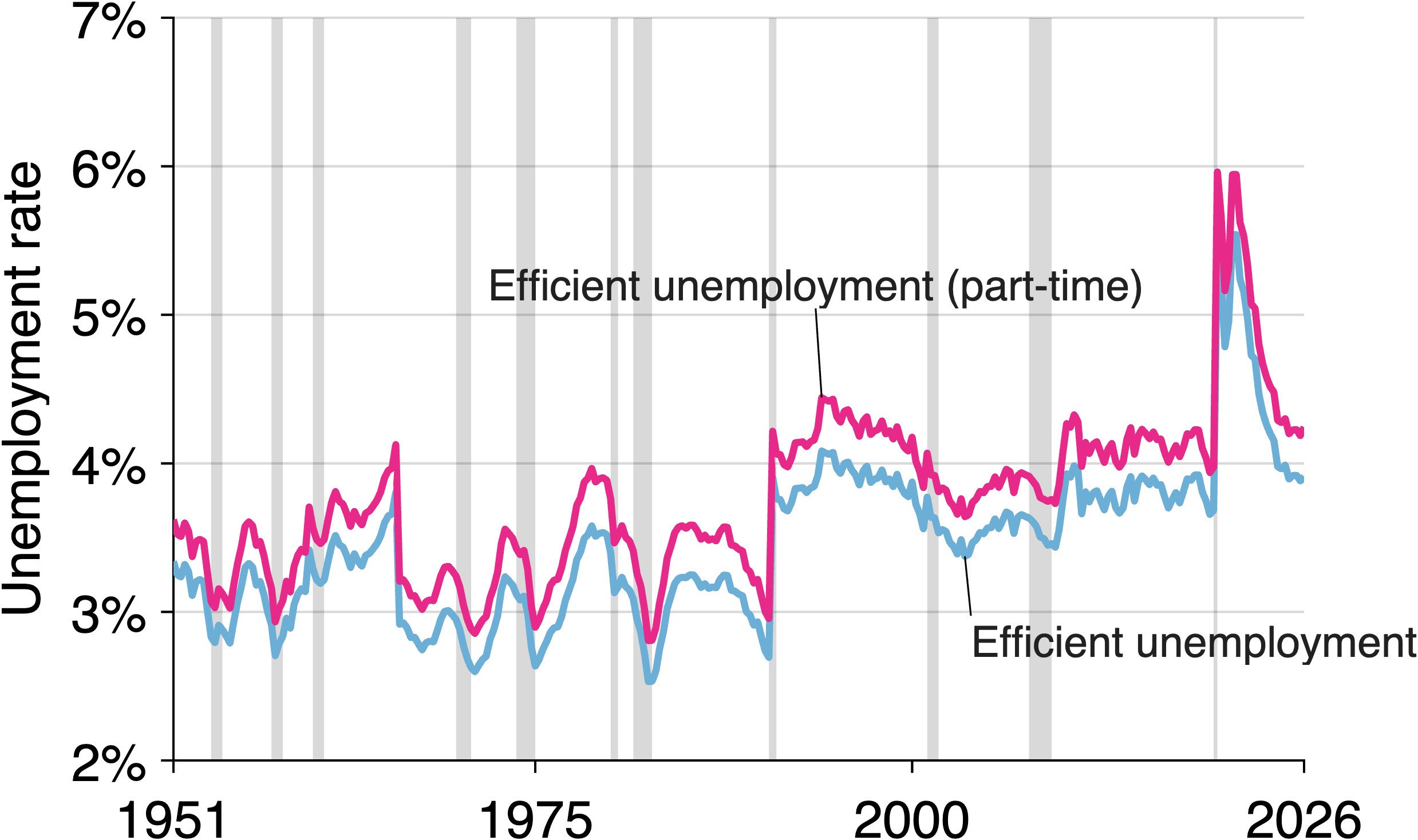}}

\vspace{0.3cm}

\subcaptionbox{Unemployment gap comparison, 1951--2026}
{\includegraphics[width=0.48\textwidth]{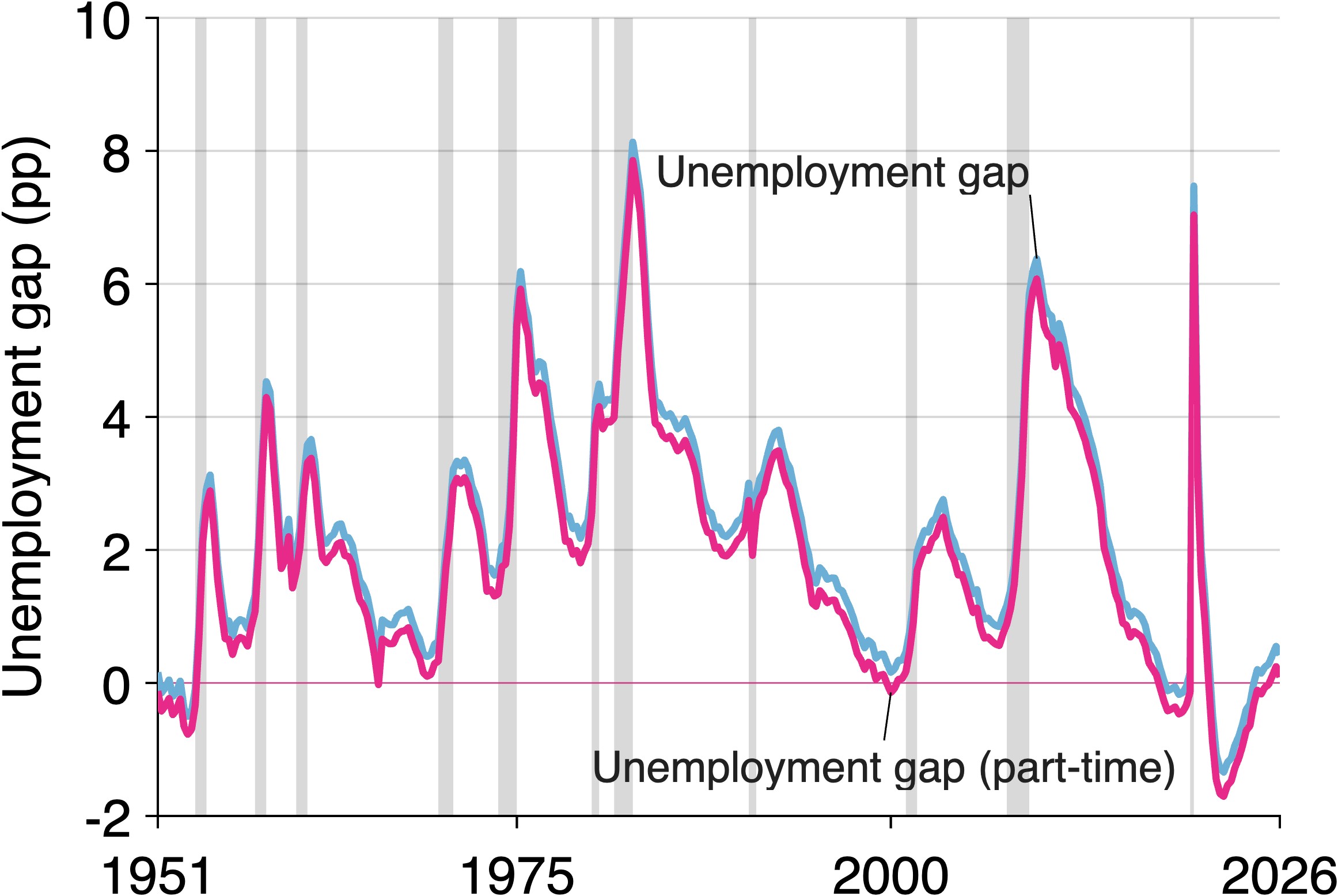}}

\caption{Comparison of efficient labor-market tightness, efficient unemployment, and unemployment gaps with and without total part-time employment in the United States, 1951--2026. \textit{Notes}: The figure compares the baseline framework of \citet{MS21b}, which treats employment as homogeneous, with the extended framework that incorporates total part-time employment.}
\label{f:graph_fullsample}

\end{figure}

\begin{figure}[H]
\centering

\subcaptionbox{Efficient labor-market tightness with involuntary part-time employment, 1951--2026}{
\includegraphics[width=0.48\textwidth]{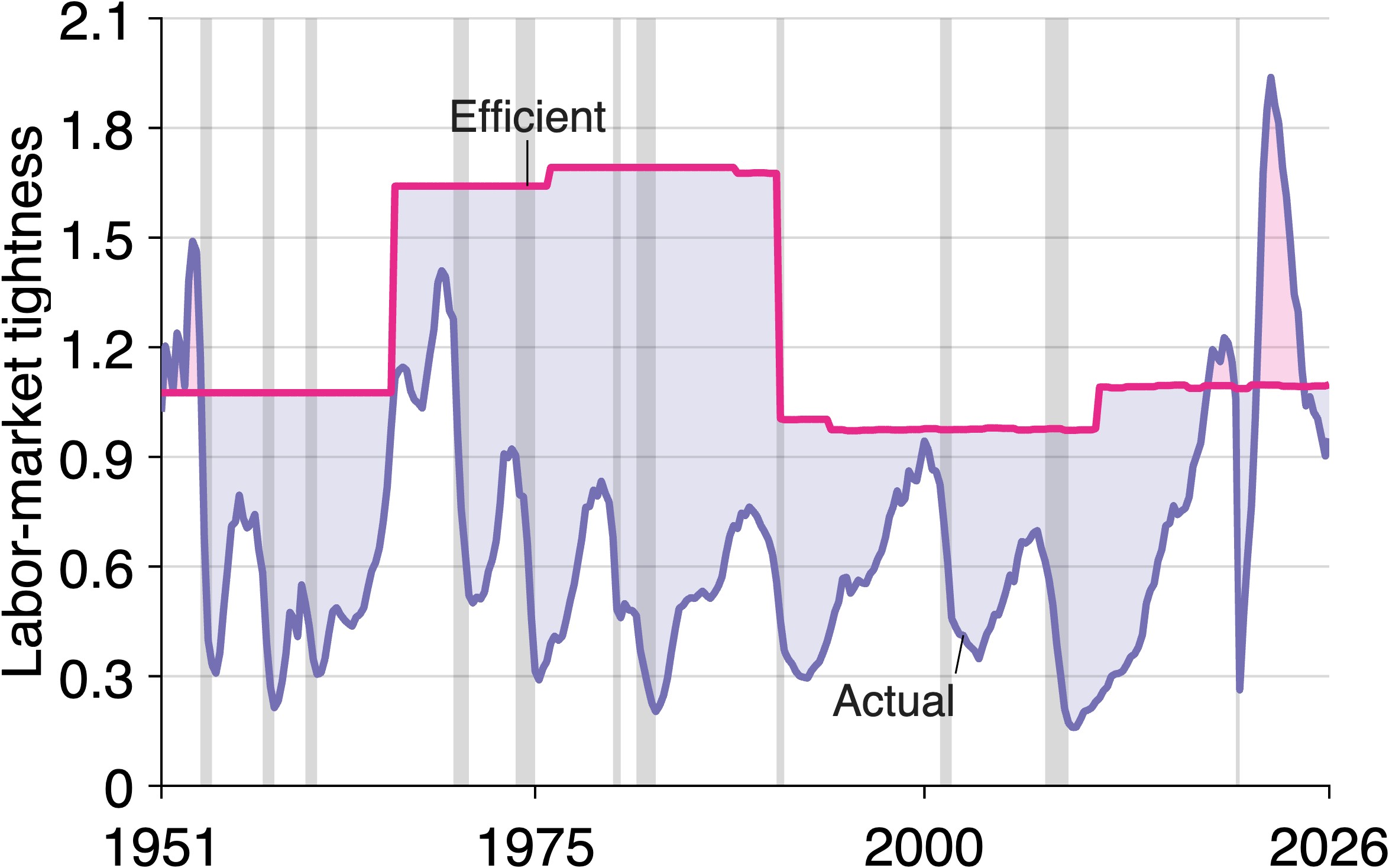}}
\hfill
\subcaptionbox{Efficient unemployment rate with involuntary part-time employment, 1951--2026}{
\includegraphics[width=0.48\textwidth]{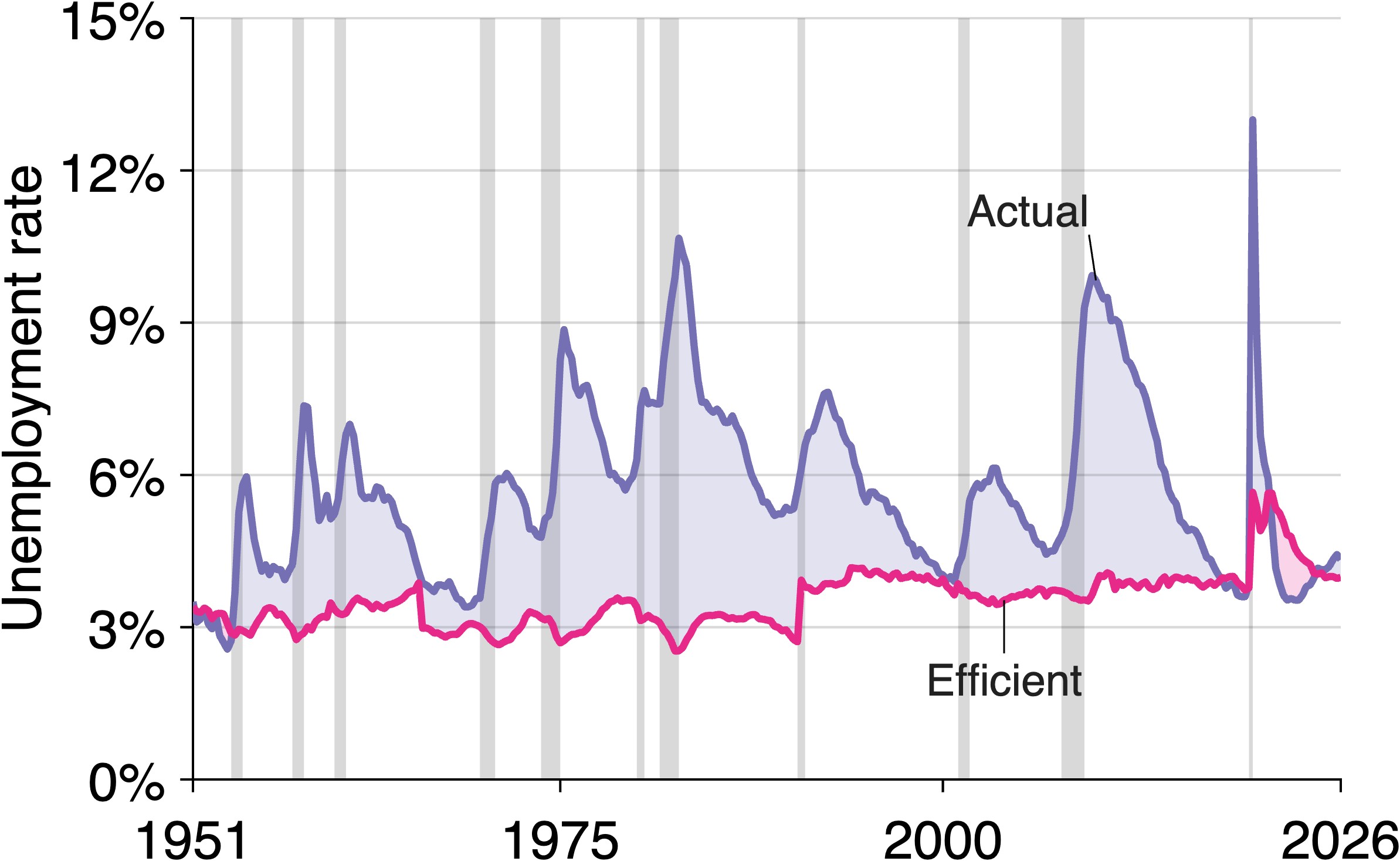}}

\vspace{0.3cm}

\subcaptionbox{Unemployment gap with involuntary part-time employment, 1951--2026}{
\includegraphics[width=0.48\textwidth]{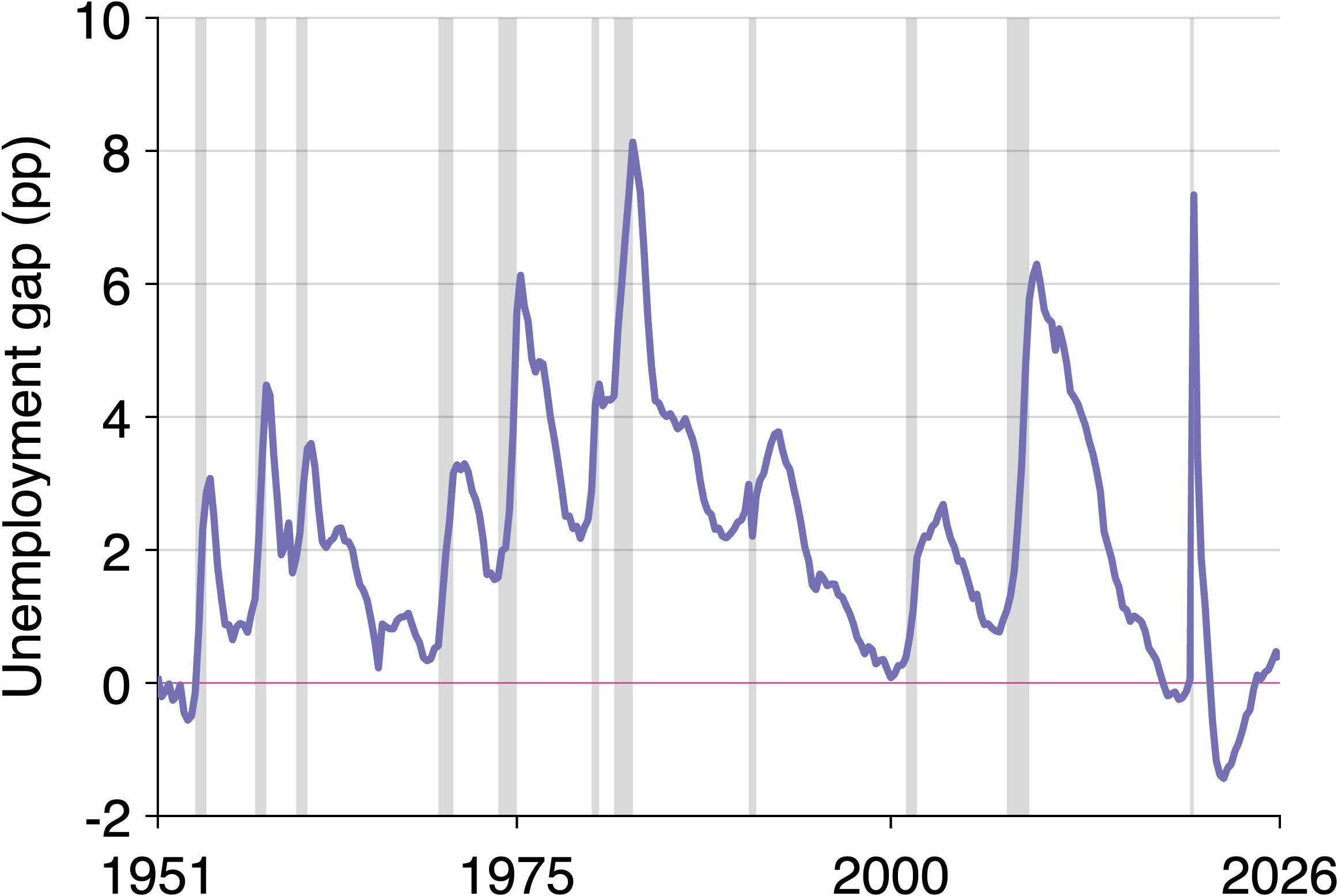}}

\caption{Efficient labor-market tightness, efficient unemployment, and unemployment gap in the United States, 1951--2026. \textit{Notes:} The figure reports efficient labor-market tightness, the efficient unemployment rate, and the unemployment gap under the extended framework that incorporates involuntary part-time employment. The unemployment gap is measured as the difference between the actual unemployment rate and the efficient unemployment rate implied by the model.}
\label{f:graph2}

\end{figure}

\begin{figure}[H]
\centering

\subcaptionbox{Efficient labor-market tightness comparison, 1951--2026}
{\includegraphics[width=0.48\textwidth]{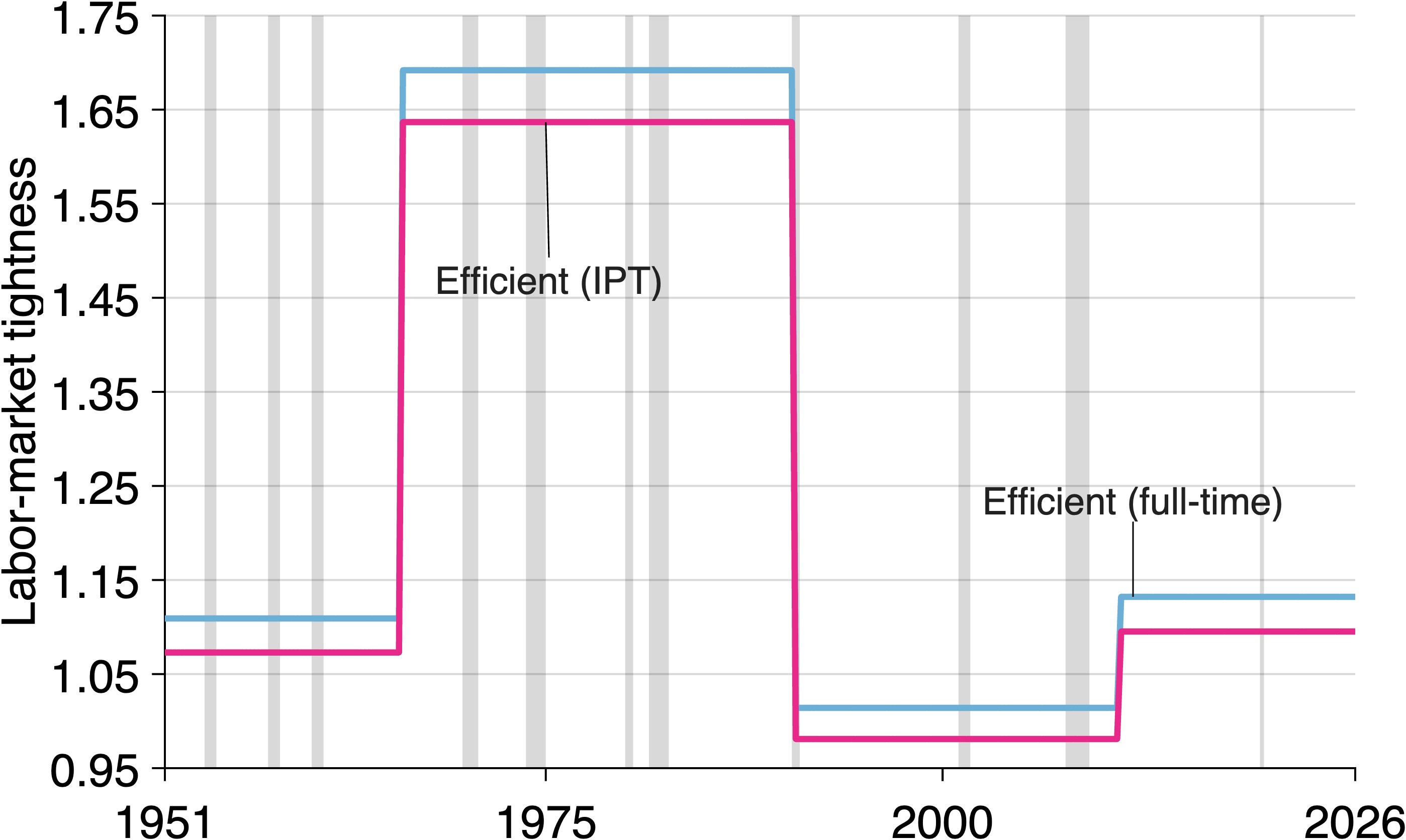}}
\hfill
\subcaptionbox{Efficient unemployment rate comparison, 1951--2026}
{\includegraphics[width=0.48\textwidth]{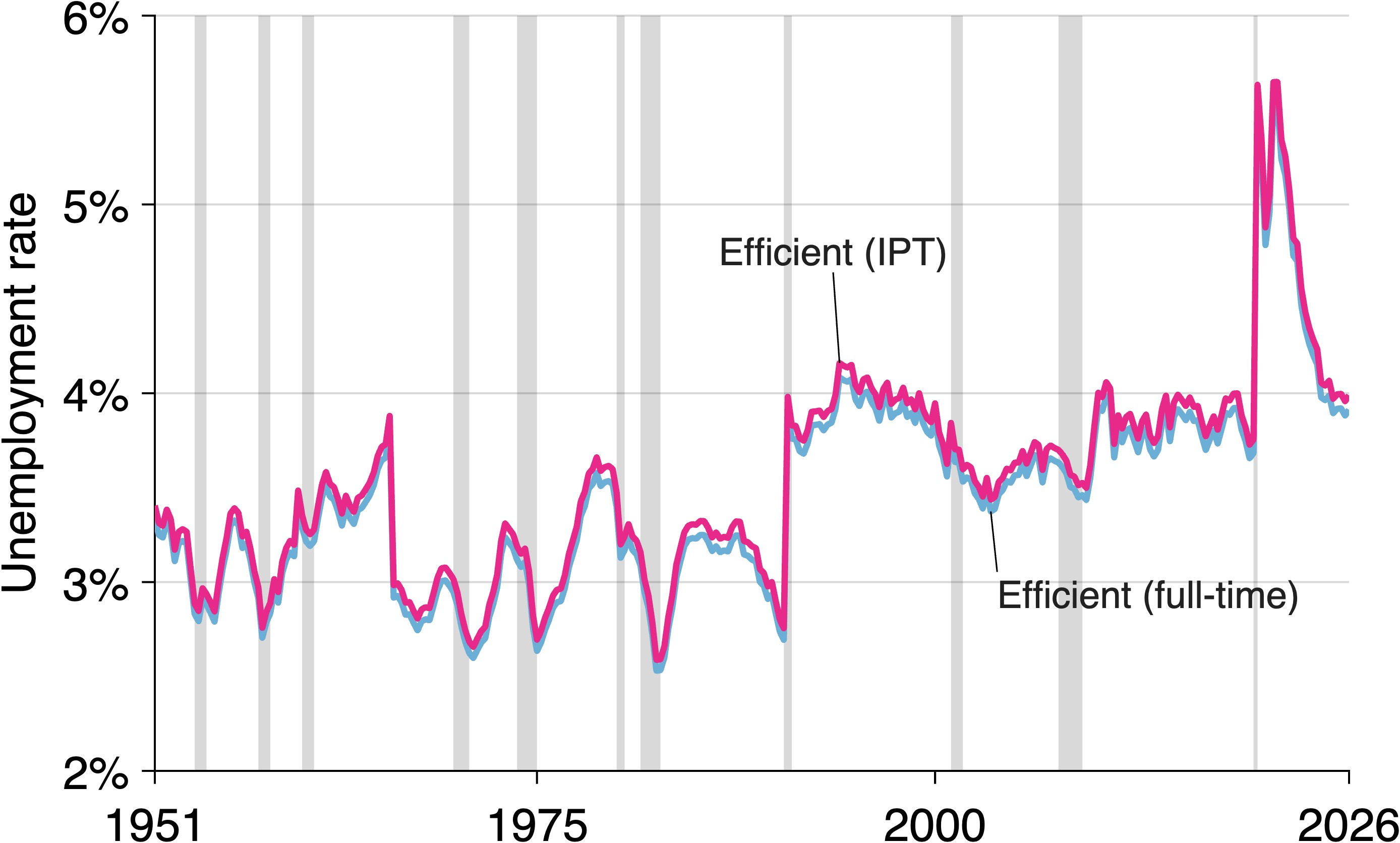}}

\vspace{0.3cm}

\subcaptionbox{Unemployment gap comparison, 1951--2026}
{\includegraphics[width=0.48\textwidth]{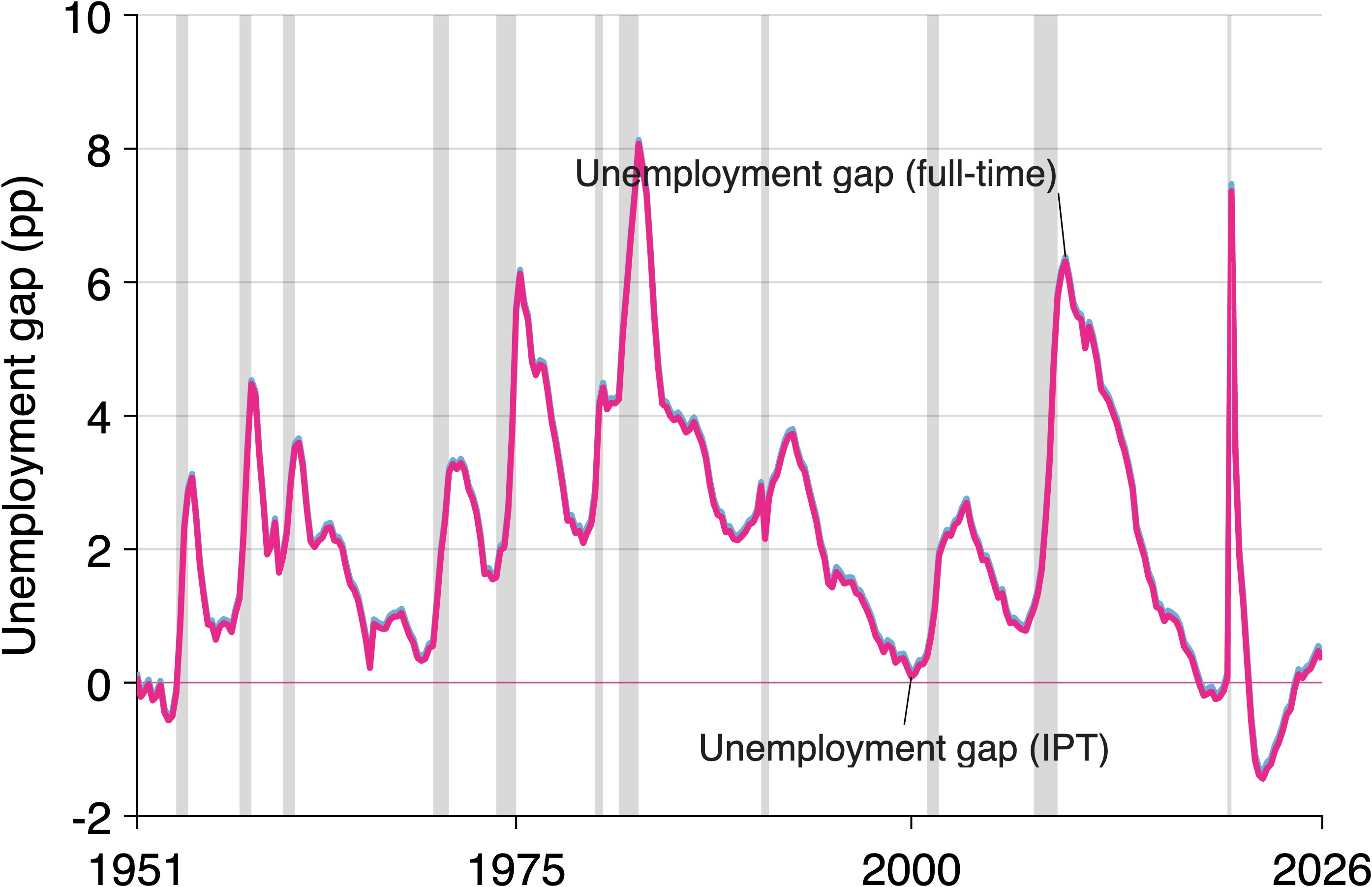}}

\caption{Comparison of efficient labor-market tightness, efficient unemployment, and unemployment gaps with and without involuntary part-time employment in the United States, 1951--2026. \textit{Notes}: The figure compares the baseline framework of \citet{MS21b}, which treats employment as homogeneous, with the extended framework that incorporates involuntary part-time employment.}
\label{f:graph_fullsample}

\end{figure}

\subsection{Additional Japanese results}

\begin{figure}[H]
\centering

\subcaptionbox{Efficient labor-market tightness\label{fig:jp_alpha_theta}}{
\includegraphics[width=0.48\textwidth]{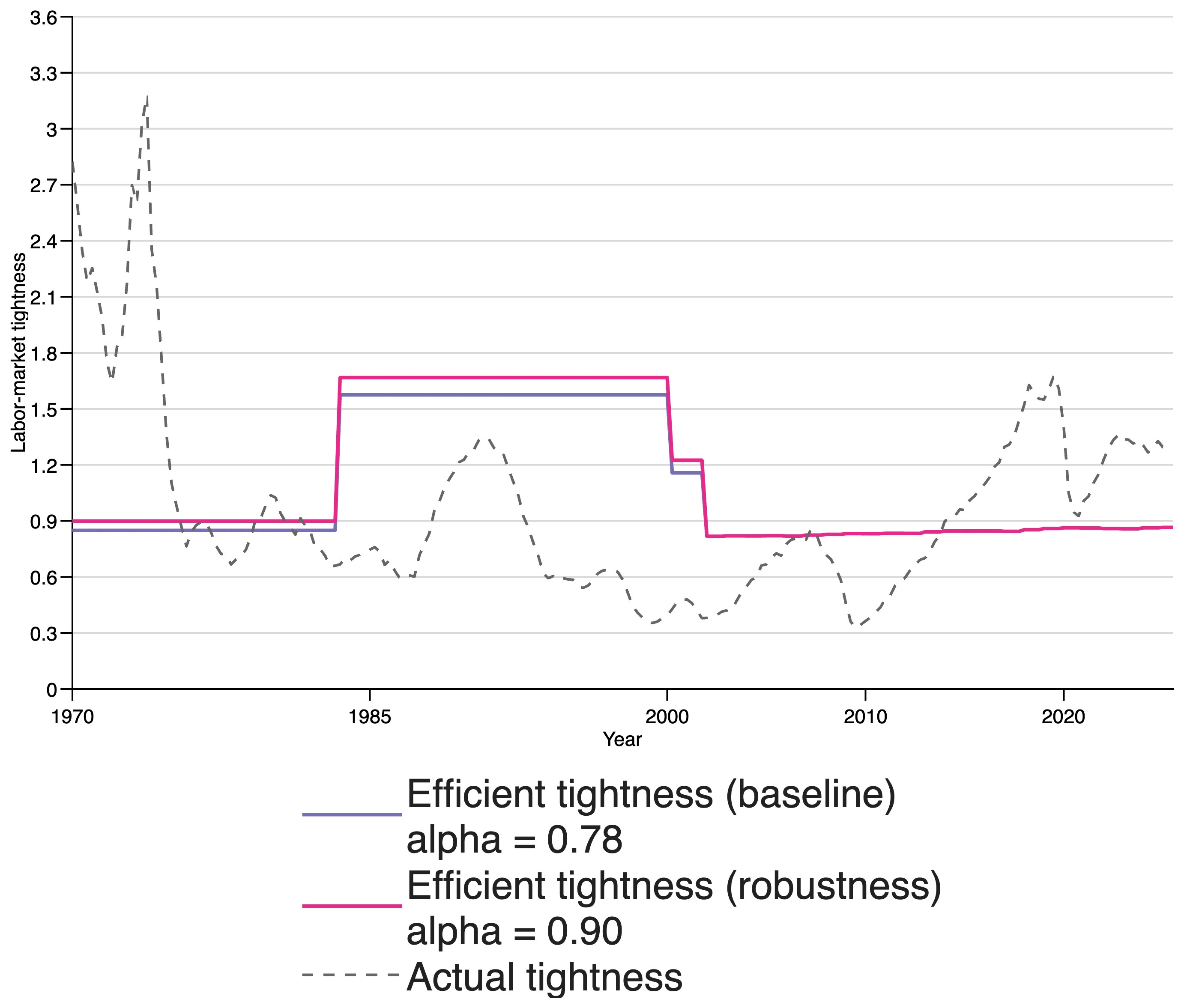}}
\hfill
\subcaptionbox{Efficient unemployment rate\label{fig:jp_alpha_ustar}}{
\includegraphics[width=0.48\textwidth]{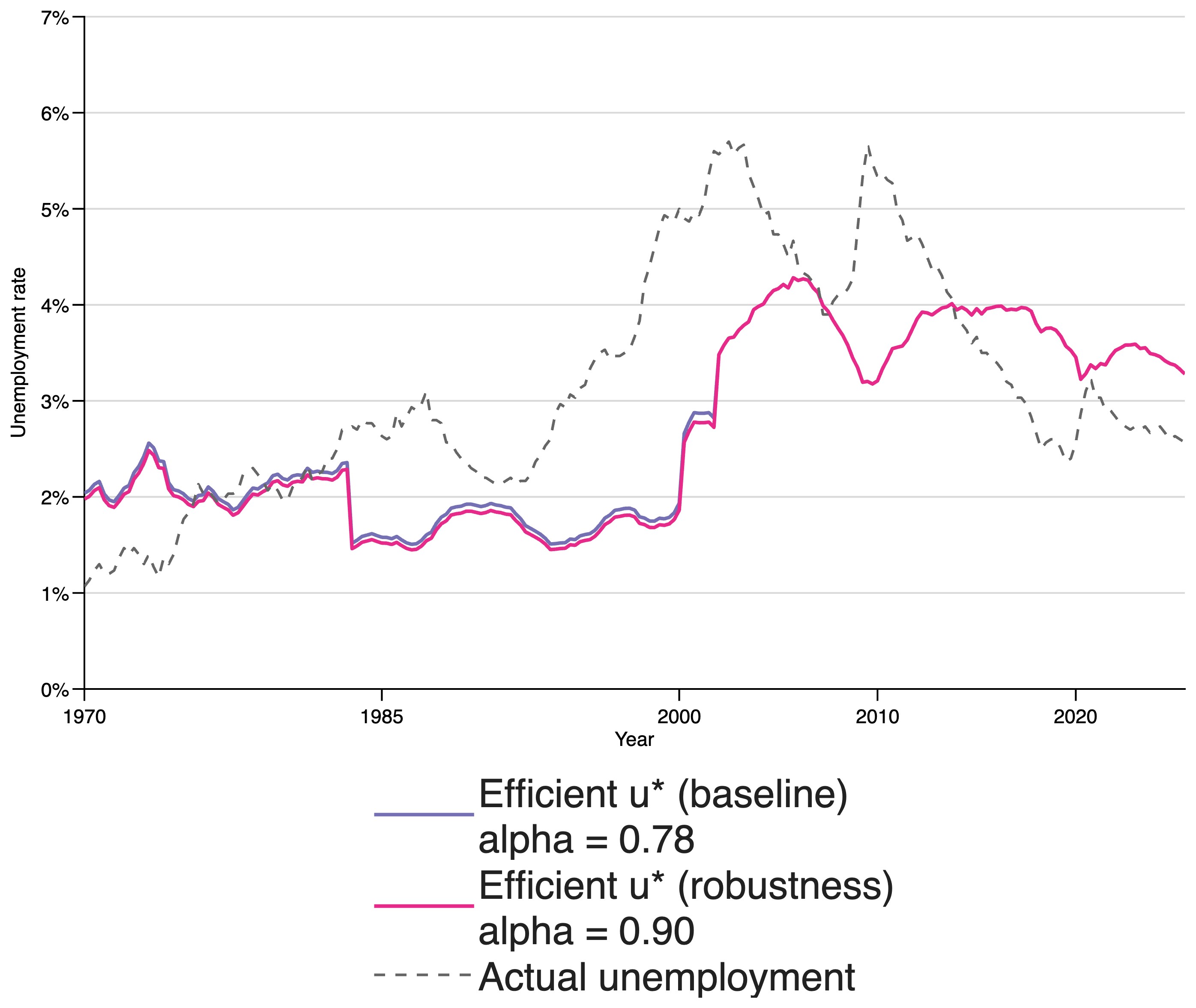}}
\hfill
\subcaptionbox{Unemployment gap\label{fig:jp_alpha_gap}}{
\includegraphics[width=0.48\textwidth]{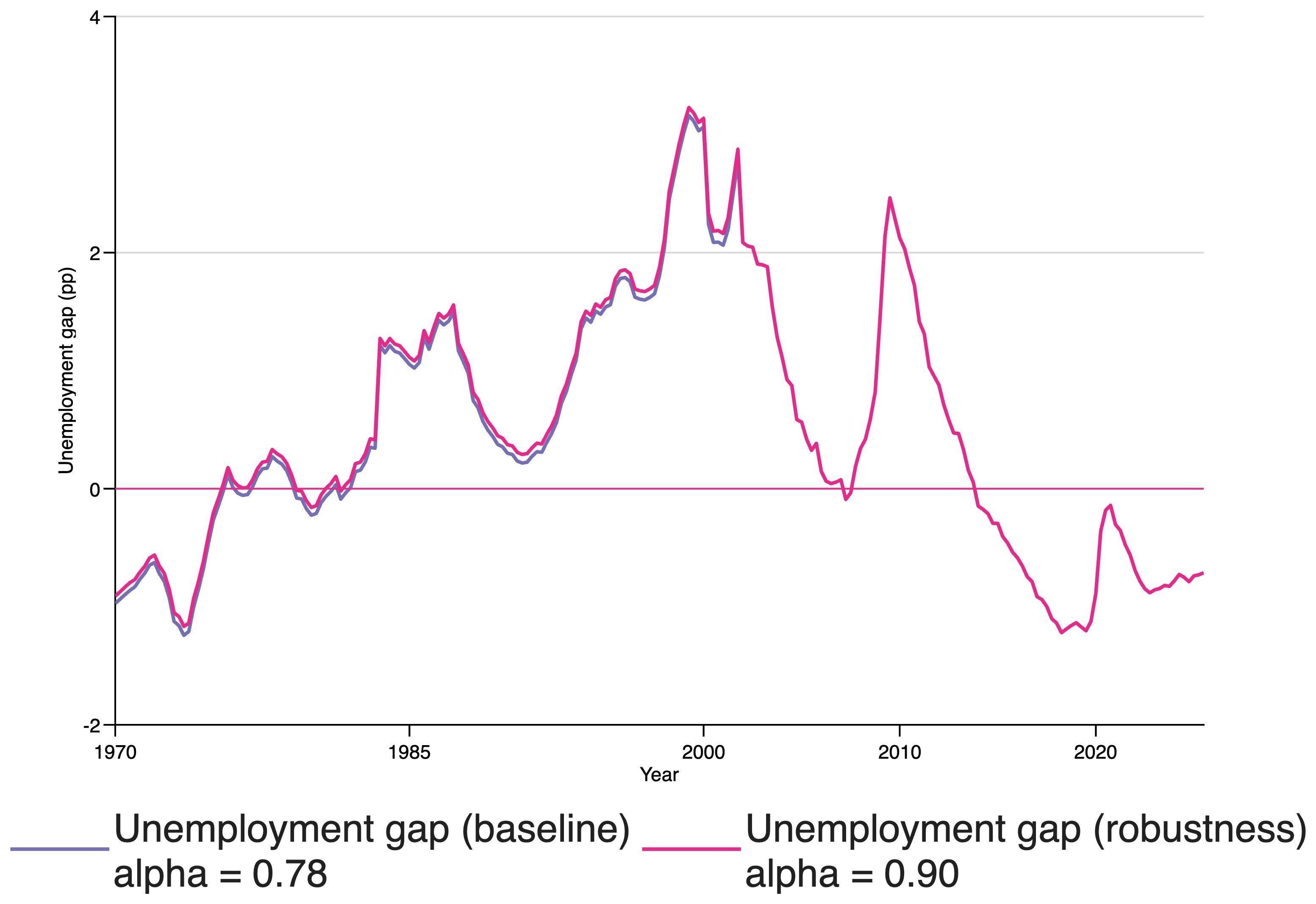}}

\caption{Robustness to the pre-2002 full-time employment share in Japan.
\textit{Notes:} The figure compares the baseline calibration, which sets $\alpha=0.78$ before 2002, with an alternative calibration using $\alpha=0.90$ before 2002. From 2002 onward, both calibrations use the observed OECD part-time employment share.}
\label{f:japan_alpha090_robustness}
\end{figure}

\begin{table}[H]
\centering
\caption{Summary Statistics of Labor-Market Tightness in Japan}
\label{tab:jpn_tightness}
\begin{tabular}{lccc}
\hline
Statistic & Actual & Efficient (Full-time) & Efficient (Part-time) \\
\hline
Mean & 0.965 & 1.338 & 1.073 \\
SD   & 0.525 & 0.299 & 0.334 \\
Min  & 0.329 & 0.940 & 0.818 \\
Max  & 3.179 & 1.744 & 1.575 \\
N    & 223   & 223   & 223   \\
\hline
\end{tabular}

\vspace{0.2cm}
\footnotesize
\textit{Notes}: The sample covers 223 quarterly observations from 1970Q1 to 2025Q3. The full-time specification assumes $\alpha=1$. The part-time specification assumes $\alpha=0.78$ before 2002 and uses the OECD part-time employment share thereafter. Parameters are $z=0.26$, $c=0.92$, and $\gamma=0.56$.

\textit{Source}: OECD Database, Statistics Bureau of Japan, and author's calculations.
\end{table}

\begin{table}[H]
\centering
\caption{Summary Statistics of Actual and Efficient Unemployment Rates in Japan}
\label{tab:jpn_ustar}
\begin{tabular}{lccc}
\hline
Statistic & Actual & Efficient (Full-time) & Efficient (Part-time) \\
\hline
Mean & 3.16\% & 2.28\% & 2.71\% \\
SD   & 1.20\% & 0.58\% & 0.95\% \\
Min  & 1.07\% & 1.40\% & 1.51\% \\
Max  & 5.70\% & 3.26\% & 4.28\% \\
N    & 223    & 223    & 223    \\
\hline
\end{tabular}

\vspace{0.2cm}
\footnotesize
\textit{Notes}: The sample covers 223 quarterly observations from 1970Q1 to 2025Q3. The full-time specification assumes $\alpha=1$. The part-time specification assumes $\alpha=0.78$ before 2002 and uses the OECD part-time employment share thereafter. Parameters are $z=0.26$, $c=0.92$, and $\gamma=0.56$.

\textit{Source}: OECD Database, Statistics Bureau of Japan, and author's calculations.
\end{table}

\begin{table}[H]
\centering
\caption{Summary Statistics of the Unemployment Gap in Japan}
\label{tab:jpn_gap}
\begin{tabular}{lcc}
\hline
Statistic & Gap (Full-time) & Gap (Part-time) \\
\hline
Mean & 0.88 pp & 0.45 pp \\
SD   & 1.07 pp & 1.09 pp \\
Min  & -1.11 pp & -1.24 pp \\
Max  & 3.28 pp & 3.16 pp \\
N    & 223 & 223 \\
\hline
\end{tabular}

\vspace{0.2cm}
\footnotesize
\textit{Notes}: The unemployment gap is defined as the difference between the actual unemployment rate and the efficient unemployment rate. The sample covers 223 quarterly observations from 1970Q1 to 2025Q3. The full-time specification assumes $\alpha=1$. The part-time specification assumes $\alpha=0.78$ before 2002 and uses the OECD part-time employment share thereafter. Parameters are $z=0.26$, $c=0.92$, and $\gamma=0.56$.

\textit{Source}: OECD Database, Statistics Bureau of Japan, and author's calculations.
\end{table}

\subsection{Robustness test tables}

\begin{table}[H]
\centering
\caption{Summary Statistics for Efficient Unemployment and Beveridge-Elasticity Uncertainty}
\label{tab:fig8a_summary}
\small
\begin{tabular}{lcc}
\hline
 & United States & Japan \\
 & (1951Q1--2019Q4; $N=276$) & (1970Q1--2025Q2; $N=223$) \\
\hline
\multicolumn{3}{l}{\textit{Point estimate $u^*$ (percent)}} \\
\quad Mean  & 4.66 & 2.72 \\
\quad Min   & 3.25 & 1.51 \\
\quad Max   & 5.84 & 4.30 \\
\hline
\multicolumn{3}{l}{\textit{Sensitivity to 95\% confidence interval for $\varepsilon$}} \\
\quad Avg.\ gap, lower bound vs.\ baseline (pp) & 0.6 & 0.33 \\
\quad Avg.\ gap, upper bound vs.\ baseline (pp) & 0.5 & 0.28 \\
\quad Max.\ deviation from baseline (pp)        & 1.3 & 0.55 \\
\quad Full envelope: min--max (percent)         & 2.85--6.55 & 1.11--4.65 \\
\hline
\end{tabular}

\vspace{0.2cm}
\footnotesize
\textit{Notes}: The shaded band varies the Beveridge elasticity within its 95\% confidence interval at each quarter; all other sufficient statistics are held at baseline ($z=0.26$, $c=0.92$).

\textit{Source}: Author's calculations.
\end{table}

\begin{table}[H]
\centering
\caption{Summary Statistics for Efficient Unemployment and Social Value of Nonwork Uncertainty}
\label{tab:fig8b_summary}
\small
\begin{tabular}{lcc}
\hline
 & United States & Japan \\
 & (1951Q1--2019Q4; $N=276$) & (1970Q1--2025Q2; $N=223$) \\
\hline
\multicolumn{3}{l}{\textit{Point estimate $u^*$ (percent)}} \\
\quad Mean  & 4.66 & 2.72 \\
\quad Min   & 3.25 & 1.51 \\
\quad Max   & 5.84 & 4.30 \\
\hline
\multicolumn{3}{l}{\textit{Sensitivity to $\zeta \in [0.03,\,0.49]$}} \\
\quad Avg.\ gap, lower bound vs.\ baseline (pp) & 0.6 & 0.42 \\
\quad Avg.\ gap, upper bound vs.\ baseline (pp) & 1.0 & 0.70 \\
\quad Max.\ deviation from baseline (pp)        & 1.3 & 1.11 \\
\quad Full envelope: min--max (percent)         & 2.81--7.15 & 1.25--5.41 \\
\hline
\end{tabular}

\vspace{0.2cm}
\footnotesize
\textit{Notes}: The shaded band varies $z$ over its calibration range; $\epsilon$ and $c$ are held at their baseline estimates.
\textit{Source}: Author's calculations.
\end{table}

\begin{table}[H]
\centering
\caption{Summary Statistics for Efficient Unemployment and Recruiting-Cost Uncertainty}
\label{tab:fig8c_summary}
\small
\begin{tabular}{lcc}
\hline
 & United States & Japan \\
 & (1951Q1--2019Q4; $N=276$) & (1970Q1--2025Q2; $N=223$) \\
\hline
\multicolumn{3}{l}{\textit{Point estimate $u^*$ (percent)}} \\
\quad Mean  & 4.66 & 2.72 \\
\quad Min   & 3.25 & 1.51 \\
\quad Max   & 5.84 & 4.30 \\
\hline
\multicolumn{3}{l}{\textit{Sensitivity to $c \in [\tfrac{2}{3}\times 0.92,\,\tfrac{4}{3}\times 0.92]$}} \\
\quad Avg.\ gap, lower bound vs.\ baseline (pp) & 0.9 & 0.60 \\
\quad Avg.\ gap, upper bound vs.\ baseline (pp) & 0.8 & 0.53 \\
\quad Max.\ deviation from baseline (pp)        & 1.2 & 0.95 \\
\quad Full envelope: min--max (percent)         & 2.61--6.83 & 1.14--5.13 \\
\hline
\end{tabular}

\vspace{0.2cm}
\footnotesize
\textit{Notes}: The shaded band varies $c$ over its calibration range; $\epsilon$ and $z$ are held at their baseline estimates. 
\textit{Source}: Author's calculations.
\end{table}

\begin{table}[H]
\centering
\caption{Summary Statistics for Inverse-Optimum Beveridge Elasticity}
\label{tab:fig9a_summary}
\small
\begin{tabular}{lcc}
\hline
 & United States & Japan \\
 & (1951Q1--2019Q4; $N=276$) & (1970Q1--2025Q3; $N=223$) \\
\hline
Mean of $\epsilon^*$                     & 1.42 & 0.84 \\
Min of $\epsilon^*$                      & 0.48 & 0.23 \\
Max of $\epsilon^*$                      & 4.35 & 2.06 \\
Mean of calibrated $\epsilon$            & 0.91 & 0.63 \\
Inside 95\% CI (\% of quarters)             & 23.0 & 24.2 \\
Below CI (\% of quarters)                   & 16.0 & 25.6 \\
Above CI (\% of quarters)                   & 60.9 & 50.2 \\
\hline
\end{tabular}
\vspace{0.2cm}

\footnotesize
\textit{Notes}: $\epsilon^*$ is the inverse-optimum Beveridge elasticity implied by observed labor-market tightness, holding $z=0.26$, $c=0.92$, and part-time parameters at baseline. The calibrated U.S. elasticity reported here, $\epsilon=0.91$, is the benchmark value from the five-break pre-COVID calibration in \citet{MS21b}, used for comparison with the original framework. It differs from the regime-specific elasticities reported in Table~\ref{tab:beveridge-elasticity}, which are estimated using the full sample and a different break structure.

\textit{Source}: Author's calculations.
\end{table}

\begin{table}[H]
\centering
\caption{Summary Statistics for Inverse-Optimum Social Value of Nonwork}
\label{tab:fig9b_summary}
\small
\begin{tabular}{lcc}
\hline
 & United States & Japan \\
 & (1951Q1--2019Q4; $N=276$) & (1970Q1--2025Q3; $N=223$) \\
\hline
Mean of $z^*$                           & 0.41 & 0.27 \\
Min of $z^*$                            & $-$0.48 & $-$1.77 \\
Max of $z^*$                            & 0.86 & 0.83 \\
Calibrated $z$                            & 0.26 & 0.26 \\
Inside $[0.03,\,0.49]$ (\% of quarters)     & 42.0 & 42.2 \\
Below range (\% of quarters)                & 12.0 & 22.0 \\
Above range (\% of quarters)                & 46.0 & 35.9 \\
$z^* < 0$ (\% of quarters)              & 12.0 & Negative $z^*$ \\
\hline
\end{tabular}
\vspace{0.2cm}

\footnotesize
\textit{Notes:} Negative values of $z^*$ occur when observed labor-market tightness is high enough that efficiency would require nonwork to have negative social value. The U.S. minimum occurs in 1969Q1 and the Japan minimum occurs in 1973Q4. U.S. maximum in 2009Q4; Japan maximum in 1999Q2.

\textit{Source}: Author's calculations.
\end{table}

\begin{table}[H]
\centering
\caption{Summary Statistics for Inverse-Optimum Recruiting Cost}
\label{tab:fig9c_summary}
\small
\begin{tabular}{lcc}
\hline
 & United States & Japan \\
 & (1951Q1--2019Q4; $N=276$) & (1970Q1--2025Q3; $N=223$) \\
\hline
Mean of $c^*$                          & 1.46 & 1.34 \\
Min of $c^*$                           & 0.46 & 0.25 \\
Max of $c^*$                           & 4.75 & 4.11 \\
Calibrated $c$                           & 0.92 & 0.92 \\
Inside $[\tfrac{2}{3}\times 0.92,\,\tfrac{4}{3}\times 0.92]$ (\% of quarters) & 42.0 & 46.2 \\
Below range (\% of quarters)                & 7.0  & 13.9 \\
Above range (\% of quarters)                & 51.0 & 39.9 \\
\hline
\end{tabular}
\vspace{0.2cm}

\footnotesize
\textit{Notes}: $c^*$ is the inverse-optimum recruiting cost, holding $\epsilon$, $z$, and part-time parameters at baseline.

\textit{Source}: Author's calculations.
\end{table}

\end{document}